\pdfoutput=1
\documentclass[11pt]{article}

\usepackage[T1]{fontenc}

\usepackage{epsfig,latexsym,amsfonts,amsmath,amsthm,amssymb,amsbsy,multirow,slashed,wasysym,textcomp,cite,comment,subfigure,wrapfig,datetime,mathtools,cancel,mathrsfs,pict2e}
\usepackage{datetime2}
\usepackage[font={footnotesize,sl},bf]{caption}

\usepackage{braket}
\usepackage{setspace}
\usepackage{tikz}
\usetikzlibrary{math}
\usetikzlibrary{calc}
\usetikzlibrary{decorations.pathreplacing}
\usetikzlibrary{decorations.pathmorphing}
\usetikzlibrary{decorations.markings}
\usetikzlibrary{arrows.meta}
\usetikzlibrary{positioning}

\usetikzlibrary{shapes.geometric}
\usetikzlibrary{patterns}

\usepackage{xcolor}
\definecolor{oldmauve}{rgb}{0.4, 0.19, 0.28}
\definecolor{pansypurple}{rgb}{0.47, 0.09, 0.29}
\definecolor{burgundy}{rgb}{0.5, 0.0, 0.13}
\definecolor{carminepink}{rgb}{0.92, 0.3, 0.26}
\definecolor{blue(pigment)}{rgb}{0.2, 0.2, 0.6}
\definecolor{darkseagreen}{rgb}{0.56, 0.74, 0.56}
\definecolor{darkspringgreen}{rgb}{0.09, 0.45, 0.27}
\definecolor{ceruleanblue}{rgb}{0.16, 0.32, 0.75}

\definecolor{navyblue}{RGB}{0,0,128}

\numberwithin{equation}{section}

\def\bea{\begin{eqnarray}}
\def\eea{\end{eqnarray}}
\newcommand{\beq}{\begin{eqnarray}}
\newcommand{\eqq}{\end{eqnarray}}
 \newcommand{\badat}{\begin{alignedat}}
 \newcommand{\eadat}{\end{alignedat}}

\newcommand{\eal}[1]{\be \begin{aligned} #1 \end{aligned}\end{equation}} 
\newcommand{\eqn}[1]{\be #1 \end{equation}} 
\newcommand{\eqa}[1]{\bea  #1\end{eqnarray}}

\newcommand{\bz}{\bar{z}}

\newcommand{\scri}{\mathscr{I}}

\newcommand{\sq}[1]{[#1]}
\newcommand{\an}[1]{\left\langle#1\right\rangle}
\newcommand{\mo}{\mathcal{O}}
\newcommand{\mc}{\mathcal{C}}
\newcommand{\mA}{\mathcal{A}}

\newcommand{\zb}{\bar{z}}
\newcommand{\Zb}{\bar{Z}}
\newcommand{\hb}{\bar{h}}

\newcommand{\D}{\Delta}
\newcommand{\e}{\epsilon}
\newcommand{\om}{\omega}

\long\def\new#1\endnew{{\bf #1}}		
\long\def\del#1\enddel{}

\def\del{\partial}

\newcommand{\be}{\begin{eqnarray}}
\newcommand{\en}{\end{eqnarray}}

\def\bz{{\bar z}}

\def\cH{{\mathcal H}}

\numberwithin{equation}{section} 

\usepackage{pict2e}
\makeatletter
\newcommand{\loplus}{\mathbin{\mathpalette\dog@lsemi{+}}}
\newcommand{\dog@lsemi}[2]{\dog@semi{#1}{#2}{270,90}}
\newcommand{\dog@semi}[3]{%
  \begingroup
  \sbox\z@{$\m@th#1#2$}%
  \setlength{\unitlength}{\dimexpr\ht\z@+\dp\z@\relax}%
  \makebox[\wd\z@]{\raisebox{-\dp\z@}{%
    \begin{picture}(1,1)
    \linethickness{\variable@rule{#1}}
    \roundcap
    \put(0.5,0.5){\makebox(0,0){\raisebox{\dp\z@}{$\m@th#1#2$}}}
    \put(0.5,0.5){\arc[#3]{0.5}}
    \end{picture}%
  }}%
  \endgroup
}
\newcommand{\variable@rule}[1]{%
  \fontdimen8  
  \ifx#1\displaystyle\textfont3\else
    \ifx#1\textstyle\textfont3\else
      \ifx#1\scriptstyle\scriptfont3\else
        \scriptscriptfont3\relax
  \fi\fi\fi
}
\makeatother

\newcommand{\pthorn}{{\text{\th}}}

\author{}
\numberwithin{equation}{section} 
\usepackage{hyperref}
\usepackage{orcidlink}

\begin{document}

\begin{titlepage}

  \thispagestyle{empty}

 \begin{flushright}
 \end{flushright}


  \begin{center}  

{\LARGE\textbf{Carrollian Physics and Holography}}




\vskip1cm 
Romain Ruzziconi\footnote{\fontsize{8pt}{10pt}\selectfont\ \href{mailto:romainruzziconi@fas.harvard.edu}{romainruzziconi@fas.harvard.edu}
} 
\vskip0.5cm

\normalsize
\medskip

\textit{Center for the Fundamental Laws of Nature, Harvard University \\
17 Oxford Street, Cambridge, MA 02138, USA} \\

\vspace{2mm}

\textit{Black Hole Initiative, Harvard University \\
20 Garden Street, Cambridge, MA 02138, USA}

\vspace{2mm}

\end{center}

\vskip0.5cm

\begin{abstract}
\noindent This report reviews key developments in Carrollian physics with an emphasis on their role in the emerging framework of holography in asymptotically flat spacetimes. We begin by introducing the Carrollian limit, understood as a contraction of the Poincaré group obtained by formally taking the speed of light to zero. The geometric structures associated with this limit are described and argued to arise naturally on null hypersurfaces, most notably on null infinity, as well as black hole and cosmological horizons. Building on this, we examine the relation between the Bondi–Metzner–Sachs symmetries governing asymptotically flat gravity and the conformal Carrollian symmetries. Explicit examples of Carrollian field theories are constructed by implementing the limit on well-known relativistic field theories, with particular attention to Carrollian CFTs. We then present the Carrollian holography proposal, according to which gravity in asymptotically flat spacetimes is dual to a Carrollian CFT living at null infinity in one lower dimension. In this framework, the massless $\mathcal{S}$-matrix written in position space at null infinity is naturally reinterpreted in terms of boundary Carrollian CFT correlators, called Carrollian amplitudes. We highlight their relation to celestial amplitudes and show how they naturally emerge from holographic CFT correlators through a correspondence between the flat space limit in the bulk and the Carrollian limit at the boundary. Using this correspondence, we provide strong evidence that flat space holography arises from a controlled and consistent limiting procedure applied to both sides of the AdS/CFT duality. We conclude by outlining future directions and open questions in the program.

\end{abstract}

\vspace{0.5cm}

\begin{center}
\textit{\large Invited Review for Physics Reports}
\end{center}

\end{titlepage}
\setcounter{page}{2}

\setcounter{tocdepth}{2}
\tableofcontents

\section{Introduction}

Poincaré symmetries constitute the foundation for most of our physical theories. It is well known that in regimes where the characteristic velocities of the system are much smaller than the speed of light, $v \ll c$, obtained in practice by formally taking $c \to \infty$, the Poincaré symmetries reduce to the Galilean symmetries, which underpin Newtonian mechanics and non-relativistic quantum mechanics. It was only later that another non-relativistic limit\footnote{Following \cite{Levy1965}, for reasons that will become clear in the text, we also use the terminology of “non-relativistic limit’’ for the Carrollian limit. This avoids confusing it with the so-called “ultra-relativistic limit”, which instead consists of taking $v \to c$.} of the Poincaré symmetries, the Carrollian limit, was discovered \cite{Levy1965,SenGupta:1966qer}. In contrast to the Galilean limit, the Carrollian limit arises when formally taking $c \to 0$. The name “Carrollian,” suggested by Lévy-Leblond, refers to Lewis Carroll, the author of Alice in Wonderland, for reasons that will become clear later in the text.

Although this limit was originally seen as a mathematical curiosity rather than of direct physical relevance, it was soon realized that the associated Carrollian geometry \cite{Henneaux:1979vn,1977asst.conf....1G, Duval:2014lpa} provides the correct structure for describing null hypersurfaces in Lorentzian spacetimes. The canonical example of such a null hypersurface is null infinity, the boundary of asymptotically flat spacetime obtained via Penrose's conformal compactification \cite{Penrose:1962ij,Penrose:1964ge,Penrose:1986uia}. Crucially, the isomorphism between the conformal Carrollian algebra, emerging from this unusual $c \to 0$ limit, and the Bondi–van der Burg–Metzner–Sachs (BMS) algebra \cite{Bondi:1962px,Sachs:1962zza,Sachs:1962wk}, which encodes the symmetries of gravity in asymptotically flat spacetimes, was explicitly realized in \cite{Duval:2014uva}. Carrollian geometry also arises on bulk null surfaces \cite{Vogel1965,Jankiewicz,Dautcourt1967}, such as black hole and cosmological horizons, providing a unified physical and mathematical framework for their analysis.

The importance of understanding the spacetime boundary structure in the development of quantum gravity has culminated in the discovery of the holographic principle \cite{tHooft:1993dmi, Susskind:1994vu}. The latter suggests that gravity in a given spacetime region can be encoded on the lower-dimensional boundary of that region. Explicit realizations of this principle have been achieved within the framework of string theory for spacetimes with negative cosmological constant, through the advent of the AdS/CFT correspondence \cite{Maldacena:1997re, Witten:1998qj, Aharony:1999ti}. This framework proposes that quantum gravity in asymptotically AdS spacetimes in $d+1$ dimensions is dual to a conformal field theory living on the $d$-dimensional boundary. In this setup, the asymptotic symmetries of the bulk coincide with the global conformal symmetries of the dual theory.

Interestingly, the cosmological constant of our universe has been measured to be positive; our spacetime is therefore closer to de Sitter than to anti–de Sitter, which means that AdS/CFT is not directly applicable in this context. For most physical models used in practice—from those describing collider physics to astrophysical systems below cosmological scales, gravitational waves, or black hole physics—the cosmological constant can be neglected, and one can adopt the framework of asymptotically flat spacetimes. This motivates the development of the holographic principle in this setup, referred to as flat space holography, which would allow us to capture structures relevant to the real world and ultimately shed light on quantum gravity.

Early attempts towards flat space holography were made \cite{Susskind:1998vk,Polchinski:1999ry,Giddings:1999jq,deBoer:2003vf,Arcioni:2003xx,Mann:2005yr}, most of them starting from AdS/CFT and implementing a flat space limit, which formally consists of taking the cosmological constant $\Lambda$ to zero. However, important technical and conceptual obstructions were encountered in this endeavour, due both to the singular nature of the flat space limit and to the fact that the boundary of flat spacetime is null, unlike the timelike boundary of AdS. It became clear that a top-down construction of flat space holography from string theory, in the spirit of AdS/CFT, was out of reach, and that the quest for a flat space holographic principle would require us to reshape entirely the way we think about holography.

Instead, one could use bottom-up constructions to address this important problem, starting from what we know about the bulk theory to constrain the dual theory and attempting to bootstrap it. It is interesting to note that, had the original derivation of AdS/CFT from string theory not been discovered, this would likely have been the way holography developed in AdS: using our knowledge of gravity and field theories in the bulk to bootstrap a putative dual CFT. See, e.g., \cite{Penedones:2016voo, Kaplan_AdSCFT_from_bottom_up} for recent lecture notes on AdS/CFT following this approach.

Adopting this philosophy, and assuming a similar pattern as in the AdS/CFT correspondence, we expect that the dual theory to flat space should live on the co-dimension one boundary, null infinity. Furthermore, by identifying the symmetries, we expect that, if a dual theory exists, it should exhibit the BMS or conformal Carrollian symmetries as global spacetime symmetries. Hence, the theory dual to gravity in $4d$ flat space should be a $3d$ BMS field theory, also called a conformal Carrollian field theory, or just a Carrollian CFT. This bottom-up approach to flat space holography is referred to as Carrollian holography.

The early successes of Carrollian holography have mostly occurred in the context of $3d$ gravity in asymptotically flat spacetime. Indeed, in this setup, the gravity theory is topological, and one can easily implement the flat space limit on asymptotically AdS$_3$ solutions of Einstein gravity to obtain asymptotically flat solutions \cite{Barnich:2012aw}. Many results have been obtained, analogous to those in AdS$_3$/CFT$_2$: the discovery of the BMS$_3$ central charge \cite{Barnich:2006av}, which is the analogue of the Brown-Henneaux central charge in AdS$_3$ \cite{Brown:1986nw}; the ``Cardyology'' of flat space cosmology entropy \cite{Barnich:2012xq,Bagchi:2012xr}, analogous to the derivation of BTZ black hole entropy via the Cardy formula \cite{Strominger:1997eq}; the construction of an effective action for the Carrollian CFT at null infinity \cite{Barnich:2013yka,Barnich:2012rz}, analogous to the Liouville action obtained by Hamiltonian reduction in AdS$_3$ \cite{Coussaert:1995zp}; the construction of a boundary dual to conformal gravity with asymptotically flat boundary conditions \cite{Bagchi:2012yk}, analogous to the discussion of the right-moving boundary CFT dual to chiral gravity in AdS$_3$ \cite{Li:2008dq}; formulae for entanglement entropy in flat space using swing surfaces \cite{Bagchi:2014iea,Jiang:2017ecm}, analogous to extremal surfaces in the Ryu-Takayanagi prescription \cite{Ryu:2006bv}; the computation of holographic stress tensor correlators via the renormalized on-shell action \cite{Detournay:2014fva,Bagchi:2015wna,Hartong:2015usd}, as in AdS \cite{deHaro:2000vlm,Skenderis:2002wp}; and a holographic anomaly in flat space \cite{Campoleoni:2022wmf}, analogous to the anomaly appearing in AdS/CFT \cite{Henningson:1998gx}, among others.

Interestingly, in parallel, a seemingly completely different approach to flat space holography was developed mostly in the context of scattering theory in $4d$ Minkowski space, known as celestial holography \cite{deBoer:2003vf,He:2015zea,Pasterski:2016qvg,Stieberger:2018onx} (see also \cite{Pasterski:2021rjz,Raclariu:2021zjz,Pasterski:2021raf} for reviews). The origin of this proposal lies in the rich infrared structure of gravity and gauge theories, and in the interplay between asymptotic symmetries, soft theorems, and memory effects \cite{Strominger:2017zoo}. It suggests that gravity in $4d$ asymptotically flat spacetime is dual to a $2d$ CFT living on the celestial sphere—the celestial CFT. The central idea is based on the observation that scattering amplitudes can be rewritten as correlation functions on the celestial sphere via a Mellin transform, satisfying conformal Ward identities.

Celestial holography has been highly successful in expressing scattering amplitude statements in the bulk in terms of a $2d$ CFT language on the celestial sphere: soft theorems correspond to Ward identities (“conformally soft theorems”) \cite{Strominger:2013jfa, Kapec:2016jld,Adamo:2019ipt, Donnay:2018neh, Puhm:2019zbl}, and collinear limits map to OPEs in the dual theory \cite{Fan:2019emx, Pate:2019lpp}. In this framework, soft OPE algebras reveal rich structures; in gravity, they form the $Lw_{1+\infty}$ algebra \cite{Guevara:2021abz, Strominger:2021mtt}. The discovery of these symmetries arguably represents one of the most important successes of the celestial holography program. Intriguingly, these celestial symmetries already appeared in the context of twistor theory in Penrose's non-linear graviton construction of self-dual spacetimes \cite{Penrose:1976js}. This construction relates $4d$ asymptotically flat (complexified) self-dual spacetimes to a complex three-dimensional projective space, called twistor space. This relation, which is holographic by nature \cite{Newman:1976gc,Hansen:1978jz}, is a non-local map, allowing integral expressions on spacetime to be reformulated as simple algebraic statements on twistor space. While the action of the $Lw_{1+\infty}$ algebra is non-local on spacetime and does not have a clear interpretation in terms of diffeomorphisms \cite{Freidel:2021ytz,Geiller:2024bgf}, its action is completely transparent on twistor space and is interpreted as Poisson diffeomorphisms \cite{Adamo:2021lrv,Bu:2022iak,Kmec:2024nmu}.

However, despite notable progress — including approaches based on twistor methods \cite{Costello:2022wso, Costello:2022jpg, Costello:2023hmi, Bittleston:2024efo} or analyses restricted to particular MHV subsectors of Yang–Mills amplitudes \cite{Stieberger:2022zyk, Stieberger:2023fju, Melton:2024akx} — an intrinsic definition of the dual celestial CFT that incorporates bulk translation invariance and captures Einstein gravity, at least in some regime, remains an open question. One of the main obstructions to an independent definition of the dual theory in the celestial holography approach is that the properties of the celestial CFT, obtained by Mellin-transforming bulk amplitudes, indicate that the putative dual CFT is not a standard $2d$ Euclidean CFT: its low-point functions are distributional \cite{Pasterski:2017ylz}, its complex continuous spectrum \cite{Pasterski:2017kqt} indicates that it is probably not reflection-positive in the usual sense, and the OPEs \cite{Fan:2019emx, Pate:2019lpp}
can fail to be associative at one loop \cite{Costello:2022upu}. These facts suggest that standard bootstrap methods to determine the dynamics of this holographic $2d$ CFT cannot be directly applied, and that there is no systematic way to construct a holographic dual.

Remarkably, although the two approaches to flat space holography described above—Carrollian and celestial holography—look completely different, both in the dimension and the nature of their dual theories, they are in fact related \cite{Donnay:2022aba, Bagchi:2022emh, Donnay:2022wvx}, creating a beautiful interplay between Carrollian physics, celestial amplitudes, and gravity in $4d$ asymptotically flat spacetime. Appreciating all the consequences of this matching is an active area of research. In particular, this connection gave rise to the concept of ``Carrollian amplitude'' \cite{Mason:2023mti}, which reinterprets the flat space $\mathcal{S}$-matrix in terms of Carrollian CFT correlators at null infinity—see \cite{Banerjee:2019prz,Salzer:2023jqv,Saha:2023hsl,Bagchi:2023fbj,Nguyen:2023miw,Bagchi:2023cen,Ruzziconi:2024zkr,Liu:2024nfc,Have:2024dff,Stieberger:2024shv,Adamo:2024mqn,Alday:2024yyj,Banerjee:2024hvb,Kraus:2024gso,Jorstad:2024yzm,Ruzziconi:2024kzo,Kulp:2024scx,Kraus:2025wgi,Nguyen:2025sqk,Kulkarni:2025qcx,Adamo:2025bfr,Marotta:2025qjh,Long:2026cpq} for developments in this rapidly growing field. These correlators naturally arise from the Carrollian partition function at null infinity, identified with the bulk on-shell action \cite{Kim:2023qbl,Kraus:2024gso,Kraus:2025wgi,Ammon:2025avo}, using a flat space analogue of the GKP/W dictionary \cite{Gubser:1998bc,Witten:1998qj}.

Explicit examples of Carrollian field theories can be systematically constructed by taking the $c \to 0$ limit of standard relativistic field theories, both at the level of the Lagrangians \cite{Henneaux:2021yzg, deBoer:2021jej, Hansen:2021fxi, Baiguera:2022lsw, Bergshoeff:2023vfd, Miskovic:2023zfz, Bagchi:2024efs} and of the correlation functions \cite{Alday:2024yyj}. In particular, bulk translation invariance and the appearance of distributional low-point functions are built in, reflecting the ultra-local nature of physics on null hypersurfaces. Hence, Carrollian holography offers hope for a systematic path to tackle the important problem of defining the dual theory of gravity in asymptotically flat spacetimes and calls for further investigation.

Probably one of the most promising features of Carrollian holography is its relation with the AdS/CFT correspondence. As mentioned earlier, taking the flat space limit of AdS/CFT is not a new idea and was already explored in the early days of holography. However, it is only recently that we have understood that the flat space limit in the bulk ($\Lambda \to 0$) corresponds to a Carrollian limit at the boundary ($c \to 0$), connecting the holographic CFT at the boundary of AdS to a Carrollian CFT at the boundary of flat space. This flat space/Carrollian limit correspondence was first hinted at by studying the flat space limit of Einstein’s equations in $3d$ and $4d$ in Bondi-type gauges and investigating the resulting boundary structures \cite{Barnich:2012aw,Ciambelli:2018wre,Compere:2019bua,Campoleoni:2023fug,Geiller:2022vto}. For instance, the covariant conservation of the stress tensor leads to the BMS flux-balance laws, which can be reinterpreted as Carroll-covariant conservation laws \cite{Fiorucci:2025twa,Hartong:2025jpp}. This limit is also understood at the level of the boundary geometry and the phase space \cite{Campoleoni:2018ltl,Compere:2020lrt,Fiorucci:2020xto,Ciambelli:2020eba,Ruzziconi:2020wrb,Campoleoni:2022wmf,Geiller:2024amx,McNees:2024iyu,Ciambelli:2024kre, Arenas-Henriquez:2025rpt}. 

Using a similar setup, the flat space/Carrollian limit correspondence has been extended to holographic correlators \cite{Alday:2024yyj}: the flat space limit of Witten diagrams in position space yields Feynman diagrams for Carrollian amplitudes, and this matches the corresponding Carrollian limit of holographic CFT correlators. We also refer to \cite{deGioia:2022fcn,deGioia:2023cbd,Bagchi:2023fbj,Bagchi:2023cen,Marotta:2024sce,Kraus:2024gso,deGioia:2024yne,deGioia:2025mwt,Kulkarni:2025qcx,Adamo:2025bfr,Marotta:2025qjh} for recent and closely related discussions, and to \cite{Kraus:2024gso,Poulias:2025eck} for developments at the level of the partition function. Altogether, this provides a systematic pathway to derive flat space holography from the flat limit of AdS/CFT, and these ideas were recently applied to the explicit realization of AdS$_4$/CFT$_3$, which relates M-theory on AdS$_4 \times S^7 / \mathbb{Z}_k$ to ABJM theory in $3d$ \cite{Aharony:2008ug}, hence reproducing flat space amplitudes through a limit taken in the CFT \cite{Lipstein:2025jfj}.

An important question of this program is to make sense intrinsically of Carrollian CFTs, and more broadly of Carrollian field theories, and not just to define them through a limit. Indeed, although many examples of Carrollian Lagrangian theories have been written in the literature, their quantization rules are still under investigation and constitute an important research area \cite{deBoer:2023fnj,Chen:2023pqf,Chen:2024voz,Cotler:2024xhb,Cotler:2025dau,Ekiz:2025hdn}. One could also adopt a more pragmatic approach and try to define Carrollian CFTs without resorting to a particular Lagrangian, as is usually done for standard CFTs. This program has been carried out in $2d$ Carrollian CFTs \cite{Campoleoni:2016vsh,Bagchi:2016geg,Bagchi:2017cpu}, as well as in $3d$ Carrollian CFTs, where it has led to constraints on the correlation functions \cite{Chen:2021xkw,Donnay:2022wvx,Nguyen:2023miw,Bagchi:2023fbj}, the structure of primaries and descendants \cite{Bagchi:2016bcd,Bagchi:2019xfx,Nguyen:2023vfz}, and the general form of the OPEs \cite{Mason:2023mti,Nguyen:2025sqk}, constituting the premises of a Carrollian bootstrap program.

The goal of this report is to provide a comprehensive and self-consistent introduction to Carrollian physics, as well as to review its recent applications in flat space holography. Let us emphasize that, beyond its central role in this context, Carrollian physics has also been shown to emerge in many other settings, including the following.
\begin{itemize}
    \item As mentioned earlier, Carrollian physics and geometry naturally arise on null hypersurfaces, including black hole and cosmological horizons. In the spirit of the membrane paradigm \cite{Damour:1978cg,1986bhmp.book.....T,Price:1986yy}, which interprets the Einstein equations projected onto a stretched timelike horizon as fluid equations, the Raychaudhuri \cite{PhysRev.98.1123} and Damour \cite{Damour:1978cg,Damour:1979wya} equations governing null hypersurfaces can be understood as Carrollian conservation equations in the limit where the stretched horizon approaches the null surface. We will discuss some of these aspects in this report and refer the reader to \cite{Chandrasekaran:2018aop,Ciambelli:2019lap,Donnay:2019jiz,Adami:2021nnf,Freidel:2022bai,Freidel:2022vjq,Freidel:2024emv,Riello:2024uvs} for recent works along these lines.  
    \item In cosmology, it has been proposed that the causal structure of spacetime during inflation, and in the regime relevant for dark energy, can be understood in terms of Carrollian physics \cite{deBoer:2021jej}. More specifically, requiring Carrollian symmetry for a perfect fluid stress tensor enforces an equation of state $\mathcal{E} + P = 0$, corresponding to $w = -1$. This suggests a remarkable connection whereby dark energy can effectively behave as a Carrollian fluid, thereby linking the $c \to 0$ contraction of the Poincaré group to cosmological acceleration.

\item Carrollian structures emerge naturally in the study of gravitational waves. As shown in \cite{Duval:2017els}, plane gravitational waves admit a description in terms of Carrollian geometry: their isometries form a Carrollian group, and the associated null hypersurfaces inherit a Carrollian structure. In this sense, the Carroll framework provides a geometrically transparent way to characterise the symmetries and kinematics of plane wave spacetimes.

 \item Carrollian physics also appears in condensed matter systems. For instance, fractons---quasiparticles with restricted mobility that typically arise in theories with higher-moment conservation laws, such as dipole symmetry---can be interpreted as Carrollian particles \cite{Bidussi:2021nmp,Figueroa-OFarrill:2023vbj,Figueroa-OFarrill:2023qty,Perez:2023uwt,Baig:2023yaz, Hartong:2024hvs,Figueroa-OFarrill:2025yum,Ahmadi-Jahmani:2025iqc}. Moreover, systems featuring flat bands \cite{Bagchi:2022eui,Ara:2024fbr} or certain types of phase transitions \cite{Biswas:2025dte} have been shown to exhibit Carrollian symmetries.

  \item Fluids in certain limiting regimes have also been shown to admit a Carrollian description \cite{deBoer:2017ing,Ciambelli:2018xat,Petkou:2022bmz}. For instance, shallow water hydrodynamics can be reformulated in terms of Carrollian geometry and dynamics \cite{Bagchi:2024ikw}. Moreover, Bjorken flows, which describe the boost-invariant longitudinal expansion of relativistic fluids (notably in heavy-ion collisions), have been shown to exhibit Carrollian features in the ultra-relativistic limit \cite{Bagchi:2023ysc, Kolekar:2024cfg}.

\end{itemize} 
Finally, let us mention that several reviews on Carrollian physics have recently appeared \cite{Bagchi:2025vri,Ciambelli:2025unn,Nguyen:2025zhg,Fiorucci:notes}, covering complementary aspects to those discussed here, including some of the points listed above.

The rest of this report is organized as follows. In Section \ref{sec:Carrollian limit}, we review the Carrollian limit of the Poincaré symmetries at the level of the group and the algebra, and define the notion of Carrollian symmetries. We compare this limit to the standard Galilean limit, highlighting the key differences. We also take the limit at the level of the conformal algebra and define the conformal Carrollian algebra. In Section \ref{sec:Carrollian geometry and symmetries}, we discuss Carrollian geometry from an intrinsic perspective, including the relevant notions of connection. We also analyze the isometries of this geometry and connect them to the Carrollian symmetries introduced in the previous section. In Section \ref{sec:Geometry of null hypersurfaces}, we show that the geometry induced on a null hypersurface in a Lorentzian manifold is Carrollian. We comment on the notion of the stretched horizon, considering black hole and cosmological horizons as limits of timelike hypersurfaces. We also discuss the example of null infinity and explain the isomorphism between the conformal Carrollian algebra and the BMS algebra. In Section \ref{sec:Carrollian field theories}, we introduce the notion of a Carrollian stress tensor. We then provide several examples of Carrollian field theories obtained by taking the $c \to 0$ limit of relativistic field theories, including the Carrollian scalar field, Maxwell and Yang-Mills theories, and Einstein gravity. We also briefly mention other models, including those with fermions or based on BMS geometric action constructions. In Section \ref{sec:Elements of Carrollian CFTs}, we focus on Carrollian CFTs and provide a comprehensive description of these theories, analogous to the treatment of standard CFTs, without relying on a Lagrangian formulation. This includes the definition of Carrollian primary fields, the general form of correlation functions implied by Ward identities, the stress tensor correlators, and the general structure of Carrollian OPEs consistent with the symmetries. In Section \ref{sec:Carrollian and celestial amplitudes}, we present the general setup for Carrollian holography in four-dimensional flat space. This includes the holographic identification between bulk fields and boundary Carrollian primaries, the definition of Carrollian amplitudes encoding the bulk $\mathcal{S}$-matrix in terms of Carrollian CFT correlators, the interpretation of soft theorems as Carrollian stress tensor Ward identities, the relation between bulk collinear limits of scattering amplitudes and Carrollian OPEs, and differential equations constraining the correlators. In Section \ref{sec:From Carrollian to celestial holography}, we give a brief overview of celestial holography and explain the relation between Carrollian and celestial amplitudes, providing examples of applications of this correspondence. In Section \ref{sec:Flat space/Carrollian limit of AdS/CFT}, we discuss the flat space / Carrollian limit correspondence at the level of correlators. More precisely, we show that the flat limit of Witten diagrams in AdS and the Carrollian limit of the corresponding holographic correlators at the boundary yield the Carrollian amplitudes living at null infinity. We apply this correspondence to the explicit realization of AdS$_4$/CFT$_3$, relating M-theory on AdS$_4 \times S^7 / \mathbb{Z}_k$ to ABJM theory in three dimensions, and show that it provides a path toward a top-down realization of flat space holography. Finally, in Section \ref{sec:Through the looking glass}, we present some open questions in the field and discuss future directions.

\section{Carrollian limit}
\label{sec:Carrollian limit}

\subsection{Another non-relativistic limit}

As emphasized in the introduction, while the Galilean limit is a well-studied and understood contraction of the Poincaré symmetries, interest in its Carrollian counterpart is quite recent. In this section, we closely follow the original reference by Lévy-Leblond \cite{Levy1965} and provide a pedagogical introduction to the Carrollian limit as a new non-relativistic contraction, comparing it to the standard Galilean limit. As we shall see, the limit can be discussed without explicitly introducing the speed of light; we set $c=1$ for now.

\begin{figure}[h]
    \centering
\includegraphics[width=0.8\textwidth]{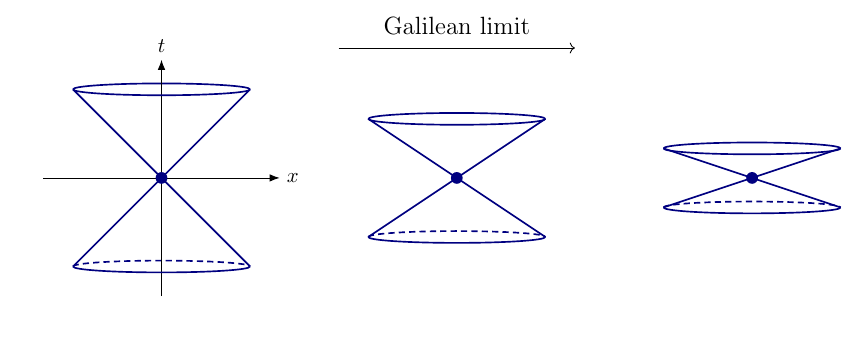}
\caption{The causal structure can be represented by light cones. The effect of taking the Galilean limit is to open the light cones. In the limit, all spacetime events become causally connected.}
\label{fig:galileanlimit}
\end{figure}

Let us start from a Poincaré boost in two dimensions of parameter $v$ acting on a spacetime interval $(\Delta t, \Delta x)$ as
\begin{equation} \label{relboost}
    \left\{
    \begin{array}{ll}
        \Delta t' &=  \frac{\Delta t + v \Delta x}{\sqrt{1-v^2}} \\
        \Delta x' &= \frac{\Delta x + v \Delta t}{\sqrt{1-v^2}}  
    \end{array}
\right.
\end{equation} There are two possible non-relativistic limits. First, the Galilean limit is defined by taking $v \ll 1$ and $|\Delta x| \ll |\Delta t|$. The Galilean boost is then given by
\begin{equation}
    \left\{
    \begin{array}{ll}
        \Delta t' &=  \Delta t  \\
        \Delta x' &= \Delta x + v \Delta t  
    \end{array}
\right.
\end{equation} Hence in a Galilean spacetime, the time does not transform under boost and becomes absolute. Notice that the second condition defining this limit, $|\Delta x| \ll |\Delta t|$, requiring the spacetime interval $(\Delta t, \Delta x)$ be timelike, is often forgotten, but is crucial to define this limit and keep the $v \Delta t$ term relative to $\Delta x$. At the level of the causal structure, this limit implies that the light cones open, making all events causally connected, see Figure \ref{fig:galileanlimit}.

\begin{figure}[h]
    \centering
\includegraphics[width=0.8\textwidth]{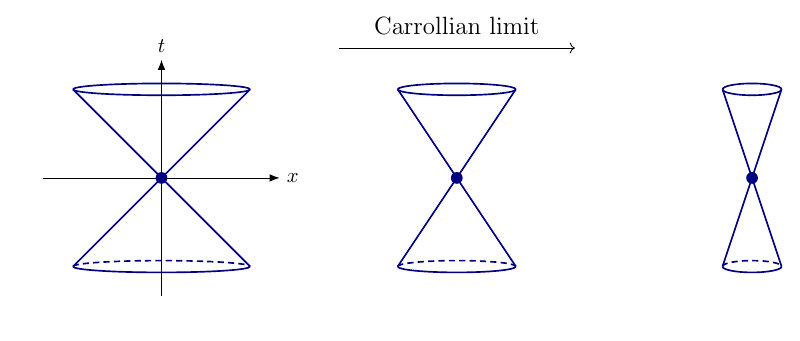}
\caption{The effect of taking the Carrollian limit is to close the light cones. In the limit, all spacetime events become causally disconnected unless they are at the same spatial point. This reflects the ultra-local nature of Carrollian physics.}
\label{fig:carrollianlimit}
\end{figure}

Surprisingly, there is another non-relativistic limit, which also corresponds to small boost parameter $v \ll 1$, but is defined by considering spacelike intervals instead of timelike intervals in \eqref{relboost}, i.e. $|\Delta t| \ll |\Delta x|$. This limit, called the Carrollian limit \cite{Levy1965}, yields
\begin{equation} \label{Carroll boost 2d}
    \left\{
    \begin{array}{ll}
        \Delta t' &=  \Delta t + v \Delta x \\
        \Delta x' &= \Delta x 
    \end{array} 
\right.
\end{equation} These Carrollian boosts act only on time and not on space. Hence, as opposed to the Galilean limit, in a Carrollian spacetime, the time is relative and the space is absolute. At the level of the causal structure, this limit implies that the light cones shrink into lines, see Figure \ref{fig:carrollianlimit}. 

The name “Carrollian limit” was introduced by Lévy-Leblond in a playful reference to Lewis Carroll, the author of \textit{Alice’s Adventures in Wonderland}. While the Galilean limit corresponds to the familiar non-relativistic physics of everyday life, the Carrollian limit exhibits a highly counterintuitive behaviour: as we see from \eqref{Carroll boost 2d}, motion in space effectively freezes and only evolution along time remains meaningful. The name Carrollian thus evokes the topsy-turvy nature of this limit, reminiscent of the strange and fantastical world imagined by Lewis Carroll.

Before moving to the next section, we comment on the interplay between the above limits and the formal limits taken on the speed of light. To do so, we restore $c$ in the spacetime interval discussion: 
\begin{itemize}
    \item Galilean limit: $\frac{|\Delta x|}{|\Delta t|} \ll c$. Hence, this limit can be obtained by keeping the spacetime interval fixed, while formally taking $c \to \infty$. 
    \item Carrollian limit: $\frac{|\Delta x|}{|\Delta t|} \gg c$. In practice this limit can be obtained by keeping the spacetime interval fixed, while formally taking $c \to 0$.
\end{itemize}
As we shall see, in practice, to implement the Carrollian limit at the level of relativistic field theories (see e.g. Section \ref{sec:Carrollian field theories}), it is convenient to restore the speed of light $c$ using dimensional arguments, and then formally take the $c\to 0$ limit to obtain explicit realizations of field theories exhibiting Carrollian symmetries.

\subsection{Carrollian and Galilean groups}
\label{sec:Carrollian and Galilean groups}

We now discuss the Galilean and Carrollian limits of the Poincaré group in a more systematic way in general dimension, and define the notion of Galilean and Carrollian groups. We work in a spacetime of dimension $d$ with coordinates $(x^0, \vec{x})$, where $x^0$ is the time coordinate and $\vec{x} = (x^A)$, $A = 1, \ldots, d-1$, are the spatial coordinates.

The Poincaré group in $d$ dimensions ($(d+1) d/2$ parameters) is composed of (spatial) rotations $R$, boosts $\vec{\beta}$, and translations $(a^0, \vec{a})$. The coordinates $(x^0 ,  \vec{x})$ of a spacetime event transform as 
\begin{equation}
    \left\{
    \begin{array}{ll}
         {x^0}' &=  \gamma (  x^0 + \vec{\beta} \cdot R \vec{x}) + a^0  \\
         \vec{x}' &= R \vec{x} + \frac{\gamma^2}{\gamma+1} (\vec{\beta}\cdot R\vec{x}) \vec{\beta} + \gamma \vec{\beta} x^0 + \vec{a}
    \end{array}
\right.
\end{equation} where $\gamma = (1 - \vec{\beta}^2)^{-\frac{1}{2}}$ is the Lorentz factor. This non-covariant notation is adopted to define the non-relativistic limits. First, we choose 
\begin{equation}
    x^0 = c u , \qquad \vec{\beta} = \frac{1}{c}\vec{v} , \qquad a^0 = c b 
\label{eq:Galilean rescaling}
\end{equation}
and take the $c \to \infty$ limit (keeping $u$, $\vec{v}$, and $b$ fixed). We find  
\begin{equation}
    \left\{
    \begin{array}{ll}
         u' &=   u + b  \\
         \vec{x}' &= R \vec{x} +  \vec{v} u + \vec{a}
    \end{array}
\right.
\end{equation} which coincides with a general Galilean transformation. These transformations form a group, the Galilean group, whose composition law can be written as
\begin{equation}
    (b_1, \vec{a}_1, \vec{v}_1, R_1  ) \circ (b_2, \vec{a}_2, \vec{v}_2, R_2  ) = (b_1+ b_2 , \vec{a}_1 + \vec{v}_1 b_2 + R_1 \vec{a}_2, \vec{v}_1 + R_1 \vec{v}_2 , R_1 R_2) .
    \label{eq:Galilean group}
\end{equation} The identity is $(0,\vec{0}, \vec{0}, \mathbb{I})$ and the inverse of $(b, \vec{a}, \vec{v}, R  )$ is $(-b, -R^{-1}(\vec{a} - \vec{v} b), -R^{-1} \vec{v}, R^{-1}  )$. 

Instead of \eqref{eq:Galilean rescaling}, one could choose 
\begin{equation}
     x^0 = c u, \qquad \vec{\beta} = c \vec{v}  , \qquad a^0 = c b
    \label{Carroll contraction}
\end{equation}
and take the $c \to 0$ limit (keeping $u$, $\vec{v}$, and $b$ fixed). We find
\begin{equation}
    \left\{
    \begin{array}{ll}
         u' &=   u + \vec{v} \cdot R \vec{x} + b   \\
         \vec{x}' &= R \vec{x} +  \vec{a}
    \end{array}
\right.
\end{equation} which corresponds to a general Carrollian transformation. Similarly to the Poincar\'e and the Galilean transformations, the Carrollian transformations ($(d+1)d/2$ parameters) are generated by spatial rotations $R$, Carrollian boosts $\vec{v}$, spatial translations $\vec{a}$ and time translation $b$. Again, these transformations form a group, the Carrollian group, whose composition is given by
\begin{equation}
     (b_1, \vec{a}_1, \vec{v}_1, R_1  ) \circ (b_2, \vec{a}_2, \vec{v}_2, R_2  ) = (b_1+ b_2 + \vec{v}_1 \cdot R_1 \vec{a}_2   , \vec{a}_1 + R_1 \vec{a}_2, \vec{v}_1 + R_1 \vec{v}_2 , R_1 R_2) .
     \label{eq:Carroll group}
\end{equation} The identity is $(0,\vec{0}, \vec{0}, \mathbb{I})$ and the inverse of $(b, \vec{a}, \vec{v}, R  )$ is $(-b + \vec{a} \cdot \vec{v}, -R^{-1}\vec{a}, -R^{-1} \vec{v}, R^{-1}  )$.

\subsection{Carrollian and Galilean algebras}
\label{sec:Carrollian and Galilean algebras}

The above non-relativistic limits can also be implemented at the level of the Poincar\'e algebra, $\mathfrak{iso}(d-1,1)$. Denoting by $J_{AB} = - J_{BA}$ the spatial rotation generators ($A,B = 1, \ldots, d-1$), $B_A$ the boost generators, $P_A$ the spatial translation generators, and $H$ the time translation generator, the Poincar\'e algebra reads as
\begin{equation}
\begin{array}{ll} 
    {[}J_{AB}, J_{CD}] = - i (\delta_{AC} J_{BD} + \delta_{BD} J_{AC} - \delta_{AD} J_{BC} - \delta_{BC} J_{AD}), \quad & {[}B_{A}, B_B] = i J_{AB}, \\
    {[}J_{AB}, B_{C}] = - i (\delta_{AC} B_{B} -  \delta_{BC} B_{A}), & {[}B_{A}, P_B] = i \delta_{AB} H,\\
    {[}J_{AB}, P_{C}] = - i (\delta_{AC} P_{B} -  \delta_{BC} P_{A}),  & {[}B_{A}, H] = i P_A,\\
    {[}J_{AB}, H] = 0, & {[}P_A, P_B] = 0,\\
    {[}P_A, H] = 0  .
\end{array} \label{Poincare algebra}
\end{equation} A representation of the Poincaré algebra is given by
\begin{equation}
    J_{AB} = i(x_A \partial_B - x_B \partial_A) , \quad B_A = i(x_0 \partial_A - x_A \partial_0 ), \quad P_A = - i \partial_A, \quad H = - i \partial_0 .
    \label{Poincare generators}
\end{equation}

Redefining the generators as
\begin{equation}
    B_A \to c B_A, \qquad P_A \to c P_A
    \label{rescaling generators}
\end{equation} in the algebra \eqref{Poincare algebra} and taking the (singular) limit $c \to \infty$, which corresponds to an \.{I}n\"{o}n\"{u}-Wigner contraction, one obtains the Galilean algebra
\begin{equation}
\begin{array}{ll} 
    {[}J_{AB}, J_{CD}] = - i (\delta_{AC} J_{BD} + \delta_{BD} J_{AC} - \delta_{AD} J_{BC} - \delta_{BC} J_{AD}), \quad & {[}B_{A}, B_B] =0 , \\
    {[}J_{AB}, B_{C}] = - i (\delta_{AC} B_{B} -  \delta_{BC} B_{A}), & {[}B_{A}, P_B] = 0,\\
    {[}J_{AB}, P_{C}] = - i (\delta_{AC} P_{B} -  \delta_{BC} P_{A}),  & {[}B_{A}, H] = i P_A ,\\
    {[}J_{AB}, H] = 0, & {[}P_A, P_B] = 0,\\
    {[}P_A, H] = 0   .
\end{array} \label{Galilean algebra}
\end{equation} This algebra, denoted by $\mathfrak{Gal}_{d}$, could have been deduced by differentiation of the Galilean group presented in \eqref{eq:Galilean group}. A representation is given by
\begin{equation}
     J_{AB} = i (x_A \partial_B - x_B \partial_A) , \quad B_A = -i u \partial_A , \quad P_A = - i \partial_A, \quad H = - i \partial_u .\end{equation} These generators can be obtained by performing spacetime contractions of their parent relativistic generators \eqref{Poincare generators}. For instance, restoring the speed of light $c$ in the time coordinate as in \eqref{eq:Galilean rescaling} in the boost generator, and then rescaling it as in \eqref{rescaling generators}, we find
     \begin{equation} 
         B_A = i(x_0 \partial_A - x_A \partial_0 ) \xrightarrow{\text{\eqref{eq:Galilean rescaling}}} i (-c u \partial_A - x_A c^{-1} \partial_u)  \xrightarrow{\text{\eqref{rescaling generators}}} B_A = i (-u \partial_A -  x_A c^{-2}  \partial_u) \xrightarrow{c \to \infty} - i u \partial_A
    \label{contraction carroll boost}
     \end{equation} and similarly for the other generators.

Alternatively, redefining the Poincaré generators as
\begin{equation}
    B_A \to \frac{1}{c} B_A, \qquad H \to \frac{1}{c} H
\label{Carrollian contraction generators}
\end{equation} in the algebra \eqref{Poincare algebra} and taking the $c \to 0$ limit, one finds the Carrollian algebra
\begin{equation}
\begin{array}{ll} 
    {[}J_{AB}, J_{CD}] = - i (\delta_{AC} J_{BD} + \delta_{BD} J_{AC} - \delta_{AD} J_{BC} - \delta_{BC} J_{AD}), \quad & {[}B_{A}, B_B] = 0, \\
    {[}J_{AB}, B_{C}] = - i (\delta_{AC} B_{B} -  \delta_{BC} B_{A}), & {[}B_{A}, P_B] = i \delta_{AB} H,\\
    {[}J_{AB}, P_{C}] = - i (\delta_{AC} P_{B} -  \delta_{BC} P_{A}),  & {[}B_{A}, H] = 0,\\
    {[}J_{AB}, H] = 0, & {[}P_A, P_B] = 0,\\
    {[}P_A, H] = 0  .
\end{array} \label{Carrollian algebra}
\end{equation} This algebra, denoted by $\mathfrak{Carr}_{d}$, could have been deduced by differentiation of the Carrollian group \eqref{eq:Carroll group}. A representation is given by
\begin{equation}
     J_{AB} = i  (x_A\partial_B - x_B \partial_A) , \quad B_A = - i x_A \partial_u , \quad P_A = - i \partial_A, \quad H = - i \partial_u .
\label{Carrollian generators repres}     
\end{equation} These generators can be obtained by spacetime contractions of \eqref{Poincare generators}. For instance, rescaling the time coordinate with $c$ as in \eqref{Carroll contraction} in the boost generator, and then rescaling it as in \eqref{Carrollian contraction generators}, we find
     \begin{equation} 
         B_A = i(x_0 \partial_A - x_A \partial_0 ) \xrightarrow{\text{\eqref{Carroll contraction}}} i (-c u \partial_A - x_A c^{-1} \partial_u)  \xrightarrow{\text{\eqref{Carrollian contraction generators}}} B_A = i ( -c^{2} u \partial_A -  x_A  \partial_u) \xrightarrow{c \to 0} -i x_A  \partial_u
     \end{equation} and similarly for the other generators. A particular feature of the Carrollian algebra \eqref{Carrollian algebra} is that the boosts commute among themselves, and the time translation generator $H$ is a central element. The homogeneous subalgebra of $\mathfrak{Carr}_d$ (i.e. the subalgebra preserving the spacetime origin) is isomorphic to the Euclidean algebra in $d-1$ dimensions, where spatial rotations and Carrollian boosts are identified with Euclidean rotations and translations, respectively. Furthermore, in the particular case of $d= 2$, the Galilean and Carrollian algebras are isomorphic, $\mathfrak{Gal}_2 \simeq \mathfrak{Carr}_2$. This can easily be seen by comparing \eqref{Galilean algebra} with \eqref{Carrollian algebra} and exchanging the role of space and time ($H \leftrightarrow P_1$). In the rest of this report, we focus on the Carrollian limit.

\subsection{Conformal Carrollian algebra}
\label{sec:Conformal Carrollian algebra}

In this section, we discuss the limit at the level of the conformal algebra and define the conformal Carrollian algebra, which plays a major role in holography. To write this section, we used \cite{Bagchi:2016bcd,Bagchi:2019xfx,Nguyen:2023vfz}.

\subsubsection{Global subalgebra}

The Poincaré algebra \eqref{Poincare algebra} can be extended into the conformal algebra in $d$ dimensions ($(d+2)(d+1)/2$ generators), $\mathfrak{so}(d,2)$, by adding the dilatation generator, $D$, and the special conformal transformations, $K_A$ and $K$. The non-vanishing brackets involving these additional generators are given by 
\begin{equation}
\begin{array}{lclcl} 
    {[}D, H] = i  H, & \hspace{1cm} & [D, P_A] =i P_A , & \hspace{1cm} & [D,  K] = -i K ,  \\  
    {[}D, K_A] = -i K_A, & & [K, H] =  2i D,  & & [K, P_A] = 2i B_A, \\  
    {[}K_A, H] = - 2 i B_A, & & [K_A, P_B] = -2i (\delta_{AB} D - J_{AB}), & & [B_A, K] = i K_A,  \\
    {[}B_A, K_B] = i \delta_{AB} K,  &  & [J_{AB}, K_C] = - i (\delta_{AC} K_B - \delta_{BC} K_A).  & &
\end{array} \label{commutation relation KD}
\end{equation}
A representation of $\mathfrak{so}(d,2)$ is given by \eqref{Poincare generators} together with 
\begin{equation}
\begin{split}
    &D = - i (x^0 \partial_0 + x^A \partial_A) , \quad K = i ( x_0 x^0 \partial_0 + 2 x_0 x^A \partial_A  - x^A x_A \partial_0), \\
    &K_A =  i (2 x_A x^0 \partial_0 + 2 x_A x^B \partial_B - x_0 x^0 \partial_A - x^B x_B \partial_A) .
\end{split} \label{D and K generators}
\end{equation} Redefining the Poincaré generators as in \eqref{Carrollian contraction generators} together with
\begin{equation}
    K \to \frac{1}{c} K
    \label{rescaling K}
\end{equation} and taking the $c \to 0$ limit, the commutation relations \eqref{commutation relation KD} become
\begin{equation} \label{conformal Carroll algebra}
\begin{array}{lclcl} 
    {[}D, H] = i  H, & \hspace{1cm}  &  [D, P_A] =i P_A , & \hspace{1cm}  & [D,  K] = -i K , \\  
    {[}D, K_A] = -i K_A , & & [K, H] = 0 , &  & [K, P_A] =  2i B_A , \\  
    {[}K_A, H] = - 2 i B_A , & & [K_A, P_B] = -2i (\delta_{AB} D - J_{AB}) , &  & [B_A, K] = 0 , \\
    {[}B_A, K_B] = i \delta_{AB} K ,    & & [J_{AB}, K_C] =- i (\delta_{AC} K_B - \delta_{BC} K_A) .  & &
\end{array} 
\end{equation} These brackets together with \eqref{Carrollian algebra} define the global conformal Carrollian algebra ($(d+2)(d+1)/2$ generators), $\mathfrak{CCarr}^{\text{glob}}_{d}$. A representation of this algebra is given by \eqref{Carrollian generators repres} together with 
\begin{equation}
\begin{split}
    &D = - i (u \partial_u + x^A \partial_A) , \quad K =  - i   x^A x_A \partial_u, \quad K_A =  i (2 x_A u \partial_u + 2 x_A x^B \partial_B - x^B x_B \partial_A)
\end{split} \label{repres Carroll K D}
\end{equation}
As before, the above generators can be obtained from their parent relativistic generators \eqref{D and K generators} by redefining the time $x^0 = c u$ as in \eqref{Carroll contraction}, performing the rescaling as in \eqref{rescaling K}, and taking $c \to 0$, see e.g. \eqref{contraction carroll boost}.  

An important observation is that $\mathfrak{CCarr}^{\text{glob}}_{d}$ is isomorphic to the Poincaré algebra in $d+1$ dimensions, $\mathfrak{iso}(d,1)$, i.e. 
\begin{equation}
  \label{isomorphism poincare conf carroll}\mathfrak{CCarr}^{\text{glob}}_{d} \simeq \mathfrak{iso}(d,1) \simeq \mathfrak{so}(d,1) \loplus \mathbb{R}^{d+1} .
\end{equation}  This isomorphism can be seen explicitly through the identification 
\begin{equation} \label{ident Poinc Carr}
    \begin{array}{llll}
         \tilde{J}_{AB} = J_{AB},  & \tilde{J}_{A0} = - \frac{1}{2} (P_A + K_A),  & \tilde{J}_{Ad} = \frac{1}{2} (P_A - K_A), & \tilde{J}_{0d} = - D \\
         \tilde{P}_0 = \frac{1}{\sqrt{2}} (H+K),  &\tilde{P}_A = - \sqrt{2} B_A, & \tilde{P}_d = \frac{1}{\sqrt{2}} (K-H) &
    \end{array}
\end{equation}
where $\{ \tilde{J}_{\mu\nu}, \tilde{P}_\mu \}$ satisfy the $\mathfrak{iso}(d,1)$ algebra, and $x^\mu = (x^0, x^A, x^d)$ with $A=1, \ldots, d-1$ \cite{Donnay:2022wvx,Nguyen:2023vfz}. This statement is already a hint towards Carrollian holography: the global conformal Carrollian symmetries of a theory in $d$ dimensions coincide with the symmetries of a Poincaré-invariant theory in $d+1$ dimensions.

\subsubsection{Infinite-dimensional enhancement}
\label{sec:Infinite-dimensional enhancement}

The $\mathfrak{CCarr}^{\text{glob}}_{d}$ algebra admits an infinite-dimensional enhancement with supertranslations in any dimension $d$. To see this, let us introduce the supertranslation generator:
\begin{equation}
    M_\mathcal{T} = -i \mathcal{T}(\vec{x}) \partial_u
    \label{supertranslation}
\end{equation} where $\mathcal{T}(\vec{x})$ is a smooth arbitrary function of the spatial coordinates $\vec{x} =(x^A)$. In particular, we have $M_\mathcal{T} = H, B_A, K$ for $\mathcal{T} = 1, x_A, x^A x_A$, respectively, in their representation \eqref{Carrollian generators repres} and \eqref{repres Carroll K D}. Furthermore, one can check that \eqref{supertranslation} forms an algebra with the rest of the generators in \eqref{Carrollian generators repres} and \eqref{repres Carroll K D}, whose commutators involving $M_\mathcal{T}$ are given by
\begin{equation} 
    \begin{array}{ll}
      [P_A, M_\mathcal{T} ] = - i  M_{\partial_A \mathcal{T}} , \quad    & [J_{AB},M_\mathcal{T}] = -i M_{-2x_{[A} \partial_{B]}\mathcal{T}} , \\
      {[} D, M_\mathcal{T}] = i M_{(-x^A\partial_A \mathcal{T} + \mathcal{T})}, & [K_A, M_\mathcal{T}] = -i M_{2x_A \mathcal{T} + x^B x_B \partial_A \mathcal{T} - 2 x_A x^B \partial_B \mathcal{T}} .
    \end{array}
\end{equation} The algebra generated by $\{M_\mathcal{T} , P_A, J_{AB}, D, K_A\}$ is called the conformal Carrollian algebra, $\mathfrak{CCarr}_d$, and the supertranslations form an abelian ideal, denoted by $\mathfrak{s}$. It forms an infinite-dimensional enhancement of $\mathfrak{CCarr}_{d}^{\text{glob}}$, hence the terminology ``global'', by analogy with two-dimensional conformal symmetries, where the conformal algebra Witt$\oplus$Witt is an infinite-dimensional extension of the global conformal algebra $\mathfrak{so}(2,2)$. In the following, it will be useful to expand the supertranslation generators in terms of polynomials in the spatial coordinates $x^A = (x^1, \ldots , x^{d-1})$:
\begin{equation}
    M^{m_1, \ldots, m_{d-1}} = -i (x^1)^{m_1} \ldots (x^{d-1})^{m_{d-1}} \partial_u
\label{super expansion}
\end{equation} with $m_i \in \mathbb{N}.$\footnote{The polynomial expansion \eqref{super expansion} assumes $\mathcal{T}(x^A)$ to be an analytic function on $\mathbb{R}^{d-1}$. As we shall see later, other topologies can be considered for the transverse space, such as the sphere $S^{d-1}$. In that case, a convenient basis is provided by spherical harmonics.} In these notations, we have 
\begin{equation}
    H = M^{0,\ldots ,0}, \quad B_1 =  M^{1, 0\ldots, 0}, \quad \ldots \quad , \quad B_{d-1} =  M^{ 0\ldots, 0, 1}, \quad K = M^{2, 0\ldots, 0} + \ldots + M^{0,\ldots,0, 2}. 
\label{translation Mm}
\end{equation}  
As we shall discuss in Section \ref{sec:Relation between Conformal Carroll and BMS}, the conformal Carrollian algebra can be further enhanced to include the superrotations.

\section{Carrollian geometry and symmetries}
\label{sec:Carrollian geometry and symmetries}

In this section, we describe the intrinsic geometry naturally associated with the Carrollian limit. We discuss the notion of Carrollian geometry, the Ehresmann connection, and several proposed formulations of a Carrollian connection. We then show how the isometries of this geometry correspond to the symmetry algebras derived in the previous section.

\subsection{Carrollian geometry}
\label{sec:Carrollian geometry}

We begin by providing some intuition about the relation between the $c\to 0$ limit and the definition of Carrollian geometry. Consider a manifold $\mathscr{S}$ of dimension $d$ with coordinates $x^a = (u, x^A)$, where $u$ is the time coordinate and $x^A = (x^1 , \ldots , x^{d-1})$ are the spatial coordinates. Restoring the speed of light $c$ in the Minkowski line element and taking $c\to 0$, we have 
\begin{equation} \label{Carrollian limit on the metric}
    ds^2 = \eta_{ab} dx^a dx^b = -c^2 du^2 + \delta_{AB} dx^A d x^B \quad \xrightarrow{c\to 0} \quad 0 du^2 + \delta_{AB} dx^A d x^B \equiv q_{ab} dx^a dx^b
\end{equation} where $q_{ab}$ is a degenerate metric with signature $(0+\ldots +)$. This metric cannot be inverted on spacetime, so it is crucial to consider the limit of the inverse Lorentzian metric with an appropriate rescaling, which contains additional information:
\begin{equation}
    -c^2 \eta^{ab} \partial_a  \otimes \partial_b =   \partial_u \otimes \partial_u - c^2 \delta^{AB} \partial_A \otimes \partial_B \quad \xrightarrow{c\to 0} \quad \partial_u \otimes \partial_u  \equiv  n^a n^b  \partial_a  \otimes \partial_b 
\end{equation} where $n^a \partial_a = \partial_u$ generates the degenerate direction of $q_{ab}$, i.e. $q_{ab} n^b = 0$.

This simple example motivates the more general definition: a Carrollian geometry \cite{Henneaux:1979vn,1977asst.conf....1G, Duval:2014lpa} on a manifold $\mathscr{S}$ is a pair
\begin{equation}
    (q_{ab}, n^c) \qquad \text{satisfying} \qquad     q_{ab} n^b = 0
\label{Carrollian geometry def}
\end{equation} where $q_{ab}$ is a degenerate metric of rank $d-1$ and $n^a$ is a nowhere-vanishing vector field in the kernel of this metric. The Carrollian geometry provides $\mathscr{S}$ with a fiber bundle structure \cite{Ciambelli:2019lap}, $\pi: \mathscr{S} \to \mathcal{S}$, whose one-dimensional fibers $\pi^{-1}(x)$, $x \in \mathcal{S}$, are the integral curves of $n^a$, and whose base space $\mathcal{S}$ is $(d-1)$-dimensional (see Figure \ref{fig:bundle}). 


\subsection{Ehresmann connection}

Given a fiber bundle structure, one can then introduce an Ehresmann connection. This allows us to single out a choice of horizontal space $H_k$ such that the tangent bundle of $\mathscr{S}$ can be split as $T \mathscr{S} = \text{ker}\, (d\pi) \oplus H_k$ unambiguously. The Ehresmann connection is encoded in the $1$-form $\boldsymbol{k} = k_a dx^a$ on $\mathscr{S}$ satisfying \cite{Henneaux:1979vn,Hartong:2015xda,Ciambelli:2019lap}
\begin{equation}
    k_a n^a =- 1 .
    \label{geometry conditions}
\end{equation} Using this, we can introduce a projector on the horizontal space $H_k = \ker \boldsymbol{k}$ as
\begin{equation}
    {q^a}_b  = \delta^a_b + n^a k_b .
    \label{projector def}
\end{equation} One can check that this object has the expected properties for a projector:
\begin{equation}
    {q^a}_b {q^b}_c =  {q^a}_c, \qquad {q^a}_b n^b = 0 , \qquad k_a {q^a}_b = 0, \qquad {q^a}_c q_{ab} = q_{ac} .
    \label{projector properties}
\end{equation} In particular, it annihilates all the vertical vectors. The identity operator can be expressed as
\begin{equation}
    \delta^a_b = {q^a}_b - n^a k_b .
    \label{identity}
\end{equation} In addition, we can define the ``inverse metric'' $q^{ab}$ through the conditions
\begin{equation}
    q^{ab} q_{bc} = \delta^a_c + n^a k_c, \qquad q^{ab} k_b = 0  
    \label{def inverse q}
\end{equation} which, by identification with \eqref{projector def}, implies 
\begin{equation}
    {q^a}_b = q^{ac} q_{cb} .
\end{equation} 




Given a Carrollian geometry $(q_{ab}, n^c)$ as defined in \eqref{Carrollian geometry def}, the freedom to choose an Ehresmann connection is encoded in the ``Carrollian boost'' or ``shift transformation'' acting as 
\begin{equation}
    \delta_\lambda k_a = \lambda_a , \qquad \delta_\lambda n^a = 0, \qquad \delta_\lambda q_{ab} = 0  \label{shift}
\end{equation} with the condition 
\begin{equation}
    \lambda_a n^a = 0
\end{equation} so that \eqref{geometry conditions} is preserved. We also deduce from \eqref{projector def} and \eqref{def inverse q} that
\begin{equation} \label{variation shifts}
    \delta_\lambda {q^a}_b =  n^a \lambda_b, \qquad \delta_\lambda q^{ab} = q^{ac}\lambda_c n^b + q^{bc} \lambda_c n^a .
\end{equation} By consistency, one can check that this does not violate the properties of the projector \eqref{projector properties}.

From the above discussion, given a Carrollian geometry $(q_{ab}, n^c)$, the pair $(q^{ab}, k_c)$ is determined only up to a shift transformation \eqref{shift}. However, given the pair $(q_{ab}, k_c)$, the conditions \eqref{Carrollian geometry def} and \eqref{geometry conditions} uniquely fix the pair $(q^{ab}, n^b)$ and vice versa. Therefore, in practice, it is convenient to consider the pairs $(q_{ab}, k_c)$ or $(q^{ab}, n^b)$ to capture the full geometry, including the Ehresmann connection.

\subsection{Carrollian frame}
\label{sec:Carrollian frame}

The above discussion has been presented in second order formalism, where the basic ingredients involve a degenerate metric $q_{ab}$. However, it is suggestive and extremely convenient in practice to work in first order formalism \cite{Hartong:2015xda,Campoleoni:2022wmf, Campoleoni:2023fug} and use a vielbein, or frame, to have all the pieces of the geometry on the same footing, i.e. either vectors or $1$-forms.

Consider $\{ m^a_I \}$ a basis for the horizontal bundle $H_k$, where $I=1, \ldots , d-1$ (see Figure \ref{fig:bundle}). We can add the Carrollian vector $n^a$ to build a basis of the tangent bundle $T\mathscr{S}$ called a Carrollian frame. This frame $e^a_i = (n^a,  m_1^a, \ldots m^a_{d-1} )$ and its associated co-frame $e_a^i = (-k_a,  m^1_a, \ldots , m_a^{d-1} )$, with $i \in \{ 0,I \} =  \{ 0,1, \ldots , d-1 \}$ the internal indices, satisfy the conditions $e_a^i e^a_j = \delta^i_j$ and $e_a^i e^b_i = \delta^b_a $ or, equivalently,
\begin{equation}
     n^a k_a = -1, \quad m_a^I m_J^a = \delta^{I}_J, \quad m^a_I m^I_b = {q^a}_b, \quad n^a m_a^I = 0 = k_a m^a_I .
     \label{normalization}
\end{equation} In terms of the frame, the metric and its inverse can be decomposed as 
\begin{equation}
    q_{ab} = e_a^i e_b^j q_{ij} = m^I_a m^J_b q_{IJ}, \qquad q^{ab} = e^a_i e_j^b q^{ij} = m^a_I m^b_J q^{IJ}
\label{sec:metric in Carrollian frame}
\end{equation} 
where $q_{0i} = 0 = q^{0i}$, $q_{IJ}$ is usually taken to be the Euclidean flat space metric, $\text{diag} (1, \ldots ,1)$, and $q^{IJ}$ its inverse. We can also define the volume form on $\mathscr{S}$ as 
\begin{equation}
    \boldsymbol{\varepsilon}_\mathscr{S} = \boldsymbol{k} \wedge  \boldsymbol \epsilon, \qquad \boldsymbol\epsilon = \boldsymbol{m}^1 \wedge \ldots \wedge \boldsymbol{m}^{d-1} 
\label{volume form def}
\end{equation} where $\boldsymbol{m}^I = m_a^I dx^a$.

In the following, we will be interested in different types of transformations of the frame. First, the subset of internal local $GL(d, \mathbb{R})$ transformations preserving the degenerate metric $q_{ij} = \text{diag} (0, 1 , \ldots 1)$ forms the homogeneous Carrollian group discussed in Section \ref{sec:Carrollian and Galilean groups}. They act infinitesimally on the Carrollian frame as
\begin{equation}
    \delta_{(r,\beta)} n^a = 0, \qquad \delta_{(r, \beta)} m^a_I = {r_I}^J m^a_J + \beta_I n^a , \qquad \delta_{(r,\beta)} k_a  =  \beta_I m^I_a , \qquad \delta_{(r, \beta)} m_a^I = {r^I}_J m_a^J
    \label{eq:variation frame}
\end{equation} where $r_{IJ} = - r_{JI}$ and $\beta_I$ are the infinitesimal parameters associated with spatial rotations and Carrollian boosts, respectively. In particular, an infinitesimal Carrollian boost acts exactly as a shift transformation \eqref{shift}, provided one makes the identification $\lambda_a =  \beta_I m^I_a$. It will also be useful to consider (homogeneous) Weyl rescalings on the Carrollian geometry, which are induced by the frame transformations
\begin{equation} \label{weyl rescaling frame}
    \delta_{\omega} n^a = \omega n^a, \qquad \delta_{\omega} m^a_I = \omega m^a_I, \qquad   \delta_{\omega} k_a =  - \omega k_a, \qquad \delta_{\omega} m_a^I = - \omega m_a^I .
\end{equation} 
Finally, spacetime diffeomorphisms on $\mathscr{S}$ simply act as Lie derivatives on the frame  
\begin{equation}
     \delta_{\xi} n^a = \mathcal{L}_\xi n^a, \qquad \delta_{\xi} m^a_I = \mathcal{L}_\xi m^a_I , \qquad  \delta_{\xi} k_a = \mathcal{L}_\xi k_a, \qquad \delta_{\xi} m_a^I = \mathcal{L}_\xi m_a^I
     \label{diffeos}
\end{equation} where $\xi^a$ are the infinitesimal diffeomorphism parameters.

\subsection{Ehresmann curvature, acceleration, vorticity and shear}
\label{sec:Ehresmann curvature}

\begin{figure}[h]
    \centering
\includegraphics[width=0.8\textwidth]{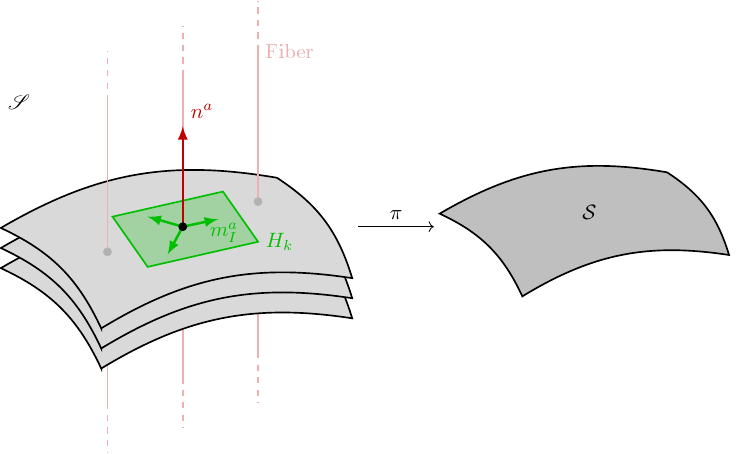}
\caption{This figure illustrates the fiber bundle structure $\pi : \mathscr{S} \to \mathcal{S}$ of a Carrollian manifold $(\mathscr{S}, q_{ab}, n^a)$. The fibers are the integral curves of the vertical vector field $n^a$, which lies in the kernel of the rank $d-1$ metric $q_{ab}$. The Ehresmann connection $k_a$ fixes a choice of horizontal bundle $H_k = \ker \boldsymbol{k}$, and the horizontal frame is denoted by $\{ m^a_I \}$. If the distribution $H_k$ is integrable, then the spacetime admits a foliation by co-dimension $1$ leaves.}
\label{fig:bundle}
\end{figure}

The curvature associated with the $1$-form Ehresmann connection $\boldsymbol{k}$ is given by $\boldsymbol{F} = d \boldsymbol{k}$ (the fibre being one dimensional, there is no quadratic term since $\boldsymbol{k} \wedge \boldsymbol{k} = 0$). It can be decomposed into the Carrollian co-frame as \cite{Ciambelli:2019lap,Freidel:2022bai}
\begin{equation}
    d \boldsymbol{k} = \boldsymbol{\varphi} \wedge \boldsymbol{k} + \boldsymbol{\varpi}
 =   {\varphi}_I \boldsymbol{m}^I  \wedge  \boldsymbol{k} + \varpi_{IJ} \boldsymbol{m}^I \wedge \boldsymbol{m}^J  .
 \label{decomposition ehresmann}
\end{equation} where the purely horizontal piece $\boldsymbol{\varpi}$ is the Carrollian vorticity and the vertical-horizontal mixed piece $\boldsymbol{\varphi}$ is called the acceleration. Assuming that the horizontal frame is obtained by taking the pullback of a holonomic frame on the base $\mathcal{S}$, $\boldsymbol{m}^I = \pi^* dx^I$ (where $x^I$ are coordinates on the base), so that $d\boldsymbol{m}^I = 0$, we have the simple commutator formulae
\begin{equation}
    [ m_I^a \partial_a , m^b_J \partial_b ] = 2  \varpi_{IJ} n^c \partial_c, \qquad  [ n^a \partial_a , m_I^b \partial_b ] = {\varphi}_I n^c \partial_c .
\label{commutatots m and n}
\end{equation} We refer to \cite{Campoleoni:2023fug,Vilatte:2024jjr, Fiorucci:2025twa} for generalizations. Given an Ehresmann connection $\boldsymbol{k}$, the Frobenius integrability condition reads as 
\begin{equation}
    \boldsymbol{k} \wedge d \boldsymbol{k} = 0 \qquad \Longleftrightarrow \qquad \boldsymbol{\varpi} = 0
\end{equation} i.e. there is no vorticity. This condition is satisfied if and only if the distribution $H_k$ is integrable, meaning there exists a co-dimension one spacetime foliation of $\mathscr{S}$ whose tangent bundle coincides with $H_k$, see Figure \ref{fig:bundle}. In that case, assuming in addition that the acceleration $\varphi$ vanishes, we can locally choose a convenient set of coordinates $(u,x^A)$ where $u$ is the coordinate along the fibres such that $n^a\partial_a=\partial_u$, and $x^A$ are coordinates on the leaves of the foliation, so that $k_A=0$. Furthermore, if the horizontal metric $q_{AB}$ is flat, one can choose the coordinates $x^A$ such that $q_{AB}=\delta_{AB}$. For such a choice of coordinates, the Carrollian geometry reads as
\begin{equation} \label{flat carrollian background}
    n^a \partial_a  = \partial_u, \quad q_{ab} dx^a dx^b = \delta_{AB} dx^A dx^B, \quad k_a dx^a = - du, \quad q^{ab} \partial_a \partial_b = \delta^{AB} \partial_A \partial_B .
\end{equation} Working with this ``flat'' or ``canonical'' Carrollian background is the analogue of working with the flat space metric $\eta_{ab} = \text{diag}(-1, 1, \ldots, 1)$ in the relativistic case. This flat background can also be expressed in terms of the frame as
\begin{equation}
    n^a \partial_a = \partial_u, \quad m^a_I \partial_a  = \partial_I, \quad k_a dx^a = - du, \quad m_a^I dx^a =  d x^I 
    \label{flat Carrollian frame}
\end{equation} which is the standard holonomic basis.

Given a Carrollian geometry $(q_{ab},n^c)$, one can define the extrinsic curvature 
\begin{equation} \label{expansion tensor}
    \theta_{ab} = \frac{1}{2}\mathcal{L}_n q_{ab} 
\end{equation} which, by definition, is symmetric and satisfies $n^a \theta_{ab} =0$ and ${q^a}_c \theta_{ab}=\theta_{cb}$. Since it is purely horizontal, we can rewrite it as
\begin{equation}
    \theta_{IJ} = m^a_I m^b_J \theta_{ab} . 
\label{expansion tensor IJ}
\end{equation} The expansion is obtained by taking the trace with respect to the inverse metric \eqref{def inverse q}:
\begin{equation} \label{expansion Carroll}
    \theta = q^{ab} \theta_{ab} = q^{IJ} \theta_{IJ} .
\end{equation} It measures how horizontal area varies along the fibers. The shear is the trace-free part of the extrinsic curvature
\begin{equation} \label{Carrollian shear}
    \sigma_{ab} = \theta_{ab} - \frac{1}{d-1} q_{ab} \theta
\end{equation} i.e. it satisfies $q^{ab} \sigma_{ab} = 0$. It measures the area-preserving deformations of the horizontal space along the fibers.

\subsection{Carrollian connections}
\label{sec:Carrollian connection}

There are several types of Carrollian connections discussed in the literature that require different compatibility conditions and constraints on the torsion. We will first discuss general considerations about connections and then specify the analysis to two relevant subcases of Carrollian connections. We refer to Section 3.1.2 of \cite{Ruzziconi:2020cjt} for a general discussion on affine connections in Riemannian geometry. We will essentially repeat the steps of that reference, adapting the discussion to the Carrollian geometry.  

\subsubsection{General considerations}

Given a Carrollian frame $e^a_i = (n^a, m^a_I)$, the coefficients of an affine connection $D$ are defined as 
\begin{equation}
   D_j e^a_i = {\Gamma^k}_{ij} e^a_k \quad \Longleftrightarrow \quad  D_j n^a = {\Gamma^0}_{0j} n^a +  {\Gamma^K}_{0j} m^a_K, \quad D_j m^a_I = {\Gamma^0}_{Ij} n^a +  {\Gamma^K}_{Ij} m^a_K  .
\end{equation} The upper index and the first lower index of the connection coefficients are internal indices, while the second lower index is naturally interpreted as a form index through ${\Gamma^k}_{ia} = {\Gamma^k}_{ij} e^j_a$. Equivalently, in terms of the co-frame $e^i_a = (-k_a, m^I_a)$, we have
\begin{equation}
     D_j e_a^i = -{\Gamma^i}_{kj} e_a^k \quad \Longleftrightarrow \quad  D_j k_a = -{\Gamma^0}_{0j} k_a +  {\Gamma^0}_{Ij} m_a^I, \quad D_j m_a^I = {\Gamma^I}_{0j} k_a -  {\Gamma^I}_{Kj} m_a^K .
\end{equation} For a vector $V^a = V^i e_i^a$ and a $1$-form $\omega_a = \omega_i e^i_a$ on $\mathscr{S}$, we deduce 
\begin{equation}
  D_j V^i = \partial_j V^i + {\Gamma^i}_{kj} V^k, \qquad D_j \omega_i = \partial_j \omega_i - {\Gamma^k}_{ij} \omega_k .
\end{equation} The structure constants of the Carrollian frame will be denoted by
\begin{equation}
    [ e_i , e_j  ] = {C^k}_{ij} e_k \quad \Longleftrightarrow \quad d \boldsymbol{e}^i = - \frac{1}{2} {C^i}_{jk} \boldsymbol{e}^j \wedge \boldsymbol{e}^k
\end{equation} where $e_i = e_i^a \partial_a$ and $\boldsymbol{e}^i = e^i_a dx^a$. Some of these encode the Carrollian vorticity and acceleration, see \eqref{commutatots m and n}. The torsion tensor is defined by
\begin{equation}
    \boldsymbol{T}^i = \frac{1}{2} {T^i}_{jk} \boldsymbol{e}^j \wedge \boldsymbol{e}^k =  d \boldsymbol{e}^i + {\boldsymbol{\Gamma}^i}_{j} \wedge \boldsymbol{e}^j , 
\end{equation} where ${\boldsymbol{\Gamma}^i}_j = {\Gamma^i}_{ja} dx^a$, and we have 
\begin{equation}
{{T}^i}_{jk} = 2 {\Gamma^i}_{[kj]} - {C^i}_{jk} .
\label{torsion versus gammas}
\end{equation} Moreover, the curvature tensor is defined by 
\begin{equation}
    {\boldsymbol{R}^i}_j = \frac{1}{2} {R^i}_{jkl}  \boldsymbol{e}^k \wedge \boldsymbol{e}^l  =  d {\boldsymbol{\Gamma}^i}_j+{\boldsymbol{\Gamma}^i}_k \wedge {\boldsymbol{\Gamma}^k}_j .
\end{equation} The commutator of covariant derivatives can be expressed as 
\begin{equation}
    [ D_i,D_j ] \omega_k = - {R^l}_{kij} \omega_l - {T^l}_{ij} D_l \omega_k  .
\end{equation} We express the degenerate metric in the frame as in \eqref{sec:metric in Carrollian frame}, and keep $q_{IJ}$ general for now. The non-metricity tensor is defined as 
\begin{equation}
    \Xi_{ijk} = - D_k q_{ij} =  -\partial_k q_{ij} + \Gamma_{ijk} + \Gamma_{jik}  .
\label{nonmetricity}
\end{equation} Finally, we define the Christoffel symbols
\begin{equation}
    \{_{ijk}\} = \frac{1}{2} (\partial_j q_{ik} + \partial_k q_{ij} - \partial_i q_{jk}) .
\end{equation}

\subsubsection{Fundamental theorem of Carrollian geometry}
\label{sec:Fundamental theorem}

The fundamental theorem of Riemannian geometry tells us that there is a unique connection which is torsion free and compatible with the metric, which is the Levi-Civita connection. As we shall now explain, this theorem no longer holds for a Carrollian geometry \cite{Vogel1965,Jankiewicz,Dautcourt1967,Henneaux:1979vn,Bekaert:2015xua, Hartong:2015xda}. To see this, let us follow the usual steps (see e.g. Section 3.1.2 of \cite{Ruzziconi:2020cjt}) and see where the differences arise. Computing $D_i q_{jk} + D_j q_{ki} - D_k q_{ij}$ and solving for $\Gamma_{ijk} = q_{il} {\Gamma^l}_{jk}$ ($\Gamma_{0jk} = 0$ for the Carrollian case since $q_{0l} = 0$), we have the usual result 
\begin{equation}
    \Gamma_{ijk} = \{_{ijk}\} + \frac{1}{2} ( \Xi_{ijk} + \Xi_{ikj} - \Xi_{jki}) + \frac{1}{2}(T_{jik}+ T_{kij}- T_{ijk}) + \frac{1}{2} (C_{jik} + C_{kij} - C_{ijk}) . 
\label{fundamental relation connection}
\end{equation} In Riemannian geometry, one can raise the index $i$ in this equation using the inverse metric to recover the Levi-Civita connection when metricity ($\Xi_{ijk} =0$) and torsion free ($T_{ijk} = 0$) conditions are imposed. In a holonomic basis ($C_{ijk}= 0)$, the connection coefficients then reduce to the Christoffel symbols. However, in the Carrollian case, the metric is degenerate and one can no longer raise the index $i$ to recover all the connection coefficients ${\Gamma^l}_{jk}$. More precisely, one can still use the inverse metric to obtain ${\Gamma^{L}}_{jk} = q^{LI} \Gamma_{Ijk}$, but ${\Gamma^0}_{jk}$ cannot be obtained in this way. Let us focus on the $i = 0$ component of this equation, which now gives a constraint
\begin{equation}
    T_{(J|0|K)} =  \theta_{JK} + \frac{1}{2} \Xi_{JK0} -  \Gamma_{(J|0|K)}
\label{fundamental relations Carroll conection}
\end{equation} and the other components do not give extra conditions. On the right-hand side, we have the extrinsic curvature defined in \eqref{expansion tensor IJ}, the non-metricity tensor \eqref{nonmetricity}, and symmetrized spin coefficients encoding the transverse deformation of $n^a$ along the horizontal directions. This relation generalizes those discussed in \cite{Vogel1965,Jankiewicz,Dautcourt1967,Henneaux:1979vn,Bekaert:2015xua, Hartong:2015xda} and expresses the well-known fact that some part of the torsion is determined by the geometry, which has no analogue in the Riemannian case. Let us study two relevant subcases of Carrollian connections.

\paragraph{Compatible Connection} Let us first discuss the closest analogue of the Levi-Civita connection \cite{Bekaert:2015xua,Figueroa-OFarrill:2020gpr}, which has notably been used in \cite{Hansen:2021fxi} to discuss the Carrollian limit of general relativity. It is also related to the connection induced at null infinity from the bulk Levi-Civita connection in asymptotically flat spacetimes \cite{Ashtekar:2014zsa}. This connection consists in imposing the compatibility with the geometry, namely
\begin{equation}
    D_a q_{bc} = 0 , \qquad D_{a} n^b = 0 .
\label{compatible connection}
\end{equation} In frame indices, this implies $\Xi_{ijk} = 0$ and ${\Gamma^i}_{0j} =0$, so that \eqref{fundamental relations Carroll conection} reduces to
\begin{equation}
    T_{(I|0|J)} = \theta_{IJ} .
\end{equation} Hence, in general, one cannot set all the components of the torsion to zero because of this geometric obstruction. In other words, a Carrollian connection compatible with the geometry and torsion free exists if and only if the extrinsic curvature vanishes \cite{Figueroa-OFarrill:2020gpr}. Once the compatibility conditions \eqref{compatible connection} are imposed, the only undetermined connection coefficients are given by ${\Gamma^0}_{(IJ)}$. Indeed, as stated below \eqref{fundamental relation connection}, ${\Gamma^0}_{ij}$ cannot be directly obtained by raising the index in this equation, but ${\Gamma^0}_{[ij]}$ are completely fixed through \eqref{torsion versus gammas}, and the coefficients ${\Gamma^i}_{0j}$ are fixed by the compatibility conditions.

\paragraph{Unconditionally torsion-free connection} As we shall discuss in Section \ref{sec:Induced Carrollian connection}, a different Carrollian connection is needed to describe Carrollian physics induced on null hypersurfaces embedded in an ambient spacetime \cite{Mars:1993mj,Chandrasekaran:2021hxc,Freidel:2022vjq,Ciambelli:2023mir,Blitz:2025spc}. We will require this connection to be torsion free, 
\begin{equation}
    {T^i}_{jk} = 0 .
\end{equation} Equation \eqref{fundamental relations Carroll conection} leads to 
\begin{equation}
     \theta_{JK}  =   - \frac{1}{2} \Xi_{JK0} +  \Gamma_{(J|0|K)}
\end{equation} and one can no longer require full compatibility with the Carrollian geometry \eqref{compatible connection}. A convenient choice \cite{Mars:1993mj,Chandrasekaran:2021hxc,Freidel:2022vjq,Ciambelli:2023mir} is to demand 
\begin{equation}
   \Xi_{JK0} = 0 \qquad\text{and} \qquad \Xi_{IJK} = 0
\end{equation} so that 
\begin{equation}
    \theta_{JK}  =      \Gamma_{(J|0|K)}, \qquad D_I q_{JK} = 0 .
\end{equation} This choice ensures the closest compatibility with full metricity without constraining the geometry encoded in $\theta_{JK}$. Putting all these assumptions together and expressing everything in terms of spacetime indices, we find 
\begin{equation}
    D_a \varepsilon_{\mathscr{S}} = -\omega_a \varepsilon_{\mathscr{S}}
    \label{derivative volume form}
\end{equation} where $\varepsilon_{\mathscr{S}}$ is the volume form \eqref{volume form def} and the $1$-form $\omega_a$ can be decomposed into a vertical and horizontal part as $\omega_a = \kappa k_a + \pi_a$, where $\kappa =- {\Gamma^0}_{00} = n^b D_b n^a \, k_a$ is the inaffinity of $n^a$ and $\pi_a = q_a{}^cD_c n^b k_b$ is the H\'aji\v{c}ek connection. Furthermore, we have 
\begin{equation} \label{connection null}
        (D_a - \omega_a) n^b = {\theta_a}^b, \qquad D_a q_{bc} = - k_b \theta_{ac} - k_c \theta_{ab} .
\end{equation} In the first expression, the (horizontal) index was raised with $q^{ab}$ defined in \eqref{def inverse q}. Notice that these expressions are covariant with respect to the Carrollian boost transformations introduced in \eqref{shift}, and reproduce the defining conditions for the connection used in \cite{Mars:1993mj,Chandrasekaran:2021hxc,Freidel:2022vjq,Ciambelli:2023mir}.

\subsection{Symmetries}
\label{sec:Symmetries}

We now discuss the symmetries of the geometry described above. We will see that the algebras discussed in Section \ref{sec:Carrollian limit}, obtained through the $c\to 0$ limit, are naturally interpreted in this framework. This further justifies why the Carrollian geometry is indeed related to the $c\to 0$ limit. We closely follow \cite{Duval:2014lpa, Duval:2014uva}.

\subsubsection{Carrollian isometries} 

In the relativistic case, the Poincaré algebra \eqref{Poincare algebra} is the symmetry algebra of the flat space metric $\eta_{ab} dx^a dx^b$ with $\eta_{ab} = \text{diag}(-1, 1, \ldots 1)$, i.e. the diffeomorphisms generated by vector fields $\xi$ satisfying the Killing equation $\mathcal{L}_\xi  \eta_{ab} = 0$. On this flat background metric, the Levi-Civita connection coefficients vanish, and this remains true under Poincar\'e transformations. The Carrollian case is a bit more subtle. To see this, consider the flat Carrollian geometry \eqref{flat carrollian background}, which we repeat here:
\begin{equation}
    n^a \partial_a = \partial_u, \quad q_{ab} dx^a dx^b = \delta_{AB} dx^A dx^B .
\label{flat Carroll geometry}
\end{equation} The Killing vectors are obtained 
by solving 
\begin{equation}
    \mathcal{L}_\xi n^a = 0, \qquad \mathcal{L}_\xi q_{ab} = 0  .
\label{Killing Carroll}
\end{equation} Denoting the components of the vector field $\xi$ as
\begin{equation}
    \xi = f \partial_u + \mathcal{Y}^A \partial_A
\label{ansatz xi}
\end{equation} the second condition implies $\partial_u \mathcal{Y}^A = 0$ and\footnote{Indices $A, B \ldots$ are lowered and raised by $q_{AB} = \delta_{AB}$ and its inverse.} $\partial_{(A} \mathcal{Y}_{B)} = 0$. The first condition further imposes $\partial_u f = 0$. Hence, the general solution is given by
\begin{equation}
    \xi = f(x^A) \partial_u + (a^A + {r^A}_B x^B )   \partial_A
\label{extension Carroll alg}
\end{equation} where $a^A$ are the $d-1$ spatial translation parameters and $r_{AB} = r_{[AB]}$ are the spatial rotation parameters. The symmetry algebra generated by \eqref{extension Carroll alg}, called the Carrollian isometry algebra, $\mathfrak{Carriso}_d$, is infinite dimensional since $f$ is an arbitrary function of the spatial coordinates $x^A$. It is a semi-direct sum 
\begin{equation}
    \mathfrak{Carriso}_d \simeq \mathfrak{iso}(d-1) \loplus \mathfrak{C}^{\infty}(\mathcal{S})
\end{equation} where $\mathfrak{iso}(d-1) = \mathfrak{so}(d-1) \loplus \mathbb R^{d-1}$ is the Euclidean algebra in $d-1$ dimensions and $\mathfrak{C}^{\infty}(\mathcal{S})$ is the abelian algebra of smooth functions $f$ in $d-1$ dimensions transforming as scalars under $\mathfrak{iso}(d-1)$. 

\subsubsection{Carrollian symmetries} 
\label{sec:Carrollian isometry}

As discussed in Section \ref{sec:Carrollian connection}, by contrast with the relativistic case, the Carrollian connection is not entirely fixed by requiring the conditions \eqref{compatible connection}. For the flat background \eqref{flat Carroll geometry}, the only free and a priori non-vanishing connection coefficients are given by $\Gamma^u_{(AB)}$. Hence, in addition to \eqref{flat Carroll geometry}, we will impose the condition
\begin{equation}
    \Gamma^u_{(AB)} = 0 .
\label{vanishing connection}
\end{equation} Diffeomorphisms preserving this condition satisfy 
\begin{equation}
\label{connection preservation}
    \delta_{\xi} \Gamma^u_{(AB)} = 0 
\end{equation}
so that
\begin{equation}
\label{reduc carroll}
    \partial_A \partial_B f = 0 .
\end{equation} This equation admits $d$ independent solutions, $f = 1, x^A$, corresponding to the time translation and the $d-1$ Carrollian boosts. In summary, the Lie bracket of infinitesimal diffeomorphisms preserving \eqref{flat Carroll geometry} together with \eqref{vanishing connection} forms the $\mathfrak{Carr}_d$ algebra defined in Equation \eqref{Carrollian algebra}.

\subsubsection{Conformal Carrollian symmetries}
\label{sec:Conformal Carrollian isometry}

In the relativistic case, the conformal algebra \eqref{commutation relation KD} is the algebra of conformal Killing vectors for the flat space metric $\eta_{ab}dx^a dx^b$, i.e. the vector fields $\xi$ obtained by solving $\mathcal{L}_{\xi} \eta_{ab} = 2 \alpha \eta_{ab}$, where the multiplicative function in the right-hand side is determined by taking the trace of this equation, $\alpha = \frac{1}{d} \partial_a \xi^a$. Analogously, in the Carrollian case, the conformal Killing vectors of the canonical Carrollian geometry \eqref{flat Carroll geometry} are obtained by solving
\begin{equation}
    \mathcal{L}_\xi n^a = - \frac{2 \alpha}{N}  n^a, \qquad \mathcal{L}_\xi q_{ab} = 2 \alpha  q_{ab}
\label{conforaml Killing equations}
\end{equation} where $N$ is an integer called the conformal Carrollian algebra level. It reflects the ambiguity of relative scaling between the degenerate metric and the Carrollian vector field. Equation \eqref{conforaml Killing equations} implies 
\begin{equation}
    \mathcal{L}_\xi (q_{ab} \otimes n^{c_1} \otimes \ldots \otimes  n^{c_N} ) =0
    \label{conformal Killing Carroll N}
\end{equation} hence providing a geometric definition of $N$. Starting with the general vector field \eqref{ansatz xi}, the second condition in \eqref{conforaml Killing equations} implies $\partial_u \mathcal{Y}^A = 0$ and 
$\partial_A \mathcal{Y}_B +\partial_B \mathcal{Y}_A = 2 \alpha \delta_{AB}$. By taking the trace of the latter, we find $\alpha = \frac{1}{(d-1)} \partial_A \mathcal{Y}^A$. Taking this into account, the first of \eqref{conforaml Killing equations} leads to $\partial_u f = \frac{2\alpha}{N}$, so that $f = \mathcal{T}(x^A) + \frac{2}{N (d-1)} u \partial_A \mathcal{Y}^A$. In summary, 
\begin{equation}
    \xi = \left( \mathcal{T} + \frac{2}{N} u \alpha \right) \partial_u + \mathcal{Y}^A \partial_A, \quad  \alpha = \frac{1}{(d-1)} \partial_A \mathcal{Y}^A, \quad \partial_A \mathcal{Y}_B +\partial_B \mathcal{Y}_A = 2 \alpha \delta_{AB}
\label{conformal Carrollian vector}
\end{equation} where $\mathcal{T}(x^A)$ is an arbitrary function of the spatial coordinates, and $\mathcal{Y}^A(x^B)$ depends on the spatial coordinates and satisfies the conformal Killing equation for the $(d-1)$-dimensional flat space metric. From this expression, the generator associated with $\mathcal{Y}^A = x^A$ reads as $z u  \partial_u + x^A \partial_A$, where we identify the quantity $z= \frac{2}{N}$ with the dynamical exponent. The level $N = 2$ is particularly interesting: space and
time are then equally dilated so that the
dynamical exponent is $z = 1$ (we refer to \cite{Afshar:2024llh,Despontin:2025dog} for recent discussions on anisotropic scaling). Notice that if $\alpha = 0 \, \Leftrightarrow \, \partial_A \mathcal{Y}^A=0$, the conformal Killing equation for the Carrollian structure \eqref{conforaml Killing equations} reduces the Killing equation \eqref{Killing Carroll}, and the solution \eqref{conformal Carrollian vector} consistently reduces to \eqref{extension Carroll alg}. The Lie bracket of the conformal Carrollian Killing vectors \eqref{conformal Carrollian vector} is given by 
\begin{equation}
    [\xi (\mathcal{T}_1, \mathcal{Y}^A_1), \xi (\mathcal{T}_2, \mathcal{Y}^A_2)] = \xi (\mathcal{T}_{12}, \mathcal{Y}^A_{12})
\end{equation} where 
\begin{equation}
    \begin{cases}
       \mathcal{T}_{12} = \mathcal{Y}_1^A \partial_A \mathcal{T}_2 - \frac{2}{N(d-1)} \partial_A \mathcal{Y}_1^A \mathcal{T}_2 - (1 \leftrightarrow 2)  , \\
      \mathcal{Y}_{12}^A = \mathcal{Y}_1^B \partial_B \mathcal{Y}_2^A - (1 \leftrightarrow 2)   .
    \end{cases}     
\end{equation} The algebra of these vector fields is a semi-direct sum between the conformal algebra in $d-1$ dimensions parametrized by $\mathcal{Y}^A$, $\mathfrak{so}(d,1)$, and the abelian algebra of densities with weight $- \frac{2}{N(d-1)}$, parametrized by $\mathcal{T}$, and denoted by $\mathfrak{s}_{N}$. This corresponds to the conformal Carrollian algebra of level $N$ in $d$ dimensions: 
\begin{equation}
    \mathfrak{CCarr}_{d,N} \simeq \mathfrak{so}(d,1) \loplus \mathfrak{s}_{N} .
\end{equation} In the particularly interesting case $N=2$ where space and time scale homogeneously, the above algebra coincides with the conformal Carrollian algebra $\mathfrak{CCarr}_{d,2} \equiv \mathfrak{CCarr}_d$ defined in Section \ref{sec:Conformal Carrollian algebra}. 

The finite-dimensional global conformal Carrollian subalgebra $\mathfrak{CCarr}^{\text{glob}}_d$, obtained from the Carrollian contraction of $\mathfrak{so}(d,2)$ (see Section \ref{sec:Conformal Carrollian algebra}), is obtained by imposing the following constraint on the supertranslation parameters:
\begin{equation}
    \left(\partial_A \partial_B - \frac{1}{(d-1)} \delta_{AB} \partial^C \partial_C \right) \mathcal{T} = 0 \quad (d> 2), \qquad \partial_{x}^3 \mathcal{T} = 0 \quad (d=2). 
\end{equation} A general solution is given by $\mathcal{T}= a+  b_A x^A + k x^2$ where $a$, $b_A$ and $k$ are real parameters for time translation, Carrollian boosts and temporal special conformal Carrollian transformation, respectively. Moreover, the solution of the conformal Killing equation in $d-1$ dimensions for $\mathcal{Y}^A$ in \eqref{conformal Carrollian vector} is given by $\mathcal{Y}^A = a^A + {r^A}_B x^B + \lambda x^A + k^A x^B x_B - 2 k_B x^B x^A$, where $a^A$, ${r}_{AB} = r_{[AB]}$, $\lambda$ and $k^A$  are real parameters for spatial translations, rotations, dilatations and spatial special conformal Carrollian transformations, respectively (the cases $d=2,3$ admit an infinite-dimensional enhancement of solutions, see Section \ref{sec:Relation between Conformal Carroll and BMS}). Injecting these explicit parametrizations into \eqref{conformal Carrollian vector} (with $N=2$), we find the generators for $\mathfrak{CCarr}^{\text{glob}}_d$:
\begin{equation}
    \xi = \left[ a+  b_A x^A + k x^2 + u (\lambda - 2 k^A x_A) \right] \partial_u +  \left[ a^A +  {r^A}_B x^B + \lambda x^A + k^A x^2 - 2 k_B x^B x^A \right] \partial_A .
\label{global conf carr Killing vector}
\end{equation} In terms of the explicit representation of the operators \eqref{Carrollian generators repres} and \eqref{repres Carroll K D}, we have
\begin{equation}
 \xi = i a H + i b^A B_A  +i  k K  + i a^A P_A + \frac{i}{2} r^{AB} J_{AB} + i\lambda D + i k^A K_A  .
\end{equation}

\subsubsection{Carrollian diffeomorphisms} 

As explained at the end of Section \ref{sec:Carrollian geometry} a manifold $\mathscr{S}$ endowed with a Carrollian geometry \eqref{Carrollian geometry def} naturally exhibits a structure of fiber bundle, with projection $\pi :\mathscr{S} \to \mathcal{S}$. On the total space, there is a preferred coordinate system $(u,x^A)$ such that $u$ is a coordinate along the one-dimensional fibers, i.e. for which $\partial_u$ aligns with the direction of the Carrollian vector $n^a$:
\begin{equation}
    n^a = F(u,x^A) \partial_u
\label{Carrollian vector F}
\end{equation} where $F$ is a nowhere-vanishing function of $(u,x^A)$. The coordinate transformations preserving the fact that $u$ is the coordinate along the fibers are called Carrollian diffeomorphisms. They are generated by vector fields \eqref{ansatz xi} satisfying 
\begin{equation}
    \mathcal{L}_{\xi} n^a \propto n^a
    \label{propto n}
\end{equation} or, equivalently $\mathcal{L}_{\xi} n^A = 0$. This implies $\partial_u \mathcal{Y}^A = 0$, so that 
\begin{equation}
    \xi = f(u, x^A) \partial_u + \mathcal{Y}^A(x^B) \partial_A .
\label{carroll diffeos expression}
\end{equation} We denote by $\mathfrak{Carrdiff}_d$ the algebra of Carrollian diffeomorphisms. The corresponding finite transformations read as
\begin{equation}
    u' = u'(u,{x}^A), \qquad {x'}^{A} = {x'}^{A}({x}^B) .
\end{equation} Choosing a coordinate system that is compatible with the fiber bundle structure is very convenient to make explicit computations in practice, see e.g. \cite{Ciambelli:2019lap,Petkou:2022bmz,Rivera-Betancour:2022lkc}. Indeed, besides the particular form of \eqref{Carrollian vector F}, the degenerate metric is of the form $q_{AB} dx^A dx^B$ (no component along $du$), and the Ehresmann connection reads as $\boldsymbol{k} =- F^{-1} du + G_A dx^A$. Horizontal vectors $V^a$ are in the kernel of the connection, $k_a V^a = 0$, and can be decomposed into a basis $F^{-1}\partial_A + G_A \partial_u$.    

\subsubsection{Newman-Unti algebra} 

We start with the flat Carrollian geometry \eqref{flat Carroll geometry}. The Newman-Unti algebra is defined as the algebra of vector fields $\xi^a$ satisfying 
\begin{equation}
    \mathcal{L}_\xi q_{ab} = 2 \alpha q_{ab}  
\label{NewmanUnti}
\end{equation}  without requiring any further constraint involving the Carrollian vector field. Starting from $0 = \mathcal{L}_{\xi} (q_{ab} n^b) = \mathcal{L}_{\xi} q_{ab} n^b + q_{ab} \mathcal{L}_{\xi} n^b$ and using \eqref{NewmanUnti}, we immediately deduce \eqref{propto n}. Hence, the Newman-Unti transformations form a subset of the aforementioned Carrollian diffeomorphisms. The constraint implied by \eqref{NewmanUnti} on \eqref{ansatz xi} has been studied below \eqref{conformal Killing Carroll N}. We get 
\begin{equation}
    \xi = f(u,x^A) \partial_u + \mathcal Y^A (x^B) \partial_A , \quad  \alpha = \frac{1}{(d-1)} \partial_A \mathcal{Y}^A, \quad \partial_A \mathcal{Y}_B +\partial_B \mathcal{Y}_A = 2 \alpha \delta_{AB}
\label{solution nu algebra}
\end{equation} where $f$ is an arbitrary function of $(u,x^A)$. Hence, the Newman-Unti algebra is as semi-direct sum of the conformal algebra in $d-1$ dimensions, $\mathfrak{so}(d,1)$, with the Lie algebra of smooth
sections of the vertical bundle,  $\Gamma\!\left(\ker(d\pi)\right) = \{f(u,x^A)\partial_u\}$:
\begin{equation}
    \mathfrak{NU}_d
    \simeq
    \mathfrak{so}(d,1)\ltimes\Gamma\!\left(\ker(d\pi)\right),
\end{equation}
where
$[f_1\partial_u,f_2\partial_u]
=(f_1\partial_u f_2-f_2\partial_u f_1)\partial_u$.

We summarize the above symmetry algebras and the associated defining conditions in terms of the geometric structure in Table \ref{tab:my_label}. We have the following inclusions:
\begin{equation}
    \mathfrak{Carr}_d \subset \mathfrak{Carriso}_d  \subset \mathfrak{CCarr}_{d, N} \subset  \mathfrak{NU}_d \subset \mathfrak{Carrdiff}_d .
\end{equation}

\begin{table}[]
    \centering
    \begin{tabular}{c|c|c|c|c}
         Name & Notation & Defining conditions & Generators & Dimension\\ \hline
         Carrollian algebra & $\mathfrak{Carr}_d$ & \eqref{Killing Carroll} with \eqref{connection preservation}  & \eqref{extension Carroll alg} with \eqref{reduc carroll} & $d(d+1)/2$ \\ \hline
         Carrollian isometry algebra & $\mathfrak{Carriso}_d$ & \eqref{Killing Carroll} & \eqref{extension Carroll alg} & $\infty$ \\ \hline
         Conformal Carrollian algebra of level $N$ & $\mathfrak{CCarr}_{d,N}$ & \eqref{conforaml Killing equations} & \eqref{conformal Carrollian vector} & $\infty$ \\ \hline
         Newman-Unti algebra & $\mathfrak{NU}_d$ & \eqref{NewmanUnti} & \eqref{solution nu algebra} & $\infty$ \\ \hline
         Carrollian diffeomorphism algebra & $\mathfrak{Carrdiff}_d$ & \eqref{propto n} & \eqref{carroll diffeos expression} & $\infty$ 
    \end{tabular}
    \caption{Summary of the symmetry algebras.}
    \label{tab:my_label}
\end{table}

\section{Geometry of null hypersurfaces}
\label{sec:Geometry of null hypersurfaces}

In Lorentzian spacetimes, null hypersurfaces play a fundamental role. In particular, within the framework of conformal compactification \cite{Penrose:1962ij,Penrose:1964ge,Newman:1966ub,Penrose:1986uia}, the boundary of asymptotically flat spacetimes, referred to as null infinity, provides a canonical example of a null hypersurface. Other notable examples include black hole and cosmological horizons. In this section, we show that the geometry induced on a null hypersurface \cite{Vogel1965,Jankiewicz,Dautcourt1967} is Carrollian, i.e. it coincides with the geometry introduced in Section \ref{sec:Carrollian geometry}. We then define the notion of asymptotic flatness, and illustrate the above statement by explaining why the geometry of null infinity \cite{1977asst.conf....1G, Ashtekar:2014zsa} is Carrollian. We also briefly comment on the Carrollian geometry of a horizon, as well as the notion of stretched horizon \cite{Chandrasekaran:2018aop,Ciambelli:2019lap,Donnay:2019jiz,Adami:2021nnf,Freidel:2022bai,Freidel:2022vjq,Freidel:2024emv,Riello:2024uvs}.

\subsection{Induced Carrollian geometry}

Consider a $(d+1)$-dimensional spacetime with coordinates $X^\mu$, $\mu = 0,1,\ldots,d$ and a Lorentzian metric $g_{\mu\nu}$ with signature $(-+\ldots+)$. A function $f = f(X^\mu)$ such that $n_\mu = \partial_\mu f \neq 0$ in a neighbourhood of the locus $f=0$ defines a foliation with $f=\text{constant}$ leaves. 

\subsubsection{Timelike and spacelike hypersurfaces}
\label{sec:Timelike and spacelike hypersurfaces}

The hypersurface $\mathcal{N}$ at $f= 0$ is timelike or spacelike if 
\begin{equation}
    n_\mu n^\mu|_{\mathcal{N}} = \alpha 
\end{equation}
 with $\alpha > 0$ or $\alpha <0$, respectively (see Figure \ref{fig:hypersurfaces}). One can define a projector on $\mathcal{N}$ as 
\begin{equation}
\label{tsprojector}
    {\Pi^\mu}_\nu = \delta^\mu_\nu - \frac{1}{\alpha} n^\mu n_\nu .
\end{equation} It satisfies ${\Pi^\mu}_\nu n^\nu = 0 = n_\mu {\Pi^\mu}_\nu$ and ${\Pi^\mu}_\nu {\Pi^\nu}_\rho = {\Pi^\mu}_\rho$. The induced metric 
\begin{equation}
    h_{\mu\nu} = g_{\rho\sigma}{\Pi^\rho}_\mu {\Pi^\sigma}_\nu = g_{\mu\nu} - \frac{1}{\alpha}n_\mu n_\nu  
\end{equation} is Lorentzian/Riemannian if $n^\mu$ is spacelike/timelike, i.e. if $\mathcal{N}$ is timelike/spacelike. The embedding of the hypersurface $\mathcal{N}$ inside of the spacetime $\mathcal{M}$ is denoted by 
\begin{equation}
\label{embedding hyp}
    p: \mathcal{N} \mapsto \mathcal{M}, \quad  x^a \to X^\mu = X^\mu(x^a)
\end{equation} where $x^a$ are now intrinsic coordinates on $\mathcal{N}$.\footnote{In the above discussion, we have used the same notation $\mathcal{N}$ for the hypersurface and its image $p(\mathcal{N
})$.} It is natural to define the maps
\begin{equation}
\label{pullpush}
\begin{split}
    &{\Pi^a}_\mu = {(p_*^{-1})^a}_\rho \circ {\Pi^\rho}_\mu : T(\mathcal{M}) \mapsto T (\mathcal{N}) , \quad V^\mu \to V^a = {\Pi^a}_\mu V^\mu =  \frac{\partial x^a}{\partial{X^\rho}} {\Pi^\rho}_\mu V^\mu , \\
    &{\Pi^\mu}_a = {(p^*)^\rho}_a \circ {\Pi^\mu}_\rho : T^*(\mathcal{M}) \mapsto T^* (\mathcal{N}) , \quad \omega_\mu \to \omega_a = \omega_\mu {\Pi^\mu}_\rho \frac{\partial X^\rho}{\partial{x^a}} .
\end{split}
\end{equation}  where $p^* : T^*(p(\mathcal{N})) \to T^*(\mathcal{N}) $ and $p_*: T(\mathcal{N}) \to T(p(\mathcal{N}))$ denote the pullback and pushforward of the embedding map, respectively.\footnote{The notation $p_*^{-1}: T(p(\mathcal{N})) \to T(\mathcal{N})$ denotes the inverse of $p_*$, understood after restricting $T(\mathcal{M})$ to the tangential subbundle $T(p(\mathcal{N}))$.} Using the above, we can map the Lorentzian/Riemannian geometry $(\mathcal{N}, h_{ab})$ to the timelike/spacelike hypersurface embedded in $(\mathcal{M},g_{\mu\nu})$
\begin{equation} \label{induced timelike spacelike geometry}
    h_{ab} = g_{\mu\nu} {\Pi^\mu}_a {\Pi^\nu}_b .
\end{equation}  In summary, we have
\begin{equation}
\begin{split}
&\text{Lorentzian/Riemannian manifold } (\mathcal{N}, h_{ab}) \\
   \xLongrightarrow{p}\,
   &\text{Timelike/spacelike hypersurface embedded in } (\mathcal{M}, g_{\mu\nu}) .
\end{split}
\end{equation}

\subsubsection{Null hypersurface} 

\begin{figure}[h]
    \centering
\includegraphics[width=1.1\textwidth]{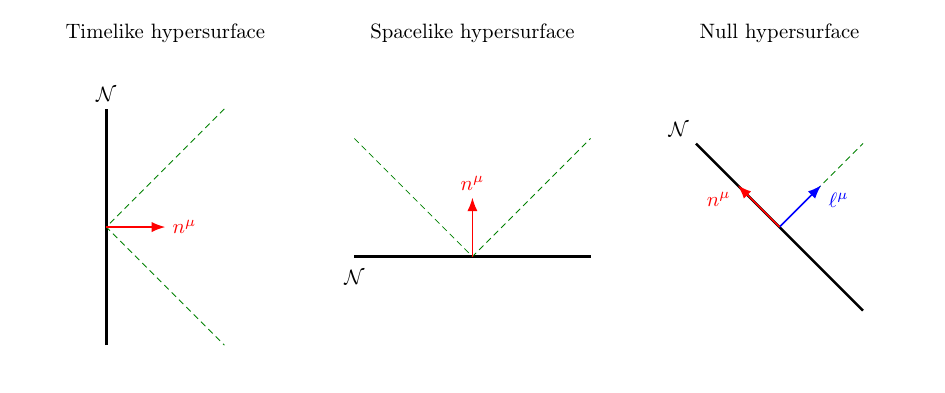}
\caption{The green dashed lines represent null directions. For a timelike/spacelike/null hypersurface $\mathcal{N}$, the normal vector $n^\mu$ is spacelike/timelike/null, respectively. In the null case, the normal is also tangent since $n^\mu n_\mu = 0$, and it is then necessary to introduce a null transverse vector $\ell^\mu$ in order to project onto $\mathcal{N}$.
}
\label{fig:hypersurfaces}
\end{figure}

We now repeat the above discussion for a null hypersurface and show that the induced geometry is Carrollian. The hypersurface $\mathcal{N}$ is null if $n_\mu = \partial_\mu f$ satisfies 
\begin{equation}
    n_\mu n^\mu |_{\mathcal{N}} = 0
\end{equation} i.e. $n^\mu = g^{\mu\nu} n_\nu$ is both tangent and orthogonal to $\mathcal{N}$ (see Figure \ref{fig:hypersurfaces}). The first issue compared to the timelike/spacelike case occurs when trying to define the projector as in \eqref{tsprojector} since $\alpha = 0$ here. To circumvent this, we introduce the null rigging vector \cite{Mars:1993mj} $\ell^\mu$ satisfying the properties 
\begin{equation}
    n_\mu \ell^\mu = -1, \qquad \ell_\mu \ell^\mu = 0  .
\end{equation} The associated rigging projector on the null hypersurface is 
\begin{equation}
\label{projector pi}
    {\Pi^\mu}_\nu = \delta^\mu_\nu + \ell^\mu n_\nu
\end{equation} so that ${\Pi^\mu}_\nu \ell^\nu = 0 = n_\mu {\Pi^\mu}_\nu$, ${\Pi^\mu}_\nu n^\nu = n^\mu$, $\ell_\mu {\Pi^\mu}_\nu = \ell_\nu$, and ${\Pi^\mu}_\nu {\Pi^\nu}_\rho = {\Pi^\mu}_\rho$. From this, we see that the vector $n^\mu$ and the $1$-form $\ell_\mu$ are tangent to $\mathcal{N}$, while the $1$-form $n_\mu$ and the vector $\ell^\mu$ are transverse. Furthermore, the induced metric 
\begin{equation} \label{metric decomposition}
    q_{\mu\nu} = g_{\rho\sigma}{\Pi^\rho}_\mu {\Pi^\sigma}_\nu = g_{\mu\nu} + n_\mu \ell_\nu + \ell_\mu n_\nu 
\end{equation} is degenerate on $\mathcal{N}$ since the tangent vector $n^\mu$ satisfies $q_{\mu\nu}n^\mu = 0$. The embedding of the hypersurface $\mathcal{N}$ \eqref{embedding hyp} and the corresponding maps \eqref{pullpush} are defined in the same way as above, but now using the projector \eqref{projector pi}. It is easy to check that $(q_{ab}, n^a, k_a)$ defined intrinsically on $\mathcal{N}$ as 
\begin{equation} \label{induced carroll}
    q_{ab} = g_{\mu\nu} {\Pi^\mu}_a {\Pi^\nu}_b , \qquad n^a = {\Pi^a}_\mu n^\mu, \qquad k_a = \ell_\mu {\Pi^\mu}_a
\end{equation} defines a Carrollian geometry together with an Ehresmann connection, in the sense of Section \ref{sec:Carrollian geometry and symmetries}. Hence, we have shown that a null hypersurface is naturally endowed with a Carrollian geometry. This is summarized as follows: 
\begin{equation}
\begin{split}
&\text{Carrollian manifold } (\mathcal{N}, q_{ab}, n^a, k_a) \\
   \xLongrightarrow{p} \,
   &\text{Null hypersurface embedded in } (\mathcal{M}, g_{\mu\nu}) .
\end{split}
\end{equation}

\subsubsection{Induced Carrollian connection}
\label{sec:Induced Carrollian connection}

In Section \ref{sec:Carrollian connection}, we have discussed different types of Carrollian connections and their ambiguities. In this section, we show that the Levi-Civita connection in the ambient Lorentzian spacetime will induce on the null hypersurface the unconditionally torsion-free Carrollian connection discussed in Section \ref{sec:Fundamental theorem} \cite{Mars:1993mj,Chandrasekaran:2021hxc}.

We denote by $\nabla_\mu$ the Levi-Civita connection in the ambient Lorentzian spacetime $\mathcal{M}$. It is easy to show that the connection induced on a timelike/spacelike hypersurface is the Levi-Civita connection for the induced Lorentzian/Riemannian metric \eqref{induced timelike spacelike geometry}. Let us now consider a null hypersurface. Using \eqref{induced carroll}, we compute
\begin{equation} \label{induced conn}
    \begin{split}
        D_a q_{bc} = {\Pi^\mu}_a {\Pi^\nu}_b {\Pi^\rho}_c \nabla_\mu q_{\nu \rho} =  -k_b \theta_{ac} - k_c \theta_{ab}, \qquad  D_a n^b \equiv {\Pi^\mu}_a {\Pi^b}_\nu \nabla_\mu n^\nu = \omega_a n^b + {\theta_a}^b
    \end{split}
\end{equation} where $D_a$ is the induced connection, $\theta_{ab}$ is the extrinsic curvature defined in \eqref{expansion tensor} and $\omega_a$ is defined in \eqref{derivative volume form}. Hence the connection $D_a$ induced on null hypersurfaces from the Levi-Civita connection is equivalent to the Carrollian unconditionally torsion-free connection defined intrinsically in Section \ref{sec:Carrollian connection}, see e.g. Equation \eqref{connection null}.

\subsubsection{Relation with Newman-Penrose formalism}

The study of null hypersurfaces has a long history, and the language of Carrollian geometry is simply a modern treatment of them, with a nice interplay with the physics arising in the $c\to 0$ limit. In this section, we briefly comment on the relation between Carrollian geometry and the Newman–Penrose formalism \cite{Newman:1961qr}, which is arguably the most adapted and sophisticated framework for dealing with null hypersurfaces in Lorentzian geometry.

The foundation of the Newman-Penrose formalism consists in trading the Lorentzian metric $g_{\mu\nu}$ for a vielbein $(\ell^\mu, n^\mu , m_I^\mu)$, $I= 1, \ldots, d-1$, such that
\begin{equation}
    g_{\mu\nu} = -\ell_\mu n_\nu - n_\mu \ell_\nu + \eta_{IJ} m^I_\mu m^J_\nu, \qquad n^\mu \ell_\mu = - 1, \qquad \ell^\mu m_\mu^I = 0 = n^\mu m_\mu^I .
\end{equation} This formalism contains from the get-go the null rigging vector $\ell^\mu$, the Carrollian vector $n^\mu$, and the degenerate metric $q_{\mu\nu} = \eta_{IJ} m^I_\mu m^J_\nu$. In particular, it is completely adapted to the use of the Carrollian frame induced on the null hypersurface, as described in Section \ref{sec:Carrollian frame}. This formalism was developed mostly in $4d$ spacetimes, where the elements of the dyad $m^\mu_I = (m^\mu , \bar m^\mu)$ are taken to be complex and null, 
\begin{equation}
    m^\mu m_\mu = 0 = \bar{m}^\mu \bar{m}_\mu, \qquad m^\mu \bar{m}_\mu = 1
\end{equation} and the vielbein $(\ell^\mu, n^\mu,  m^\mu , \bar m^\mu)$ is called a null tetrad. These conditions can be generalized in higher even dimensions. For the use of this formalism in $3$ dimensions, see \cite{Barnich:2016lyg}.

In addition, the Newman-Penrose spin coefficients encode all the pieces of the Carrollian geometry described above, such as the shear, the expansion, the acceleration, and the vorticity discussed in Section \ref{sec:Ehresmann curvature}. Finally, the Geroch-Held-Penrose differential operators \cite{Geroch:1973am, Penrose:1985bww}, defined in terms of those spin-coefficients, precisely contain the Carrollian connection induced on $\mathcal{N}$ \eqref{induced conn}:
\begin{equation}
    \pthorn' = n^a D_a, \qquad \eth = m^a D_a, \qquad {\eth}' =  \bar m^a D_a
\end{equation} ($\pthorn'$, $\eth$ and $\eth'$ are the components of the connection tangent to $\mathcal{N}$).

\subsection{Null Brown-York stress tensor}

In this section, we show that Einstein's vacuum equations governing the gravitational field, when restricted to a hypersurface (via the Gauss–Codazzi equations), reduce to stress tensor conservation equations, which are naturally interpreted as hydrodynamic conservation equations. For timelike or spacelike hypersurfaces, this is described by a relativistic fluid, while for a null hypersurface, it is described by a Carrollian fluid.

\subsubsection{Timelike and spacelike hypersurfaces}

On a timelike/spacelike hypersurface with unit normal $n^\mu$, the extrinsic curvature is given by 
\begin{equation}
    {K_{ab}} = \frac{1}{2} \mathcal{L}_n g_{\mu\nu} {\Pi^\mu}_a  {\Pi^\nu}_b =  \nabla_{\mu} n_{\nu} {\Pi^\mu}_a  {\Pi^\nu}_b  .
\end{equation}
The Brown-York stress tensor \cite{Brown:1992br} can then be defined as 
\begin{equation}
    T_{ab} = \frac{1}{8\pi G} (K_{ab} - h_{ab} K) 
\end{equation} where $K = h^{ab} K_{ab}$ is the mean curvature. This object can be extracted from the variation of the gravitational action with respect to the induced metric. By construction, this stress tensor satisfies
\begin{equation} \label{relativistic fluid eq}
    \nabla^a T_{ab} =0, \qquad T_{ab} = T_{ba} , \qquad {T^a}_a = h^{ab} T_{ab} = \frac{1-d}{8\pi G} K .
\end{equation} As we will review in Section \ref{sec:Reminder on the relativistic stress tensor}, these are the typical properties of a relativistic stress tensor. The first equation follows from Einstein's vacuum equations induced on the hypersurface. This equation is naturally interpreted as a conservation equation for a relativistic fluid. In the context of AdS/CFT, where the hypersurface is taken to be the timelike boundary of AdS, this object plays a fundamental role: it corresponds to the vacuum expectation value of the stress tensor in the dual theory \cite{Balasubramanian:1999re,deHaro:2000vlm}. In the hydrodynamic picture, this gives rise to the fluid/gravity correspondence \cite{Bhattacharyya:2007vjd}, which allows one to reconstruct the bulk AdS metric from a relativistic fluid at the boundary.

\subsubsection{Null hypersurfaces}

The analogue of the Brown-York stress tensor on null hypersurfaces can be constructed as follows. First, introduce the null Weingarten tensor
\begin{equation}
    {W^a}_b = D_b n^a = n^a \omega_b  + {\theta^a}_b
\end{equation} where the second equality follows from \eqref{induced conn}. This is the analogue of the extrinsic curvature ${K^a}_b$, with one index raised. The null Brown-York stress tensor is then defined as \cite{Chandrasekaran:2021hxc} 
\begin{equation}
    {T^a}_b = \frac{1}{8\pi G} ({W^a}_b - \delta^a_b W)
\end{equation} where $W = D_a n^a = \theta - \kappa$. Here $\theta$ is the expansion defined in \eqref{expansion Carroll} and $\kappa$ is the inaffinity. One can check that this stress tensor satisfies
\begin{equation}
    D_a {T^a}_b = 0, \qquad T_{ab} = T_{ba}, \qquad m_a^I {T^a}_b n^b = 0, \qquad    {T^a}_a = \frac{1-d}{8\pi G} W
\end{equation} which are typical properties of a Carrollian stress tensor, see Section \ref{sec:Carrollian case}. The first equation follows from Einstein’s vacuum equations induced on the null hypersurface. Projecting this conservation equation along the Carrollian vector $n^b$ generating the null hypersurface, we find
\begin{equation} \label{raychau}
    n^b D_a {T^a}_b = 0 \quad \Longleftrightarrow \quad (\mathcal{L}_n + \theta ) \theta - \mu \theta + {\sigma^a}_b  {\sigma^b}_a = 0 .
\end{equation} This equation is called the Raychaudhuri equation \cite{PhysRev.98.1123} and controls how the expansion $\theta$ evolves along the null congruence generating the hypersurface. The shear $\sigma_{ab}$ was defined in \eqref{Carrollian shear}. Now projecting on the horizontal direction yields
\begin{equation}
    {q^b}_a D_c {T^c}_b = 0 \quad \Longleftrightarrow \quad (\mathcal{L}_n + \theta ) \pi_a - {q^b}_a D_b \mu + (\theta - \mu) \varphi_a + {q^c}_b {q^d}_a (D_c + \varphi_c) {\sigma^b}_d = 0
\label{eq:Damour}
\end{equation} where $\pi_a$ is the H\'aji\v{c}ek connection (see below \eqref{derivative volume form}), $\varphi_a$ is the acceleration defined in \eqref{decomposition ehresmann}, and we introduced the combination 
\begin{equation}
   \mu=-\kappa+\frac{d-2}{d-1}\theta .
\end{equation}  This second equation \eqref{eq:Damour} is called the Damour equation \cite{Damour:1978cg,Damour:1979wya} and describes the evolution of a Carrollian momentum density along the null hypersurface. These are naturally interpreted as conservation equations for a Carrollian fluid \cite{deBoer:2017ing,Ciambelli:2018xat,Ciambelli:2018wre,Donnay:2019jiz,Adami:2021nnf,Adami:2021kvx,Petkou:2022bmz,Freidel:2022bai,Adami:2023fbm,Adami:2023wbe,Adami:2024mtu,Adami:2024rkr}.

\subsection{Membrane paradigm, black hole and cosmological horizons}

One of the most important examples of null hypersurfaces is the notion of a horizon. A horizon is defined as a null hypersurface with a causal boundary property. For example, the event horizon of a black hole is defined as the boundary of the causal past of future null infinity (see Section \ref{sec:Asymptotically flat spacetimes}). It corresponds to the null hypersurface separating regions of spacetime that can communicate with an observer at infinity from those that cannot. Another famous example is the cosmological horizon, which is a null hypersurface that separates events that can influence (or be observed by) a given observer from those that never can, due to the global causal structure of spacetime. By contrast with general null hypersurfaces, which generically admit caustics (i.e., points where the null generators intersect), future event horizons have no future endpoints under the usual global assumptions, although their generators may have past endpoints forming caustic or crease sets, as in the case of black hole formation. Furthermore, starting from the Raychaudhuri equation \eqref{raychau}, one can show, under the null energy condition and the usual assumptions of the area theorem, that the expansion of their future-directed generators is non-negative, $\theta\geq 0$, in agreement with the area law \cite{Hawking:1971vc,Chrusciel:2000cu}.

\begin{figure}[h]
    \centering
\includegraphics[width=0.4\textwidth]{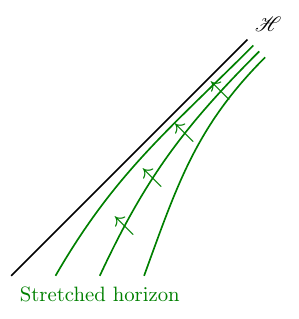}
\caption{This figure illustrates the membrane paradigm. The null hypersurface $\mathscr{H}$ represents the horizon, while the green hypersurfaces are timelike slices approaching it. As these slices move closer to the horizon, their induced geometry and dynamics undergo a Carrollian limit: the Lorentzian structure degenerates into a Carrollian one, and the induced Einstein equations reduce accordingly to the Carrollian fluid conservation equations.}
\label{fig:stretched}
\end{figure}

An interesting framework to study horizons is provided by the membrane paradigm \cite{Damour:1978cg,1986bhmp.book.....T,Price:1986yy}, originally applied to black hole horizons. In this setup, the horizon is effectively replaced by a timelike hypersurface, called the “stretched horizon,” located just outside the true event horizon (see Figure \ref{fig:stretched}). As discussed in the previous section, Einstein's equations induced on this timelike hypersurface take the form of hydrodynamic conservation laws, and the stretched horizon then serves as a fictitious membrane whose evolution equations act as boundary conditions for the exterior spacetime, making the framework particularly useful for astrophysical applications. From an intrinsic perspective, taking the stretched horizon closer and closer to the actual horizon corresponds to taking a Carrollian limit, and the relativistic fluid equations \eqref{relativistic fluid eq} reduce consistently to the Carrollian fluid conservation equations encoding the Raychaudhuri \eqref{raychau} and Damour \eqref{eq:Damour} equations in the limit. The geometry of the stretched horizon has been recently discussed in Carrollian terms in \cite{Donnay:2019jiz,Freidel:2024emv}, and has been applied to null infinity in \cite{Riello:2024uvs}. Stretched horizons can also be used to study cosmological horizons; see, e.g., \cite{Anninos:2024wpy,Anninos:2025zgr} for recent discussions.

\subsection{Asymptotically flat spacetimes and BMS symmetries}
\label{sec:Asymptotically flat spacetimes}

In gravity, four-dimensional asymptotically flat spacetimes are used as models to describe isolated sources of the gravitational field, which can emit radiation, including gravitational waves. These systems are such that, very far away from the sources, the curvature becomes a small perturbation around Minkowski spacetime. Naively, in this asymptotic region one might expect that general relativity reduces to special relativity, and that the Poincaré group is recovered as the asymptotic symmetry group. However, as we shall explain, when we move far from gravitational sources, an infinite-dimensional enhancement of the Poincaré group, known as the Bondi–Metzner–Sachs (BMS) group \cite{Bondi:1962px,Sachs:1962zza,Sachs:1962wk}, emerges, and this will make the whole story very interesting.

The literature on $4d$ gravity in asymptotically flat spacetime is vast. Besides the first references \cite{Bondi:1962px,Sachs:1962zza,Sachs:1962wk} discussing the Bondi-Sachs expansion and the BMS symmetries, the conformal compactification formalizing the notion of null infinity has been introduced in \cite{Penrose:1962ij, Penrose:1964ge, Penrose:1986uia}, see also \cite{Newman:1966ub,1977asst.conf....1G,Ashtekar:2014zsa}. The complete characterization of the solution space and characteristic initial value problem has been discussed in \cite{Newman:1962cia,Tamburino:1966zz}, and the construction of the radiative phase space and surface charges has been performed in \cite{Ashtekar:1981bq,Wald:1999wa,Barnich:2011mi}. Weaker definitions of asymptotic flatness, obtained by relaxing some of the falloff conditions and leading to enhanced asymptotic symmetry groups, have become an active area of research, see e.g. \cite{Barnich:2010eb,Barnich:2011ct,Flanagan:2015pxa,Campiglia:2014yka,Campiglia:2015yka,Compere:2018ylh,Campiglia:2020qvc,Ruzziconi:2020cjt,Freidel:2021fxf,Geiller:2022vto,Fuentealba:2022xsz,Campoleoni:2023fug,Geiller:2024amx,Fiorucci:2024ndw,Geiller:2024ryw,Geiller:2025dqe}, see also \cite{Compere:2018aar,Ruzziconi:2019pzd} for reviews.

\subsubsection{Minkowski spacetime in Bondi coordinates}
\label{sec:Minkowski spacetime in Bondi coordinates}

Let us first discuss Minkowski spacetime and use Bondi coordinates $(u_\circ,r_\circ,x^A_\circ)$, $x^A_\circ = (z_\circ, \bar z_\circ)$, which are particularly suited to approach null infinity. We follow Appendix A of \cite{Donnay:2022wvx} for notations and conventions. Here $u_\circ \in \mathbb{R}$ is a retarded time coordinate ($u_\circ = \text{constant}$ labels some null hypersurface) and $r_\circ >0$ is a radial coordinate. The future conformal boundary of the spacetime, $\mathscr I^+ = \mathbb{R} \times S^2 \equiv \{r_\circ\to+\infty\}$, is a null hypersurface called future null infinity, and each of its cuts at constant $u_\circ$ is a sphere, called the celestial sphere. The latter is parametrized by stereographic coordinates $(z_\circ,\bar z_\circ)$, where $z_\circ = e^{i\varphi}\tan\frac{\theta}{2}$. In Bondi coordinates, the Minkowski line element reads as 
\begin{equation}
    d s^2 = -d u_\circ^2 - 2 d u_\circ d r_\circ + \frac{4\,r_\circ^2}{(1+z_\circ \bar z_\circ)^2}d z_\circ d\bar z_\circ. \label{Mink usu}
\end{equation}
 The change of coordinates from the Cartesian chart $X^\mu$ is given by
\begin{equation}
    X^\mu = u_\circ \, \delta_0^\mu + r_\circ\,\frac{\sqrt{2}}{1+z_\circ\bar z_\circ}\,q^\mu(z_\circ,\bar z_\circ), \qquad q^\mu(z,\bar z) \equiv \frac{1}{\sqrt{2}} \Big(1+z\bar z,z+\bar z,-i(z-\bar z),1-z\bar z\Big)
\end{equation}
where $q^\mu (z, \bar z)$
is the standard parametrization of a null direction pointing toward $(z,\bar z)$ on the celestial sphere. In a similar way, one defines advanced Bondi coordinates $(v_\circ, r_\circ, z_\circ, \bar z_\circ)$, suited to approach past null infinity, $\mathscr{I}^-$. 

\begin{figure}[h!]
\centering
\includegraphics[scale=1.2]{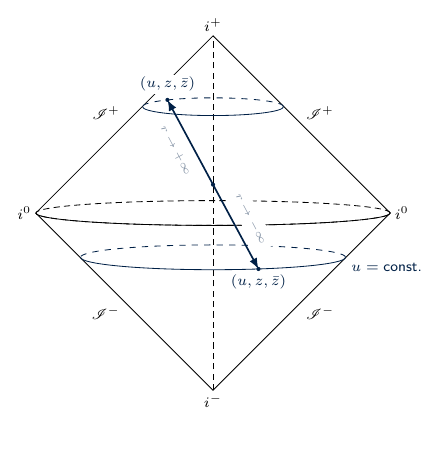}
\caption{There is a natural geometric identification between $\mathscr I^-$ and $\mathscr I^+$. Starting from a point at $\mathscr I^-$ and following a null geodesic along $\partial_r$, we end up at a point at $\mathscr I^+$ with the same $(u,z,\bar z)$ coordinates \cite{Alday:2024yyj}.}
    \label{FigureBondi}
\end{figure}

In practice, in the holographic context, it is convenient to use a different version of Bondi coordinates, sometimes referred to as ``rectangular Bondi coordinates,'' denoted by $(u,r,z, \bar z)$, with $u \in \mathbb{R}$, $r\in \mathbb{R}$, $z\in \mathbb{C}$. These coordinates trade the unit round-sphere metric on the boundary for the flat complex plane metric. The diffeomorphism that implements the boundary Weyl rescaling has been worked out e.g. in \cite{Compere:2016hzt} and reads as
\begin{equation}
    r = \frac{\sqrt{2}}{1+z_\circ\bar z_\circ}r_\circ +\frac{u_\circ}{\sqrt{2}}\,,\quad u = \frac{1+z_\circ \bar z_\circ}{\sqrt{2}}u_\circ - \frac{z_\circ\bar z_\circ u_\circ^2}{2r}\,,\quad z = z_\circ - \frac{z_\circ u_\circ}{\sqrt{2}r} \label{flatten future}
\end{equation}
for retarded coordinates. The relation with Cartesian coordinates is now given by
\begin{equation}
    X^\mu = u\,\partial_z\partial_{\bar z}q^\mu(z,\bar z) + r\, q^\mu(z,\bar z) \label{Retarded flat BMS coordinates (appendix)}
\end{equation}
and the Minkowski line element reads as
\begin{equation}
    d s^2 = -2d u d r+2r^2 d z d \bar z  . \label{Mink best}
\end{equation} Notice that, compared to \eqref{Mink usu}, the transverse part of the metric is flat, and there is no $-du^2$ term. 
These coordinates are such that $\mathscr{I}^+$ and $\mathscr{I}^-$ are respectively obtained by taking the limits $r \to +\infty$ and $r\to -\infty$. Cuts of $\mathscr{I}^+$ and $\mathscr{I}^-$ are now complex planes endowed with a flat metric. Lines generated by $\partial_r$ and obtained by keeping $(u,z,\bar z)$ fixed form a null geodesic congruence extending from past to future null infinity, see Figure \ref{FigureBondi}. The geometric matching between past and future null infinities is therefore immediate, see Figure \ref{FigureBondi}. We could have used instead the advanced version of these coordinates $(v,r, z,\bar{z})$, but it is not strictly needed in Minkowski spacetime since the coordinate system $(u,r,z,\bar z)$ is enough to discuss both $\mathscr{I}^+$ and $\mathscr{I}^-$.

\subsubsection{Asymptotically flat spacetimes in Bondi gauge}

\begin{figure}[h]
    \centering
\includegraphics[width=0.4\textwidth]{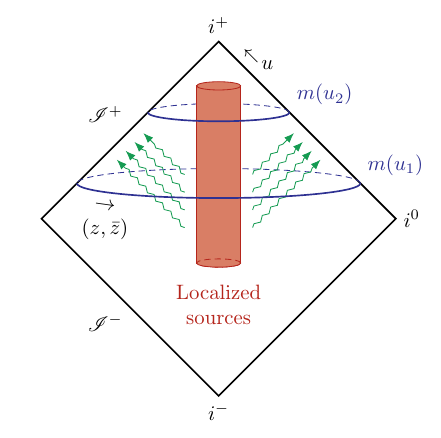}
\caption{This figure illustrates the Bondi mass loss formula: the mass of the system decreases in time due to the emission of radiation.}
\label{fig:fluxbalance}
\end{figure}

We now wish to study perturbations of the gravitational field with respect to the Minkowski metric \eqref{Mink best}, generated by localized sources of the gravitational field. The discussion will be valid in a neighbourhood of $\mathscr{I}^+$ where the radial expansion of the metric converges and no caustics are present (the discussion would be similar around $\mathscr{I}^-$). We work in Bondi gauge, where bulk metric $g_{\mu\nu}$ is required to satisfy the four following conditions:
\begin{equation}
    g_{rA} = 0, \qquad g_{rr} = 0, \qquad \det (g_{AB}) = r^4 \det q_{AB} =- r^4 . \label{Bondi gauge}
\end{equation} Here $q_{AB} dx^A dx^B = 2 dz d\bar z$ is the metric on the complex plane and $\det q_{AB} =-1$. These four gauge-fixing conditions can always be reached using the four degrees of freedom in the diffeomorphisms. In Bondi gauge, asymptotically flat spacetimes are solutions of Einstein's equations with vanishing cosmological constant, $R_{\mu\nu} - \frac{1}{2} g_{\mu\nu} R = 8 \pi G T_{\mu \nu}$, that fall off sufficiently fast in $1/r$ to recover the Minkowski metric \eqref{Mink best} in the asymptotic region when $r\to \infty$. We have 
\begin{equation}
    \begin{split}
        ds^2 = \, &\left(\frac{2M}{r}+\mathcal O(r^{-2})\right) d u^2 - 2 \left( 1+\mathcal O(r^{-2})\right) d u d r\\
        &+ \left(r^2 {q}_{AB} + r\, C_{AB} + \mathcal{O}(r^{0})\right) d x^A d x^B \\
    &+ \left(\frac{1}{2}\partial^B C_{AB} + \frac{2}{3r}(N_A + \frac{1}{4}C_{AB} \partial_C C^{BC}) + \mathcal O(r^{-2}) \right)d u d x^A .
    \end{split} \label{Bondi gauge metric}
\end{equation}
Indices $A$, $B$ are lowered and raised by $q_{AB}$ and its inverse. The asymptotic shear $C_{AB}$ is a symmetric trace-free tensor (${q}^{AB} C_{AB} = 0$). The Bondi mass aspect $M(u,z,\bar z)$ and the angular momentum aspect $N_A (u,z,\bar z)$ satisfy the time evolution/constraint equations
\begin{equation}
\begin{split}
\partial_u M &= - \frac{1}{8} N_{AB} N^{AB} + \frac{1}{4} \partial_A \partial_B N^{AB} - 4\pi G\, T_{uu}^{m(2)} , \\ 
\partial_u N_A &= \partial_A M + \frac{1}{16} \partial_A (N_{BC} C^{BC}) - \frac{1}{4} N^{BC} \partial_A C_{BC} - 8\pi G\, T_{uA}^{m(2)}\\
&\quad  -\frac{1}{4} \partial_B (C^{BC} N_{AC} - N^{BC} C_{AC}) - \frac{1}{4} \partial_B \partial^B \partial^C C_{AC}+ \frac{1}{4} \partial_B \partial_A \partial_C C^{BC} ,
\end{split}\label{EOM1} 
\end{equation} 
with $N_{AB} = \partial_u C_{AB}$ the Bondi news tensor encoding the outgoing radiation and $T_{\mu\nu}^m$ the null matter stress tensor whose expansion near null infinity is taken to be
\begin{equation}
    \begin{split}
        T_{uu}^m(u,r,z,\bar z) &= T_{uu}^{m(2)}(u,z,\bar z)\frac{1}{r^2}+\mathcal O(r^{-3}) , \\ 
        T_{uA}^m(u,r,z,\bar z) &= T_{uA}^{m(2)}(u,z,\bar z)\frac{1}{r^2}+\mathcal O(r^{-3})  .
    \end{split}
\end{equation} To gain some intuition on these equations, let us define the total mass of the system at a cut $u=\text{constant}$ of $\mathscr{I}^+$ as 
\begin{equation}
    m (u) = \int d^2 z M(u, z, \bar z) .
\end{equation} The difference of mass between two cuts $u_1$ and $u_2$ of $\mathscr{I}^+$, $u_1 < u_2$, is non-positive since
\begin{equation}
    m(u_2) - m(u_1)  = -  \int_{u_1}^{u_2} du \int d^2 z \left( \frac{1}{8}N_{AB} N^{AB} + 4 \pi G T_{uu}^{m(2)} \right) \le 0
\label{bondi mass loss formula}
\end{equation} where we used the first of \eqref{EOM1}, dropped the total derivative terms under the integral, and assumed that the matter stress tensor obeys a nonnegative averaged null energy condition on the interval $[u_1, u_2]$, $\int_{u_1}^{u_2} du T_{uu}^{m(2)} \ge 0$. Equation \eqref{bondi mass loss formula} is called the Bondi mass loss formula \cite{Trautman:1958zdi, Bondi:1962px}, and states that the mass of the system decreases in time due to the emission of gravitational waves (encoded in $N_{AB} N^{AB}$) or other forms of radiation (encoded in $T_{uu}^{m(2)}$). This situation is depicted in Figure \ref{fig:fluxbalance} and illustrates the use of the asymptotically flat spacetime framework.

\subsubsection{BMS algebra}
\label{sec:BMS algebra}

The Killing vectors $\zeta^\mu$ associated with the Minkowski metric \eqref{Mink best} are obtained by solving the Killing equation 
\begin{equation}
    \mathcal{L}_\zeta g_{\mu\nu} = 0 .
    \label{Killing equation bulk}
\end{equation} The general solution is given by 
\begin{equation}
\begin{split}
&\zeta^u =\mathcal{T} + \frac{u}{2} (\partial \mathcal{Y} + \bar \partial \bar{\mathcal{Y}}) , \qquad
    \zeta^r = - \frac{r}{2} (\partial \mathcal{Y} + \bar \partial \bar{\mathcal{Y}}) + \partial \bar \partial \mathcal{T} , \\
&\zeta^z = \mathcal{Y} - \frac{1}{r} \left(\bar{\partial} \mathcal{T} + \frac{u}{2} \bar \partial^2 \bar{\mathcal{Y}} \right) , \qquad \zeta^{\bar{z}} = \bar{\mathcal{Y}}   - \frac{1}{r} \left({\partial} \mathcal{T} + \frac{u}{2} \partial^2 {\mathcal{Y}} \right)  ,
    \label{Killing vectors eq}
\end{split}
\end{equation} where the parameters $\mathcal{T}= \mathcal{T}(z, \bar{z})$, $\mathcal{Y} = \mathcal{Y}(z)$ and $\bar{\mathcal{Y}} = \bar{\mathcal{Y}}(\bar z)$ satisfy 
\begin{equation} \label{conditions Killing}
    \bar \partial \mathcal{Y} = 0 = \partial \bar{\mathcal{Y}}  , \qquad \partial^3 \mathcal{Y} =0 =  \bar \partial^3 \bar{\mathcal{Y}}  , \qquad \partial^2 \mathcal{T} = 0 = \bar \partial^2 \mathcal{T} .
\end{equation} The Killing equation admits $10$ independent solutions, and the vector fields \eqref{Killing vectors eq} form the Poincaré algebra under the standard Lie bracket. The $6$ parameters $\mathcal{Y}= 1, z, z^2$ and $\bar{\mathcal{Y}}= 1, \bar z, \bar z^2$ generate Lorentz transformations, and the $4$ parameters $\mathcal{T}= 1, z, \bar z, z \bar z$ generate translations.

Asymptotically flat spacetimes \eqref{Bondi gauge metric} do not generically admit solutions to the Killing equation \eqref{Killing equation bulk}. However, far from the gravitational sources, one might still hope to define a notion of symmetries for the metric, called asymptotic symmetries. Since the metric asymptotes to the Minkowski metric, one would expect to recover the above Poincaré symmetries. However, as we now explain, the result will be much richer than this.

The asymptotic symmetries of asymptotically flat spacetimes are the spacetime diffeomorphisms preserving the solution space \eqref{Bondi gauge metric}. They are generated by vector fields $\zeta = \zeta^u \partial_u + \zeta^z \partial_z + \zeta^{\bar{z}} \partial_{\bar z} + \zeta^r \partial_r$ which preserve the Bondi gauge conditions \eqref{Bondi gauge}, i.e. satisfying
\begin{equation}
    \mathcal{L}_\zeta g_{rr} = 0, \qquad \mathcal{L}_\zeta g_{rA} = 0, \qquad g^{AB}\mathcal{L}_\zeta g_{AB} = 0
\end{equation} together with the falloff conditions, implying 
\begin{equation}
    \mathcal{L}_\zeta g_{uu} = \mathcal{O}(r^{-1}), \qquad \mathcal{L}_\zeta g_{ur} = \mathcal{O}(r^{-2}), \qquad \mathcal{L}_\zeta g_{uA} = \mathcal{O} (r^{0} ), \qquad \mathcal{L}_\zeta g_{AB} = \mathcal{O}(r) .
\end{equation} Solving these asymptotic Killing equations leads to the following general solution:
\begin{equation}
\begin{split}
    &\zeta^u =\mathcal{T} + \frac{u}{2} (\partial \mathcal{Y} +  \bar \partial \bar{\mathcal{Y}}) , \qquad  \zeta^r = - \frac{r}{2} (\partial \mathcal{Y} + \bar{\partial} \bar{\mathcal{Y}}) + \partial \bar \partial \mathcal{T}+ \mathcal{O}(r^{-1})  , \\
    &\zeta^z = \mathcal{Y} - \frac{1}{r} \left(\bar{\partial} \mathcal{T} + \frac{u}{2} \bar \partial^2 \bar{\mathcal{Y}} \right) + \mathcal{O}(r^{-2}) , \qquad \zeta^{\bar{z}} = \bar{\mathcal{Y}} - \frac{1}{r} \left({\partial} \mathcal{T} + \frac{u}{2} \partial^2 {\mathcal{Y}} \right)  + \mathcal{O}(r^{-2}) , 
    \end{split}
    \label{AKV Bondi}
\end{equation} where $\mathcal{T}= \mathcal{T}(z, \bar{z})$ is an arbitrary function on the celestial plane called supertranslation parameter, and $\mathcal{Y} = \mathcal{Y}(z)$, $\bar{\mathcal{Y}} = \bar{\mathcal{Y}}(\bar{z})$ are the superrotation parameters satisfying the conformal Killing equation in $2d$,
\begin{equation}
    \bar \partial \mathcal{Y} = 0 = \partial \bar{\mathcal{Y}}  .\label{CKV Y}
\end{equation} It is instructive to compare the solution for the asymptotic Killing vectors \eqref{AKV Bondi}-\eqref{CKV Y} with the Poincaré generators obtained in \eqref{Killing vectors eq}-\eqref{conditions Killing}. At leading order in $r$, these solutions look very similar, but with crucial differences: the supertranslations $\mathcal{T}(z, \bar z)$ now contain an infinite number of independent generators (which are sometimes interpreted as ``angle-dependent translations''), and the superrotations $\mathcal{Y}(z)$ and  $\bar{\mathcal{Y}}(\bar z)$ generate all the local conformal transformations in $2d$, and therefore contain an infinite number of independent parameters. The global conformal transformations generate the $6$ Lorentz transformations in $4d$ discussed above. This infinite-dimensional enhancement of the Poincaré algebra is called the Extended BMS algebra \cite{Barnich:2009se, Barnich:2010eb, Barnich:2011ct}, $\mathfrak{ebms}_4$, which further enhances the original BMS algebra \cite{Bondi:1962px,Sachs:1962zza,Sachs:1962wk} with meromorphic superrotations.

Another crucial difference compared to Minkowski space is that the asymptotic Killing vectors \eqref{AKV Bondi} now contain subleading orders in $r$, which involve explicit field dependence in the metric components. As a consequence, a technical aspect is that one can no longer use a standard bracket to compute the structure constants. Instead, one has to use the modified Lie bracket $[\zeta_1, \zeta_2]_\star \equiv [\zeta_1, \zeta_2] - \delta_{\zeta_1}\zeta_2 + \delta_{\zeta_2} \zeta_1$, where the last two terms take into account the field-dependence of \eqref{AKV Bondi} at subleading orders in $r$ \cite{Barnich:2010eb}. These asymptotic Killing vectors satisfy the commutation relations
\begin{equation}
    \big[\zeta (\mathcal{T}_1, \mathcal{Y}_1, \bar{\mathcal{Y}}_1), \zeta (\mathcal{T}_2, \mathcal{Y}_2, \bar{\mathcal{Y}}_2) \big]_\star = \zeta (\mathcal{T}_{12}, \mathcal{Y}_{12}, \bar{\mathcal{Y}}_{12}) ,
    \label{commutation relations 1}
\end{equation} with
\begin{equation}
\begin{split}
    \mathcal{T}_{12} &= \mathcal{Y}_1 \partial \mathcal{T}_2 - \frac{1}{2} \partial \mathcal{Y}_1 \mathcal{T}_2 + \text{c.c.} - (1 \leftrightarrow 2)   \, , \\
    \mathcal{Y}_{12} &= \mathcal{Y}_1 \partial \mathcal{Y}_2 - (1 \leftrightarrow 2)\,, \quad \bar{\mathcal{Y}}_{12} = \bar{\mathcal{Y}}_1 \bar \partial \bar{\mathcal{Y}}_2 - (1 \leftrightarrow 2) .
\end{split}
    \label{commutation relations 2}
\end{equation} Hence we see that the Extended BMS algebra is a semi-direct sum between Witt$\oplus$Witt superrotations and supertranslations. To read the structure constants of $\mathfrak{ebms}_4$, it is useful to use the following expansions on the complex punctured plane:
\begin{equation}
\label{expansion ebms}
    \mathcal{Y}(z) = \sum_{m \in \mathbb{Z}} L_m z^{1-m}, \qquad \bar{\mathcal{Y}}(\bar{z}) = \sum_{m \in \mathbb{Z}}  \bar{L}_m \bar{z}^{1-m}, \qquad \mathcal{T}(z, \bar{z}) = \sum_{k,l\in \mathbb{Z} + \frac{1}{2}} P_{k,l}{z}^{\frac{1}{2}-k} \bar{z}^{\frac{1}{2}-l} 
\end{equation} where $m \in \mathbb{Z}$ and $k,l \in \mathbb{Z} + \frac{1}{2}$ \cite{Barnich:2017ubf, Barnich:2021dta}. From \eqref{commutation relations 2} we obtain
\begin{equation} \label{BMS4 structure constants}
    \begin{split}
        &[L_m, L_n] = (m-n) {L}_{m+n}  , \qquad
        [\bar{L}_m, \bar{L}_n] = (m-n) {\bar{L}}_{m+n} , \\
        &[L_m, P_{k,l}] = \Big(\frac{1}{2} m - k  \Big) P_{m+k, l}, \qquad [\bar{L}_m, P_{k,l}] = \Big(\frac{1}{2} m - l  \Big) P_{k, m+l}, \\
        &[P_{k,l}, P_{r,s}] = 0
    \end{split}
\end{equation} (we dropped the $\star$ notation for the bracket).  

One can compute the action of the BMS symmetries on the metric \eqref{Bondi gauge metric}, and these can be found in \cite{Barnich:2010eb}. For future reference, we include the transformation of the asymptotic shear: 
\begin{equation} \label{shear transfo}
    \delta_\zeta C_{zz} = \left[ \left(\mathcal{T} + \frac{u}{2} (\partial \mathcal{Y} + \bar \partial \bar{\mathcal{Y}} ) \right) \partial_u + \mathcal{Y} \partial + \bar{\mathcal{Y}}\bar{\partial} + \frac{3}{2} \partial \mathcal{Y} - \frac{1}{2}\bar{\partial} \bar{\mathcal{Y}}  \right] C_{zz} - 2 \partial^2 \mathcal{T} - u \partial^3 \mathcal{Y} .
\end{equation}

\subsection{BMS and Conformal Carrollian algebras}
\label{sec:Relation between Conformal Carroll and BMS}

We now show that the geometry and the symmetries induced at $\mathscr{I}$ from the bulk analysis reproduce exactly the Carrollian geometry and symmetries discussed in the previous sections.

\subsubsection{Conformal Carrollian geometry of null infinity}
\label{sec:Conformal Carrollian geometry of null infinity}

The proper way to discuss the geometry induced at null infinity is the conformal compactification procedure \cite{Penrose:1962ij, Penrose:1964ge, Penrose:1986uia}. Starting from the metric \eqref{Bondi gauge metric}, one can perform a Weyl rescaling
\begin{equation}
    ds^2 \to \Omega^2 (x) ds^2
\end{equation} where the conformal factor $\Omega > 0$ is a spacetime function such that $\Omega \sim r^{-1}$. This guarantees the cancellation of the $\sim r^{2}$ divergence appearing in \eqref{Bondi gauge metric} when $r\to \infty$. One can then compactify the spacetime by adding a conformal boundary, $\mathscr{I}^+$ and $\mathscr{I}^-$. At null infinity, we have $\Omega = 0$ and $\nabla_\mu \Omega \neq 0$. For the canonical choice $\Omega = r^{-1}$, the metric induced at $\mathscr{I}^+$ is 
\begin{equation}
    q_{ab} dx^a dx^b = \lim_{\bar{r} \to \infty} \Omega^2 ds^2|_{r = \bar{r}}  = 0 du^2 + 2 dz d\bar{z} 
    \label{degenerate metric at scri}
\end{equation} where $r = \bar{r}$ defines a (timelike) foliation approaching $\mathscr{I}^+$. Hence the metric induced at $\mathscr{I}^+$ is degenerate, as expected for a null hypersurface. Furthermore, the vector
\begin{equation}
    n^a \partial_a = \lim_{\bar{r} \to \infty} \Omega^{-2} g^{\mu\nu} \nabla_\nu \Omega  \, \partial_\mu |_{r=\bar{r}} = \partial_u
\label{vector in the kernel}
\end{equation} is in the kernel of \eqref{degenerate metric at scri}, i.e. $q_{ab} n^b = 0$. Therefore, the pair $(q_{ab}, n^c)$ defines precisely a Carrollian geometry at $\mathscr{I}^+$, in the sense of \eqref{Carrollian geometry def}. The associated Ehresmann connection can be taken as
\begin{equation}
    k_a dx^a = -du
\end{equation} and satisfies \eqref{geometry conditions}. The Carrollian structure at $\mathscr{I}^+$ implied by the standard definition of asymptotic flatness \eqref{Bondi gauge metric} is a flat Carrollian background in the sense of \eqref{flat carrollian background}. Relaxations of the Bondi gauge conditions have been studied in \cite{Ciambelli:2020eba,Ciambelli:2020ftk,Campoleoni:2022wmf, Campoleoni:2023fug} to induce a more generic Carrollian geometry at the boundary. However, because null infinity must be compatible with the Einstein equations and asymptotic flatness, it is not the most general null hypersurface or Carrollian manifold (e.g., the intrinsic Carrollian shear defined in \eqref{Carrollian shear} vanishes). Technically, null infinity corresponds to a weakly isolated horizon \cite{Ashtekar:2024mme,Ashtekar:2024bpi}, namely a non-expanding null hypersurface whose horizon structure is time-independent along its null generators.

Of course, the above discussion relies on a particular choice of $\Omega$, and one is free to choose a different conformal factor as 
\begin{equation}
    \Omega \to \omega \Omega
\end{equation} with $\omega > 0$ \cite{Ashtekar:2014zsa}. This implies the following rescalings for \eqref{degenerate metric at scri} and \eqref{vector in the kernel}:
\begin{equation}
\label{conformal Carrollian geometry}
    q_{ab} \to \omega^2 q_{ab} , \qquad n^a \to \omega^{-1} n^a .
\end{equation} This is sometimes referred to as a conformal Carrollian geometry \cite{Duval:2014uva,Duval:2014lpa}. It corresponds to the ``universal structure'' introduced in \cite{1977asst.conf....1G, Ashtekar:2014zsa}, namely the part of the geometric data that is common to all asymptotically flat spacetimes.

Here, we restrict our discussion to the geometry, but the dynamics at null infinity described by the time evolution of the mass and angular momentum aspects \eqref{EOM1} can also be understood in terms of Carrollian geometry, analogously to the Raychaudhuri \eqref{raychau} and Damour \eqref{eq:Damour} equations at finite distance. In \cite{Donnay:2022aba, Donnay:2022wvx}, the mass and angular momentum aspects were recast in terms of a Carrollian stress tensor at null infinity. The radiation encoded in the asymptotic shear $C_{AB}$ was interpreted as an external source breaking the conservation of this stress tensor. Recently, these sources were understood as part of the Carrollian connection, which is not solely captured by the Carrollian geometry $(q_{ab}, n^c)$ \cite{Fiorucci:2025twa,Hartong:2025jpp}; see Section \ref{sec:Carrollian connection}. The flux-balance laws \eqref{EOM1} are then interpreted as Carroll-covariant conservation laws.

\subsubsection{Isomorphism between BMS and conformal Carroll}
\label{sec:Isomorphism between BMS and conformal Carroll}

The natural symmetries of the conformal Carrollian geometry at $\mathscr{I}^+$ are the conformal symmetries for the flat Carrollian geometry, i.e. the boundary diffeomorphisms ${\xi}$ preserving the geometry, up to some homogeneous scaling 
\begin{equation}
    \mathcal{L}_\xi q_{ab} = 2 \alpha q_{ab}, \qquad  \mathcal{L}_\xi n^a = -\alpha n^a
\end{equation} where $\alpha$ is a multiplicative function. The relative scaling between $q_{ab}$ and $n^a$ is fixed by \eqref{conformal Carrollian geometry}. In Section \ref{sec:Symmetries}, we have already solved these equations for the flat Carrollian background, see \eqref{flat Carroll geometry}, \eqref{conforaml Killing equations} and \eqref{conformal Carrollian vector} for $N=2$. We repeat the solution here: 
\begin{equation}
    \xi = \left( \mathcal{T} +  u \alpha \right) \partial_u + \mathcal{Y} \partial + \bar{\mathcal{Y}} \bar \partial , \qquad  \alpha = \frac{1}{2} (\partial \mathcal{Y} + \bar \partial \bar{\mathcal{Y}}), \qquad \partial \bar{\mathcal{Y}} = 0 = \bar \partial \mathcal{Y} .\label{conf Killing vector at Scri}
\end{equation} Notice that in the special case of $d=3$, the conformal Killing equation for $\mathcal{Y}^A = (\mathcal{Y}, \bar{\mathcal{Y}})$ admits an infinite set of solutions, further enhancing the conformal Carrollian algebra discussed in Section \ref{sec:Conformal Carrollian algebra} with meromorphic superrotations. Interestingly, the vector field \eqref{conf Killing vector at Scri} precisely coincides with the restriction of the asymptotic Killing vector \eqref{AKV Bondi} at $\mathscr{I}^+$:
\begin{equation}
    \xi = \lim_{\bar{r} \to \infty} \zeta |_{r=\bar{r}} .
\end{equation} One can show that the vector field \eqref{conf Killing vector at Scri} represents the $\mathfrak{ebms}_4$ algebra under the standard Lie bracket at $\mathscr{I}^+$, reproducing \eqref{commutation relations 2}. Hence we have shown the isomorphism: 
\begin{equation}
    \mathfrak{ebms}_{4} \simeq \mathfrak{eCCarr}_3
\end{equation} where $\mathfrak{eCCarr}_3$ stands for the $\mathfrak{CCarr}_3$ algebra discussed in Section \ref{sec:Conformal Carrollian algebra} enhanced with meromorphic superrotations. 

More generally, in asymptotically flat spacetimes in $d+1$ dimensions, one can define the BMS algebra as the semi-direct sum between Lorentz transformations and supertranslations: $\mathfrak{bms}_{d+1} \simeq \mathfrak{so}(d,1) \loplus \mathfrak{s}$. By contrast with the $4d$ case, the necessity of supertranslations in higher dimension is less clear, and we refer to \cite{Hollands:2003xp,Hollands:2003ie,Hollands:2004ac,Tanabe:2009va,Tanabe:2011es,Kapec:2015vwa,Capone:2019aiy,Campoleoni:2020ejn,Fuentealba:2021yvo,Capone:2021ouo,Capone:2023roc,Fuentealba:2023fwe} for discussions on this topic. We have the following isomorphism:
\begin{equation}
    \mathfrak{bms}_{d+1} \simeq \mathfrak{CCarr}_{d} .
    \label{isomorphism between BMS and Carr}
\end{equation} This is the infinite-dimensional analogue of \eqref{isomorphism poincare conf carroll}, which now includes supertranslations. This isomorphism is one of the most important points of this report, as it relates Carrollian physics in $d$ dimensions with gravity in asymptotically flat spacetimes in $d+1$ dimensions. It is the starting point of Carrollian holography, which will be discussed in Section \ref{sec:Carrollian and celestial amplitudes}.

Finally, let us conclude this section  by mentioning that BMS and Carrollian supersymmetric algebras have been considered in the literature. The Carrollian limit of the superconformal algebra has been discussed in various dimensions in \cite{Bagchi:2022owq}. In particular, for $d=2$, these results reproduce the super BMS algebra found in \cite{Barnich:2014cwa} by looking at asymptotic symmetries in $3d$ supergravity. For $d=3$, supersymmetric extensions of supertranslations have been early discussed in \cite{Awada:1985by}, then derived in \cite{Fotopoulos:2020bqj} in the context of celestial holography, and in \cite{Henneaux:2020ekh, Fuentealba:2023hzq} as asymptotic symmetries of $4d$ supergravity at spacelike infinity. Finally, supersymmetric extensions of the BMS symmetries have appeared in the derivation of flat space Carrollian holograms from AdS/CFT \cite{Lipstein:2025jfj}, see also Section \ref{sec:Towards a top-down flat space hologram from AdS/CFT}.

\section{Carrollian field theories}
\label{sec:Carrollian field theories}

In this section, we construct field theories that exhibit Carrollian symmetries by implementing the $c\to 0$ limit of relativistic field theories. We also discuss general features of these theories, including the properties of their stress tensor and some features that arise when attempting to quantize them.

\subsection{Carrollian stress tensor}
\label{sec:Carrollian stress tensor}

\subsubsection{Reminder on the relativistic stress tensor}
\label{sec:Reminder on the relativistic stress tensor}

In a relativistic (i.e. Poincar\'e-invariant) field theory, there are two ways to extract the stress tensor $T_{ab}$. The first is to use the Noether procedure and construct the conserved currents associated with translation symmetries. The canonical stress tensor obtained in this way may not have all the desired properties from the get-go, such as the symmetry in its indices, and one would need to use a Belinfante procedure to obtain an improved expression. The other way to obtain the stress tensor is to couple the theory originally living in Minkowski space with gravity by turning on a non-trivial background metric $g_{ab}$. The stress tensor can be obtained by varying the action with respect to the metric, and restricting to a flat background at the end. The resulting stress tensor is then guaranteed to be symmetric. More explicitly, we have 
\begin{equation} \label{var rel 1}
    \delta S = \frac{1}{2}\int d^{d} x \sqrt{|g|}  T_{ab} \delta g^{ab}, \qquad  T_{ab} = \frac{2}{\sqrt{|g|}} \frac{\delta S}{\delta g^{ab}} 
\end{equation} and we set $g_{ab} = \eta_{ab}$ at the end. 

This argument can be easily repeated in first order formalism by decomposing the background metric in terms of a frame as $g_{ab} = \eta_{ij} e^i_a e^j_b$. We have
\begin{equation}
    \delta S = \int d^{d} x e  T^i_a \delta e^a_i, \qquad  T^i_a = \frac{1}{e} \frac{\delta S}{\delta e^a_i}
\end{equation} where $e = \det{(e^i_a)}$. Comparing with \eqref{var rel 1}, we have $T_{ab} e^b_j \eta^{ij} = T^i_a$. To study the properties of the stress tensor, we consider the following transformation of the vielbein:
\begin{equation}
    \delta_{\xi, \Lambda, \omega} e^a_i = \mathcal{L}_{\xi} e^a_i  + {\Lambda_i}^j e^a_j + \omega e^a_i
\end{equation} where $\xi^a$, $\Lambda_{ij} = - \Lambda_{ji}$ and $\omega$ are respectively diffeomorphism, local Lorentz and Weyl rescaling parameters. Invariance under diffeomorphisms implies
\begin{equation}
    e^{-1} \partial_a \left( e T^i_b e^a_i  \right) = - T^i_a \partial_b e^a_i .
\end{equation}
Invariance under local Lorentz implies 
\begin{equation}
    T^i_a e^a_j = T^j_a e^a_i .
\end{equation}
Finally, invariance under Weyl rescalings leads to
\begin{equation}
    T^i_a e^a_i = 0 .
\end{equation} On a flat background $g_{ab} = \eta_{ab}$ and $e^a_i = \delta^a_i$, we recover $\partial^a T_{ab} = 0$ coming from translation invariance, $T_{ab} = T_{ba}$ coming from Lorentz invariance, and ${T^a}_a = 0$ coming from the possible dilatation and special conformal transformation invariance. In the next section, we extend these considerations to define the Carrollian stress tensor and study its properties.

\subsubsection{Carrollian case}
\label{sec:Carrollian case}

As in the relativistic case, there are two ways to extract the Carrollian stress tensor. The Noetherian procedure and its potential Belinfante improvement have been discussed e.g. in Appendix A of \cite{Ruzziconi:2024kzo}. Here we focus on the second method, which consists in coupling the theory to a Carrollian background, as defined in Section \ref{sec:Carrollian geometry and symmetries}. Following the discussion below Equation \eqref{variation shifts}, we consider an action defined on a background $(n^a, q^{ab})$, which fully encodes the Carrollian geometry and the Ehresmann connection. Varying the action with respect to the background and evaluating the result on a flat background at the end leads to a stress tensor corresponding to the canonical stress tensor obtained from the Noether procedure, after a possible Belinfante improvement.

The variation of the action with respect to the background yields 
\begin{equation}
    \delta S =  \int d^{d}x \varepsilon \left[ T^{(n)}_a \delta n^a + \frac{1}{2} T_{ab}^{(q)} \delta q^{ab} \right]
\end{equation} where we denoted the volume form \eqref{volume form def} as $\boldsymbol{\varepsilon}_\mathscr{I} = \varepsilon d^d x$ and introduced notations for functional derivatives:
\begin{equation}
    T^{(n)}_a  = \frac{1}{\varepsilon}\frac{\delta S}{\delta n^a}, \qquad T_{ab}^{(q)} = \frac{2}{\varepsilon} \frac{\delta S}{\delta q^{ab}} .
\label{variation second order}
\end{equation} The Carrollian stress tensor is then defined as \cite{Ciambelli:2018xat,Ciambelli:2018ojf,Rivera-Betancour:2022lkc,Baiguera:2022lsw, Campoleoni:2022wmf} 
\begin{equation}
    {T^a}_b = n^a T^{(n)}_b  + q^{ac} T_{cb}^{(q)}  .
\end{equation}

Alternatively, this discussion can be performed using the frame $e^a_i = (n^a, m^b_I)$ as a background. As emphasized in Section \ref{sec:Carrollian frame}, this formalism is natural and handy to deal with Carrollian geometry, and we will adopt this second point of view. The variation of the action in terms of the frame reads as
\begin{equation}
    \delta S =  \int d^{d}x \varepsilon \left[ T^{(n)}_a \delta n^a +  T_{a}^{(m_I)} \delta m^{a}_I \right] .
    \label{variation first order}
\end{equation} Comparing with \eqref{variation second order}, we easily derive the relation between the momenta: $T_{ab}^{(q)} m^b_J \eta^{IJ} = T_{a}^{(m_I)}$ while $T_a^{(n)}$ is the same in the two frameworks. From these, we can define the full Carrollian stress tensor as
\begin{equation}
    {T^a}_b  = n^a T_b^{(n)} + m^a_I T^{(m_I)}_b \qquad \Longleftrightarrow \qquad T_b^{(n)} = - k_a {T^a}_b , \quad T^{(m_I)}_b = m_a^I {T^a}_b .
\label{Carrollian stress tensor def}
\end{equation} Analogously to the relativistic case, we will impose invariance of the action under diffeomorphisms and homogeneous Carrollian transformations of the frame and deduce the constraints implied on the stress tensor components. Evaluating the variation of the action \eqref{variation first order} for the infinitesimal diffeomorphisms acting on the frame as in \eqref{diffeos}, and setting it to zero, we find  
\begin{equation}
\begin{split}
    \delta_{\xi} S &=  \int d^{d}x \varepsilon \left[ T^{(n)}_a \mathcal{L}_\xi n^a +  T_{a}^{(m_I)} \mathcal{L}_\xi  m_I^a \right] \\
    &= \int d^{d}x \varepsilon \xi^b \left[ T_a^{(n)} \partial_b n^a + \varepsilon^{-1} \partial_a \left( \varepsilon T_b^{(n)} n^a \right) + T_a^{(m_I)} \partial_b m^a_I + \varepsilon^{-1} \partial_a \left( \varepsilon T_b^{(m_I)} m_I^a  \right)  \right] = 0 
\end{split}
\end{equation} where, to obtain the second line, we integrated by parts and discarded total-derivative terms. Since the parameters $\xi^a$ are arbitrary functions of the coordinates, the above condition is satisfied if and only if 
\begin{equation}
    \varepsilon^{-1} \partial_a \left( \varepsilon T_b^{(n)} n^a + \varepsilon T_b^{(m_I)} m_I^a  \right)  = -T_a^{(n)} \partial_b n^a -  T_a^{(m_I)} \partial_b m^a_I  .
\end{equation} Equivalently, in terms of the full Carrollian stress tensor \eqref{Carrollian stress tensor def}, this condition reads as
\begin{equation}
     \varepsilon^{-1} \partial_a \left( \varepsilon {T^a}_b  \right)  = k_c {T^c}_a \partial_b n^a -  m^I_c {T^c}_a \partial_b m^a_I .
    \label{diffeo invariance full stress tensor}
\end{equation} Similarly, invariance of the action under infinitesimal homogeneous Carrollian transformations \eqref{eq:variation frame} leads to 
\begin{equation}
    \delta_{(r,\beta)} S =  \int d^{d}x \varepsilon \left[ T^{(m_I)}_a \beta_I n^a +  T_{a}^{(m_I)} {r_I}^J m_J^a \right] = 0 .
\end{equation} Since the symmetry parameters ${r_I}^J$ and $\beta_I$ depend arbitrarily on the spacetime coordinates, one deduces
\begin{equation}
   T_{a}^{(m_I)} m^a_J = T_{a}^{(m_J)} m^a_I, \qquad T^{(m_I)}_a n^a = 0 
\end{equation} or, in terms of the Carrollian stress tensor \eqref{Carrollian stress tensor def}, 
\begin{equation}
    m_a^I {T^a}_b m^b_J =  m_a^J {T^a}_b m^b_I, \qquad m_a^I  {T^a}_b n^b =0  .
    \label{homogeneous Carroll conditions}
\end{equation} If one further requires the invariance of the action under infinitesimal Weyl rescalings of the frame \eqref{weyl rescaling frame},
\begin{equation}
    \delta_{\omega} S =  \int d^{d}x \varepsilon \omega \left[ T^{(n)}_a  n^a +  T_{a}^{(m_I)}  m_I^a \right] = 0 .
\end{equation} we find
\begin{equation}
    T^{(n)}_a n^a + T_{a}^{(m_I)} m_I^a =0 .
\end{equation} In terms of the Carrollian stress tensor \eqref{Carrollian stress tensor def}, we have 
\begin{equation}
    {T^a}_a = 0
    \label{trace condition}
\end{equation} which is the usual trace-free condition for Weyl invariance. 

Restricting to a flat Carrollian background \eqref{flat carrollian background}-\eqref{flat Carrollian frame}, the conditions \eqref{diffeo invariance full stress tensor} and \eqref{homogeneous Carroll conditions} simplify to 
\begin{equation}
     \partial_a  {T^a}_b    =0, \qquad {T^A}_B = {T^B}_A, \qquad {T^A}_u = 0 
\label{Carrollian stress tensor conditions in flat background}
\end{equation} where in the last two conditions, we have freely replaced internal frame indices by spacetime indices $x^a = (u,x^A)$. The first condition is the standard stress tensor conservation associated with global translation invariance. The second condition tells us that the stress tensor is symmetric with respect to its spatial components as a consequence of spatial rotation invariance. Finally, the third condition follows from Carrollian boost invariance. It tells us that Carrollian particles cannot move in space, unless they have zero energy, which manifests the ultra-local nature of the theory, see Figure \ref{fig:carrollianlimit}. The condition \eqref{trace condition} manifests potential conformal invariance.

With a stress tensor at hand, one can construct conserved currents. We work on a flat Carrollian background \eqref{flat carrollian background}-\eqref{flat Carrollian frame}. Given a Carrollian stress tensor satisfying \eqref{Carrollian stress tensor conditions in flat background}, we have
\begin{equation}
    \partial_a j^a_\xi = 0, \qquad j^a_\xi = {T^a}_b \xi^b
    \label{conservation Carrollian currents}
\end{equation} if $\xi$ is a Carrollian symmetry as defined in Section \ref{sec:Carrollian isometry}, satisfying the $\mathfrak{Carr}_d$ algebra. Furthermore, if ${T^a}_b$ satisfies the trace condition \eqref{trace condition}, then the currents \eqref{conservation Carrollian currents} are conserved for all conformal Carrollian symmetries defined in Section \ref{sec:Conformal Carrollian isometry} satisfying the $\mathfrak{CCarr}_d$ algebra. Hence, Noether's theorem allows us to establish the relation between (conformal) Carrollian symmetries and symmetries of the action, through the existence of a Carrollian stress tensor.

\subsubsection{Hypermomenta}

Notice that there might be Carrollian invariant theories whose conserved Noether currents and conservation laws are not of the form \eqref{conservation Carrollian currents} \cite{Ciambelli:2017wou,Ciambelli:2018wre,Petkou:2022bmz,Rivera-Betancour:2022lkc,Armas:2023dcz,Fiorucci:2025twa, Hartong:2025jpp}. This would arise, for instance, if there is more background structure than the Carrollian frame assumed in the above discussion. For instance, a natural object to add would be the undetermined piece of the compatible Carrollian connection discussed in Section \ref{sec:Fundamental theorem}. The momenta associated with this extra background structure, called ``hypermomenta'', generically arise as the replicas of the stress tensor in the Carrollian limit of a relativistic stress tensor \cite{Petkou:2022bmz,Campoleoni:2023fug}, and contribute to the expression and conservation of the Noether currents \eqref{conservation Carrollian currents}. This situation occurs at null infinity if one tries to describe the flux-balance laws \eqref{EOM1} in terms of Carrollian geometry, see the last paragraph of Section \ref{sec:Conformal Carrollian geometry of null infinity}, and references \cite{Fiorucci:2025twa, Hartong:2025jpp} for more details.

\subsection{Scalar field}
\label{sec:Scalar field}

We now construct the Carrollian scalar field theories by implementing the $c\to 0$ limit at the level of the relativistic scalar field theory. It is instructive to present the limit both in Lagrangian and Hamiltonian formalisms. This simple example allows us to discuss general features of Carrollian field theories. The Carrollian limit of the scalar field has been discussed e.g. in \cite{deBoer:2021jej,Henneaux:2021yzg,Rivera-Betancour:2022lkc,Baiguera:2022lsw,Chen:2023pqf,Chen:2024voz, Majumdar:2025juk}.

\subsubsection{Limit in Lagrangian formalism}
\label{sec:Limit in Lagrangian formalism}

We start from the relativistic massless scalar field Lagrangian in $d$ dimensions in Lorentzian signature
\begin{equation}
    S =  \int d^{d} x \, \mathcal{L}, \qquad \mathcal{L} = -\frac{1}{2}  \partial_\mu \phi \partial^\mu \phi  =   \frac{1}{2 c^2} \dot{\phi}^2 - \frac{1}{2} \partial_A \phi \partial^A \phi 
    \label{scalar field theory}
\end{equation} where we restored the explicit dependence in the speed of light $c$ using $x^a = (c u, \vec{x})$, $\vec{x} = (x^A)$, see \eqref{Carroll contraction}, and $\dot{\phi}= \partial_u \phi$. The scalar field $\phi$ transforms as a conformal primary of dimension $\Delta = \frac{d-2}{2}$. Under Lorentz boost of parameter $\vec{\beta}$, we have 
\begin{equation}
    \delta_{\vec{\beta}} \phi = cu \vec{\beta} \cdot \vec{\partial} \phi + \frac{1}{c} \vec{\beta} \cdot \vec{x} \dot{\phi} .
    \label{transfo under boost}
\end{equation} We expand the field in powers of $c$ as \cite{deBoer:2021jej}
\begin{equation}
    \phi = c^{\Delta} ( \phi_0 + c^2 \phi_1 + c^4 \phi_2+ \ldots ) = c^{\Delta} \sum_{n = 0}^\infty \phi_n c^{2n} .
\label{expansion c}
\end{equation} Injecting this into \eqref{transfo under boost}, and rescaling the boost parameter as in \eqref{Carroll contraction}, we identify the following transformations order by order in $c$:
\begin{equation}
    \delta_{\vec{v}} \phi_0 =  \vec{v} \cdot \vec{x} \dot{\phi}_0, \qquad \delta_{\vec{v}} \phi_n = \vec{v} \cdot \vec{x} \dot{\phi}_n + u \vec{v} \cdot \vec{\partial} \phi_{n-1} \qquad (n>0) .
    \label{Carroll boost transfo}
\end{equation} The field $\phi_0$ transforms into itself, while the other fields transform into each other under boost. 

We now describe the procedure allowing us to build explicit examples of Carrollian theories. Injecting the expansion \eqref{expansion c} into \eqref{scalar field theory}, we find
\begin{equation}
    \mathcal{L} = c^{2\Delta-2} (  \mathcal{L}_0 + c^2  \mathcal{L}_1 + \mathcal{O}(c^4) ) \label{expansion L}
\end{equation} with 
\begin{equation}
    \mathcal{L}_0 = \frac{1}{2} \dot{\phi}^2_0 , \qquad \mathcal{L}_1 = \dot{\phi}_0 \dot{\phi}_1 - \frac{1}{2} \partial_A \phi_0 \partial^A \phi_0  \label{Carrollian theories by expansion}
\end{equation} and so on. One can check explicitly that $\mathcal{L}_0$ and $\mathcal{L}_1$ are invariant under translations and spatial rotations. In particular, the associated stress tensors can either be obtained by expanding the relativistic stress tensor as
\begin{equation}
    {T^a}_{b} = {(T_0)^{a}}_ b + c^2  {(T_1)^{a}}_b + \mathcal{O}(c^4)
\end{equation} and identifying the appropriate orders in the expansion, by applying the usual Noether procedure, or by turning on a background as described in Section \eqref{sec:Carrollian stress tensor}. The stress tensor associated with $\mathcal{L}_0$ is
\begin{equation}
\label{T0 stress}
    {(T_0)^u}_u = -\frac{1}{2} \dot{\phi}^2_0, \quad  {(T_0)^A}_u = 0, \quad  {(T_0)^u}_A = - \dot{\phi}_0 \partial_A \phi_0,  \quad  {(T_0)^A}_B =  \frac{1}{2} \dot{\phi}^2_0 \delta^A_B
\end{equation} while the one associated with $\mathcal{L}_1$ reads as
\begin{equation} \label{electric stress tensor}
\begin{split}
    &{(T_1)^u}_u = -\dot{\phi}_0 \dot{\phi}_1 - \frac{1}{2} \partial_A \phi_0 \partial^A \phi_0   , \quad  {(T_1)^A}_u = \dot{\phi}_0 \partial_A \phi_0, \\
    &{(T_1)^u}_A = - \dot{\phi}_1 \partial_A \phi_0 - \dot{\phi}_0 \partial_A \phi_1,  \quad  {(T_1)^A}_B = \delta^A_B \mathcal{L}_1 + \partial^A \phi_0 \partial_B \phi_0 .
\end{split}
\end{equation}

One can check explicitly that $\mathcal{L}_0$ is invariant under Carrollian boost \eqref{Carroll boost transfo}. This first example of Carroll-invariant field theory is called the electric Carrollian scalar field. By contrast, the Lagrangian $\mathcal{L}_1$ is not invariant under Carrollian boost \eqref{Carroll boost transfo}. As discussed around Equation \eqref{Carrollian stress tensor conditions in flat background}, this fact could have been directly deduced from the observation that ${(T_1)^A}_u \neq 0$. To remedy this problem and obtain a Carroll-invariant theory, one can replace $\dot{\phi}_1$, which only appears in the kinetic term, by a Lagrange multiplier $\pi$, leading to a first-order theory in time:
\begin{equation}
    \mathcal{L}'_1 = \pi \dot{\phi}_0 - \frac{1}{2} \partial_A \phi_0 \partial^A \phi_0 .
\label{magnetic action from Lagrangian}
\end{equation} It is then easy to check that this action is invariant under Carrollian boosts acting on the fields as follows
\begin{equation}
    \delta_{\vec{v}} \phi_0 =  \vec{v} \cdot \vec{x} \dot{\phi}_0, \qquad \delta_{\vec{v}} \pi =  \vec{v} \cdot \vec{x} \dot{\pi} + \vec{v} \cdot \vec{\partial} \phi_0 .
\label{transfo of the field under boost}
\end{equation} The stress tensor is given by 
\begin{equation} \label{magnetic stress tensor}
\begin{split}
    &{(T'_1)^u}_u =  - \frac{1}{2} \partial_A \phi_0 \partial^A \phi_0   , \quad  {(T'_1)^A}_u = 0, \\
    &{(T'_1)^u}_A = - \pi \partial_A \phi_0 ,  \quad  {(T'_1)^A}_B = - \frac{1}{2} \delta^A_B\partial_C \phi_0 \partial^C \phi_0 + \partial^A \phi_0 \partial_B \phi_0 
\end{split}
\end{equation}
after imposing the on-shell condition $\dot{\phi}_0 = 0$. It now satisfies the condition ${(T'_1)^A}_u = 0$ implied by Carrollian boost invariance. This second example of Carroll-invariant theory \eqref{magnetic action from Lagrangian} is called the magnetic scalar field theory. The terminology of ``electric'' and ``magnetic'' Carrollian theories was introduced in \cite{Duval:2014uoa} and will become transparent in the Carrollian limit of Maxwell theory described in Section \ref{sec:Maxwell theory}. The procedure described above can, in principle, be iterated to obtain more examples of Carrollian field theories by looking at higher order terms in the expansion \eqref{expansion L}.

\subsubsection{Limit in Hamiltonian formalism}
\label{sec:Limit in Hamilatonian formalism}

As we shall now discuss, the Carrollian Lagrangians $\mathcal{L}_0$ and $\mathcal{L}'_1$ can be directly obtained by taking the limit in the Hamiltonian formulation of the relativistic theory \eqref{scalar field theory} \cite{Henneaux:2021yzg}. The momentum $\pi$ conjugate to the field $\phi$ is given by 
\begin{equation}
    \pi = \frac{\partial \mathcal{L}}{\partial \dot{\phi}}=\frac{1}{c^2} \dot{\phi}
\end{equation} and the Hamiltonian action associated with \eqref{scalar field theory} reads as
\begin{equation}
    S_H [\phi, \pi] = \int du \, d\vec{x} \, \left[ \pi \dot{\phi} - \frac{1}{2} \partial_A \phi \partial^A \phi - \frac{1}{2} c^2 \pi^2 \right] .
    \label{Ham action scalar}
\end{equation}  There are two interesting Carrollian limits that preserve the kinetic term $\pi \dot{\phi}$. The magnetic limit is found by just taking $c\to 0$ and not rescaling the fields: $\phi \to \phi$ and $\pi \to \pi$. We find
\begin{equation}
\label{magnetic scalar action}
 S_H^M [\phi, \pi] = \int du \, d\vec{x} \, \left[ \pi \dot{\phi}  - \frac{1}{2} \partial_A \phi \partial^A \phi \right] .
\end{equation} In this case, the Hamiltonian only involves spatial derivatives of $\phi$. This theory coincides with the magnetic Carrollian scalar field theory \eqref{magnetic action from Lagrangian}. Notice that the field $\pi$ can be integrated out by injecting its associated constraint $\dot{\phi}= 0$ into the action, yielding the second order Lagrangian action
\begin{equation} 
    S_L^M = -L \int d\vec{x} \,  \frac{1}{2} \partial_A \phi \partial^A\phi
    \label{magnetic scalar}
\end{equation} where we have integrated over the $u$ coordinate and set $\int du = L$, with $L$ an IR cut-off in time. Up to the overall scale $L$, this coincides with a Euclidean CFT Lagrangian in $d-1$ dimensions. While the form of the action \eqref{magnetic scalar} is conceptually interesting, this theory is no longer equivalent to \eqref{magnetic scalar action}. In fact,
the connection between $\dot{\phi}$ and $\pi$ is lost, since $\dot{\phi} = 0$ but in general $\pi \neq 0$.

The electric limit is found by rescaling the fields as $\phi \to c \phi$ and $\pi \to c^{-1} \pi$ in \eqref{Ham action scalar}, which preserves the kinetic term, and taking $c\to 0$, yielding
\begin{equation}
     S_H^E [\phi, \pi] = \int du \, d\vec{x} \, \left[ \pi \dot{\phi}  - \frac{1}{2} \pi^2 \right] .
     \label{electic scalar}
\end{equation} By contrast, the Hamiltonian now only involves the momentum $\pi$. This theory is equivalent to the electric Carrollian scalar field theory \eqref{Carrollian theories by expansion}. To see this, we can integrate out $\pi$, which leads to
\begin{equation} \label{electric action simple}
    S_L^E = \frac{1}{2}\int du \, d\vec{x} \,  \dot{\phi}^2 .
\end{equation} This action is indeed classically equivalent to \eqref{electic scalar} as $\pi$ and $\dot{\phi}$ are still related. It reproduces exactly $\mathcal{L}_0$ in \eqref{Carrollian theories by expansion}.

\subsubsection{Coupling constants}
\label{sec:Coupling constants}

The above discussion concerns the Carrollian contraction of a massless free scalar field. This can easily be extended to massive or interacting scalar fields. The only difference is that, in addition to providing the scalings of the fields in the limit, one also needs to provide the scalings of the coupling constants. To see this, consider the theory
\begin{equation}
    S_H [\phi, \pi] = \int du \, d\vec{x} \, \left[ \pi \dot{\phi} - \frac{1}{2} \partial_A \phi \partial^A \phi - \frac{1}{2} c^2 \pi^2 - \frac{1}{2}m^2 c^2 \phi^2 - \lambda \phi^n \right] 
\end{equation} which is just \eqref{Ham action scalar} with a mass term and a general $\lambda \phi^n$ coupling. Performing the magnetic Carrollian contraction of this Lagrangian $\phi \to \phi$ and $\pi \to \pi$ without rescaling the couplings leads to a massless scalar interacting field in the limit. To obtain a massive scalar field, one also needs to rescale the mass as $m \to i c^{-1} m$,\footnote{The factor $i$ in the rescaling of the mass flips the sign of the mass term and comes from the fact that magnetic Carrollian theories are obtained from the limit of tachyonic field theories, by consistency with the dispersion relations \cite{deBoer:2021jej, Bergshoeff:2023vfd}.} yielding 
\begin{equation} \label{interacting magnetic}
    S_H^M [\phi, \pi] = \int du \, d\vec{x} \, \left[ \pi \dot{\phi} - \frac{1}{2} \partial_A \phi \partial^A \phi  + \frac{1}{2}m^2 \phi^2  -\lambda \phi^n \right] .
\end{equation} Similarly, the electric scaling on the fields alone, $\phi \to c \phi$ and $\pi \to c^{-1} \pi$, does not yield a finite mass term nor a finite coupling in the limit. To circumvent this result, one can rescale the mass and the coupling as $m \to c^{-2}m$ and $\lambda \to c^{-n}\lambda$ before taking the limit. We obtain
\begin{equation}
    S_H^E [\phi, \pi] = \int du \, d\vec{x} \, \left[ \pi \dot{\phi}  - \frac{1}{2}  \pi^2 - \frac{1}{2}m^2 \phi^2 - \lambda \phi^n \right] . \label{electric limit with coupling}
\end{equation} As we will see later, it is important to specify both the scaling of the fields and that of the coupling constants in the limit.

\subsubsection{Carroll covariant form}
\label{sec:Carroll covariant form}

The above magnetic and electric Carrollian scalar field theories admit a Carroll-covariant version when a background is turned on. In the relativistic case, the action \eqref{scalar field theory} can be made diffeomorphism invariant by substituting the flat space metric by a generic curved background $\eta_{ab} \to g_{ab}$, so that
\begin{equation}
    S[\phi] \to -\frac{1}{2} \int du \, d\vec{x} \,\sqrt{-g} g^{ab} \partial_a \phi \partial_b \phi  = -\frac{1}{2} \int du \, d\vec{x} \, e \, \eta^{ij} e^a_i e^b_j \partial_a \phi  \partial_b \phi .
\end{equation} This allows us to derive the stress tensor of the theory by varying with respect to the background as reviewed in Section \ref{sec:Reminder on the relativistic stress tensor}. Similarly, in the Carrollian case the actions can be made diffeomorphism invariant by promoting the flat space Carrollian structure to a generic Carrollian background \eqref{Carrollian geometry def}.

The covariantization of the electric Carrollian scalar field \eqref{electric action simple} reads as
\begin{equation}
    S^E_L \to \frac{1}{2}\int \boldsymbol\varepsilon_\mathscr{S} \, (n^a \partial_a \phi)^2 
\end{equation} where $n^a$ is the vector defining the Carrollian geometry \eqref{Carrollian geometry def}, $\varepsilon_\mathscr{S}$ is the Carrollian volume form defined in \eqref{volume form def}. It is easy to check that this action is invariant under diffeomorphisms, as well as local Carrollian transformations \eqref{eq:variation frame} of the frame. Taking the flat Carrollian background \eqref{flat Carrollian frame}, we recover the action \eqref{electric action simple}, as desired.

The covariantization of the magnetic Carrollian scalar field \eqref{magnetic scalar action} reads as  
\begin{equation}
\label{covariant magnetic}
    S^M_L \to \int  \boldsymbol{\varepsilon}_\mathscr{S}  \left[ \pi n^a \partial_a \phi - \frac{1}{2} q^{ab} \partial_a \phi  \partial_b \phi \right] = \int  \boldsymbol{\varepsilon}_\mathscr{S} \left[\pi n^a \partial_a \phi - \frac{1}{2} q^{IJ} m^a_I m^b_J \partial_a \phi  \partial_b \phi \right]
\end{equation} where $q^{ab}$ is the ``inverse metric'' defined in \eqref{def inverse q}, and in the second equality, we used the horizontal Carrollian frame, see Equation \eqref{sec:metric in Carrollian frame}. Again, it can be checked that this action is invariant under diffeomorphisms. While the action is manifestly invariant under spatial rotations of the frame, invariance under Carrollian boost requires the momentum $\pi$ to transform as
\begin{equation}
    \delta_\beta \pi = q^{ab} \lambda_b \partial_a \phi = q^{IJ} \beta_I m^a_J \partial_a \phi
\end{equation} with $\lambda_a = \beta_I m^I_a$ (see \eqref{eq:variation frame} and below), which is consistent with \eqref{transfo of the field under boost}. Hence, although the action \eqref{covariant magnetic} appears to involve the Ehresmann connection through $q^{ab}$ (see \eqref{def inverse q}), it is actually independent of this choice. Choosing the flat Carrollian background \eqref{flat Carrollian frame}, one recovers \eqref{magnetic scalar action}.

Finally, it is instructive to check that, following the procedure described in Section \ref{sec:Carrollian stress tensor}, one can extract the components of the Carrollian stress tensor. Evaluating the result on a flat Carrollian background \eqref{flat Carrollian frame}, one reproduces \eqref{T0 stress} and \eqref{magnetic stress tensor} in the electric and magnetic cases, respectively. 

\subsection{Maxwell, Yang-Mills and gravity}

Many other examples of Carrollian field theories have been discussed in the literature. In this section, we present some of the basic examples.

\subsubsection{Maxwell theory}
\label{sec:Maxwell theory}

Carrollian electrodynamics was first discussed in \cite{Duval:2014uoa} at the level of the equations of motion. In this section, following \cite{Henneaux:2021yzg}, we apply the same method as the one presented in Section \ref{sec:Limit in Hamilatonian formalism} (see also \cite{Bagchi:2016bcd,Basu:2018dub,deBoer:2021jej}). After restoring the dependence on the speed of light $c$, the Hamiltonian action of electrodynamics is 
\begin{equation}
    S_H[A^A, \pi_A, A_u] = \int du \int d \vec x \left[ \pi_A \dot{A}^A - \frac{c^2}{2} \pi_A \pi^A - \frac{1}{4} F_{AB} F^{AB}  + A_u \partial_A \pi^A \right]
\end{equation} where $E^A = -\pi^A$ is the electric field and, in $4d$, $B^A = \frac{1}{2} \epsilon^{ABC}  F_{BC}$ is the magnetic field. The Hamiltonian involves one quadratic term with the electric field and one with the magnetic field, as well as the Gauss constraint with Lagrange multiplier $A_u$. The magnetic Carrollian contraction is obtained by taking $c\to 0$, which yields
\begin{equation}
    S^M_H[A_A, \pi^A, A_u] = \int du \int d \vec x  \left[ \pi_A \dot{A}^A - \frac{1}{4} F_{AB} F^{AB}  + A_u \partial_A \pi^A \right] .
    \label{magnetic Maxwell}
\end{equation} In this action, the Hamiltonian involves only the magnetic field $B_A$ and the Gauss constraint, which is why we call this limit ``magnetic'' \cite{Duval:2014uoa}. To perform the electric contraction, we rescale the fields as $A_A \to c A_A$, $A_u \to c A_u$ and $\pi^A \to c^{-1} \pi^A$ and then take $c\to 0$. We find
\begin{equation}
    S_H^E[A_A, \pi^A, A_u] = \int du \int d \vec x \left[ \pi_A \dot{A}^A - \frac{1}{2} \pi_A \pi^A   + A_u \partial_A \pi^A \right] .
    \label{electric Maxwell}
\end{equation} The Hamiltonian now involves only the electric field $E_A$, hence the name ``electric'' limit. Analogously to the scalar field case discussed in Section \ref{sec:Carroll covariant form}, the Carrollian actions \eqref{magnetic Maxwell} and \eqref{electric Maxwell} can be written in a covariant form by turning on a non-trivial background, allowing to derive the stress tensor along the lines of Section \ref{sec:Carrollian case}.

In four spacetime dimensions, the relativistic Maxwell theory exhibits electric-magnetic duality, which is reflected in the invariance of Maxwell's equations under the exchange $\{ E_A, B_A \} \to \{  -B_A , E_A \}$. Interestingly, this duality is intimately related to the electric and magnetic Carrollian field theories, which can be viewed as dual to each other \cite{Bunster:2012hm,OConnor:2024rku}. One way to see this is to start from the parent relativistic action involving two potentials that manifest electric-magnetic duality, and then take the Carrollian contraction. Integrating out one of the fields in the resulting Carrollian action yields the electric Carrollian Maxwell theory, while integrating out the other leads to the magnetic one. Hence, the electric and magnetic Carrollian limits of the Maxwell action are dual to each other in the usual sense that they arise from the same parent action, depending on which field is integrated out.

\subsubsection{Yang-Mills theory}

The above considerations can be directly generalized to Yang-Mills theory \cite{Henneaux:2021yzg,Islam:2023rnc}. This will provide simple examples of interacting Carrollian field theories. Consider a non-abelian gauge field $A_a = A_a^i T_i$, where $T_i$ are some Lie algebra generators satisfying $[T_i,T_j] = {f_{ij}}^k T_k$. The Hamiltonian form of the Yang-Mills action reads as 
\begin{equation}
    S_H[A^i_A, \pi_i^A, A_u^i] = \int du \int  d \vec x \left[ \pi_A^i \dot{A}^A_i - \frac{c^2}{2} \pi_A^i \pi^A_i - \frac{1}{4} F_{AB}^i F^{AB}_i  + A_u^i D_A \pi^A_i \right]
\end{equation} where $F_{AB}^i =\partial_A A_B^i - \partial_B A_A^i + g {f_{jk}}^i A^j_A A^k_B$, $D_A \pi^A_i = \partial_A \pi^A_i + g {f_{ij}}^k A^j_B   \pi^B_k$ and $g$ is the coupling constant. The magnetic Carrollian contraction is straightforward and leads to 
\begin{equation}
     S_H^M[A^i_A, \pi_i^A, A_u^i] = \int du \int  d \vec x \left[ \pi_A^i \dot{A}^A_i  - \frac{1}{4} F_{AB}^i F^{AB}_i  + A_u^i D_A \pi^A_i \right] .
\end{equation} The electric contraction requires the following rescaling of the fields: $A_A \to c A_A$, $A_u \to c A_u$ and $\pi^A \to c^{-1} \pi^A$. To keep the coupling constant in the limit, one also has to rescale $g$ as $g\to c^{-1}g$ so that, at the end, we find 
\begin{equation}
    S_H^E[A^i_A, \pi_i^A, A_u^i] = \int du \int  d \vec x \left[ \pi_A^i \dot{A}^A_i - \frac{1}{2} \pi_A^i \pi^A_i  + A_u^i D_A \pi^A_i \right] .
\end{equation} As discussed in Section \ref{sec:Coupling constants}, in general, the limit requires giving a scaling for the fields and the coupling constants with $c$ to find a non-trivial interacting theory.

\subsubsection{Einstein's gravity}

The electric Carrollian limit of Einstein's gravity theory was discovered early in Hamiltonian formalism \cite{Henneaux:1979vn}, and revisited in \cite{Henneaux:2021yzg}. The magnetic Carrollian limit was discussed much later in \cite{Campoleoni:2022ebj}. The Carrollian limit of general relativity was also discussed in \cite{Bergshoeff:2017btm}, and systematically studied in \cite{Hansen:2021fxi} by performing a Carrollian expansion of the covariant Einstein-Hilbert action, along the lines pioneered in \cite{Dautcourt:1997hb}. This method generalizes the one presented in Section \ref{sec:Limit in Lagrangian formalism}, where the electric Carrollian field theory is found at leading order in the $c$ expansion, while the magnetic theory is found at
the next-to-leading order. In this section, we follow the presentation of \cite{Henneaux:2021yzg,Campoleoni:2022ebj}.

The canonical action for gravity is given by
\begin{equation}
    \label{eq:relativistic_ADM}
S_H[g_{AB}, \pi^{AB}, N, N^A]=\int du  d \vec x \left[\pi^{AB}\dot{g}_{AB}-N \mathcal{H}_{\perp}-N^{A}\mathcal{H}_{A}\right] 
\end{equation} where $N$ is the lapse and $N^A$ is the shift. The Hamiltonian constraints take the form
\begin{equation}
\mathcal{H}_{\perp} = \mathcal{H}_M +\mathcal{H}_E \, ,
\qquad 
\mathcal{H}_{A} = -\,2\,\nabla_{\!B}\pi_{A}{}^{B}\,,
\end{equation}
where
\begin{equation}
\label{eq:electricandmagneticH}
\mathcal{H}_M = -\frac{\sqrt{g}}{16\pi G}\left(R-2\Lambda\right) ,
\qquad
\mathcal{H}_E = \frac{16\pi G c^2}{\sqrt{g} } \left(\pi^{AB}\pi_{AB}-\frac{1}{d-2}\, \pi^{2}\right) , 
\end{equation}
with $G = c^{-4} G_N$ and $G_N$ denoting Newton's constant. Again, we have two possible Carrollian contractions. The first possibility is to take the $c \to 0$ limit (keeping $G$ fixed), effectively dropping the term $\mathcal{H}_E$ in the Hamiltonian density. One thus obtains the action of magnetic Carrollian gravity:
\begin{equation} \label{eq:magnetic_action}
S_H^M  [g_{AB}, \pi^{AB}, N, N^A] =\int du  d \vec x \left[\pi^{AB}\dot{g}_{AB}-N \mathcal{H}_{M}-N^{A}\mathcal{H}_{A}\right] .
\end{equation}
 The electric Carrollian limit is obtained instead by rescaling the fields as $g_{AB}\to c g_{AB}$, $\pi^{AB} \to c^{-1} \pi^{AB}$, $N \to c N$ and $N^A \to N^A$, and the coupling as $G \to c^{\frac{d-7}{2}} G$, so as to keep in the $c \to 0$ limit the term $\cH_{E}$ in the Hamiltonian density while dropping $\cH_{M}$, without affecting either the kinetic term nor $\cH_A$. We obtain 
\begin{equation} 
S_H^E [g_{AB}, \pi^{AB}, N, N^A] =\int du  d \vec x \left[\pi^{AB}\dot{g}_{AB}-N \mathcal{H}_{E}-N^{A}\mathcal{H}_{A}\right] .
\end{equation}

Besides their mathematical interest, Carrollian theories of gravity have also been argued to be physically relevant. In particular, the electric Carrollian limit corresponds to the strong coupling limit \cite{Isham:1975ur} or the zero-signature limit \cite{Teitelboim:1978wv}, introduced long ago, which are relevant to BKL behaviour \cite{Belinsky:1970ew}. This story has been revisited recently in \cite{Oling:2024vmq}, where it was shown that the Carrollian limit of general relativity coupled to matter captures the ultra-local BKL dynamics near spacelike singularities, including chaotic cosmological billiards, thereby providing a tractable framework for studying near-singularity physics and deep infrared dynamics in AdS/CFT. 

The Carrollian limit of gravity solutions was also used in \cite{Fontanella:2025tbs} in a holographic context. In the same spirit as non-relativistic AdS/CFT, the Carrollian contraction is implemented on both sides of the duality, leading to a Carrollian bulk theory and a correspondingly contracted boundary theory. The analysis provides important insights into the fate of the compact dimensions in AdS$_5 \times S^5$ when addressing its relation to flat-space holography (see also Section \ref{sec:Towards a top-down flat space hologram from AdS/CFT}).

Let us also mention that the Carrollian limit of two-dimensional JT gravity has been discussed in \cite{Grumiller:2020elf,Gomis:2020wxp}. These models are particularly interesting, as they admit “Carrollian black holes”, that is, solutions of Carrollian gravity which display thermal behaviour and admit a notion of extremal surface, despite the absence of a conventional Lorentzian causal structure \cite{Ecker:2023uwm}.

More speculatively, Carrollian theories of gravity could be interesting toy models for quantum gravity, as recently argued in \cite{Ecker:2024czh}. Indeed, there is an interesting triple scale limit of quantum gravity when $\hbar \to \infty$, $G_N \to 0$, $c \to 0$, keeping fixed the gravitational coupling $G_N c^{-4}$
and the combination $\hbar c$. This limit has the virtue of preserving the laws of black hole thermodynamics, which means that puzzles
related to black holes and their evaporation could be addressed more easily in this limit than
in full-fledged quantum gravity.

\subsection{Fermions, swiftons, strings, and more...}

In the previous sections, we have discussed some basic Carrollian Lagrangians. Many other examples exist in the literature. For instance, the Carrollian limit of $p$-form theories has been discussed in \cite{Henneaux:2021yzg}. Furthermore, while all the above examples of Carrollian field theories involve bosonic fields, fermions have recently been included in the discussion. Carroll-invariant fermionic actions have been constructed in \cite{Bagchi:2022eui,Banerjee:2022ocj} from an intrinsic perspective by working with modified (degenerate) Clifford algebras adapted to Carrollian geometry. Carrollian fermions have also been obtained in \cite{Bergshoeff:2023vfd,Koutrolikos:2023evq,Bergshoeff:2024ytq} from a Carrollian limit of the Dirac action, leading to distinct electric and magnetic theories. We refer to \cite{Grumiller:2025rtm} for a concise summary of these constructions and further references. These developments illustrate that Carrollian fermions are richer than the scalar field example discussed in Section \ref{sec:Scalar field}: the space of invariant fermionic actions includes multiple Carrollian sectors, distinguished both by how the limit is taken and by the choice of Carrollian Clifford representations.

In two dimensions, Carrollian Liouville theories have been derived in \cite{Barnich:2012rz} by looking at the electric and magnetic limits of Liouville theory,\footnote{In that reference, the Carrollian limit is induced by a flat space limit in the bulk of AdS$_3$, we refer to Section \ref{sec:Flat space/Carrollian limit of AdS/CFT} for a discussion on the flat space/Carrollian limit correspondence.} and provide useful tractable models for non-trivial interacting Carrollian theories. In particular, the magnetic theory has been shown to encode the asymptotic dynamics of $3d$ gravity in asymptotically flat spacetime \cite{Barnich:2013yka}. Alternatively, these Carrollian Liouville actions can be obtained via the geometric action construction \cite{Barnich:2017jgw}, which does not rely on a limit procedure. The latter is a systematic method to write actions that are invariant under a given symmetry group. This method has been successfully applied to construct BMS/conformal Carroll-invariant actions relevant for the holographic description of three- \cite{Barnich:2017jgw,Merbis:2019wgk,Cotler:2024cia} and four-dimensional \cite{Barnich:2022bni} asymptotically flat spacetimes. This construction relies on the coadjoint representation of the BMS group discussed in three \cite{Barnich:2015uva} and four dimensions \cite{Barnich:2021dta,Donnelly:2020xgu}.

Another interesting way to construct Carrollian field theories without resorting to a $c \to 0$ limit can be achieved in $2d$ via a marginal deformation of a relativistic CFT \cite{Rodriguez:2021tcz,Tempo:2022ndz,Parekh:2023xms,Banerjee:2026qyc}. The marginal operator is of the form $\sqrt{T\bar{T}}$ (where $T$ and $\bar{T}$ are the holomorphic and anti-holomorphic components of the stress tensor, respectively), and allows one to flow from the conformal algebra $\mathrm{Witt} \oplus \mathrm{Witt}$ to the two-dimensional conformal Carrollian algebra, which is isomorphic to $\mathfrak{bms}_3$.

As mentioned above, Carrollian fields are ultra-local and are naively not expected to exhibit propagating degrees of freedom. This can be seen from Figure \ref{fig:carrollianlimit}, where movement would violate causality, or from Section \ref{sec:Carrollian case}, where we have seen that Carrollian boost invariance implies ${T^A}u = 0$, forbidding spatial displacements of energy. However, this situation can be circumvented by considering interacting Carrollian field theories, called “Carrollian swiftons”. These have been discussed in detail in \cite{Ecker:2024czx}, where scalar and vector models invariant under Carrollian symmetry are constructed that support field excitations propagating with strictly non-zero speed outside the Carrollian light cone (hence the name ``swifton''), and analogous constructions are given in the presence of Carrollian gravity in two dimensions. An example of such an action is obtained by coupling two Carrollian scalar fields
\begin{equation}
S[\phi, \chi] = \frac{1}{2} \int \boldsymbol{\varepsilon}_{\mathscr{S}} \left[ (n^a \partial_a \phi)^2 + (n^a \partial_a \chi)^2 + g B_a B^a \right], \qquad B_b = n^{a} \partial_{[a} \phi \partial_{b]} \chi.
\end{equation}
This action is structurally related to continuum descriptions of fracton phases, where constrained mobility and unconventional propagation arise, although standard fracton models are usually formulated in terms of higher-rank gauge fields rather than Carrollian scalar theories \cite{Baig:2023yaz,Kasikci:2023tvs}. The fields $\phi$ and $\chi$ satisfy nontrivial hyperbolic equations of motion and therefore admit propagating degrees of freedom on a Carrollian spacetime, in contrast with the ultra-local behavior of free Carrollian scalars. 

Extended objects with Carrollian symmetries have also been studied in the context of string theory. The so-called null string, originally introduced as the tensionless limit of the relativistic string \cite{Schild:1976vq,Isberg:1993av}, provides an early example of a Carrollian theory on the worldsheet, in which the dynamics becomes ultra-local. This perspective has been revisited recently in \cite{Bagchi:2015nca,Bagchi:2020fpr,Bagchi:2023cfp,Bagchi:2024qsb}, where the tensionless limit has been related to a Carrollian limit on the worldsheet (see \cite{Bagchi:2026wcu} for a review). Notice that the null string is classically equivalent to the ambitwistor string \cite{Mason:2013sva,Casali:2016atr}. Interestingly, based on the ambitwistor string, worldsheet actions have been constructed at null infinity and shown to encode tree-level scattering formulae in flat space \cite{Adamo:2014yya,Adamo:2015fwa}. These are relevant in the holographic context discussed in Section \ref{sec:Carrollian and celestial amplitudes}. Finally, let us mention that the Carrollian limit can also be implemented directly on the target space, leading to strings propagating in Carrollian geometry \cite{Cardona:2016ytk}.

\subsection{Quantization}
\label{sec:Quantization}

In the previous sections, we have seen how to construct examples of classical Carrollian field theories. Of course, a natural question is to try to quantize these theories. In particular, as we shall explain in Section \ref{sec:Carrollian and celestial amplitudes}, the dual theory for gravity in asymptotically flat spacetimes, if it exists, is expected to be an example of such a quantum Carrollian field theory. In this section, we discuss recent attempts to quantize Carrollian field theories, starting from their Lagrangians \cite{deBoer:2023fnj,Chen:2023pqf,Chen:2024voz,Cotler:2024xhb,Cotler:2025dau}.  

\subsubsection{Electric theory}

Let us consider an electric Carrollian scalar field theory
\begin{equation}
\label{electric interacting theory}
S[\phi] = \int du\, d \vec x \left( \frac{1}{2}\dot{\phi}^2 - V(\phi) \right),
\end{equation}
where $V(\phi)=\frac{m^2}{2}\phi^2$ is a potential, which we will take Gaussian for now. The propagator is readily obtained by inverting the quadratic operator \cite{deBoer:2023fnj}:
\begin{equation} \label{feynamn propagator}
(\partial_u^2+m^2)\,\langle \phi(u,\vec x)\phi(0)\rangle
=-i \delta(u)\,\delta^{(d-1)}(\vec x)
\quad\Longrightarrow\quad
\langle \phi(u,\vec x)\phi(0)\rangle
= \frac{1}{2m}e^{-im|u|}\,\delta^{(d-1)}(\vec x).
\end{equation} A first problem occurs when computing correlators of composite operators, which are typically relevant for holography, see Section~\ref{sec:Towards a top-down flat space hologram from AdS/CFT}. Considering $\phi^k$, the connected two-point function behaves as
\begin{equation}
\langle \phi^k(u,\vec x)\phi^k(0)\rangle
\sim (\delta^{(d-1)}(\vec x))^k,
\end{equation}
which diverges for $k>1$, since it contains powers of $\delta^{(d-1)}(0)$. As suggested in~\cite{Cotler:2024xhb}, a natural regularization is to introduce a lattice scale $a$ that separates spatial points, so that $\delta^{(d-1)}(\vec x)\sim a^{-(d-1)}$. Rescaling the operators as
\begin{equation}
\phi^k \to O_k = a^{\frac{(k-1)(d-1)}{2}}\phi^k,
\end{equation}
one finds that in the continuum limit $a\to 0$,
\begin{equation}
\langle O_k(u,\vec x)\,O_k(0)\rangle \sim \delta^{(d-1)}(\vec x),
\end{equation}
which is ultra-local. However, as a result, all connected higher-point correlators become trivial in the limit. For instance,
\begin{equation}
\langle O_k(u,\vec x)\,O_k(u',\vec y)\,O_k(u'',\vec z)\rangle
\sim a^{\frac{d-1}{2}}\,
\delta^{(d-1)}(\vec x-\vec y)\,
\delta^{(d-1)}(\vec y-\vec z),
\end{equation}
which vanishes as $a\to 0$. Therefore, properly normalized composite operators retain non-trivial two-point functions, but do not generate any non-trivial higher-point structure. Hence, this regularization of electric Carrollian field theories leads to generalized free field theories. 

\begin{figure}[h]
    \centering   \includegraphics[width=0.6\textwidth]{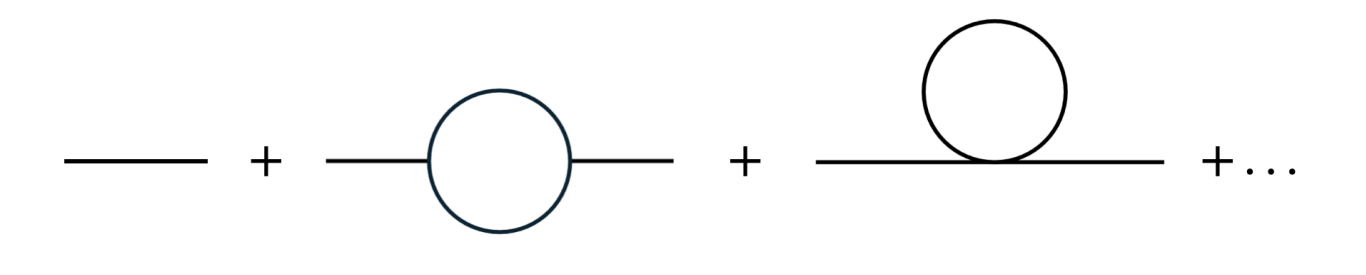}
    \caption{$1$-loop Feynman propagator.}
    \label{fig:mylabel}
\end{figure}

A second related feature occurs when looking at the loop corrections to the Feynman propagator \eqref{feynamn propagator} in presence of non-Gaussian interactions. The relativistic dispersion relation for a momentum $k^\mu = (\frac{E}{c}, \vec k)$ reads as
\begin{equation}
      E^2  =  |\vec k|^2 c^2 + m^2 c^4
\end{equation} Taking the $c\to 0$ limit of this dispersion relation, with the scaling of the mass discussed around Equation \eqref{electric limit with coupling}, $m\to c^{-2} m$, we obtain the Carrollian dispersion relation
\begin{equation}
    E^2 =  m^2
\end{equation} Notice that this could also be read directly from the quadratic term in \eqref{electric interacting theory}. This is an example of ``flat band'' dispersion relation \cite{Bagchi:2022eui,Ara:2024fbr}. In particular, the Fock space is highly degenerate, as, for fixed $n$, all the states $a^\dagger_{\vec{k}_1} \ldots a^\dagger_{\vec{k}_n} |0\rangle$ have the same energy. Let us now compute the $1$-loop correction to the Feynman propagator for a potential $V(\phi) = \frac{m^2}{2}\phi^2 + \frac{\lambda_3}{3!} \phi^3 + \frac{\lambda_4}{4!} \phi^4$. Since the $\vec k$ does not appear in the dispersion relation, we can work in position space for the spatial components, while staying in Fourier space for the energy. We find (see Figure \ref{fig:mylabel})
\begin{equation}
\begin{split}
&\langle \phi(E, \vec x)\phi(-E, \vec y)\rangle
=
\frac{i}{E^2 - m^2} \frac{\delta_{\vec{x}, \vec{y}}}{a^{d-1}} \\
&\quad+
\left[
\frac{i}{E^2 - m^2}
\right]^2
\frac{1}{2}
\int \frac{dE'}{2\pi}
\left[
(i\lambda_3)^2
\frac{i}{(E-E')^2 - m^2 }
\frac{i}{E'^2 - m^2}
+
i\lambda_4
\frac{i}{E'^2 - m^2 }
\right]  \frac{\delta_{\vec{x}, \vec{y}}}{a^{d-1}} 
+ \cdots .
\end{split}
\end{equation} where we have introduced the lattice scale $a$ to regulate the $\delta$-function, $\delta (\vec x - \vec y) \sim \delta_{\vec{x}, \vec{y}}\,  a^{-(d-1)}$, and the volume integral appearing in the $1$-loop correction, $\left(\int \frac{d\vec k}{(2\pi)^{d-1}}\right) \sim a^{-(d-1)}$. This leads to a renormalized $2$-point function $\sim \delta (\vec x - \vec y)$ in the continuum limit. However, as in the Gaussian case above, one can show that this procedure eliminates all the connected higher-point functions. For instance, we have
\begin{equation}
    \langle \phi (E_1, \vec x) \phi( E_2 , \vec y) \phi( E_3 , \vec z) \rangle \sim \lambda_3 a^{\frac{d-1}{2}} \delta^{(d-1)}(\vec x- \vec y) \delta^{(d-1)}(\vec y- \vec z) 
\end{equation} which vanishes in the continuum limit. This is a manifestation of the UV/IR mixing: regulating short-distance physics (the UV) directly controls the structure of long-distance correlations (the IR), which is unusual compared to standard relativistic QFTs. Here we focused on the electric scalar field case, but other examples have been considered in \cite{Cotler:2024xhb}, such as the quantization of the electric Carrollian Maxwell theory \eqref{electric Maxwell}. The question of finding an intrinsically Carrollian quantization procedure for electric theories with an appropriate regulator that yields non-trivial higher-point electric correlators in presence of interactions remains an open question.

\subsubsection{Magnetic theory}

The status of quantum magnetic Carrollian field theories is better. Let us consider the magnetic scalar field action 
\begin{equation}
S[\chi, \pi] = \int du\, d\vec x \left( \pi \dot{\chi} - \mathcal{H} (\chi) \right),
\end{equation} where $\mathcal{H}(\chi) = -\frac{m^2}{2}\chi^2+ \frac{1}{2} \partial_A \chi \partial^A \chi + \ldots$ is a Hamiltonian depending on $\chi$ and its spatial derivatives, see e.g. \eqref{interacting magnetic}. As discussed earlier, $\pi$ plays a role of Lagrange multiplier in this action. Integration over $\pi$ in the path integral will force time independent configurations of $\chi$. A basis of states is provided by $|\chi ( \vec x) \rangle$ and the inner product can be computed via \cite{Cotler:2024xhb}
\begin{equation}
    \langle \chi_1 ( \vec x) | \chi_2 ( \vec x) \rangle =  
\int \mathcal{D}\pi( \vec x )\,
e^{-i \int d\vec x \,
\pi( \vec x )(\chi_2( \vec x)-\chi_1( \vec x ))}  =\delta [\chi_1 ( \vec x) - \chi_2 ( \vec x) ] .
\end{equation} These orthonormal states are energy eigenstates since
\begin{equation}
    \hat{H} | \chi (\vec x) \rangle =  H [\chi (\vec x)] | \chi (\vec x) \rangle , \qquad   H [\chi (\vec x)] =  \int d \vec x \, \mathcal{H} (\chi(\vec x)) .
\end{equation} Indeed, in the Schrödinger picture, we have
\begin{equation}
    [\hat\pi (\vec{x}) , \hat \chi (\vec y)]  = - i \delta^{(d-1)} (\vec x - \vec{y} ), \qquad [\hat{H} , \hat{\chi} (\vec x)] = 0 .
\end{equation}
Our theory has a continuous spectrum of energy eigenstates, which in turn implies an IR-divergent partition function:
\begin{equation} \label{partition magnetic}
    \text{Tr}\left(e^{-\beta \hat H}\right) 
= \int \mathcal{D}\chi(\vec x)\,\langle \chi(\vec x)| e^{-\beta \hat H}|\chi(\vec x)\rangle
= \delta[0]\left( \int \mathcal{D}\chi( \vec x)\, e^{-\beta H[\chi]} \right) .
\end{equation} Hence upon stripping off this divergence $\delta [0]$, we recover a Euclidean partition function of a standard relativistic field theory in one dimension lower. This is reminiscent of what we observed at the classical level in Equation \eqref{magnetic scalar}. In the same vein, correlators of $\hat\chi$ are time-independent and can be computed in the usual way: 
\begin{equation}
    \langle \hat \chi (u_1, \vec{x}_1 ) , \ldots \hat \chi (u_n, \vec{x}_n )   \rangle = \frac{\int \mathcal{D}\chi( \vec x)\, \chi ( \vec{x}_1 ) \ldots \chi ( \vec{x}_n) e^{-\beta H[\chi]}}{\int \mathcal{D}\chi( \vec x) e^{-\beta H[\chi]}} .
\end{equation} Correlators of $\pi$ are related to those of $\chi$ via the Heisenberg equations of motion
\begin{equation}
    \frac{d \hat{\pi}}{du} = i [\hat{H}, \hat{\pi}] = -\frac{\delta H [\hat{\chi}]}{\delta \hat \chi}
\end{equation} and therefore involve linear time dependence.

\subsubsection{Electric-magnetic theory}

On the one hand, we have seen that the quantization of electric Carrollian field theories using lattice regulator leads to generalized free field theories. On the other hand, the quantization of magnetic Carrollian field theories leads to interesting, but time-independent higher-point functions. One may try to couple these two types of theories together to get an interesting model with non-trivial time-dependent higher-point functions. An example of such a theory is obtained by coupling an electric scalar with a magnetic scalar field through some Yukawa-type of coupling: 
\begin{equation}
S[\phi, \pi, \chi]
= \int du d\vec{x} \, \left(
\frac{1}{2}\dot{\phi}^2
-
\frac{m^{2}}{2}\phi^{2}
+ \pi \dot{\chi}
-
\frac{1}{2}\partial_A \chi \partial^A \chi
+
\frac{M^{2}}{2}\chi^{2}
+
\lambda \chi\phi^{2} \right) .
\label{eq:lag}
\end{equation} This model was investigated in \cite{Cotler:2024xhb}. As in the pure electric case, using the above lattice regularization, it can be shown that there are only Gaussian correlations, i.e. all the connected higher-point functions vanish.

At this stage, one may wonder whether there exists a non-trivial quantum Carrollian field theory that produces non-Gaussian, time-dependent higher-point functions. This is a central question for Carrollian physics. Indeed, as we shall explain in Section \ref{sec:Carrollian and celestial amplitudes}, the correlators of a putative holographic Carrollian CFT—which encode the bulk massless $\mathcal{S}$-matrix and are referred to as ``Carrollian amplitudes'' (see Section \ref{sec:Definition of Carrollian amplitudes})—are precisely of this type.

In the next section, we will adopt another approach and attempt to define the quantum theory without referring to a Lagrangian. We will focus on Carrollian CFTs, as these are the relevant theories for flat space holography. We will follow a treatment analogous to the standard construction of relativistic CFTs \cite{DiFrancesco:1997nk}. Finally, as we shall see in Section \ref{sec:Carrollian limit of holographic correlators}, the Carrollian limit of CFT correlators leads to non-trivial higher-point functions. This suggests that there may exist a quantization procedure, with $c$ acting as a regulator, that yields non-trivial time-dependent higher-point correlators. We will further comment on this in the discussion of Section \ref{sec:Through the looking glass}.

\section{Elements of Carrollian CFTs}
\label{sec:Elements of Carrollian CFTs}

In the previous section, we discussed several examples of classical Carrollian field theories and briefly examined their quantization using perturbative methods. In this section, we focus on conformal Carrollian field theories, or Carrollian CFTs, namely theories exhibiting the conformal Carrollian algebra (or equivalently the BMS algebra, see Section \ref{sec:Isomorphism between BMS and conformal Carroll}) as a global spacetime symmetry. A natural way to define a quantum Carrollian CFT is to use methods analogous to those developed for relativistic CFTs, which do not rely on a Lagrangian or perturbative techniques. These include the definition of primary operators, the constraints imposed by Ward identities on correlators, the state–operator correspondence, and the definition of OPEs. In what follows, we revisit parts of this framework in the Carrollian context, focusing primarily on the $d=3$ case, which is relevant for holography in four-dimensional asymptotically flat spacetimes.

\subsection{Conformal Carrollian primaries}

\subsubsection{General definition}
\label{sec:General definition}

In this section, we introduce the notion of conformal Carrollian primary in $d$ dimensions \cite{Bagchi:2019xfx, Donnay:2022wvx , Saha:2023hsl,Nguyen:2023vfz,Chen:2023pqf}, which is a field that transforms covariantly under the global conformal Carrollian algebra $\mathfrak{CCarr}_d^{\text{glob}}$ discussed in Section \ref{sec:Conformal Carrollian algebra}. As in the standard case of relativistic conformal fields \cite{Mack:1969rr}, we first look for (irreducible) representations of the stability subgroup of the
origin $x^a = 0$. By looking at \eqref{Carrollian generators repres} and \eqref{repres Carroll K D}, we see that the latter is generated by 
\begin{equation}
    J_{AB}, \,  B_A, \,  K_A, \,  K, \,  D .
\label{stabilizer}
\end{equation}  First, the spatial rotations $J_{AB}$ satisfy the $\mathfrak{so}(d-1)$ algebra, and we are thus led to consider symmetric and traceless spin-$s$ tensor fields $\Phi^{(s)} = (\Phi_{A_1 \ldots A_s})$ transforming as
\begin{equation}
    [J_{AB},{\Phi}^{(s)}(0) ] = \mathcal{S}_{AB}^{(s)}   \Phi^{(s)}(0) 
\label{rotation confCarr}
\end{equation} where $\mathcal{S}_{AB}^{(s)}$ is the $\mathfrak{so}(d-1)$ irreducible spin-$s$ hermitian representation. The boost generator $B_A$ transforms as a vector under $\mathfrak{so}(d-1)$, and its action typically mixes components of a multiplet $\boldsymbol{\Phi} = (\Phi^{(s_1)} \ldots \Phi^{(s_n)})$ with different spins. We write 
\begin{equation}
    [B_A, \boldsymbol{\Phi} (0)] = \mathcal{B}_A  \boldsymbol{\Phi} (0) .
\label{boost confCarr}
\end{equation} An important example of such a Carrollian multiplet is the Carrollian stress tensor introduced in Section \eqref{sec:Carrollian stress tensor}, see Section \ref{sec:Examples Carrollian primaries}. As in standard relativistic CFT, the action of the dilatation operator is taken to be diagonal
\begin{equation}
    [D, \boldsymbol{\Phi}(0)] =  i \Delta \boldsymbol{\Phi} (0)
\label{dilatation confCarr}
\end{equation} where $\Delta$ is the conformal Carrollian dimension. Finally, the special conformal Carrollian transformations annihilate the field and define the primary condition 
\begin{equation}
    [K, \boldsymbol{\Phi} (0)] = 0 , \qquad [K_A, \boldsymbol{\Phi} (0)] = 0 .
\label{sct confCarr}
\end{equation} A Carrollian multiplet $\boldsymbol{\Phi}$ satisfying \eqref{dilatation confCarr}, \eqref{sct confCarr}, and whose components transform as \eqref{rotation confCarr}, \eqref{boost confCarr}, is called a conformal Carrollian multiplet of conformal dimension $\Delta$, and its components are conformal Carrollian primary fields of possibly different spin $s$ but with the same conformal dimension $\Delta$.

Starting from the above representation of the subalgebra \eqref{stabilizer} stabilizing the origin, we can infer the transformation of the conformal Carrollian primary multiplet $\boldsymbol{\Phi}$ at all points $x^a = (u, x^A)$ of the spacetime by using the following transformation under spacetime translations $P_a = (H , P_A)$: 
\begin{equation}
    \boldsymbol{\Phi}(u,x^A) = U \boldsymbol{\Phi}(0) U^{-1}, \qquad U = e^{-i (u H + x^A P_A)} .
\end{equation} To work out the action of an arbitrary generator $X \in \mathfrak{CCarr}_d^{\text{glob}}$, we use 
\begin{equation}
    [X, \boldsymbol{\Phi}(x)] = U [X', \boldsymbol{\Phi}(0)] U^{-1}
\end{equation} where 
\begin{equation}
    X' = U^{-1} X U = \sum_{n=0}^\infty \frac{i^n}{n!} x^{a_1} \ldots x^{a_n} [P_{a_1},[\ldots [P_{a_n}, X]]] .
\end{equation} Using the commutation relations \eqref{Carrollian algebra} and \eqref{conformal Carroll algebra}, this series truncates at order $n=2$ at most, and leads to the following infinitesimal transformations: 
\begin{equation}
    \begin{split}
		&[H,\boldsymbol{\Phi}(u,x^A)]=i\partial_u \boldsymbol{\Phi}(u,x^A) ,  \\ 
		&[P_A,\boldsymbol{\Phi}(u,x^A)]=i\partial_A \boldsymbol{\Phi} (u,x^A) ,  \\ 
		&[J_{AB},\boldsymbol{\Phi}(u,x^A)]=i(-i\mathcal{S}_{AB}-x_A\partial_B+x_B\partial_A) \boldsymbol{\Phi}(u,x^A) ,   \\ 
		&[D,\boldsymbol{\Phi}(u,x^A)]=i(\Delta+u\partial_u+x^A\partial_A) \boldsymbol{\Phi}(u,x^A),  \\ 
		&[K,\boldsymbol{\Phi}(u,x^A)]=(ix^2\partial_u+2x^A\mathcal{B}_A) \boldsymbol{\Phi}(u,x^A) , \\ 
		&[B_A,\boldsymbol{\Phi}(u,x^A)]=(ix_A\partial_u+\mathcal{B}_A) \boldsymbol{\Phi}(u,x^A) , \\
		&[K_A,\boldsymbol{\Phi}(u,x^A)]=i(-2x_A \Delta -2ix^B\mathcal{S}_{AB}+2iu\mathcal{B}_A-2ux_A\partial_u-2x_A x^B\partial_B+x^2\partial_A) \boldsymbol{\Phi}(u,x^A) 
    \end{split}
\end{equation} where $\mathcal{S}_{AB}$ is a (possibly reducible) representation of $\mathfrak{so}(d-1)$. These transformations can be rewritten compactly in terms of the Lie derivative with respect to global conformal Carrollian Killing vectors derived in \eqref{global conf carr Killing vector}. We have 
\begin{align} 
    \delta_\xi \boldsymbol{\Phi} &\equiv - i \left[a H +  b^A B_A  +  k K  +  a^A P_A + \frac{1}{2}  r^{AB} J_{AB} + \lambda D +  k^A K_A , \boldsymbol{\Phi}  \right] \nonumber \\
    &= \left(f\partial_u + \mathcal{Y}^A \partial_A + \frac{i}{2} \partial_{[A} \mathcal{Y}_{B]} \mathcal{S}^{AB} - i \partial^A f \mathcal{B}_A + \frac{\Delta}{d} \partial_a \xi^a  \right) \boldsymbol{\Phi}  \label{transfo field final}\\
    &= \left((\mathcal{T}  + \frac{u}{d-1} \partial_A \mathcal{Y}^A ) \partial_u + \mathcal{Y}^A \partial_A + \frac{i}{2} \partial_{[A} \mathcal{Y}_{B]} \mathcal{S}^{AB} - i \partial^A ( \mathcal{T}  + \frac{u}{d-1} \partial_B \mathcal{Y}^B) \mathcal{B}_A +\frac{\Delta}{d-1} \partial_A \mathcal{Y}^A  \right) \boldsymbol{\Phi} . \nonumber
\end{align}

At this stage, the discussion focused on the global conformal Carrollian algebra. We now carry on and extend the definition of primary to the infinite-dimensional conformal Carrollian algebra $\mathfrak{CCarr}_d$ introduced in Section \ref{sec:Infinite-dimensional enhancement}. For this, we have to prescribe the action of pure supertranslations on the primary multiplet $\boldsymbol{\Phi}$ at the origin $x^a=0$, since they are part of the stability subalgebra. Using the notation \eqref{super expansion}, a natural choice is to replace the first primary condition in \eqref{sct confCarr} by 
\begin{equation}
    [ M^{m_1, \ldots , m_{d-1}} , \boldsymbol{\Phi}(0)] = 0 \qquad \text{for } \sum_{i=1}^{d-1}m_i>1 .
\end{equation} In particular, this is compatible with a non-trivial action of the boosts $B_A$ as in \eqref{boost confCarr}, see \eqref{translation Mm}. Repeating the above procedure for this definition, we can repackage the action of supertranslations as 
\begin{equation}
    [M_{\mathcal{T}}, \boldsymbol\Phi(u,x^A) ] = ( i \mathcal{T} (x^B) \partial_u + \partial^A \mathcal{T}(x^B) \mathcal{B}_A )\boldsymbol\Phi(u,x^A) .
\end{equation} All the other variations remain the same. Hence, the resulting variation of the field under the action of $\mathfrak{CCarr}_d$ is still given by \eqref{transfo field final}, where now the supertranslation parameters $\mathcal{T}$ are arbitrary functions of the spatial coordinates $x^A$, see \eqref{conformal Carrollian vector} for $N=2$. A multiplet transforming as \eqref{transfo field final} under the full $\mathfrak{CCarr}_d$ algebra is a conformal Carrollian primary multiplet. By analogy with the terminology in $2d$ CFT, a multiplet transforming as \eqref{transfo field final} under the global subalgebra $\mathfrak{CCarr}^{\text{global}}_d$ will be sometimes called a quasi-conformal Carrollian primary multiplet.

Notice that the unitary irreducible representations of the $\mathfrak{bms}_4$ algebra were studied in great detail in \cite{Mccarthy:1972ry,McCarthy1972BMS,McCarthy1973BMS,McCarthyCrampin1973BMS,McCarthyCrampin1976BMS}, and their analogues for $\mathfrak{bms}_3$ were constructed in \cite{Barnich:2014kra,Barnich:2015uva,Campoleoni:2016vsh} using the method of induced representations. It is only recently that the induced representations of $\mathfrak{ebms}_4$ were studied in \cite{Ruzziconi:2026isv}. Remarkably, the explicit construction of the wavefunction in position space associated with BMS symmetries with superrotations has been shown to naturally lead to ``extended objects'', such as strings, instead of point particles. It would be interesting to fully embrace the consequences of this in the context of our present discussion, and to include extended objects in the Carrollian CFT framework.

\subsubsection{Examples}
\label{sec:Examples Carrollian primaries}

\paragraph{Scalar field singlet} First we consider the simplest example of a scalar field singlet $\phi$, for which spatial rotations and Carrollian boosts act trivially on the field at the origin, i.e. $[J_{AB}, \phi (0)] = 0 = [B_A, \phi (0)]$. The transformation \eqref{transfo field final} simply reduces to 
\begin{equation}
    \delta_\xi \phi =  \left(\left(\mathcal{T}  + \frac{u}{d-1} \partial_A \mathcal{Y}^A\right) \partial_u + \mathcal{Y}^A \partial_A +\frac{\Delta}{d-1} \partial_A \mathcal{Y}^A  \right) \phi .
\label{scalar field singlet}
\end{equation} For $\Delta = \frac{d-2}{2}$, this coincides with the transformation of an electric Carrollian scalar field discussed in Section \ref{sec:Scalar field}. In particular, under boosts, we recover the transformation of $\phi_0$ in \eqref{Carroll boost transfo} by taking $\mathcal{T} = \vec{v} \cdot \vec{x}$.

\paragraph{Reducible indecomposable representation}
The magnetic Carrollian scalar field discussed in Section
\ref{sec:Scalar field} requires one to consider a pair of scalar fields
$(\phi,\pi)$ which does not form a primary multiplet in the strict
sense defined above. The two fields have conformal dimensions
$\Delta_\phi=\Delta$ and $\Delta_\pi=\Delta+1$ and transform
non-trivially under Carrollian boosts, i.e.
$[B_A,\pi(0)]=i\partial_A\phi(0)$ and $[B_A,\phi(0)]=0$.
Their transformations read
\begin{equation}
\label{reducible indecomposable}
\begin{split}
    \delta_\xi\phi
    &=
    \left(
        \left(\mathcal{T}
        +\frac{u}{d-1}\partial_A\mathcal{Y}^A\right)\partial_u
        +\mathcal{Y}^A\partial_A
        +\frac{\Delta}{d-1}\partial_A\mathcal{Y}^A
    \right)\phi,
    \\
    \delta_\xi\pi
    &=
    \left(
        \left(\mathcal{T}
        +\frac{u}{d-1}\partial_A\mathcal{Y}^A\right)\partial_u
        +\mathcal{Y}^A\partial_A
        +\frac{\Delta+1}{d-1}\partial_A\mathcal{Y}^A
    \right)\pi
    +\partial^A\left(
        \mathcal{T}
        +\frac{u}{d-1}\partial_B\mathcal{Y}^B
    \right)\partial_A\phi.
\end{split}
\end{equation}
For $\Delta=(d-2)/2$, so that
$\Delta_\phi=(d-2)/2$ and $\Delta_\pi=d/2$, these transformations
coincide with those of the magnetic scalar field.
In particular, under Carrollian boosts, we recover
\eqref{transfo of the field under boost}. This representation is
reducible because it contains the scalar-field singlet representation
\eqref{scalar field singlet}, but it is indecomposable because
Carrollian boosts mix $\pi$ with a descendant of $\phi$.

\paragraph{Carrollian stress tensor} Let us consider the Carrollian CFT stress tensor studied in Section \ref{sec:Carrollian stress tensor} on a flat Carrollian background \eqref{flat carrollian background}. After imposing the constraints \eqref{Carrollian stress tensor conditions in flat background} and \eqref{trace condition} implied by conformal Carrollian invariance, the components of the stress tensor can be written as
\begin{equation} \label{Carrollian stress tensor}
    {T^a}_b = \begin{pmatrix}
        \mathcal{M}  &  \mathcal{N}_B \\
        0 & {\mathcal{A}^A}_{B} 
    \end{pmatrix}
\end{equation} with ${\mathcal{A}^A}_{A} = - \mathcal{M}$. It is a conformal Carrollian primary multiplet of dimension $\Delta = d$, whose (independent) components $\mathcal{M}$, $\mathcal{N}_B$, and ${(\mathcal{A}^{TF})^A}_B = \mathcal{A^A}_{B} + \frac{1}{d-1}q_{AB}\mathcal{M}$ transform in the spin $0$, $1$ and $2$ representations of $\mathfrak{so}(d-1)$, respectively. Focusing on the multiplet $(\mathcal{M}, \mathcal{N}_B$) \cite{Barnich:2021dta, Barnich:2022bni,Ruzziconi:2024kzo, Nguyen:2025sqk}, \eqref{transfo field final} becomes 
\begin{equation}
    \begin{split}
        \delta_\xi \mathcal{M} &= \left( f \partial_u + \mathcal{Y}^A \partial_A + \frac{d}{d-1} \partial_A \mathcal Y^A  \right)  \mathcal{M} , \\
        \delta_\xi \mathcal{N}_B &=  \left(f \partial_u + \mathcal{Y}^A \partial_A   +\frac{d}{d-1} \partial_A \mathcal{Y}^A  \right) \mathcal{N}_B +  \partial_{[B} \mathcal{Y}_{C]} \mathcal{N}^C + \frac{d}{d-1} \partial_B f \mathcal{M} .  
    \end{split}
\end{equation} These transformations coincide with the coadjoint representation of the $\mathfrak{bms}_{d+1}$ algebra, see \cite{Barnich:2015uva} for $d=2$ and \cite{Barnich:2021dta} for $d=3$.

\subsubsection{Conformal Carrollian primaries in $d=3$}
\label{sec:Conformal Carrollian primary in d=3}

Let us now specify the definition of conformal Carrollian primary field to the $d=3$ case, as it will be relevant to holographically encode the scattering states in four-dimensional Minkowski space, see Section \ref{sec:Carrollian and celestial amplitudes}. We work with coordinates $x^a = (u, z ,\bar{z})$ where $u$ is the Carrollian time and $x^A = (z, \bar{z})$ are spatial coordinates on the complex plane. The flat Carrollian background \eqref{flat carrollian background} reads as 
\begin{equation}
    n^a \partial_a = \partial_u, \qquad q_{ab} dx^a dx^b = 0 du^2 + 2 dz d\bar{z}, \qquad k_a dx^a = - du .
\end{equation} The relevant symmetry algebra in that case is the extended BMS algebra, $\mathfrak{ebms}_4$, discussed in Section \ref{sec:BMS algebra}. A convenient basis was introduced in \eqref{expansion ebms}. The global conformal Carrollian subalgebra is spanned by $P_{k,l}$ with $(k,l)= (\frac{1}{2}, \frac{1}{2}), (\frac{1}{2}, -\frac{1}{2}), (-\frac{1}{2}, \frac{1}{2}), (-\frac{1}{2}, -\frac{1}{2})$, and $L_m$, $\bar{L}_m$ with $m = -1, 0,1$. These generators can be expressed in terms of those appearing in \eqref{Carrollian algebra} and \eqref{conformal Carroll algebra} through \cite{Bagchi:2016bcd, Bagchi:2023cen}
\begin{align} \label{Map}
&P_{\frac{1}{2}, \frac{1}{2}}  =i H, \quad P_{-\frac{1}{2}, \frac{1}{2}} =i B_{\bar z}, \quad P_{\frac{1}{2}, -\frac{1}{2}} =i B_z, \quad P_{-\frac{1}{2}, -\frac{1}{2}} =iK , \nonumber \\  
&L_{1}= i P_z, \quad L_0=\frac{i}{2}( J_{z \bar z}+D), \quad L_{-1}=  - \frac{i}{2} K_{\bar z} ,  \\ \nonumber
&\bar{L}_{1}= i P_{\bar z}, \quad \bar{L}_{0}=-\frac{i}{2}(J_{z\bar z}-D),\quad \bar{L}_{-1}=-\frac{i}{2} K_z .  
\end{align}
 The enhancement from $\mathfrak{CCarr}_d^{\text{glob}}$ to $\mathfrak{ebms}_4$ is obtained by considering the higher modes in $L_m$, $\bar{L}_m$ and $P_{k,l}$.

Consider a spin $s$ conformal Carrollian primary field $\Phi_{(k,\bar{k})}$, with $\mathcal{B_A}  = 0$ since it is a singlet under Carrollian boosts. This field is characterized by Carrollian weights $(k,\bar{k})$ defined through
\begin{equation} \label{primary 2d 1}
    [L_0 ,{\Phi}_{(k,\bar{k})}(0)] = -k {\Phi}_{(k,\bar{k})}(0), \qquad [\bar{L}_0 , {\Phi}_{(k,\bar{k})}(0)] = -\bar{k} {\Phi}_{(k,\bar{k})}(0) .
\end{equation} 
It satisfies the primary conditions
\begin{equation}
\label{primary 2d 2}
    [L_m , {\Phi}_{(k,\bar{k})}(0)] = 0 , \qquad [\bar{L}_m , {\Phi}_{(k,\bar{k})}(0)] = 0, \qquad \forall m <0 , 
\end{equation} together with 
\begin{equation}
    [P_{r,l}, {\Phi}_{(k,\bar{k})}(0)] = 0, \qquad \forall r,l \le \frac{1}{2}, \qquad r+ l <1 .
\end{equation} 
Here $r,l\leq\frac12$ restricts to supertranslation modes regular at the origin, while $r+l<1$ means that every nonconstant regular mode annihilates $\Phi_{(k,\bar k)}$; in particular, the degree-one modes annihilate it because they are Carrollian boosts. This definition of conformal Carrollian primary field is consistent with the definition given in Section \ref{sec:General definition} in arbitrary dimension, and we have
\begin{equation}
    \Delta = k + \bar{k}, \qquad s =  k-\bar{k}
\end{equation} where the spin $s$ representation of $\mathfrak{so}(2)$ in $d=3$ reduces to $\mathcal{S}_{AB}^{(s)} = is \epsilon_{AB}$. Notice that \eqref{primary 2d 1} and \eqref{primary 2d 2} naturally extend highest-weight representations in $2d$ CFT \cite{Bagchi:2016bcd, Bagchi:2023cen}. 

Using the same procedure as described in Section \ref{sec:General definition}, we can deduce the transformation of the conformal Carrollian primary field $\Phi$ at any point $(u,z,\bar{z})$. In the present case, the translation operator reads as 
\begin{equation}
    U = \exp \left[- u P_{\frac{1}{2},\frac{1}{2}} - z L_{1} - \bar{z} \bar{L}_{1} \right] .
\end{equation} We find 
\begin{equation}
\begin{split}
\label{primary}
	[L_m,\Phi_{(k,\bar k )}(u,z,\bar{z})]&=-z^{1-m}\partial_z \Phi_{(k,\bar k )}(u,z,\bar{z})-(1-m)\Big(k+\frac{u}{2}\partial_u \Big)z^{-m}\Phi_{(k,\bar k )}(u,z,\bar{z}), \\ 
		[\bar{L}_m,\Phi_{(k,\bar k )}(u,z,\bar{z})]&= -\bar{z}^{1-m}\partial_{\bar{z}} \Phi_{(k,\bar k )}(u,z,\bar{z})-(1-m)\Big(\bar{k}+\frac{u}{2}\partial_u \Big)\bar{z}^{-m}\Phi_{(k,\bar k )}(u,z,\bar{z}) , \\ 
		[P_{r,l},\Phi_{(k,\bar k )}(u,z,\bar{z})]&=-z^{\frac{1}{2}-r}\bar{z}^{\frac{1}{2}-l}\partial_u\Phi_{(k,\bar k )}(u,z,\bar{z}) .
	\end{split}
    \end{equation}
    In terms of the conformal Carrollian Killing vectors \eqref{conf Killing vector at Scri} whose parameters admit the mode expansion \eqref{expansion ebms}, we can recombine these transformations as
\begin{equation}
    \delta_\xi \Phi_{(k,\bar{k})} (u,z,\bar{z}) = \left[ \left(\mathcal{T} + \frac{u}{2} (\partial \mathcal{Y} + \bar \partial \bar{\mathcal{Y}} ) \right) \partial_u + \mathcal{Y} \partial + \bar{\mathcal{Y}}\bar{\partial} + k \partial \mathcal{Y} + \bar{k}\bar{\partial} \bar{\mathcal{Y}}  \right] \Phi_{(k,\bar{k})} (u,z,\bar{z})
\label{Carrollian primary 3d}
\end{equation}
which coincides with the definition of conformal Carrollian primary given in \cite{Donnay:2022aba,Donnay:2022wvx,Mason:2023mti}. Again, if $\Phi_{(k,\bar k)}$ transforms as \eqref{Carrollian primary 3d} only under the global conformal Carrollian algebra, we call it a quasi-conformal Carrollian primary. An important technical remark is that $\partial_u$-descendants of these Carrollian primaries are also primaries; indeed, one can check that $\partial_u^m \Phi_{(k,\bar{k})}$ transforms as a Carrollian primary \eqref{Carrollian primary 3d} with shifted weights $(k + \frac{m}{2}, \bar k + \frac{m}{2})$.

\subsection{Stress tensor Ward identities}

As discussed around Equation \eqref{Carrollian stress tensor}, an important primary in the theory is the stress tensor. In Section \ref{sec:Carrollian stress tensor}, we derived the general properties obeyed by a Carrollian CFT stress tensor. We now extend this discussion at a quantum level by writing the Carrollian CFT Ward identities for the stress tensor. We start from the general form of the local infinitesimal Ward identities associated with a Noether symmetry of parameter $\xi$
\begin{equation} \label{starting WI}
    \langle \partial_a j^a_\xi (x) X \rangle =  -i \sum_{j=1}^n \delta^{(3)}(x-x_j) \delta_{\xi_j}\langle X \rangle  .
\end{equation}
Here $j^a_\xi$ is the conserved current operator associated with the symmetry $\xi$, $X = \Phi_1 (x_1) \ldots \Phi_n (x_n)$ is a string of operators, and $\delta_{\xi_j}$ generates the infinitesimal transformation of $\Phi_j$ under $\xi$. We are interested in the case where $\xi$ is a (global) conformal Carrollian symmetry, as given in \eqref{conf Killing vector at Scri}, $j^a_\xi = {T^a}_b \xi^b$ is built out of the Carrollian CFT stress tensor, and $\Phi_j (x_j)$ are conformal Carrollian primaries \eqref{Carrollian primary 3d}.\footnote{For a generalization of the discussion with non-trivial boost parameter $\mathcal{B}_A$, we refer to \cite{Ruzziconi:2024kzo}. In particular, the right-hand side of \eqref{eq:3} generally has a non-trivial contact term.} Note that \eqref{starting WI} is the quantum analogue of \eqref{conservation Carrollian currents}. The Carrollian CFT stress tensor Ward identities read as \cite{Donnay:2022wvx,Saha:2023hsl,Chen:2023pqf}
\begin{align}
\partial_a\langle {T^a}_{b}(u, z , \bar z)X\rangle &=-i\sum\limits_{j=1}^n{{\partial_{b_j}}\langle X\rangle}\delta(u-u_j)\delta^{(2)}(z-z_j)\, ,\label{eq:2}\\
\langle {T^A}_{u}(u, z , \bar z)X\rangle &=0 , \label{eq:3}\\
\langle {T^a}_{a}(u, z , \bar z)X\rangle &=-i\sum\limits_{j=1}^n{\Delta}_j{\langle X\rangle}\delta(u-u_j)\delta^{(2)}(z-z_j)\, ,\label{eq:4}\\
\langle {T^z}_{z}(u, z , \bar z)X\rangle-\langle {T^{\bar{z}}}_{\bar{z}}(u, z , \bar z)X\rangle &=-i\sum\limits_{j=1}^n s_j{\langle X\rangle} \delta(u-u_j)\delta^{(2)}(z-z_j)\, .\label{eq:5}
\end{align} These are the quantum analogues of \eqref{trace condition} and \eqref{Carrollian stress tensor conditions in flat background}. Integrating \eqref{starting WI} over the Carrollian manifold and discarding the boundary term, this implies the invariance of the correlators under symmetry transformations:
\begin{equation}
    \sum_{j=1}^n  \delta_{\xi_j}\langle X \rangle = 0 .\label{integrated form}
\end{equation}

\subsection{Consequences of Ward identities}
\label{sec:Ward identities}

In this section, we present the general form of the two-, three- and four-point functions compatible with the global conformal Carrollian symmetries. Again, we restrict our discussion to three dimensions, but it can easily be extended to any dimension. Using \eqref{Carrollian primary 3d}, the infinitesimal form of the global conformal Carrollian Ward identities \eqref{integrated form} reads explicitly as 
\cite{Chen:2021xkw,Donnay:2022wvx}
\begin{multline}
\label{Carrollian WI}
    \sum_{i=1}^n \Big[ \Big( \mathcal{T}(z_i, \bar{z}_i) + \frac{u_i}{2} (\partial_{z_i}\mathcal{Y}(z_i) + {\partial}_{\bar{z}_i} \bar{\mathcal{Y}}(\bar{z}_i ) )\Big)\partial_{u_i} + \mathcal{Y}(z_i) \partial_{z_i} + \bar{\mathcal{Y}}(\bar{z}_i) {\partial}_{\bar{z}_i}
    \\
     + k_i \partial_{z_i} \mathcal{Y}(z_i) + \bar{k}_i  \partial_{\bar{z}_i} \bar{\mathcal{Y}} (\bar z_i) \Big] \langle \Phi_{(k_1,\bar k_1)}(u_1, z_1, \bar{z}_1) \ldots \Phi_{(k_n,\bar k_n)}(u_n, z_n, \bar{z}_n) \rangle = 0 
\end{multline} where the parameters $(\mathcal{T}(z,\bar z), \mathcal{Y}(z), \bar{\mathcal{Y}}(\bar z))$ are restricted to satisfy the global conditions \eqref{conditions Killing}. More explicitly, we have the three translations $\mathcal{T} = 1$, $\mathcal{Y} = 1$, $\bar{\mathcal{Y}} = 1$, the spatial rotation $(\mathcal{Y}, \bar{\mathcal{Y}})  = (z, -\bar{z})$, the two Carrollian boosts $\mathcal{T} = z, \bar{z}$, the dilatation $(\mathcal{Y}, \bar{\mathcal{Y}})  = (z, \bar{z})$, and the three special conformal Carrollian transformations $\mathcal{T} = z\bar{z}$, $\mathcal{Y} = z^2$, $\bar{\mathcal{Y}} = \bar{z}^2$.

\subsubsection{Two-point function}

The one-point function of a conformal Carrollian primary is required to vanish, $\langle \Phi_{k,\bar{k}} (u,z,\bar{z}) \rangle = 0$, by consistency with \eqref{Carrollian WI}, unless $(k,\bar{k}) = (0,0)$. 

Let us focus on the two-point function. Invariance under translation implies
\begin{equation}
     \langle \Phi_{k_1,\bar{k}_1} (u_1,z_1,\bar{z}_1) \Phi_{k_2,\bar{k}_2} (u_2,z_2,\bar{z}_2)  \rangle = F(u_{12}, z_{12}, \bar{z}_{12})  
\end{equation} where $u_{12} = u_1 - u_2$, $z_{12} = z_1 - z_2$, $\bar{z}_{12} = \bar{z}_1 - \bar{z}_2$ and $F$ is an arbitrary function. Invariance under Carrollian boosts leads to 
\begin{equation}
    z_{12} \partial_{u_{12}} F(u_{12}, z_{12}, \bar{z}_{12}) = 0 = \bar{z}_{12} \partial_{u_{12}} F(u_{12}, z_{12}, \bar{z}_{12}) 
\end{equation} which yields two possible branches of solutions:
\begin{equation}
    F(u_{12}, z_{12}, \bar{z}_{12})  = f(z_{12}, \bar{z}_{12}) + g(u_{12}) \delta^{(2)} (z_{12}).
\end{equation} The first term is time independent and is referred to as the ``magnetic branch''. The second term is time dependent but involves a $\delta$-function with respect to the spatial coordinates; it is referred to as the ``electric branch''. This terminology comes from the fact that this is the typical behaviour for propagators in magnetic and electric Carrollian field theories, respectively, see Section \ref{sec:Quantization}. We now discuss these branches separately, keeping in mind that the most general two-point function could be a linear combination of them.

Let us first look at the magnetic branch. Dilatation invariance implies
\begin{equation} 
    (z_{12} \partial_{z_{12}} + \bar{z}_{12} \partial_{\bar{z}_{12}} + \Delta_1 + \Delta_2)  f(z_{12} ,\bar{z}_{12}) = 0 \quad \Longrightarrow \quad  f(z_{12}, \bar{z}_{12}) = \frac{C^{(m)}_{12}}{z^a_{12} \bar{z}^b_{12}}, \quad a + b  = k_1 + k_2 + \bar{k}_1 + \bar{k}_2
\end{equation} where $C^{(m)}_{12}$ is a constant. Rotation invariance leads to $a - b = k_1 - \bar{k}_1 + k_2 - \bar{k}_2$. Finally, special conformal transformations imply $k_1 = k_2$ and $\bar{k}_1 = \bar{k}_2$. Putting everything together, we have 
\begin{equation}
\label{two-point magnetic}
    \langle \Phi_1 \Phi_2 \rangle = \frac{C^{(m)}_{12}}{z_{12}^{k_1+k_2} \bar{z}_{12}^{\bar{k}_1 +\bar{k}_2}} \delta_{k_1, k_2} \delta_{\bar k_1, \bar{k}_2} .
\end{equation}  Hence the magnetic branch just coincides with the $2$-point function of a $2d$ CFT. This agrees with the observation made earlier at the level of the magnetic Carrollian theories: there is a sense in which these could be reduced to relativistic field theories in one dimension lower, both at the level of Lagrangians (see discussion around Equation \eqref{magnetic scalar}) and the correlation functions (see below \eqref{partition magnetic}).

Now for the electric branch, dilatation invariance implies
\begin{equation}
   (u_{12}\partial_{u_{12}} - 2 + \Delta_1 + \Delta_2 ) g (u_{12}) = 0 \quad \Longrightarrow \quad g ({u_{12}}) = \frac{C^{(e)}_{12}}{u_{12}^{k_1 + k_2 + \bar{k}_1 + \bar{k}_2 - 2}}
\end{equation} where $C^{(e)}_{12}$ is a constant. Rotation invariance leads to $k_1 - \bar{k}_1 + k_2 - \bar{k}_2 = 0$, while special conformal Carrollian invariance does not impose further constraint. Putting all together, we find
\begin{equation}
\label{two-point electric}
     \langle \Phi_1 \Phi_2  \rangle 
     =  \frac{C^{(e)}_{12}}{u_{12}^{k_1 + k_2 + \bar{k}_1 + \bar{k}_2 - 2}} \delta^{(2)}(z_{12}) \delta_{k_1 - \bar{k}_1, \bar{k}_2 - k_2} .
\end{equation} Interestingly, the electric branch can be non zero for Carrollian primaries of different weights, which differs from the relativistic CFT expectation. Only the spins have to be of opposite signs, $s_1 = - s_2$. The forms of \eqref{two-point magnetic} and \eqref{two-point electric} of the two-point function capture the ultra-local nature of Carrollian field theories; correlations between spatially separated points can exist, but they have to be time independent. Time-dependent correlations happen at the same spatial point.

Notice that we only discuss the generic case here; particular values of the Carrollian weights may lead to different behaviours (e.g. the electric branch for $\Delta_1 + \Delta_2 - 2 = 0$ has been discussed in \cite{Donnay:2022wvx}). Furthermore, in the above discussion, we considered $z$ and $\bar{z}$ to be complex conjugate. Relaxing this condition may lead to additional branches. This is relevant in the context of the scattering theory in Kleinian signature, see Section \ref{sec:Three-point function}, where $z$ and $\bar{z}$ are real and independent. In that case a new electric branch can arise \cite{Nguyen:2025sqk}:
\begin{equation} \label{Kleinian sign 2}
     \langle \Phi_1  \Phi_2  \rangle  = \frac{C^{(e)\prime}_{12}}{z^{2 k_1}_{12}} \delta (\bar{z}_{12}) \delta_{k_1, k_2} \delta_{\bar{k}_1 + \bar{k}_2, 1}
\end{equation} together with its complex conjugate.

\subsubsection{Three-point function}

We now investigate the three-point function. As for the two-point function, there exist different branches of three-point functions solving the Ward identities: the time-independent magnetic branch will always exist, while there will be more electric branches associated with different structures of $\delta$-functions \cite{Bagchi:2023cen,Nguyen:2023miw, Nguyen:2025sqk}.

Writing 
\begin{equation}
     \langle \Phi_{k_1,\bar{k}_1} (u_1,z_1,\bar{z}_1) \Phi_{k_2,\bar{k}_2} (u_2,z_2,\bar{z}_2) \Phi_{k_3,\bar{k}_3} (u_3,z_3,\bar{z}_3)  \rangle = F(u_{i}, z_{i}, \bar{z}_{i}) 
\end{equation} where $i,j=1,2,3$, translation invariance implies $F = F (u_{ij}, z_{ij}, \bar{z}_{ij})$. Carrollian boost invariance yields 
\begin{equation}
    ( z_{12} \partial_{u_{12}} + z_{23}  \partial_{u_{23}} + z_{13}  \partial_{u_{13}} ) F = 0 = ( \bar{z}_{12} \partial_{u_{12}} + \bar{z}_{23}  \partial_{u_{23}} + \bar{z}_{13}  \partial_{u_{13}} ) F . 
\end{equation} This leads to several branches defined by their $\delta$-function structure: 
\begin{multline}
    F = f(z_{12}, z_{13}, z_{23}, \bar z_{12}, \bar z_{13}, \bar z_{23} ) + g (u_{12}, z_{23},  z_{13}, \bar{z}_{23},  \bar{z}_{13}) \delta^{(2)}(z_{12}) \\
    +  h (u_{12}, u_{23},  u_{31})  \delta^{(2)}(z_{12}) \delta^{(2)}(z_{23}) 
\end{multline} plus the possible permutation terms (i.e. $\propto \delta^{(2)} (z_{23})$, $\delta^{(2)} (z_{13})$, $\delta^{(2)}(z_{12}) \delta^{(2)}(z_{13})$, $\delta^{(2)}(z_{13}) \delta^{(2)}(z_{23})$). We did not include branches with more $\delta$-functions, such as $\propto \delta^{(2)}(z_{13}) \delta^{(2)}(z_{23}) \delta^{(2)}(z_{31})$, as they are formally divergent with $\delta (0)$. Again, let us discuss the different branches separately. 

The first time-independent branch is the magnetic branch, and a similar discussion as for the two-point function fixes it to be a three-point function of a $2d$ CFT: 
\begin{equation}
     \langle \Phi_1 \Phi_2 \Phi_3  \rangle = \frac{C^{(m)}_{123}}{z_{12}^{k_1+ k_2 - k_3} z_{23}^{k_2+ k_3 - k_1} z_{13}^{k_1+ k_3 - k_2} \bar{z}_{12}^{\bar{k}_1+ \bar{k}_2 - \bar{k}_3} \bar{z}^{\bar k_2+ \bar k_3 -\bar k_1}_{23} \bar{z}_{13}^{\bar k_1+ \bar k_3 - \bar k_2} } .
\end{equation} For the second branch, assuming 
\begin{equation}
    g (u_{12}, z_{23},  z_{13}, \bar{z}_{23},  \bar{z}_{13})  \delta^{(2)}(z_{12})   = \frac{C^{(e,1)}_{123}}{u_{12}^a z_{23}^b  z_{13}^c \bar{z}_{23}^d  \bar{z}_{13}^e}  \delta^{(2)}(z_{12}) =  \frac{C^{(e,1)}_{123}}{u_{12}^a z_{23}^{b+c} \bar{z}_{23}^{d+e} }  \delta^{(2)}(z_{12}) , 
\end{equation} dilatation invariance leads to $a+(b+c)+(d+e)= \Delta_{1}+\Delta_2 + \Delta_3 - 2$. Spatial rotation invariance implies $(b   + c) -( d+ e) = k_1 - \bar{k}_1 + k_2 - \bar{k}_2 + k_3 - \bar{k}_3$. Finally, special conformal Carrollian invariance gives $a = k_1 + \bar{k}_1 + k_2 + \bar{k}_2 -k_3 - \bar{k}_3 -2$, $(b+c) = 2 k_3$, $(d+e) = 2 \bar{k}_3$. Putting all together, we find 
\begin{equation}
   \langle \Phi_1 \Phi_2 \Phi_3  \rangle =  \frac{C^{(e,1)}_{123}}{u_{12}^{k_1+\bar{k}_1 + k_2 + \bar{k}_2 - k_3 - \bar{k}_3-2} z_{23}^{2k_3} \bar{z}_{23}^{2 \bar{k}_3} }  \delta^{(2)}(z_{12}) \delta_{k_3 - \bar{k}_3, k_1 - \bar{k}_1 + k_2 - \bar{k}_2} . 
\end{equation} Finally, similar steps can be applied for the third branch, and we find explicitly 
\begin{equation}
    \langle \Phi_1 \Phi_2 \Phi_3  \rangle = \frac{C^{(e,2)}_{123}}{u_{12}^a u_{23}^b u_{31}^c} \delta^{(2)}(z_{12}) \delta^{(2)}(z_{23}) \label{third branch}
\end{equation} with 
\begin{equation}
    a+b+c = \Delta_1 + \Delta_2 + \Delta_3 - 4  \quad \text{and} \quad s_1 + s_2 + s_3 = 0 . 
\end{equation} 

An important difference compared to relativistic CFTs is that the form of the low-point functions is not completely fixed by symmetries; there is some freedom in the choice of branch. Furthermore, there can also be extra freedom within a given branch, as can be seen from \eqref{third branch} where the powers in the denominator are not completely fixed. 

Of course, as in the two-point function, if one instead considers $z$ and $\bar{z}$ as independent and real variables, additional $\delta$-function structures will be allowed and lead to new branches. For instance, we have
\begin{equation} \label{3-point z bz indep}
    \langle \Phi_1 \Phi_2 \Phi_3 \rangle   =\frac{C^{(e,3)}_{123}\, \delta(\bz_{12})\delta(\bz_{23})}{ (z_{12})^{s_1+s_2-\Delta_3+2}\, (z_{23})^{s_2+s_3-\Delta_1+2}\, (z_{13})^{s_1+s_3-\Delta_2+2}\, (F_{123})^{2(\bar k_1+\bar k_2+\bar k_3-2)}}\,,
\end{equation}
\begin{equation}
    \langle \Phi_1 \Phi_2 \Phi_3 \rangle=C^{(e,4)}_{123}\,\frac{\delta(z_{12})\delta(\zb_{12})\, \delta_{s_3,s_1+s_2}}{u^{\Delta_1+\Delta_2-\Delta_3-2}_{12} z^{2k_3}_{23}\bz^{2\bar k_3}_{23}} ,
\end{equation}
\begin{equation}
    \langle \Phi_1 \Phi_2 \Phi_3 \rangle=C^{(e,5)}_{123}\, \frac{\delta(z_{12})\delta(\bz_{12})\delta(z_{13})\, \delta_{s_1-s_2+\Delta_3,1}}{u^{2(-2+k_1+k_2+k_3)}_{12}\bz^{2\bar k_3}_{13}} .
\end{equation}
In the above expression, we introduced the quantity
\begin{equation}
    F_{123} = u_1 z_{23} + u_2 z_{31} + u_3 z_{12}
\end{equation} which is translation and Carroll boost invariant on the support of $\bar{z}_1 = \bar{z}_2 = \bar{z}_3$.

\subsubsection{Four-point function}

As for the two- and three-point functions, several branches of four-point functions are allowed by the symmetries. In this section, we focus on two of them, which will be relevant within the next sections. Of course, there is the usual magnetic branch, which coincides with the four-point function of a $2d$ CFT: 
\begin{equation} \label{mangetic 4pt}
    \langle \Phi_1 \Phi_2 \Phi_3 \Phi_4  \rangle = g (z, \bar{z}) \prod_{i<j} \frac{1}{z^{a_{ij}}_{ij} {\bar{z}}_{ij}^{\bar{a}_{ij}}}
\end{equation} where $i,j = 1, 2, 3, 4$,  
\begin{equation} \label{aijcft}
    a_{ij} = k_i + k_j - \sum_l k_l /3, \qquad  \bar{a}_{ij} = \bar k_i + \bar k_j - \sum_l \bar k_l /3, 
\end{equation} and $g$ is a function of the $2d$ cross ratios
\begin{equation}
    z = \frac{z_{12}z_{34}}{z_{13}z_{24}}, \qquad  \bar{z} = \frac{\bar z_{12} \bar z_{34}}{\bar z_{13} \bar z_{24}}. 
\end{equation} Another interesting branch, which is of electric type, arises on the support $z= \bar{z}$ and will be of direct relevance for holography, see Section \ref{sec:Four-point function}. The most general ansatz we can make, which involves $\delta (z-\bar{z})$ and is compatible with translation and Carroll boost invariance is provided by
\begin{equation} \label{electric 4pt}
    \langle \Phi_1 \Phi_2 \Phi_3 \Phi_4  \rangle =  \delta(z-\bz) G(z)\prod_{i<j} \frac{1}{(z_{ij})^{a_{ij}} (\bz_{ij})^{\bar{a}_{ij}}(F_{1234})^c}
\end{equation} where $G$ is a function of the $2d$ cross ratio $z$, and 
\begin{equation}
	\label{F1234}
	\begin{split}
	F_{1234}&\equiv u_4-u_1 z \left| \frac{z_{24}}{z_{12}}\right|^2+u_2 \frac{1-z}{z}\left| \frac{z_{34}}{z_{23}}\right|^2-u_3 \frac{1}{1-z}\left| \frac{z_{14}}{z_{13}}\right|^2\\
	&=u_4-u_1\, \frac{z_{34} \bz_{24}}{z_{13}\bz_{12}}+u_2\, \frac{z_{14} \bz_{34}}{z_{12}\bz_{23}}-u_3\, \frac{z_{24}\bz_{14}}{z_{23}\bz_{13}} \\
    &= -u_{14} z \left| \frac{z_{24}}{z_{12}}\right|^2+u_{24} \frac{1-z}{z}\left| \frac{z_{34}}{z_{23}}\right|^2-u_{34} \frac{1}{1-z}\left| \frac{z_{14}}{z_{13}}\right|^2
	\end{split}
	\end{equation} is a combination invariant under translations and Carrollian boosts on the support $z= \bar{z}$ \cite{Mason:2023mti, Nguyen:2025sqk, Kulkarni:2025qcx}. Notice that permutations of $1,2,3,4$ do not lead to independent combinations on the support $z=\bar{z}$. Dilatation and rotation invariance further yield the constraints
\begin{equation}
    \sum_{i<j} (a_{ij}+\bar{a}_{ij})+c=\sum_i \Delta_i\,, \qquad \sum_{i<j} (a_{ij}-\bar{a}_{ij})=\sum_i s_i\,.
\end{equation} Finally, invariance under special conformal Carrollian transformations leads to 
\begin{equation}
	\sum_{i\neq j} a_{ij}=2k_j \quad (j\neq 4)\,, \qquad 
	\sum_{i \neq 4}a_{i4}+c=2k_4\,,
	\end{equation}
	together with the complex conjugate relations. Solving the constraints leads to
\begin{align}
	\label{aij}
	\begin{split}
	a_{ij}&=k_i+k_j- \sum_l k_l /3+c/6\,, \quad (i,j \neq 4)\,,\\
	a_{i4}&=k_i+k_4- \sum_l k_l /3-c/3\,,
	\end{split}
	\end{align}
	together with the complex conjugate relations. Hence, we are left with one free parameter $c$, and all the $u$-dependence in \eqref{electric 4pt} appears through the combination \eqref{F1234}. If $c=0$, the expression \eqref{electric 4pt} reduces to the magnetic branch \eqref{mangetic 4pt} evaluated on the support $z=\bar{z}$.

\subsection{Operator product expansions}

\subsubsection{Reminder on CFT OPEs}

The notion of operator product expansion (OPE) in CFT is absolutely central to identifying the data that define the dynamics of the theory \cite{DiFrancesco:1997nk}. It is the idea that a product of two local operators can be expanded into a sum of local operators 
	\begin{equation}
	\mathcal O_1(x)\, \mathcal O_2 (0)= \sum_{primaries~O_i} C_{12i}(x, \partial)\, \mathcal O_i (0) \,
	\end{equation}
	where the sum is over primary operators and their descendants, encoded in the notation $C_{12i}(x, \partial)$. This statement is made particularly robust in CFT, thanks to the state-operator correspondence, which states that any quantum state can be created from linear combinations of states created by insertions of local operators. Using conformal invariance, the OPE takes the simple form
	\begin{equation}
	\label{standard OPE coincident}
	\mathcal O_1( x)\, \mathcal O_2(0)  = \sum_{i} \frac{c_{12i}}{| x|^{\Delta_1+\Delta_2-\Delta_i}}\,\mathcal O_i(0)+ subleading\,,
	\end{equation}
	where the subleading terms in the $x \to 0$ limit contain derivatives of the primary operators and therefore account for their descendants. The latter are actually completely fixed by conformal symmetry, such that the set of coefficients $\{c_{12k}\}$ carry all the independent data.

\subsubsection{Uniform OPE limit}

One may wonder whether there is a similar notion of OPE in Carrollian CFT. As we have seen above, the structure of Carrollian CFT correlation functions is more involved because of the presence of contact terms that lead to different branches. The same will be true for the OPEs. There are three different types of uniform OPE limits $(u_1, z_1, \bar z_1) \to  (u_2, z_2, \bar z_2)$ compatible with scale invariance:
\begin{itemize}
    \item The first one is just the standard OPE limit where there is no contact term: 
    \begin{equation}
        \Phi_1 (u,z,\bar{z}) \Phi_2 (0,0,0) = \sum_i \frac{c_{12i}}{z^{k_1 + k_2 - k_i -a} \bar z^{\bar k_1 + \bar k_2 -\bar k_i - a} u^{2a}} \Phi_{i}(0) + \ldots
    \end{equation} One can show that invariance under Carrollian boost implies $a=0$ \cite{Nguyen:2025sqk}, hence reducing this expression to a $2d$ CFT OPE at leading order. This is a magnetic-type of Carrollian OPE; 
    \item The second one is obtained on the support $(z_1, \bar z_1) = (z_2, \bar z_2) = (0,0)$:
    \begin{equation}
        \Phi_1 (u,z,\bar{z}) \Phi_2 (0,0,0) = \sum_i \frac{c_{12i} \,  \delta^{(2)}(z)}{u^{\Delta_1 + \Delta_2 - \Delta_i - 2}} \Phi_{i}(0) + \ldots 
    \end{equation} This is an electric-type of Carrollian OPE; 
    \item Considering $z$ and $\bar z$ as real and independent, one can consider additional branches, in the same way we had an additional $2$-point function in \eqref{Kleinian sign 2}: 
    \begin{equation}
        \Phi_1 (u,z,\bar{z}) \Phi_2 (0,0,0) = \sum_i \frac{c_{12i} \,  \delta(\bar z)}{ z^{s_1 + s_2 - s_i + 1} u^{2\bar k_1 + 2 \bar{k}_2  - 2 \bar{k}_i -2}} \Phi_{i}(0) + \ldots
    \end{equation}
\end{itemize} The ``$\ldots$'' terms in the above OPE expansions have been discussed in \cite{ Nguyen:2025sqk} and fixed by requiring consistency with symmetries. The upshot is that Carrollian descendants appear together with Carrollian ascendants, and that primary operators other than those in \eqref{Carrollian primary 3d} also appear in the OPE decomposition. These include, for instance, primary multiplets transforming in reducible but indecomposable representations, which can potentially involve massive Casimirs \cite{Kulp:2024scx,Nguyen:2025sqk}.

Notice that there is no established notion of a state-operator correspondence in Carrollian CFT. Hence, the discussion of Carrollian OPEs is on less solid ground than in standard relativistic CFT. Establishing a state-operator correspondence in Carrollian CFT is an important open question.

\subsubsection{Holomorphic OPE limit}
\label{sec:Holomorphic OPE limit}

In addition, to discuss holographic OPEs in Section~\ref{sec:Collinear limit and holomorphic Carrollian OPEs}, it will be useful to consider the holomorphic OPE limit, namely the limit in which $z_1 \to z_2$ without taking limits on the other coordinates, i.e. keeping $u_{12} \neq 0$ and $\bar{z}_{12} \neq 0$. Again, this is in the same spirit as treating $z$ and $\bar{z}$ as real and independent variables, on which different limits can be taken.

A natural ansatz for the holomorphic limit is provided by \cite{Nguyen:2025sqk}
\begin{equation}
\label{integral OPE ansatz}
\Phi_1(u_1, z_1, \bar z_1) \Phi_2(u_2, z_2, \bar z_2) \stackrel{z_{12}\sim 0}{\approx} \int_0^1 dt \int_0^1 ds F(u_{12}, z_{12}, \bar{z}_{12};t,s) \Phi_3(u_2+t u_{12},z_2,\bz_2+s \bz_{12}),
\end{equation} where the wavy equality means we keep only the most divergent term in the $z_1 \to z_2$ limit. This ansatz can be readily checked to be consistent with translations. Setting $(u_2, z_2, \bar z_2)=(0,0,0)$ without loss of generality, we have
\begin{equation}
\Phi_1(u, z, \bar z) \Phi_2(0)\stackrel{z\sim 0}{\approx} \frac{1}{z^\alpha} \int^1_0 dt \int_0^1 ds F(u,\bz;t,s)\Phi_3(ut,0,s\bz),
\end{equation}
where we have assumed that the leading singularity when $z_1 \to z_2$ goes like $\sim z^{-\alpha}$, where $\alpha$ has to be determined. We did not add terms with derivatives $\partial_u$ or $\bar \partial$ acting on $\Phi_3$ since one can use integration by parts and redefine the function $F(u,\bz;t,s)$ to reabsorb them.

We now constrain the above ansatz with global conformal Carrollian invariance. Dilatation and rotation invariance implies the conditions
\begin{equation}
\label{L0 constraint}
u\partial_u F+2(k_1+k_2-k_3-\alpha)F=0, \qquad
u\partial_u F+2\bz \bar{\partial}F+2(\bar k_1+\bar k_2-\bar k_3)F=0,
\end{equation} Carrollian boost invariance yields the equation
\begin{equation}
\int^1_0 dt \int_0^1 ds\left(\partial_u F \Phi_3+u^{-1}(t-s)F \partial_t \Phi_3 \right)=0
\end{equation} for which a solution is provided by $F=\delta(t-s) f(\bz;t)$. Hence the first equation in \eqref{L0 constraint} fixes $\alpha= k_1 + k_2 - k_3$. Consistency with special conformal Carrollian transformations then implies
\begin{equation}
\int^1_0 dt \left(\bz \bar{\partial}f \Phi_3 +t(1-t) f \frac{d}{d t}\Phi_3+2 \Phi_3 f(\bar{k}_1-t \bar{k}_3)\right)=0,
\end{equation}
which, after integration by parts, gives the following differential equation:
\begin{equation}
\label{diff eq}
\frac{d}{dt}\left(f t(1-t)\right)-2f(\bar{k}_1-t \bar{k}_3)+(\bar{k}_1+\bar{k}_2-\bar{k}_3)f=0.
\end{equation}
Taking into account the constraints \eqref{L0 constraint}, the solution reads as
\begin{equation}
f(\bz;t)=c_{123} \bz^{\bar k_3-\bar k_2-\bar k_1}t^{\bar k_3-\bar k_2+\bar k_1-1}(1-t)^{\bar k_3+\bar k_2-\bar k_1-1}. 
\end{equation}
Notice that the boundary terms in the integration by parts to obtain the differential equation \eqref{diff eq} can be discarded provided $\bar k_3-\bar k_2+\bar k_1>0$ and $\bar k_3+\bar k_2-\bar k_1>0$, which we will assume here. We will also assume that $\bar k_3-\bar k_2-\bar k_1>0$, so that the result is nonsingular as $\bar z\to0$. In summary, the leading term in the holomorphic OPE limit is given by
\begin{equation}
\label{eq:holomorphicOPE}
\Phi_1(u, z, \bar{z})\Phi_2(0)\stackrel{z\sim 0}{\approx} c_{123} \frac{\bar{z}^{\bar{k}_3 - \bar{k}_1 - \bar{k}_2}}{{z}^{k_1 + k_2 - k_3}}\int^1_0 dt \, t^{\bar{k}_3-\bar{k}_2+\bar{k}_1-1}(1-t)^{\bar{k}_3+\bar{k}_2-\bar{k}_1-1} \Phi_3(tu,0,t\bz),
\end{equation}
which is regular in $\bar z$, and possibly singular in $z$. Similar arguments could in principle be used to fix the subleading terms in the OPE limit.

\section{Carrollian holography}
\label{sec:Carrollian and celestial amplitudes}

In the previous section, we discussed the general properties of a Carrollian CFT, which preclude the identification of the key data required to properly define such a theory and to develop a corresponding bootstrap program. In this section, we explain how massless scattering amplitudes and their associated properties in four-dimensional Minkowski spacetime can be recast in terms of data in a three-dimensional Carrollian CFT living at null infinity. In doing so, we present the basic entries of the Carrollian holographic dictionary.

\subsection{Boundary operators}
\label{sec:Boundary operators}

\subsubsection{From bulk to boundary: the large-\textit{r} expansion}
\label{sec:From bulk to boundary}

First, following closely \cite{Donnay:2022wvx}, we show that the boundary value at $\mathscr{I}^\pm$ of a bulk massless spin-$s$ field $\phi^{(s)}_I (X)$ in $4d$ Minkowski spacetime corresponds to a Carrollian primary, as defined in Section \ref{sec:Conformal Carrollian primary in d=3}. Here $I = (\mu_1\mu_2\dots\mu_s)$ is a symmetrized multi-index notation, $\mu_i\in \{0,\dots, 3\}$, and $X^\mu$ are standard Cartesian coordinates. Working in De Donder gauge defined by $\partial^\nu \phi^{(s)}_{\nu\mu_2\dots\mu_s}(X) =0$ and $\eta^{\mu\nu}\phi^{(s)}_{\mu\nu\mu_3\dots\mu_s}(X)=0$, the equations of motion in the free theory reduce to
\begin{equation}
    \partial^\mu \partial_\mu \phi^{(s)}_I(X)=0 .\label{Alembert}
\end{equation} It is convenient to expand the field in Fourier modes. Each mode is labelled by an on-shell null $4$-momentum vector parametrized as 
\begin{equation}
    p^\mu(\omega,w,\bar w) = \epsilon \omega q^\mu(w,\bar w) ,\quad   q^\mu(w,\bar w) = \frac{1}{\sqrt{2}} \Big(1+w\bar w,w+\bar w,-i(w-\bar w),1-w\bar w\Big) 
    \label{q in terms of w bar w future}
\end{equation}
where $\omega>0$ is the light-cone energy, $(w,\bar w)$ the coordinates on the celestial complex plane, and $\epsilon = \pm 1$ if the momentum is outgoing/incoming. The polarization vectors are parametrized as
 \begin{equation}
    \begin{split}
        \varepsilon^{+}_{\mu}(\vec q) &= \partial_w q_\mu = \frac{1}{\sqrt{2}}\big(-\bar w,1,-i,-\bar w\big), \qquad
        \varepsilon^{-}_{\mu}(\vec q) = [\varepsilon^{+}_{\mu}(\vec q)]^* = \partial_{\bar w} q_\mu = \frac{1}{\sqrt{2}}\big(-w,1,i,-w\big).
    \end{split}
    \label{epsilon pola}
\end{equation} and satisfy $\varepsilon_+^\mu \varepsilon^-_\mu = 1$, $q^\mu \varepsilon^\pm_\mu = 0$. The Fourier mode expansion of the spin-$s$ field reads as
\begin{equation}
    \phi^{(s)}_I(X) = \frac{K_{(s)}}{16\pi^3} \sum_{\alpha=\pm}  \int \omega\,d\omega\,d^2 w \Big[a^{(s)}_\alpha(\omega, w,\bar w) \varepsilon_I^{*\alpha}(w,\bar w)\, e^{i \omega q^\mu X_\mu} + a^{(s)}_\alpha(\omega,w,\bar w)^\dagger \varepsilon_I^{\alpha}(w,\bar w)\, e^{-i \omega q^\mu X_\mu} \Big] \label{pre-fourier}
\end{equation}
after choosing a Lorentz-invariant measure in momentum space, using the parametrization \eqref{q in terms of w bar w future}-\eqref{epsilon pola} with convention $\epsilon = 1$, and defining the polarization tensors $\varepsilon^\pm_{\mu_1\dots \mu_s}(w, \bar{w}) = \varepsilon^\pm_{\mu_1}(w, \bar{w})\varepsilon^\pm_{\mu_2}(w, \bar{w})\dots \varepsilon^\pm_{\mu_s}(w, \bar{w})$, which are fully symmetric and transverse tensors, i.e. $q^{\mu_i}\varepsilon^\pm_{\mu_1\dots\mu_i\dots\mu_s} = 0$ for any $i=1,\dots,s$. The overall normalization constant $K_{(s)}$ may depend on the coupling constant for the relevant spin (e.g. $K_{(1)} = e\sqrt{\hbar}$, $K_{(2)} = \sqrt{32\pi G\hbar}$). Ladder operators obey the usual commutation relations
\begin{equation}
    \left[ a^{(s)}_\alpha(\omega,w,\bar w),a^{(s)}_{\alpha'} (\omega',w',\bar w')^\dagger\right] = 16\pi^3 \, \omega^{-1}\delta(\omega-\omega')\,\delta^{(2)}(w-w')\,\delta_{\alpha,\alpha'} \label{commu a adagger}
\end{equation} as induced by the canonical commutation relations of the field \eqref{pre-fourier}. Here $a^{(s)}_+(\omega, w,\bar w)^\dagger$ (resp. $a^{(s)}_-(\omega, w,\bar w)^\dagger$) creates a massless particle of spin $s$ and helicity $J = +s$ (resp. $J=-s$), with energy $\omega$ and a null momentum pointing towards the direction $q^{\mu}(w,\bar w)$.

In Minkowski spacetime, it is convenient to introduce the rectangular Bondi coordinates $\{u , r, z, \bar{z}  \}$ ($u,r \in \mathbb{R}$, $z \in \mathbb{C}$), which are related to the Cartesian coordinates through \eqref{Retarded flat BMS coordinates (appendix)}, and where the Minkowski line element reads as in \eqref{Mink best}. Here, future null infinity $\mathscr I^+$ is reached in the limit $r\to + \infty$ while past null infinity $\mathscr I^-$ is obtained in the limit $r\to - \infty$ (see Figure \ref{FigureBondi}). In particular, there is a natural geometric (but importantly not dynamical) identification between points at $\mathscr I^-$ and points at $\mathscr I^+$ which have the same boundary coordinates $x^a = (u,z,\bar z)$.

Using these coordinates, we wish to push the bulk field \eqref{pre-fourier} to $\mathscr{I}^\epsilon$ by taking $r\to \epsilon \infty$ with $\epsilon = \pm 1$. Using \eqref{Retarded flat BMS coordinates (appendix)} and \eqref{q in terms of w bar w future}, one has $q^\mu X_\mu = -u-r|z-w|^2$. Using the stationary phase approximation
\begin{equation}
\label{stationary phase}
   \lim_{r\to \infty} |z-w| e^{-ir \omega |z-w|^2} =  - \frac{i}{2 r \omega } \delta (|z-w|) + \mathcal{O}(r^{-2})
\end{equation} (see \cite{Donnay:2022wvx} for details), the limit $r\to \epsilon \infty$ implies $z \to w$, i.e. the momentum and the light ray pointing to $\mathscr{I}^{\epsilon}$ become collinear. The leading components of $\phi^{(s)}_I(X)$ near $\mathscr I^\epsilon$ are given by
\begin{equation}
\phi^{(s)\epsilon}_{z\dots z}(X)
=
-\frac{iK_{(s)}}{8\pi^2}r^{s-1}
\int_{0}^{+\infty}d\omega
\left[
a_+^{(s)\epsilon}(\omega,z,\bar z)e^{-i\omega u}
-
a_-^{(s)\epsilon}(\omega,z,\bar z)^\dagger e^{+i\omega u}
\right]
+\mathcal O(r^{s-2}),
\label{Phi s final}
\end{equation}
its complex-conjugate component $\phi^{(s)\epsilon}_{\bar z\dots\bar z}(X)$, as well as components of the form $\phi^{(s)\epsilon}_{ r z \dots z}(X)$, $\phi^{(s)\epsilon}_{r rz\dots z}(X)$, $\ldots$, which are all generically of order $\mathcal{O}(r^{s-1})$. One can extract the boundary value of the field through
\begin{equation}
    \begin{split}
        &\bar{\phi}^{(s)\epsilon}_{z\dots z} (u, z, \bar z)\, d z \otimes \cdots \otimes d z + \bar{\phi}^{(s) \epsilon}_{\bar z\dots \bar z} (u, z, \bar z)\, d \bar z \otimes \cdots \otimes d \bar z \\
        &\equiv \lim_{\bar{r} \to \epsilon \infty} \bar{r}^{1-s} \left( \phi^{(s)}_{\mu_1\dots\mu_s}dx^{\mu_1}\otimes \dots \otimes d x^{\mu_s}\right)\Big|_{r = \bar{r}}  ,
    \end{split}
\label{Boundary value}
\end{equation} 
where the restriction to the hypersurface $r = \bar{r}$ before taking the limit eliminates the radial components. Hence the boundary value at $\mathscr{I}^\epsilon$ can be expressed in terms of the creation/annihilation operators of the bulk field as \cite{He:2014laa,Strominger:2017zoo}
\begin{equation}
\bar\phi^{(s)\epsilon}_{z\dots z} (u,z,\bar z) = -\frac{i K_{(s)}}{8\pi^2} \int_{0}^{+\infty}d \omega  \left[ a_+^{(s)\epsilon}(\omega,z,\bar z) e^{ -i  \omega u} - a_-^{(s)\epsilon}(\omega,z,\bar z)^\dagger e^{+i  \omega u}\right], \label{outgoing spin s}
\end{equation}  while the modes for $\bar \phi^{(s)}_{\bar z \dots \bar z} = (\bar \phi^{(s)}_{z \dots z})^{\dagger}$ are obtained from  \eqref{outgoing spin s} by exchanging $a^{(s)}_\pm \to a^{(s)}_\mp$. We added an index $\epsilon$ on the creation/annihilation operators in \eqref{outgoing spin s}. Indeed, although the incoming/outgoing operators are the same in the free theory, they will differ in the interacting theory. It will be crucial to keep track of this when defining the Carrollian amplitudes.

\subsubsection{From boundary to bulk: the Kirchhoff-d'Adh\'emar formula}

Up to this stage, we have reviewed how a bulk free field induces a boundary field by taking the large-$r$ limit. We now describe how to reconstruct the bulk field from the boundary value, using the Kirchhoff-d'Adh\'emar formula \cite{Penrose1980GoldenON,Penrose:1985bww}. 

Inverting the Fourier transform \eqref{outgoing spin s}, we get
\begin{equation}
    \begin{split}
        a_{+}^{(s)\epsilon}(\omega,z,\bar z) & = \frac{4\pi i}{K_{(s)}} \int_{-\infty}^{+\infty}du \, e^{i  \omega u}\, \bar\phi^{(s)\epsilon}_{z\dots z}(u, z, \bar z) ,\\
        a_{-}^{(s) \epsilon}(\omega,z,\bar z)^{\dagger} & = - \frac{4\pi i}{K_{(s)}} \int_{-\infty}^{+\infty} d u\, e^{-i  \omega u}\, \bar\phi^{(s)\epsilon}_{z\dots z}(u, z, \bar z),
    \end{split} \label{fourier transform formula at scri}
\end{equation}
for $\omega >0$. Inserting these formulae into \eqref{pre-fourier} yields 
\begin{align}
    \phi_I^{(s)}(X) &= \frac{i}{4\pi^2}\int \omega\, d\omega\, d^2 w\, \varepsilon^{*+}_I(w,\bar w)\int_{-\infty}^{+\infty} d\tilde u\left[e^{i \omega(q^\mu X_\mu+\tilde u)} - e^{-i \omega(q^\mu X_\mu+\tilde u)}\right]\bar\phi_{z\dots z}^{(s)\epsilon}(\tilde u,w,\bar w) + \text{h.c.} \nonumber \\
    &= \frac{i}{4\pi^2}\int d^2 w\, \varepsilon^{*+}_I(w,\bar w) \int_{-\infty}^{+\infty} d\tilde u \int_{-\infty}^{+\infty}d\omega\,\omega \, e^{i \omega(q^\mu X_\mu+\tilde u)}\, \bar\phi_{z\dots z}^{(s)\epsilon}(\tilde u,w,\bar w) + \text{h.c.} 
\end{align}
The integrand can be rewritten as a derivative of $\delta$-function, leading to
\begin{align}
    \phi^{(s)}_I (X) &=  \frac{1}{2\pi} \int\,d^2 w\, d \tilde u   \;  \varepsilon^{*+}_I(w,\bar w) \Big[   \partial_{\tilde u}\delta\left(\tilde u+q^\mu X_\mu\right) \bar\phi^{(s)\epsilon}_{z\dots z}( \tilde u, w,\bar w) \Big] + \text{h.c.}
\end{align}
 Integrating out the $\delta$-function, we obtain the Kirchhoff-d'Adh\'emar formula \cite{Penrose:1985bww,Penrose1980GoldenON}
\begin{equation}\label{KA}
    \phi^{(s)}_I (X) =  -\frac{1}{2\pi} \int\,d^2 w  \; \varepsilon^{*+}_I(w,\bar w)   \;  \partial_{\tilde u}\bar\phi_{z\dots z}^{(s)\epsilon} (\tilde u = -q^\mu X_\mu, w,\bar  w)  + \text{h.c.}
\end{equation} 
which allows us to reconstruct the bulk field $\phi^{(s)}_I (X)$ from its boundary value at $\mathscr{I}^+$ or $\mathscr{I}^-$. This free theory statement is holographic by nature, and already points towards a description of the bulk theory in terms of data at null infinity.

\subsubsection{Transformation laws}

In this section, we show that the boundary value of bulk massless field transforms as a conformal Carrollian primary, as defined in Section \ref{sec:Conformal Carrollian primary in d=3}.

Poincaré transformations $X'^\mu = {\Lambda^\mu}_\nu X^\nu + t^\mu$ act on the gauge field as
\begin{equation}
    \phi^{(s)}_{\mu_1\mu_2\dots\mu_s}(X)\mapsto \phi'^{(s)}_{\mu_1\mu_2\dots\mu_s}(X') = {\Lambda_{\mu_1}}^{\nu_1}{\Lambda_{\mu_2}}^{\nu_2}\dots {\Lambda_{\mu_s}}^{\nu_s}\phi^{(s)}_{\nu_1\nu_2\dots \nu_s}(X) \label{lorentz rep}.
\end{equation} A Poincaré transformation in the bulk induces the following transformation at the boundary
\begin{equation} \label{boundary action of poincare}
    u \to u' = \left|\frac{\partial z'}{\partial z}\right| (u - q^\mu (z,\bar z) {\Lambda_\mu}^\nu t_\nu ) , \qquad  z \to z' = \frac{a z +b}{cz + d}
\end{equation} and the complex conjugate transformation for $\bar{z}$. Here $ad-bc = 1$, so that $(z,\bar{z})$ transforms under the Möbius group on the celestial plane. The Jacobian reads explicitly as $\left|\frac{\partial z'}{\partial z}\right| = \frac{1}{|cz+d|^2}$ and we used the parametrization \eqref{q in terms of w bar w future}. The explicit relation between the Lorentz matrix ${\Lambda^\mu}_\nu$ and the Möbius parameters $(a,b,c,d)$ can be found e.g. in \cite{Oblak:2015qia}. Since $p^\mu$ is a Lorentz vector, one has
\begin{equation}
\omega' = \left|\frac{\partial w'}{\partial w}\right|^{-1}\omega,\quad q^\mu(w',\bar w') = \left|\frac{\partial w'}{\partial w}\right|{\Lambda^\mu}_\nu q^\nu(w,\bar w). \label{transfo omega et q}
\end{equation}
 Owing to \eqref{epsilon pola} and \eqref{transfo omega et q}, one can show that, up to an inhomogeneous term that can be re-absorbed by gauge transformation, we have 
\begin{equation}
\varepsilon'^\pm_\mu(w',\bar w') = \left(\frac{\partial w'}{\partial w}\right)^{\mp\frac{1}{2}}\left(\frac{\partial \bar w'}{\partial \bar w}\right)^{\pm\frac{1}{2}}
{\Lambda_\mu}^\nu \varepsilon_\nu^\pm(w,\bar w) . \label{transfo pola conformal}
\end{equation}
 For the expansion \eqref{pre-fourier}, recalling that the integration measure and the plane wave are Lorentz-invariant, we deduce from \eqref{transfo pola conformal} that the transformation of ladder operators under the Poincaré group is  
\begin{equation}
    a_\pm'^{(s)}(\omega',w',\bar w') = \left(\frac{\partial w'}{\partial w}\right)^{-\frac{J}{2}}\left(\frac{\partial \bar w'}{\partial \bar w}\right)^{\frac{J}{2}}\,e^{-i\omega q^{\mu}(w, \bar w) {\Lambda^{\nu}}_{\mu} t_{\nu}}\,a_\pm^{(s)}(\omega,w,\bar w). \label{transfo a poincare}
\end{equation}
In particular, one recovers that the ladder operators are eigenvectors of translations, which is expected since they are assumed to create/annihilate energy eigenstates. The infinitesimal version of \eqref{transfo a poincare} can be obtained by setting $X'^\mu = X^\mu - \epsilon\xi^\mu$, with $\xi^\mu = {\varpi^\mu}_\nu X^\nu +\tau^\mu$ ($\varpi_{\mu\nu} = \varpi_{[\mu\nu]}$) and $w'(w) = w-\epsilon \mathcal Y(w)$, and retaining only the linear terms in $\epsilon$. One concludes that
\begin{equation}
    \begin{split}
        &\delta_{\xi(\mathcal T,\mathcal Y, \bar{\mathcal{Y}})}a_\pm^{(s)}(\omega,w,\bar w) \\
        &= \left[ -i\omega\mathcal T + \mathcal Y\partial + \bar{\mathcal{Y}} \bar{\partial} + \frac{J}{2}\partial \mathcal Y - \frac{J}{2} \bar{\partial} \bar{\mathcal{Y}} -\frac{\omega}{2} (\partial \mathcal Y + \bar{\partial} \bar{\mathcal{Y}})\partial_\omega  \right] a_\pm^{(s)}(\omega,w,\bar w)
    \end{split}
    \label{transfo a poincare infinitesimal}
\end{equation}
where $\xi(\mathcal T,\mathcal Y, \bar{\mathcal{Y}})$ is now parametrized by the function $\mathcal T(w,\bar w) = -q^{\mu}(w,\bar w)\tau_\mu$ and the vector $\mathcal Y(w)\partial + \bar{\mathcal Y} (\bar w)\bar \partial$ on the Riemann sphere, while acting in momentum space covered by coordinates $(\omega,w,\bar w)$.

Using \eqref{transfo omega et q}, the Fourier transform of \eqref{transfo a poincare} gives 
\begin{equation}
\bar\phi'^{(s)\epsilon}_{z\dots z}(u',z',\bar z') = \left(\frac{\partial z'}{\partial z}\right)^{-\frac{1+ J}{2}}\left(\frac{\partial \bar z'}{\partial \bar z}\right)^{-\frac{1- J}{2}} \bar\phi^{(s)\epsilon}_{z\dots z}(u,z,\bar z) \label{bar phi transfo lorentz}
\end{equation}
 where $J=s$. The complex conjugated component $\bar\phi^{(s)}_{\bar z\dots \bar z}$ transforms in the same way but with the flipped helicity $J=-s$. This translates infinitesimally into
\begin{equation}
\begin{split}
    &\delta_{\xi(\mathcal T,\mathcal Y, \bar{\mathcal{Y}})} \bar\phi_{z...z}^{(s)\epsilon}(u, z, \bar z) = \left(\mathcal T + \frac{u}{2}\left(\partial \mathcal{Y} + \bar \partial \bar{\mathcal{Y}}  \right) \right)\partial_u \bar\phi_{z...z}^{(s)\epsilon}(u,z,\bar z) \\
    &\quad + \left(\mathcal{Y} \partial +\bar{\mathcal{Y}} \bar \partial + \frac{1+ J}{2} \partial \mathcal{Y}  + \frac{1- J}{2} \bar \partial \bar{\mathcal{Y}} \right) \bar\phi_{z...z}^{(s)\epsilon}(u, z, \bar z ).
\end{split} \label{Poincare on boundary value}
\end{equation} This reproduces exactly the transformation of a conformal Carrollian primary \eqref{Carrollian primary 3d} under the global conformal Carrollian algebra, with Carrollian weights given by $k = \frac{1+ J}{2}$, $\bar k = \frac{1- J}{2}$. Hence, we see that the boundary values of the bulk fields are naturally identified with operators in a putative dual Carrollian CFT living at null infinity, and we write
\begin{equation}
    \Phi^{\epsilon = +1}_J (u,z,\bar z) \equiv \bar\phi_{z...z}^{(s)\epsilon = + 1}(u, z, \bar z), \qquad  \Phi^{\epsilon = -1}_J (u,z,\bar z) \equiv \bar\phi_{z...z}^{(s)\epsilon = - 1}(u, z, \bar z)^\dagger = \bar\phi_{\bar z... \bar z}^{(s)\epsilon = - 1}(u, z, \bar z)
\label{identification operators}
\end{equation} where the Carrollian weights of $\Phi^\epsilon_J(u,z, \bar z)$ are fixed by 
\begin{equation} \label{fixed weight}
    k = \frac{1+ \epsilon J}{2} , \qquad \bar k = \frac{1- \epsilon J}{2} .
\end{equation} The conformal dimension of these operators is $\Delta = k + \bar{k} =  1$, and the spin is $\epsilon J = k-\bar{k}$, which matches the helicity of the bulk field. An example of such a quasi-primary field with $J=2$ is given by the asymptotic shear $C_{zz}(u, z, \bar z)$ in Section \ref{sec:Asymptotically flat spacetimes}, see in particular Equation \eqref{shear transfo} where the weights can be read. Notice that the inhomogeneous terms in the transformation \eqref{shear transfo} vanish for the global subalgebra, see \eqref{conditions Killing}. In the next section, we show that correlation functions of the Carrollian operators identified in \eqref{identification operators} encode the bulk massless $\mathcal{S}$-matrix.

\subsection{Definition of Carrollian amplitudes}
\label{sec:Definition of Carrollian amplitudes}

In momentum space, we can construct an incoming or outgoing state, respectively denoted as $\ket{\omega,z,\bar z,s,\alpha}$ and $\bra{\omega,z,\bar z,s,\alpha}$, representing a massless particle of light-cone energy $\omega$, spin $s$ and helicity $J = \alpha s=\pm s$ coming from or heading to the point $(z,\bar z)$ on the celestial sphere, by acting on the respective vacua with the ladder operators $a^{(s)\,\text{in/out}}_\alpha(\omega,z,\bar z)$, i.e.
\begin{equation}
\bra{\omega,z,\bar z,s,\alpha} = -  \frac{i K_{(s)}}{4\pi} \bra{0} a_\alpha^{(s)\,\text{out}} (\omega, z, \bar{z} ) , \qquad \ket{\omega,z,\bar z,s,\alpha} =  \frac{iK_{(s)}}{4\pi} a_\alpha^{(s)\,\text{in}}(\omega,z,\bar z)^\dagger |0\rangle .
 \label{energy eigenstates}
\end{equation}
The choice of normalization of energy eigenstates in \eqref{energy eigenstates} is inspired by the relations \eqref{fourier transform formula at scri} and will be justified a posteriori. The scattering amplitudes in momentum space involving $n$ massless particles, $m$ of them being outgoing, are defined as
\begin{equation} \label{usual momentum space amplitude}
    \mathcal A_n (\{ \omega_1, z_1, \bar z_1 \}^{\epsilon_1}_{J_1} ,  \ldots , \{ \omega_n, z_n, \bar z_n \}^{\epsilon_n}_{J_n}) =\braket{\text{out}|\text{in}}_{\text{mom}}
\end{equation} where $\epsilon_1 = \ldots = \epsilon_m = +1$ label the $m$ outgoing states, and $\epsilon_{m+1} = \ldots = \epsilon_n = -1$ label the $n-m$ incoming states. The out/in states are given explicitly by
\begin{equation}
    \bra{\text{out}} = \bra{\omega_{1},z_{1},\bar z_{1},s_{1},\alpha_{1}}\otimes \dots\otimes \bra{\omega_m,z_m,\bar z_m,s_m,\alpha_m}
\end{equation}
and
\begin{equation}
    \ket{\text{in}} = \ket{\omega_{m+1},z_{m+1},\bar z_{m+1},s_{m+1},\alpha_{m+1}}\otimes \dots\otimes |\omega_n,z_n,\bar z_n,s_n,\alpha_n\rangle.
\end{equation}

Alternatively, we can define the asymptotic quantum states directly in position space at null infinity. Outgoing states at $\mathscr{I}^+$ are naturally defined as \cite{Donnay:2022wvx}
\begin{equation}
\begin{aligned}
\bra{u,z,\bar z,s,+} &= \bra{0}\bar\phi^{(s)\epsilon = +1}_{z\dots z}(u,z,\bar z) = \frac{1}{2\pi} \int_{0}^{+\infty} d \omega\, e^{-i\omega u} \langle \omega,z,\bar z,s,+| ,\\
\bra{u,z,\bar z,s,-}  &= \bra{0}\bar\phi^{(s)\epsilon = +1}_{\bar z\dots \bar z}(u,z,\bar z)= \frac{1}{2\pi} \int_{0}^{+\infty} d \omega\, e^{-i\omega u} \langle \omega,z,\bar z,s,-|,
\end{aligned} \label{direct to fourier boundary}
\end{equation} 
after using \eqref{outgoing spin s}. The boundary field $\bar\phi^{(s)}_{z\dots z}(u,z,\bar z)$ creates outgoing spin-$s$ particles with positive helicity and destroys outgoing spin-$s$ particles with negative helicity, while $\bar\phi^{(s)}_{\bar z\dots\bar z}(u,z,\bar z)$ $=$ $\bar\phi^{(s)}_{z\dots z}(u,z,\bar z)^{\dagger}$ acts in the opposite way. The normalization of \eqref{energy eigenstates} allows to represent the creation and annihilation operators in position space \eqref{direct to fourier boundary} without extra normalization factors. Similarly, at $\mathscr{I}^-$, we construct the incoming states in position space as 
\begin{equation}
\begin{aligned}
\ket{u,z,\bar z,s,+} &=  \bar\phi^{(s)\epsilon = -1}_{z\dots z}(u,z,\bar z)^{\dagger}\ket{0} = \frac{1}{2\pi}  \int_{0}^{+\infty} d \omega\, e^{i\omega u} | \omega,z,\bar z,s,+\rangle ,\\
\ket{u,z,\bar z,s,-} &=  \bar\phi^{(s)\epsilon = -1}_{\bar z\dots \bar z}(u,z,\bar z)^\dagger\ket{0} = \frac{1}{2\pi}  \int_{0}^{+\infty} d \omega\, e^{i\omega u} |\omega,z,\bar z,s,-\rangle .
\end{aligned} \label{direct to fourier boundary PAST}
\end{equation}
We then introduce ``position space amplitudes'' at $\mathscr{I}^\pm$, denoted by $\mathcal C_n = \braket{\text{out}|\text{in}}_{\text{pos}}$, which are obtained from the usual momentum representation of the $\mathcal S$-matrix as \cite{Donnay:2022wvx, Mason:2023mti}
\begin{equation}
    \begin{split}
        &\mathcal C_n( \{ u_{1},z_{1},\bar z_{1} \}^{\epsilon_1}_{J_1},\dots ,\{ u_{n},z_{n},\bar z_{n} \}^{\epsilon_n}_{J_n}) \\
        &= \frac{1}{(2\pi)^n}\prod_{k=1}^n \int_0^{+\infty} d\omega_k \, e^{-i \epsilon_k \omega_k u_k}    \mathcal A_n(\{ \omega_1, z_1, \bar z_1 \}^{\epsilon_1}_{J_1} ,  \ldots , \{ \omega_n, z_n, \bar z_n \}^{\epsilon_n}_{J_n}) .
        \label{Carrollian S-matrix}
    \end{split}
\end{equation} Consistently with the identification between the boundary value of the bulk field with Carrollian primaries in the dual theory \eqref{identification operators}, we can re-interpret the position space amplitude as a correlator of these primaries: 
\begin{equation} \label{eq:Carrollian dictionary}
    \langle \Phi_{J_1}^{\epsilon_1} (u_1, z_1, \bar z_1) \ldots \Phi_{J_n}^{\epsilon_n} (u_n, z_n, \bar z_n)    \rangle \equiv  \mathcal C_n( \{ u_{1},z_{1},\bar z_{1} \}^{\epsilon_1}_{J_1},\dots ,\{ u_{n},z_{n},\bar z_{n} \}^{\epsilon_n}_{J_n}) .
\end{equation} By construction, these correlators satisfy the Carrollian CFT Ward identities discussed in Section \ref{sec:Ward identities}. For this reason, we call these amplitudes in position space at $\mathscr{I}$ ``Carrollian amplitudes'' \cite{Mason:2023mti}. 

At this stage, we have shown that the massless $\mathcal{S}$-matrix in flat space can be fully encoded in terms of correlators of Carrollian primaries with conformal dimension $\Delta = 1$. Starting from the scattering amplitude in momentum space, there is a natural way to generalize the formula \eqref{Carrollian S-matrix} to produce correlators of Carrollian primaries with any conformal dimension $\Delta \in \mathbb C$:
\begin{equation}
\begin{split}
    &\langle \Phi_{\Delta_1, J_1}^{ \epsilon_1} (u_1, z_1, \bar z_1) \ldots \Phi_{\Delta_n, J_n}^{\epsilon_n} (u_n, z_n, \bar z_n)    \rangle \\
    &= \frac{1}{(2\pi)^n}\prod_{k=1}^n \int_0^{+\infty} d\omega_k \, \omega^{\Delta_k - 1} e^{-i \epsilon_k \omega_k u_k}    \mathcal A_n(\{ \omega_1, z_1, \bar z_1 \}^{\epsilon_1}_{J_1} ,  \ldots , \{ \omega_n, z_n, \bar z_n \}^{\epsilon_n}_{J_n}). 
\end{split} \label{modified mellin transform}
\end{equation} Indeed, it can be easily shown that this expression transforms as a Carrollian CFT correlator. The above integral transform is called the modified Mellin transform, and was originally proposed in \cite{Banerjee:2018gce,Banerjee:2018fgd,Banerjee:2019prz} to regulate the UV divergences of celestial graviton amplitudes. In this setup, $u_k$ was interpreted as a regulator, and setting $u_k=0$ reproduces the usual Mellin transform defining celestial amplitudes, see Section \ref{sec:From Carrollian to celestial holography}. However, let us emphasize that, as it stands, \eqref{modified mellin transform} is a redundant encoding of the $\mathcal{S}$-matrix, since we have shown that operators with $\Delta = 1$ are sufficient to encode it. For the integer values $\Delta = 1, 2, 3, \ldots$, one can simply recover \eqref{modified mellin transform} from \eqref{Carrollian S-matrix} by taking $\partial_{u_k}$-derivatives. Indeed, from the comment below \eqref{Carrollian primary 3d}, we know that $\Phi_{\Delta, J}^\epsilon = \partial_u^m \Phi_{\Delta = 1, J}^\epsilon$ is still a Carrollian primary, with conformal dimension $\Delta = 1 +m$. In practice, it will be convenient to use the modified Mellin transform \eqref{modified mellin transform}, notably to obtain IR-finite expressions in position space, but keeping in mind this redundancy.

Equations \eqref{identification operators}, \eqref{eq:Carrollian dictionary}, and \eqref{modified mellin transform} constitute the main entries of the Carrollian holography dictionary. Notice that this is a flat-space “extrapolate-type” dictionary, in the sense that boundary correlators are obtained from bulk correlators by pushing the insertions to infinity; see Section~\ref{sec:From bulk to boundary}. This terminology is inspired by AdS/CFT, where CFT correlators can be obtained by extrapolating bulk insertions to the boundary \cite{Susskind:1998dq,Banks:1998dd}, sometimes referred to as the BDHM dictionary (see also \cite{Polchinski:1999ry,Polchinski:2010hw}). However, in AdS/CFT there exists another holographic dictionary based on the identification between bulk and boundary partition functions, in which boundary correlators are obtained by differentiating with respect to boundary sources \cite{Witten:1998qj,Gubser:1998bc}. This is sometimes referred to as the GKP/W dictionary, and the equivalence between these two dictionaries was established in \cite{Harlow:2011ke}. One may then wonder whether there exists an analogue of the GKP/W dictionary in flat space. This question was answered in \cite{Kim:2023qbl,Kraus:2024gso,Kraus:2025wgi}, using a formulation of the $\mathcal{S}$-matrix in terms of a path integral with specified asymptotic boundary data, as originally proposed by Arefeva, Faddeev, and Slavnov \cite{Arefeva:1974jv}. It was shown that this prescription, when expressed in position space, reproduces the Carrollian amplitudes defined in this section, thereby establishing the equivalence between the two dictionaries. Finally, a similar approach was taken in \cite{Ammon:2025avo}, using holographic renormalisation of the scalar field in flat space and a timelike foliation to approach null infinity, and it was also shown to reproduce Carrollian amplitudes.

\subsection{Examples of Carrollian amplitudes}
\label{sec:Example of Carrollian amplitudes}

In this section, we give some examples of tree-level Carrollian amplitudes, and show that they correspond to some electric branches of solutions for the Carrollian CFT Ward identities discussed in Section \ref{sec:Ward identities}. This section is mostly based on \cite{Mason:2023mti,Nguyen:2025sqk}. 

\subsubsection{Two-point function}

The two-point function in momentum space reads as
\begin{equation}
    \mathcal A_2(\{\omega_1,z_1,\bar z_1\}^{-}_{J_1},\{\omega_2,z_2,\bar z_2\}^{+}_{J_2}) = K^2_{J_1,J_2}\,\pi\,\frac{\delta(\omega_1-\omega_2)}{\omega_1}\,\delta^{(2)}(z_1-z_2)\,\delta_{J_1,J_2},
\label{two point amplitude in fourier}
\end{equation} 
where we took the first and the second particle as incoming and outgoing, respectively. Applying two integral transforms as prescribed in \eqref{modified mellin transform}, we find 
\begin{equation}
   \langle \Phi_{\Delta_1, J_1}^{-1} (u_1, z_1, \bar z_1) \Phi_{\Delta_2, J_2}^{+1} (u_2, z_2, \bar z_2) \rangle =     \frac{K^2_{J_1,J_2}}{4\pi}   \frac{\Gamma (\Delta_1 + \Delta_2 - 2)}{(u_{12} + i \varepsilon)^{\Delta_1 + \Delta_2 - 2}}\delta^{(2)}(z_{12})\,\delta_{J_1,J_2} 
\label{2point News}
\end{equation} where the infinitesimal regulator $\varepsilon >0$ was introduced to compute the second integral over $\omega$. This result matches the electric branch of solution of the Carrollian CFT Ward identities for the two-point function, see \eqref{two-point electric}. Notice that for $\Delta_1 + \Delta_2 = 2$, which is the case for operators identified with boundary values as in \eqref{identification operators}, then the Carrollian $2$-point function diverges due to the $\Gamma$ function. This particular case has been discussed in great details in \cite{Donnay:2022wvx}, and requires introducing a $u$-independent IR regulator, which can be checked to be consistent with the Ward identities. 

\subsubsection{Three-point function}
\label{sec:Three-point function}

In the Lorentzian signature, the three-point function vanishes, up to collinear contributions of the form \eqref{third branch}. Instead, to find more interesting branches, it is useful to work in Klein space, which is a split signature spacetime with metric $\eta = \text{diag} \left(-,+,-,+\right)$ and whose conformal boundary is $\scri=\mathbb{R}\times \mathcal{LT}_2$ where now $\mathcal{LT}_2 = S^1 \times S^1 /\mathbb{Z}_2$ is a Lorentzian torus, referred to as the ``celestial torus'' \cite{Atanasov:2021oyu}. Null momenta in Klein space can be parameterized in a manner similar to \eqref{q in terms of w bar w future} as\footnote{This parameterization is reached by Wick rotating the third component of \eqref{q in terms of w bar w future}.}
\begin{equation}
    \label{eq:mompar22}
    p_i^\mu = \e_i \omega_i q_i^{\mu} = \frac{\e_i \om_i}{\sqrt{2}} \left(1+z_i \zb_i, z_i + \zb_i, z_i - \zb_i, 1 - z_i \zb_i \right).
\end{equation}
Here $(z_i, \zb_i)$ are coordinates on a Poincar{\'e} patch of $\mathcal{LT}_2$, $\om_i$ is the energy and $\e_i = \pm 1$ labels the Poincar{\'e} patches which are now connected. For more details on how they relate to global coordinates on $\mathcal{LT}_2$, we refer to \cite{Atanasov:2021oyu,Mason:2022hly}. Null momenta admit a decomposition into real spinor helicity variables 
\begin{align}
    p_{\alpha \dot{\alpha}} \equiv \sigma^{\mu}_{\alpha \dot{\alpha}}p_{\mu} = \sqrt{2}\kappa_{\alpha} \tilde{\kappa}_{\dot{\alpha}}. 
\end{align}
These are defined up to a little group scaling $\kappa \to t \kappa, \tilde{\kappa} \to \frac{1}{t} \tilde{\kappa}$ ($t\in \mathbb R \setminus \lbrace 0 \rbrace$) and for the parametrization in \eqref{eq:mompar22}, we can set
\begin{align}
    \label{eq:shvars}
    \kappa_i = \sqrt{\omega_i} \begin{pmatrix}
        1\\ z_i
    \end{pmatrix}, \qquad  \tilde{\kappa}_i = \sqrt{ \omega_i} \e_i \begin{pmatrix}
        1\\ \zb_i
    \end{pmatrix}.
\end{align} 
Using the standard notation $\mathcal A_3(\{\omega_1,z_1,\bar z_1\}^{\epsilon_1}_{J_1},\{\omega_2,z_2,\bar z_2\}^{\epsilon_2}_{J_2}, \{\omega_3,z_3,\bar z_3\}^{\epsilon_3}_{J_3}) \equiv \mathcal{A}_3(1^{J_1} 2^{J_2} 3^{J_3})$, the form of the three-point scattering amplitude is entirely fixed by the little group scalings and locality of the interaction, which is most conveniently displayed in spinor-helicity variables \cite{Elvang:2015rqa,Badger:2023eqz} as
	\begin{align}
	\label{spinhel3pt}
	\mathcal{A}_3(1^{J_1}2^{J_2}3^{J_3})=\!\!\begin{cases} \braket{12}^{J_3-J_1-J_2}\braket{31}^{J_2-J_1-J_3}\braket{23}^{J_1-J_2-J_3}\delta^{(4)}(\Sigma_i\, p_i),\!\!\! &J_1+J_2+J_3<0\,,\\
	[12]^{-J_3+J_1+J_2}[31]^{-J_2+J_1+J_3}[23]^{-J_1+J_2+J_3}\, \delta^{(4)}(\Sigma_i\, p_i),\!\!\! &J_1+J_2+J_3>0\,,\end{cases}
	\end{align}
	up to an overall coefficient. Using the parametrization \eqref{eq:shvars}, we have $[ij] = \tilde{\kappa}_{i\dot{\alpha}}  \tilde{\kappa}_j^{\dot{\alpha}} = -\epsilon_i \epsilon_j\sqrt{\omega_i\omega_j}\, \zb_{ij}$ and $ \langle i j \rangle = \kappa_i^\alpha  \kappa_{j\alpha} = \sqrt{\omega_i\omega_j}\,z_{ij}$. We disregard amplitudes with $J_1+J_2+J_3 = 0$ as they do not lead to consistent four-particle interactions\footnote{The only exception is the $\phi^3$ interaction, see below.} \cite{McGady:2013sga}. For some combinations of integer spins, it will be necessary to introduce an extra colour index for these amplitudes to be compatible with spin-statistics. For instance, in Yang-Mills theory with gauge group $SU(N)$, corresponding to $J_1=J_2=-J_3 = 1$, each particle transforms in the adjoint representation and the amplitude must be multiplied by the structure constant $f^{abc}$.

    Now, applying the three integral transforms over $\omega$, as prescribed by \eqref{modified mellin transform}, for $J_1+J_2+J_3<0$, and rewriting the momentum conserving $\delta$-function as 
    \begin{equation}
    \delta^{(4)}\left(p_1+p_2+p_3\right) = \frac{1}{\left|\zb_{12}\zb_{13}\right|\omega_1^2} \delta\left(z_{12}\right)\delta\left(z_{23}\right) \delta\left(\om_{2} + \frac{\zb_{13}}{\zb_{23}}\e_1 \e_2 \om_1\right)\delta\left(\om_{3} - \frac{\zb_{12}}{\zb_{23}}\e_1 \e_3 \om_1\right),
\end{equation} we find 
	\begin{equation}
	\label{eq:3ptwithTheta}
	\begin{split}
	\langle \Phi_1 \Phi_2 \Phi_3 \rangle &=\Gamma[2\Sigma_i \bar k_i-4]\, \Theta\left(-\frac{z_{13}}{z_{23}}\epsilon_1\epsilon_2\right)\Theta\left(\frac{z_{12}}{z_{23}}\epsilon_1\epsilon_3\right)\\
	&\times \frac{\delta(\bz_{12}) \delta(\bz_{13}) (z_{12})^{\Delta_3-J_1-J_2-2} (z_{23})^{\Delta_1-J_2-J_3-2} (z_{13})^{\Delta_2-J_1-J_3-2}}{\left(z_{23}\, u_1-z_{13}\, u_2+z_{12}\, u_3 \right)^{2\Sigma_i \bar k_i-4}}\,,
	\end{split}
	\end{equation}
	again up to an overall constant. Here $\Theta(x)$ denotes the step function. The above expression is indeed of the general form \eqref{3-point z bz indep} derived from the Ward identities. The expression for $J_1+J_2+J_3>0$ is obtained by the replacement $z_i \to \bz_i$ and $k_i \to \bar k_i$.
	
It is also useful to provide the explicit expression for the scalar field case ($J_1 = J_2 = J_3 = 0$). Considering a $\phi^3$ interaction, the tree-level Carrollian $3$-point function associated with $\mathcal{A}_3 =\delta^{(4)} (p_1+p_2+p_3)$ reads as 
\begin{equation}
	\label{eq:3ptwithThetascal}
	\begin{split}
	\langle \Phi_1 \Phi_2 \Phi_3 \rangle &=\Gamma[2\Sigma_i \bar k_i-4]\, \Theta\left(-\frac{z_{13}}{z_{23}}\epsilon_1\epsilon_2\right)\Theta\left(\frac{z_{12}}{z_{23}}\epsilon_1\epsilon_3\right)  \frac{\delta(\bz_{12}) \delta(\bz_{13}) (z_{12})^{\Delta_3-2} (z_{23})^{\Delta_1-2} (z_{13})^{\Delta_2-2}}{\left(z_{23}\, u_1-z_{13}\, u_2+z_{12}\, u_3 \right)^{2\Sigma_i \bar k_i-4}}\,. 
	\end{split}
	\end{equation}

\subsubsection{Four-point function}
\label{sec:Four-point function}

At tree-level the only non-zero $4$-point gluon and graviton amplitudes are the MHV (``maximal helicity violating'') amplitudes, which are given by\footnote{For Yang-Mills theory, we are working with colour-ordered amplitudes and as in the $3$-particle case, we are suppressing the colour indices on all the particles. For more details, we refer the reader to the review \cite{Mangano:1990by} and references therein.} \cite{Elvang:2015rqa}
	\begin{align}
	\mathcal A^{\text{YM}}_4(1^{+1}2^{-1}3^{-1}4^{+1})&=\frac{\langle 23 \rangle^4}{\langle 12 \rangle \langle 23\rangle \langle 34\rangle \langle 41\rangle} \delta^{(4)}(\Sigma_i\, p_i) =\frac{\omega_2\omega_3}{\omega_1\omega_4} \frac{z_{23}^3}{z_{12}z_{34}z_{41}} \delta^{(4)}(\Sigma_i\, p_i)\,,\\
	\mathcal A_4^{\text{GR}}(1^{+2}2^{-2}3^{-2}4^{+2})&=\frac{\langle 23 \rangle^7 [23]}{\langle 13\rangle \langle 34\rangle \langle 12 \rangle \langle 24 \rangle \langle 14 \rangle^2} \delta^{(4)}(\Sigma_i\, p_i) =\frac{(\omega_2 \omega_3)^3}{(\omega_1\omega_4)^2}\frac{z_{23}^7 \bz_{23}}{z_{13}z_{34}z_{12}z_{24}z_{14}^2} \delta^{(4)}(\Sigma_i\, p_i)\,.
	\end{align}
Writing the momentum conserving $\delta$-function as 
\begin{multline}
\label{delta function explicit}
      \delta^{(4)}\left(p_1+p_2+p_3+p_4\right) = \frac{1}{4\om_4\left|z_{24}\zb_{13}\right|^2} \delta\left(\om_1 + z \left|\frac{z_{24}}{z_{12}}\right|^2 \e_1 \e_4 \om_4 \right)\\
      \times \delta\left(\om_2 - \frac{1-z}{z} \left|\frac{z_{34}}{z_{23}}\right|^2 \e_2 \e_4 \om_4 \right) \delta\left(\om_3 + \frac{1}{1-z} \left|\frac{z_{14}}{z_{13}}\right|^2 \e_3 \e_4 \om_4 \right) \delta\left(z-\zb\right)
\end{multline}
where $z = \frac{z_{12}z_{34}}{z_{13}z_{24}}$ is the $2d$ cross ratio and $\left|z_{ij}\right|^2 = z_{ij} \zb_{ij}$, we can apply the four integral transforms as in \eqref{modified mellin transform} to obtain the general form 
\begin{multline}
	\label{C4general}
 \langle \Phi_1 \Phi_2 \Phi_3 \Phi_4 \rangle =   \Theta\left(-z\left|\frac{z_{24}}{z_{12}}\right|^2 \epsilon_1 \epsilon_4  \right)\Theta\left(\frac{1-z}{z} \left|\frac{z_{34}}{z_{23}}\right|^2\epsilon_2 \epsilon_4  \right)\Theta\left(-\frac{1}{1-z}\left|\frac{z_{14}}{z_{13}}\right|^2\epsilon_3 \epsilon_4 \right)\\ \times \Gamma[c]\, \prod_{i<j} \frac{G(z)\delta(z-\bz)}{(z_{ij})^{a_{ij}} (\bz_{ij})^{\bar{a}_{ij}}(iF_{1234})^c} .
\end{multline} This four-point function is of the general form of the electric branch \eqref{electric 4pt} satisfying the Carrollian CFT Ward identities. The exact expressions of $G(z)$ and $c$ (and therefore of $a_{ij}$ and $\bar{a}_{ij}$ through \eqref{aij}) depend on the theory. For MHV gluon and graviton amplitudes, we find
\begin{equation} \label{Gz for gluon and gravitons}
\begin{split}
	G^{\text{YM}}_{+--+}(z)&=z^{-1/3} (1-z)^{5/3}\,, \qquad c^{\text{YM}}=\Sigma_i \Delta_i -4\,,\\
	G^{\text{GR}}_{+--+}(z)&=z^{-2/3} (1-z)^{10/3}\,,   \qquad c^{\text{GR}}=\Sigma_i \Delta_i -2\,,
\end{split}
	\end{equation} respectively. For the scalar field involving a $\phi^4$ interaction, the Carrollian amplitude associated with the tree-level contact diagram $\delta^{(4)} (\sum_i p_i)$ is given explicitly by 
\begin{equation} \label{4pt contact diagram Carroll}
	\begin{split}
	\langle \Phi_1 \Phi_2 \Phi_3 \Phi_4 \rangle_c &=\delta(z-\bz)\Theta\left(-z\left|\frac{z_{24}}{z_{12}}\right|^2 \epsilon_1 \epsilon_4  \right)\Theta\left(\frac{1-z}{z} \left|\frac{z_{34}}{z_{23}}\right|^2\epsilon_2 \epsilon_4  \right)\Theta\left(-\frac{1}{1-z}\left|\frac{z_{14}}{z_{13}}\right|^2\epsilon_3 \epsilon_4 \right)\\
	&\times \frac{z^{\Delta_1-\Delta_2} (1-z)^{\Delta_2-\Delta_3}}{|z_{13}z_{24}|^2} \left|\frac{z_{24}}{z_{12}} \right|^{2(\Delta_1-1)}\left|\frac{z_{34}}{z_{23}} \right|^{2(\Delta_2-1)} \left|\frac{z_{14}}{z_{13}} \right|^{2(\Delta_3-1)}\frac{\Gamma[\Sigma \Delta-4]}{(iF_{1234})^{\Sigma \Delta-4}}
	\end{split}
    \end{equation}
    so that 
    \begin{equation}
	G(z)=\left[z(1-z)\right]^{2/3}\,, \qquad c=\Sigma_i \Delta_i -4\,
	\end{equation} in \eqref{C4general}. Similarly, the Carrollian amplitude associated with the four-point $s$-channel exchange diagram $\mathcal{A}_{4,e} =  \frac{1}{s} \delta^{(4)}( p_1 + p _2 + p_3 + p_4)$ is 
\begin{equation}
    \begin{split}
	\langle \Phi_1 \Phi_2 \Phi_3 \Phi_4 \rangle_e &=\delta(z-\bz)\Theta\left(-z\left|\frac{z_{24}}{z_{12}}\right|^2 \epsilon_1 \epsilon_4  \right)\Theta\left(\frac{1-z}{z} \left|\frac{z_{34}}{z_{23}}\right|^2\epsilon_2 \epsilon_4  \right)\Theta\left(-\frac{1}{1-z}\left|\frac{z_{14}}{z_{13}}\right|^2\epsilon_3 \epsilon_4 \right)\\
	&\times \frac{z^{\Delta_1-\Delta_2} (1-z)^{\Delta_2-\Delta_3-1}}{|z_{13}|^2 |z_{24}|^4} \left|\frac{z_{24}}{z_{12}} \right|^{2(\Delta_1-1)}\left|\frac{z_{34}}{z_{23}} \right|^{2(\Delta_2-2)} \left|\frac{z_{14}}{z_{13}} \right|^{2(\Delta_3-1)}\frac{\Gamma[\Sigma \Delta-6]}{(iF_{1234})^{\Sigma \Delta-6}} .
	\end{split} \label{four-point exchange}
\end{equation} The $t$- and $u$-channels are obtained in a similar way ($s = (p_1 + p_2)^2$, $t = (p_1 + p_3)^2$ and $u = (p_1 + p_4)^2$ are the Mandelstam variables).

Finally, let us mention that $n$-point MHV gluon and graviton amplitudes have been written as Carrollian correlators in \cite{Mason:2023mti}. Superstring scattering amplitudes have been rewritten as Carrollian amplitudes in \cite{Stieberger:2024shv}. A generalization of the above discussion in general dimension was recently presented in \cite{Kulkarni:2025qcx}.

\subsection{Soft limit and Carrollian CFT stress tensor Ward identities}
\label{sec:Soft limit and Carrollian CFT stress tensor Ward identities}

\subsubsection{Soft theorems for Carrollian amplitudes}

It is interesting to reformulate statements about massless scattering amplitudes in the bulk in the language of a Carrollian CFT defined at null infinity. Let us begin with the soft graviton theorems, which relate $(n+1)$-point scattering amplitudes involving a graviton that becomes soft (i.e. with energy $\omega \to 0$) to $n$-point amplitudes without the graviton insertion. More explicitly, the amplitudes satisfy the leading \cite{Weinberg:1965nx} and subleading \cite{Cachazo:2014fwa} soft graviton theorems which, in the above parametrization, read as 
\begin{equation} \label{leading soft omega}
  \lim_{\omega \to 0} \omega \mathcal{A}_{n+1} (\{\omega, z, \bar{z}\}, \{\omega_i, z_i, \bar{z}_i\}^{\epsilon_i}_{J_i})  = -\left( \sum_{i=1}^n \frac{\epsilon_i \omega_i 
(\bar z -  \bar z_i)}{z - z_i}  \right) \mathcal{A}_{n}  ( \{\omega_i, z_i, \bar{z}_i\}^{\epsilon_i}_{J_i} ) 
\end{equation}
and
\begin{multline} \label{subleading soft omega}
\lim_{\omega \to 0} (1 + \omega \partial_\omega) \mathcal{A}_{n+1} (\{\omega, z, \bar{z}\}, \{\omega_i, z_i, \bar{z}_i\}^{\epsilon_i}_{J_i})  \\
= -\left( \sum_{i=1}^n \frac{(\bar z- \bz_i)^2 \bar \partial_i+ ( \epsilon_i J_i + \omega_i \partial_{\omega_i}) (\bz -\bz_i) }{z - z_i}  \right)  \mathcal{A}_{n}  ( \{\omega_i, z_i, \bar{z}_i\}^{\epsilon_i}_{J_i} ) ,
\end{multline} respectively. We now rewrite these statements in position space at null infinity by applying the integral transforms \eqref{Carrollian S-matrix} or \eqref{modified mellin transform}. Defining the leading and the subleading soft graviton operators \cite{Strominger:2017zoo}, 
\begin{equation}
    {S}_0^+(z,\bar{z}) = \lim_{\omega \to 0^+} \frac{\omega}{2} [ a^{\text{out}}_+ (\omega) - a^{\text{in}}_- (\omega)^\dagger  ] \, , \quad S_1^+(z,\bar{z}) = \lim_{\omega \to 0^+} \frac{i}{2} (1 + \omega \partial_{\omega}) [a^{\text{out}}_+ (\omega) + a^{\text{in}}_- (\omega)^\dagger] \, ,
\label{leading and subleading soft operator}
\end{equation}
the leading soft theorem reads as 
\begin{equation}
    \langle i S^+_0 (z,\bar{z}) \prod_{i=1}^n \Phi^{\epsilon_i}_{(k_i,\bar{k}_i)} (u_i, z_i, \bz_i)   \rangle = \left(\sum_{i=1}^n    \frac{\bar z- \bar z_i}{z- z_i} \partial_{u_i} \right)  \langle  \prod_{i=1}^n \Phi^{\epsilon_i}_{(k_i,\bar{k}_i)} (u_i, z_i, \bz_i)  \rangle  
    \label{leading soft them position}
\end{equation} and the subleading soft theorem as
\begin{multline} \label{subleading soft graviton theorem}
    \langle i S_1^+ (z,\bar{z})  \prod_{i=1}^n \Phi^{\epsilon_i}_{(k_i,\bar{k}_i)} (u_i, z_i, \bz_i)   \rangle  \\
    = \left( \sum_{i=1}^n \frac{(\bar z-\bar z_i)^2 \bar \partial_i+ (2 \bar k_i + u_i \partial_{u_i}) (\bar z_i- \bar z) }{z - z_i}  \right) \langle \prod_{i=1}^n \Phi^{\epsilon_i}_{(k_i,\bar{k}_i)} (u_i, z_i, \bz_i)   \rangle . 
\end{multline}  In \cite{Ruzziconi:2024kzo}, it was shown that the above relations follow from the Carrollian CFT stress tensor Ward identities \eqref{eq:2}-\eqref{eq:5} in the soft limit, which in position space at $\mathscr{I}$, and for fixed values of Carrollian weights \eqref{fixed weight} for the inserted primaries, corresponds to the $|u| \to \infty$ limit. Let us briefly repeat the steps here.

\subsubsection{Leading soft theorem}

Let us first reproduce \eqref{leading soft them position}. Subtracting the spatial divergence of \eqref{eq:3} from \eqref{eq:2}$_{b=u}$, we obtain: 
\begin{align}
\partial_u\langle {T^u}_{u}(u,z, \bz)X\rangle=-i\sum\limits_{j=1}^n\delta(u-u_j)\delta^{(2)}(z-z_j)\partial_{u_j}\langle X\rangle .
\end{align} 
Choosing the `retarded' initial condition\footnote{\label{foot:adv}Alternatively, one could choose the `advanced' initial condition $\langle {T^u}_{u}(u\rightarrow+\infty,z, \bz)X\rangle=0$, which would affect the signs.} $
\langle {T^u}_{u}(u\rightarrow-\infty,z, \bz)X\rangle=0$,
the solution to the above temporal partial differential equation is found to be
\begin{align}
\langle {T^u}_{u}(u,z, \bar z)X\rangle=-i\sum\limits_{j=1}^n\text{ }\theta(u-u_j)\delta^{(2)}(z-z_j)\partial_{u_j}\langle X\rangle \, \label{eq:7}
\end{align} where $\theta(u)$ is the temporal  Heaviside function. We wish to convert all the contact-term singularities in \eqref{eq:7} into pole singularities while avoiding branch-cuts. For this purpose, we note that
\begin{align}
\langle {T}^u_{\hspace{1.5mm}u}(u,z, \bar z)X\rangle&=-\frac{i}{\pi}\sum\limits_{j=1}^n\text{ }\theta(u-u_j)\text{ }{\bar{\partial}}^2\left[\frac{\bar{z}-\bar{z}_j}{z-z_j}\right]\partial_{u_j}\langle X\rangle\label{eq:8}\\
&=-\frac{i}{\pi}\sum\limits_{j=1}^n\text{ }\theta(u-u_j)\text{ }{{\partial}}^2\left[\frac{z-z_j}{\bar{z}-\bar{z}_j}  \right] \partial_{u_j} \langle X\rangle\label{eq:9}
\end{align}
where we used $\partial (\frac{1}{\bar{z}}) = \pi \delta^{(2)} (z)$ and its complex conjugate. It is then suggestive to introduce the Carrollian operators $S^\pm_0 (u,z, \bar z)$ \cite{Saha:2023hsl} (where $\pm$ will coincide later with the bulk helicity) by respectively inverting the ${\bar{\partial}}^2$ operator in \eqref{eq:8} and the $\partial^2$ operator in \eqref{eq:9} as
\begin{align}
&S_0^+(u,z,\bar{z}):=\int d^2 z^\prime\frac{\bar{z}-\bar{z}^\prime}{z-z^\prime}\text{ }{T^u}_{u}(u,z^\prime , \bar{z}^\prime)\hspace{2.5mm}\Longrightarrow\hspace{2.5mm}\bar{\partial}^2S_0^+=\pi {{T}^u}_{u} \, ,\label{Tuu component}\\
&S_0^-(u,z,\bar{z}):=\int d^2 z^\prime \frac{z-z^\prime}{\bar{z}-\bar{z}^\prime}\text{ }{T}^u_{\hspace{1.5mm}u}(u,z^\prime , \bar{z}^\prime)\hspace{2.5mm}\Longrightarrow\hspace{2.5mm}{\partial}^2S_0^-=\pi {T^u}_{u} \, .
\end{align}
The scaling dimension $\Delta$ and the spin $s$ of the fields $S^\pm_0$ are $(\Delta,s)=(1,\pm 2)$. By construction, the fields $S^\pm_0$ are the $2d$ shadow transforms (on $S^2$) of each other. Since a field and its shadow
cannot both be treated as local fields in a theory \cite{Banerjee:2022wht, Saha:2023hsl}, only one among $S^\pm_0$ is to be chosen as a local field. Following \cite{Ruzziconi:2024kzo}, we opt to treat $S^+_0$ as the local field while relegating $S^-_0$ as its non-local shadow. This amounts to focus on the holomorphic sector of the $3d$ Carrollian CFT. As a consequence of \eqref{eq:8}, we have 
\begin{align}
&\langle S^+_0(u,z,\bar{z})X\rangle=-i\sum\limits_{j=1}^n\theta(u-u_j)\frac{\bar{z}-\bar{z}_j}{z-z_j}\partial_{u_j}\langle X\rangle\label{3}
\\
\Longrightarrow\hspace{2.5mm}& {\partial_u}\langle S^+_0(u,z,\bar{z})X\rangle=\left[\text{temporal contact terms}\right] \, . \label{2}
\end{align}
Taking $u\to+\infty$ in \eqref{3}, we reproduce exactly the leading soft graviton theorem \eqref{leading soft them position} written in position space at $\mathscr{I}$, provided 
\begin{equation}
    \lim_{u \to +\infty} S^+_0 (u,z,\bar{z}) = {S}_0^+ (z,\bar{z})
    \label{identific1}
\end{equation} where the leading soft graviton operator $S_0^+ (z,\bar{z})$ was defined in \eqref{leading and subleading soft operator}. A similar relation would hold for $S^-_0$. This discussion shows explicitly that the soft limit $\omega \to 0$ coincides with the $u\to \infty$ limit via the integral transform \eqref{modified mellin transform} at fixed value of $\Delta_k$.

\subsubsection{Subleading soft theorem}

We now show that the large time behaviour of the Carrollian CFT stress tensor Ward identities also reproduces \eqref{subleading soft graviton theorem}. The Ward identities \eqref{eq:4}, \eqref{eq:5} and \eqref{eq:7} can be linearly combined into the following form: 
\begin{equation}
\langle {T^z}_{z}(u,z,\bz)X\rangle+\frac{1}{2}\langle {T^u}_{u}(u,z,\bz)X\rangle=-i\sum\limits_{j=1}^n {k}_j{\langle X\rangle}\delta(u-u_j)\delta^{(2)}(z-z_j) \, ,
\end{equation} which implies
\begin{align}
 i\langle {T^z}_{z}(u,z,\bz)X\rangle 
&=\sum\limits_{j=1}^n\left[\delta(u-u_j)\delta^{(2)}(z-z_j)k_j-\frac{\theta(u-u_j)}{2}\delta^{(2)}(z-z_j)\partial_{u_j}\right]{\langle X\rangle}  .
\end{align}
Thus, subtraction of $\partial_z\langle T^z_{\hspace{1.5mm}z}(\mathbf{x})X\rangle$ from \eqref{eq:2}$_{b=z}$ leads to
\begin{align}
&\partial_{\bar{z}}\langle {T^{\bar{z}}}_{z}(u,z,\bz)X\rangle+\partial_u\langle {T^u}_{z}(u,z,\bz)X\rangle 
\\&=-i\sum\limits_{j=1}^n\Big[\delta(u-u_j)\left\{\delta^{(2)}(z-z_j)\partial_{z_j}-k_j\partial_{z}\delta^{(2)}(z-z_j)\right\} +\frac{\theta(u-u_j)}{2}\partial_z\delta^{(2)}(z-z_j)\partial_{u_j}\Big]{\langle X\rangle} \, . \nonumber
\end{align}
Choosing a `retarded' initial condition, $\langle {T^u}_{z}(u\rightarrow-\infty,z, \bar z)X\rangle=0$, we obtain the following solution to the above temporal partial differential equation:
\begin{equation}
\begin{split}
&\langle {T^u}_{z}(u,z,\bz)X\rangle+\int^u_{-\infty}du^\prime\partial_{\bar{z}}\langle {T^{\bar{z}}}_{z}(u^\prime,z,\bz)X\rangle \\
&=-i\sum\limits_{j=1}^n\theta(u-u_j)\Big[\delta^{(2)}(z-z_j)\partial_{z_j}-k_j\partial_{z}\delta^{(2)}(z-z_j)+\frac{u-u_j}{2}\partial_z\delta^{(2)}(z-z_j)\partial_{u_j}\Big]{\langle X\rangle} \, . \label{a13}
\end{split}
\end{equation}
The complex-conjugate version can be found in a similar way. Next, we extract a $\partial^3$-derivative from the right-hand side of \eqref{a13} to convert contact terms into pole singularities. The conformal Carrollian fields $S^\pm_1$ were defined by inverting these derivatives in \cite{Saha:2023hsl} as below: 
\begin{align}
&S_1^-(u,z,\bar{z})=\int d^2{z^\prime}\frac{{(z-z^\prime)}^2}{\bar{z}-\bar{z}^\prime}\left[{T^u}_{z}(u,z^\prime , \bar{z}^\prime)+\int_{-\infty}^u du^\prime\partial_{\bar{z}^\prime}{T^{\bar{z}}}_{z}(u^\prime,z^\prime , \bar{z}^\prime)\right]\label{47} \, ,\\
&S_1^+(u,z,\bar{z})=\int d^2{z^\prime}\frac{{(\bar{z}-\bar{z}^\prime)}^2}{{z}-{z}^\prime}\left[{T^u}_{\bar{z}}(u,z^\prime , \bar{z}^\prime)+\int_{-\infty}^u du^\prime\partial_{{z}^\prime} {T^{{z}}}_{\bar{z}}(u^\prime,z^\prime , \bar{z}^\prime)\right] \, .\label{46}
\end{align}
The dimensions of the fields $S^\pm_1$ are $(\Delta,s)=(0,\pm2)$. It was shown in \cite{Saha:2023hsl}, following \cite{Banerjee:2022wht}, that the fields $S^+_0$ and $S^-_1$ are not mutually local, while $S^+_0$ and $S^+_1$ can be treated as local simultaneously. Then the $S^+_1$ Ward identity is:
\begin{align}
&\langle S_1^+(u,z,\bar{z}) X\rangle=-i\sum\limits_{j=1}^n\theta(u-u_j)\Big[\frac{(\bar{z}-\bar{z}_j)^2}{z-{z}_j}\partial_{\bar{z}_j}-2\bar{k}_j\frac{\bar{z}-\bar{z}_j}{z-{z}_j} +(u-u_j)\frac{\bar{z}-\bar{z}_j}{z-{z}_j}\partial_{u_j}\Big]\langle X\rangle \label{10}\\
\Longrightarrow\hspace{5mm}& {\partial_u}\langle S^+_1(u,z,\bar{z})X\rangle-\langle S^+_0(u,z,\bar{z})X\rangle=\left[\text{temporal contact terms}\right] \, . \label{11}
\end{align}
To isolate the subleading soft graviton theorem, and motivated by \eqref{2} and \eqref{11}, we re-express the $S^+_1$ field inside the correlator as 
\begin{align}
&S^+_1(u,z,\bar{z})=S^+_{1e}(u,z,\bar{z})+u S^+_0(u,z,\bar{z})\label{12}\\
\Longrightarrow\hspace{5mm}&{\partial_u}\langle S^+_{1e}(u,z,\bar{z})X\rangle=\left[\text{temporal contact terms}\right]
\end{align} Using this definition for $S_{1e}^+$, we deduce from \eqref{3} and \eqref{10} that 
\begin{equation}
    \langle S_{1e}^+(u,z,\bar{z}) X\rangle=-i\sum\limits_{j=1}^n\theta(u-u_j)\Big[\frac{(\bar{z}-\bar{z}_j)^2}{z-{z}_j}\partial_{\bar{z}_j}-2\bar{k}_j\frac{\bar{z}-\bar{z}_j}{z-{z}_j} 
-u_j\frac{\bar{z}-\bar{z}_j}{z-{z}_j}\partial_{u_j}\Big]\langle X\rangle \, . 
\end{equation} Taking the soft limit in position space at $\mathscr{I}$, corresponding to $u \to  \infty$, one reproduces exactly the subleading soft graviton theorem \eqref{subleading soft graviton theorem}, provided one makes the following identification: 
\begin{equation}
    \lim_{u \to +\infty} S_{1e}^+ (u, z , \bar{z}) = S_{1}^+ (z , \bar{z})
\label{identif2}
\end{equation} where the subleading soft graviton operator $S^+_{1} (z , \bar{z})$ was defined in \eqref{leading and subleading soft operator}. 

In \cite{Ruzziconi:2024kzo}, the identifications \eqref{identific1} and \eqref{identif2} allowed to deduce the expression of the ${T^u}_u$ and ${T^u}_A$ components of the Carrollian stress tensor, respectively, in terms of the bulk radiative modes.

\subsection{Collinear limit and holomorphic Carrollian OPEs}
\label{sec:Collinear limit and holomorphic Carrollian OPEs}

The collinear limit of scattering amplitudes is another important aspect of tree-level scattering amplitudes \cite{Mangano:1990by,Berends:1988zn}. When two external momenta, say $p_1$ and $p_2$, become collinear, then the amplitude factorizes. Let us be more explicit. We will consider the collinear limit of two incoming particles, i.e. $\e_1 = \e_2 = -1$. The other cases can be handled similarly. At tree-level, scattering amplitudes have collinear poles and the corresponding residues factorize. This can be written as\footnote{The notion of collinear factorization is well defined in both Minkowski and Klein spacetime. In Minkowski spacetime, where the three-point amplitudes involving massless particles vanish, this is defined by considering amplitudes with one leg slightly off shell. We will work in split signature for convenience.}
\begin{align} \label{eq:collinear limit}
    \mA_n \left(1^{J_1}, 2^{J_2}, 3^{J_3}, \dots , n^{J_n} \right) \xrightarrow[]{1 || 2} \sum_{J}\mA_3 \left(1^{J_1}, 2^{J_2}, -P^{-J} \right) \frac{1}{\an{12}\sq{21}} \mA_{n-1} \left(P^{J}, 3^{J_3}, \dots ,n^{J_n}\right) .
\end{align}
Here $J$ corresponds to the helicity of the exchanged particle. In general every massless particle that can be exchanged leads to a collinear pole.
 Since $P$ is a null momentum, we can write $P =  \frac{\om}{\sqrt 2} \left(1+z \zb, z + \zb, z - \zb , 1 - z \zb \right)$. A convenient change of variables is $\om_1 = t \om, \om_2 = \left(1-t\right) \om$. We can now  solve $p_1 + p_2 - P = 0$ for $z, \zb$ in terms of $z_1, z_2, \zb_1, \zb_2, t$, giving $z = z_1$ and $\zb = t \zb_1 + (1-t)\zb_2$.

We wish to re-interpret this statement about scattering amplitudes in terms of a holomorphic Carrollian OPE, as defined in Section \ref{sec:Holomorphic OPE limit}. To do so, let us apply the integral transforms \eqref{Carrollian S-matrix} defining Carrollian amplitudes on \eqref{eq:collinear limit}. We have
\begin{multline}
  \mc_n  \xrightarrow[]{1 || 2}  \frac{1}{4\pi^2} \int_0^{+\infty} d\om \, \int_0^1 dt \,\om e^{i\left(t u_1+ (1-t)u_2\right)\om}  \\ \times \sum_{J}\mA_3 \left(1^{J_1}, 2^{J_2}, -P^{-J}\right) \frac{1}{\an{12}\sq{21}} \prod_{i=3}^n  \left(\int_0^{+\infty}  \frac{d\om_i}{2\pi} e^{-i\epsilon_i \omega_i u_i} \right)
   \mA_{n-1} \left(P^{J}, 3^{J_3}, \dots , n^{J_n}\right)   .
\end{multline}
For the holomorphic collinear limit, $z_{12} \to 0$, the only non-zero three-particle amplitudes are the anti-holomorphic ones. Thus, we must have $J_1+J_2-J >0$ and plugging in the appropriate three-point amplitude from \eqref{spinhel3pt}, we get
\begin{multline}
\label{eq:carrollcollinear}
    \mc_n  \xrightarrow[]{1 || 2} -  \sum_p\frac{\kappa_{J_1, J_2, -J}}{4\pi^2} \frac{\zb_{12}^{p}}{z_{12}}\int_0^{+\infty} d\om \, \om^{p}  \left[\int_0^1 dt\,  e^{it\om u_{12}} t^{J_2-J-1} (1-t)^{J_1-J-1}\right] e^{i\omega u_2} \\
    \times \prod_{i=3}^n  \left(\int_0^{+\infty}  \frac{d\om_i}{2\pi} e^{- i \epsilon_i \om_i u_i}   \right)   \mA_{n-1} \left(\left\lbrace \om, z_2, \zb_2 + t \zb_{12}\right\rbrace, 3^{J_3}, \dots , n^{J_n}\right) 
\end{multline}
where $p = J_1+J_2-J-1 \geq 0$. Hence the sum over $p$ is implicitly a sum over $J$. It runs over all allowed values of $p$, which are deduced by solving the inequalities  
\begin{align}
    \label{eq:prestrictions}
    p\geq 0, \qquad \left|J_1+J_2-p-1\right| \leq  2 \quad \text{and} \quad \left|J_1\right| \leq 2, \qquad \left|J_2\right| \leq 2 . 
\end{align}
The first of these is just taking into account the fact that only anti-holomorphic three-point amplitudes  contribute to the holomorphic collinear limit, while the remaining ones neglect massless higher spins. The right-hand side of \eqref{eq:carrollcollinear} depends non trivially on $\zb_1, \zb_2, u_1$ and $u_2$. This reflects the fact that we have considered the holomorphic collinear limit and left the remaining variables at generic values. Following the Carrollian holography dictionary \eqref{eq:Carrollian dictionary}, we can write \cite{Mason:2023mti}
\begin{align}
    \label{eq:carrollOPEblock}
    &\Phi_{J_1} \left(u_1, z_1, \zb_1 \right) \Phi_{J_2} \left(u_2, z_2, \zb_2 \right) \\
    &\nonumber \qquad \sim  - \sum_p\frac{\kappa_{J_1, J_2, -J}}{4\pi^2}   \frac{\zb_{12}^{p}}{z_{12}} \int_0^{+\infty} d\om \, \om^{p}  \left[\int_0^1 dt \,  e^{it\om u_{12}} t^{J_2-J-1} (1-t)^{J_1-J-1}\right]  e^{i u_2 \om }\,  \Phi_{J}\left(\om, z_2, \zb_2 + t \zb_{12}\right) .
\end{align}
Since we use the integral transform \eqref{Carrollian S-matrix}, the weights of the Carrollian primaries are fixed through \eqref{fixed weight}. The above expression constitutes a holomorphic Carrollian OPE, in the sense defined in Section \ref{sec:Holomorphic OPE limit}.

We first work out a compact expression for the Carrollian OPE that includes all the $u$- and $\zb$-descendants in one integral formula. To obtain this, we perform the $\om$ integral by making the identification 
\begin{equation}
    \frac{1}{2\pi}\int_0^{+\infty} d\om \, \om^{p} e^{i \left(u_2 + t u_{12} \right) \,\om }\,  \Phi_{J}\left(\om, z_2, \zb_2+t \zb_{12}\right)  = \partial_u^{p}\Phi_{J}\left(u, z_2, \zb_2+t \zb_{12}\right)\rvert_{u = u_2 + t u_{12}} .
\end{equation}
  We land on 
\begin{align}
\label{eq:carrollOPEformula1}
    &\Phi_{J_1} \left(u_1, z_1, \zb_1 \right) \Phi_{J_2} \left(u_2, z_2, \zb_2 \right) \\
    &\nonumber \qquad\qquad \sim  -\sum_p\frac{\kappa_{J_1, J_2, -J}}{2\pi}\frac{\zb_{12}^{p}}{z_{12}} \int_0^1 dt\,  t^{J_2-J-1} (1-t)^{J_1-J-1} \,  \left(\frac{\partial}{\partial u}\right)^{p}\Phi_{J}\left(u, z_2,\zb_2 +t \zb_{12}\right)\rvert_{u = u_2 + t u_{12}}.
\end{align}
The integral compactly encodes the contributions from all $u$- and $\zb$-descendants. This formula agrees with the general holomorphic OPE formula derived in \eqref{eq:holomorphicOPE} by identifying $\partial_u^{p}\Phi_{J} \equiv \Phi_{\Delta, J}$ in the right-hand side, with $\Delta = 1 + p$. This is consistent with the fact that this operator indeed transforms as a Carrollian primary, see paragraph below \eqref{modified mellin transform}. The weights of the Carrollian primaries $\Phi_1$ and $\Phi_2$ are fixed through \eqref{fixed weight}. The pole $z^{-1}$ in \eqref{eq:carrollOPEformula1} fixes $k_3 = k_1 + k_2 - 1$, consistently with $k_3 = \frac{1+ p +J}{2}$.

We can make this OPE formula even more explicit by expanding in Taylor series in $\bar{z}$ and $u$, and performing the integral over $t$. We refer to \cite{Mason:2023mti} for details. At the end, we find the expression:
\begin{align}
\label{eq:carrollOPEformula3}
   &\Phi_{J_1} \left(u_1, z_1, \zb_1 \right) \Phi_{J_2} \left(u_2, z_2, \zb_2 \right) \\
   &\qquad\sim -\sum_p\frac{\kappa_{J_1, J_2, -J}}{2\pi z_{12}} \sum_{m,n=0}^{\infty} B(J_2-J+m+n, J_1-J) \frac{\zb_{12}^{p+m}u_{12}^n}{m!n!}\left(\frac{\partial}{\partial \zb_2}\right)^m\left(\frac{\partial}{\partial u_2}\right)^{p+n}\Phi_{J}\left(u_2, z_2, \zb_2 \right). \nonumber
\end{align} where $B(x,y)$ is the Euler Beta function. This last expression makes the presence of $u$ and $\bar{z}$ descendants explicit.

\subsection{Feynman rules for Carrollian amplitudes}
\label{sec:Feynman rules for Carrollian amplitudes}

In this section, we show that the Carrollian amplitudes can be directly computed via Feynman diagrams in position space, where the insertion points are at null infinity. The latter are the flat space analogue of the well-known Witten diagrams in AdS. We mostly follow \cite{Alday:2024yyj}; see also \cite{Liu:2024nfc,Long:2026cpq}.

\subsubsection{Propagators}
\label{sec:Propagators}

The building blocks of the diagrams are the bulk-to-bulk and bulk-to-boundary propagators. In flat spacetime, the bulk-to-bulk propagator for a massless scalar field, which solves the sourced Klein-Gordon equation 
\begin{equation}
 \Box_{X_{1}} \mathcal{G}_{BB}^{Flat}(X_1,X_2) = \frac{\delta^{(4)}(X_{12})}{\sqrt{-g}} \ , \qquad \Box= \Big[  \frac{2}{r^2} \partial_z {\partial}_{\bar z} - \frac{2}{r} \partial_u - 2 \partial_u \partial_r   \Big]
 \label{box equation flat}
\end{equation} and is relevant for scattering is given by the Feynman propagator
\begin{equation}
    \mathcal{G}_{BB}^{Flat}(X_1,X_2)  = \frac{-i}{\left(2\pi\right)^2} \frac{1}{\xi^{Flat}_{12} + i \varepsilon}
    \label{bulktobulkflat} \ ,
\end{equation} with
\begin{equation}
    \xi_{12}^{Flat} =  (X_{12})^\mu (X_{12})_\mu  =  -2  r_{12} u_{12} + 2 r_1 r_2 |z_1 - z_2|^2 
    \label{distance flat}
\end{equation} where we used the parametrization of the Cartesian coordinates in terms of Bondi coordinates \eqref{Retarded flat BMS coordinates (appendix)}. The infinitesimal regulator $\varepsilon >0$ in position space ensures that the solution is well-defined on the physical spacetime and can be obtained by analytic continuation in Cartesian coordinates from Euclidean signature \cite{Duffin}. The bulk-to-bulk propagator admits an integral representation using Fourier transform:
\begin{equation}
\begin{split}
     \mathcal{G}_{BB}^{Flat}(X_1,X_2)   &= -\int \frac{d^4p}{\left(2\pi\right)^4} \frac{e^{-i p \cdot X_{12}}}{p^2 - i \varepsilon_p} 
    \end{split}
    \label{fourierBulktoBulk}
\end{equation} where $\varepsilon_p > 0$ is a momentum space regulator. We can derive an explicit expression for the bulk-to-boundary propagators in Minkowski
space by taking a large-$r$ limit of one of the insertion points of the bulk-to-bulk propagator, say $X_1$, and rescaling by an appropriate power of $r_1$ to obtain a finite result. This computation is in the same spirit as the one we presented in Section \ref{sec:From bulk to boundary} to extract the boundary value of the bulk field. The outgoing bulk-to-boundary propagator associated with positive energy particle is defined as 
\begin{equation}
    \mathcal{G}_{Bb,+1}^{Flat}(x_1;X_2)  \equiv \lim_{r_1 \to + \infty } r_1 \mathcal{G}_{BB}^{Flat}(X_1,X_2) =  \frac{-i}{2\left(2\pi\right)^2} \frac{1}{-u_{1}-q_1 \cdot X_2 + i \varepsilon } \ ,
    \label{Bbout}
\end{equation} where $x_1 = (u_1, z_1, \bar z_1)$ is a point at $\mathscr{I}^+$ and we used $q_1 \cdot X_2 = -u_2-r_2|z_{12}|^2$ in the parametrization \eqref{Retarded flat BMS coordinates (appendix)} and \eqref{q in terms of w bar w future} to write the last expression. Similarly, the incoming bulk-to-boundary propagator associated with positive energy particle is defined as
\begin{equation}
\mathcal{G}_{Bb,-1}^{Flat}(x_2;X_1)  \equiv \lim_{r_2 \to - \infty } r_2 \mathcal{G}_{BB}^{Flat}(X_1,X_2) =  \frac{-i}{2\left(2\pi\right)^2} \frac{1}{-u_{2}-q_2 \cdot X_1 - i \varepsilon } \ ,
\label{Bbin}
\end{equation} where we rescaled $-\varepsilon / r_2 \to \varepsilon$ to keep $\varepsilon > 0$. A comprehensive discussion on the extraction of positive energy bulk-to-boundary propagators from this limit can be found in Appendix A of \cite{Alday:2024yyj}. These bulk-to-boundary propagators can be nicely rewritten as 
\begin{equation}
\mathcal{G}_{Bb,\epsilon}^{Flat}(x;X) = \frac{-\epsilon}{2\left(2\pi\right)^2} \int_{0}^{+\infty} {d\omega} e^{-\varepsilon \omega} e^{-i \epsilon \omega u_x } e^{-i \epsilon \om q^\mu_x X_\mu} = \frac{-i}{2\left(2\pi\right)^2} \frac{1}{-u_x - q_x \cdot X +i \epsilon \varepsilon} \ ,
\label{bulktoboundaryflat}
\end{equation} which corresponds to the Fourier transform of a plane wave. The integral expression can be obtained from taking the large-$r$ limit at the level of the integral expression \eqref{fourierBulktoBulk} and using the stationary phase approximation \eqref{stationary phase}, as was done in Section \ref{sec:From bulk to boundary}. It is consistent with the change of basis from momentum space to position space in \eqref{Carrollian S-matrix} to define Carrollian amplitudes. Analogously to the discussion around Equation \eqref{modified mellin transform}, it is convenient to consider the modified Mellin transform to define the following bulk-to-boundary propagator \cite{Bagchi:2023fbj,Bagchi:2023cen, Alday:2024yyj}: 
\begin{equation}
  \label{Bb propagator with D}  \mathcal{G}_{Bb,\epsilon}^{Flat,\Delta}(x;X) = \frac{-i (-i\epsilon)^\Delta}{2\left(2\pi\right)^2} \int_{0}^{+\infty} {d\omega} \omega^{\Delta - 1} e^{-\varepsilon \omega} e^{-i \epsilon \omega u } e^{-i \epsilon \om q^\mu X_\mu} = \frac{-i}{2\left(2\pi\right)^2} \frac{\Gamma (\Delta)}{(-u - q \cdot X +i \epsilon \varepsilon)^{\Delta}}
\end{equation} In particular, the above propagator with integer values of $\Delta = 1, 2, 3 \ldots$ can be obtained from \eqref{bulktoboundaryflat} by taking $\partial_u$-derivatives, which we use to fix the overall factor.

\subsubsection{Conformal Carrollian primary wave functions}
\label{sec:Conformal Carrollian primary wave functions}

As discussed in Section \ref{sec:Boundary operators}, the boundary value at $\mathscr I^\pm$ of a bulk massless field with radiative falloffs is identified with a Carrollian primary operator in the putative Carrollian CFT. For a bulk scalar field $\phi (X)$, we have explicitly
\begin{equation}
    \Phi^{\epsilon =\pm1}_{(\frac{1}{2}, \frac{1}{2})}(x)  =  \lim_{r\to \pm \infty} r \, \phi (X)  ,
    \label{boundary values}
\end{equation} where the Carrollian weights have been fixed according to \eqref{fixed weight} with $J = 0$. The Feynman propagator is defined as the time-ordered correlator 
\begin{equation}
     \mathcal{G}_{BB}^{Flat}(X_1,X_2) = \left\langle T (\phi (X_1) \phi (X_2) )\right\rangle  .
\end{equation} It is therefore natural to rewrite the bulk-to-boundary propagators \eqref{Bbout} and \eqref{Bbin} as 
\begin{equation}
\mathcal{G}_{Bb,+1}^{Flat}(x;X) = \left\langle \Phi^{\epsilon =+1}_{(\frac{1}{2}, \frac{1}{2})}(x) \phi (X) \right\rangle , \qquad \mathcal{G}_{Bb,-1}^{Flat}(x;X) = \left\langle \phi (X) \Phi^{\epsilon =-1}_{(\frac{1}{2}, \frac{1}{2})}(x)  \right\rangle 
\end{equation} where the time order is automatically implied. Similarly, the bulk-to-boundary propagator \eqref{Bb propagator with D} can be written as
\begin{align}
\label{descendantsbB}
&\mathcal{G}_{Bb,+1}^{Flat,\Delta}(x;X) = \left\langle \Phi^{\epsilon =+1}_{(\frac{\Delta}{2}, \frac{\Delta}{2})}(x) \phi (X) \right\rangle \ , \qquad \mathcal{G}_{Bb,-1}^{Flat,\Delta}(x;X) = \left\langle  \phi (X) \Phi^{\epsilon =-1}_{(\frac{\Delta}{2}, \frac{\Delta}{2})}(x)  \right\rangle  .
\end{align} Using \eqref{box equation flat} and the definition of a Carrollian primary in Section \ref{sec:Conformal Carrollian primary in d=3}, this notation makes it manifest that $\mathcal{G}_{Bb,\epsilon}^{Flat,\Delta}(x;X)$ is a conformal Carrollian primary wavefunction, i.e. $(i)$ it satisfies the Klein-Gordon equation $\Box_X \mathcal{G}_{Bb,\epsilon}^{Flat,\Delta}(x;X) =0$, $(ii)$ it satisfies the primary conditions $K\mathcal{G}_{Bb,\epsilon}^{Flat,\Delta}(0;X) = 0 = K_A \mathcal{G}_{Bb,\epsilon}^{Flat,\Delta}(0;X)$, where $K$ and $K_A$ are the scalar operators given in \eqref{repres Carroll K D} and solely acting on the boundary coordinates $x = (u, x^A) =  (u, z, \bar z)$ ($x=0$ is at $\mathscr{I}^\epsilon$), $(iii)$ it transforms as a scalar conformal Carrollian primary of dimension $\Delta$ under the action of a boundary dilatation corresponding to a bulk isometry generated by $\tilde{J}_{0d}$ (see \eqref{ident Poinc Carr}), $(iv)$ it is annihilated by boundary rotations and Carrollian boosts $J_{AB} \mathcal{G}_{Bb,\epsilon}^{Flat,\Delta}(0;X) = 0 = B_A \mathcal{G}_{Bb,\epsilon}^{Flat,\Delta}(0;X)$.

\subsubsection{Diagrams}
\label{sec:Diagrams}

The bulk-to-bulk and bulk-to-boundary propagators introduced above allow us to express the Carrollian amplitudes directly in terms of Feynman diagrams with insertion points at $\mathscr{I}$ \cite{Liu:2024nfc, Alday:2024yyj}. First, the two-point Carrollian amplitude \eqref{2point News} can be obtained from \eqref{Bb propagator with D} by simply taking $\epsilon = \pm 1$, $r \to \mp \infty$, and using the stationary phase approximation. Alternatively, we can compute the two-point Carrollian amplitude by integrating over Minkowski space a product of bulk-to-boundary propagators:
\begin{equation} \label{2point carrollian diagram}
    \langle \Phi_{\Delta_1}^{\epsilon = +1} (x_1)\Phi_{\Delta_2}^{\epsilon = -1} (x_2)   \rangle = \int_{Flat} d^4 X \, \mathcal G_{Bb, +1}^{Flat , \Delta_1} (x_1; X) \mathcal G_{Bb, -1}^{Flat , \Delta_2} (x_2; X) 
\end{equation} To show this, we can use the integral expression \eqref{Bb propagator with D} of the bulk-to-boundary propagator and reconstruct the momentum conserving $\delta$-function. We have 
\begin{align}
        \langle \Phi_{\Delta_1}^{\epsilon = +1} (x_1)\Phi_{\Delta_2}^{\epsilon = -1} (x_2)   \rangle &=  \int_{Flat} d^4 X \int_0^{+\infty}  \frac{d \omega_1}{2\pi} \int_0^{+\infty}  \frac{d \omega_2}{{2\pi}} \, \omega_1^{\Delta_1-1} \omega_1^{\Delta_2-1} e^{-i\omega_1 u_1 + i \omega_2 u_2}  e^{-i \omega_1 q_1 X + i \omega_2 q_2 X} \nonumber \\
        &=  \left( \prod_{j=1}^2 \int_0^{+\infty}  \frac{d \omega_j}{2\pi} \, \omega_j^{\Delta_j -1} e^{-i \epsilon_j\omega_j u_j}  \right) \int_{Flat} d^4 X \, e^{-i ( \omega_1 q_1  - \omega_2 q_2 ) X} \nonumber \\
        &=    \left( \prod_{j=1}^2 \int_0^{+\infty} \frac{d\omega_j}{2\pi} \, \omega_j^{\Delta_j -1} e^{-i \epsilon_j\omega_j u_j}  \right) (2\pi)^4 \delta^{(4)} (p_1+p_2) ,
\end{align} up to an overall constant (we refer to \cite{Alday:2024yyj} for the global factor conventions). The last line coincides with the definition of the $2$-point Carrollian amplitude for a scalar field, see \eqref{modified mellin transform}. The momentum is parametrized as in \eqref{q in terms of w bar w future}.

Similarly, the tree-level Carrollian three-point function associated with a vertex $\phi^3$ can be computed as the product of three bulk-to-boundary propagators integrated over Minkowski space:
\begin{equation}
    \langle \Phi_{\Delta_1}^{\epsilon_1} (x_1) \Phi_{\Delta_2}^{\epsilon_2} (x_2) \Phi_{\Delta_3}^{\epsilon_3} (x_3)  \rangle  =  \int_{Flat} d^4 X \mathcal{G}^{Flat, \Delta_1}_{Bb,\epsilon_1}(x_1;X) \mathcal{G}^{Flat, \Delta_2}_{Bb,\epsilon_2} (x_2;X)\mathcal{G}^{Flat, \Delta_3}_{Bb,\epsilon_3}(x_3;X) .
\label{3pointdiagram}
\end{equation} Indeed, using again the integral representation of the bulk-to-boundary propagator in \eqref{Bb propagator with D}, we obtain 
\begin{equation} \label{3ptdiagramflat}
\begin{split}
     \langle \Phi_{\Delta_1}^{\epsilon_1} (x_1) \Phi_{\Delta_2}^{\epsilon_2} (x_2) \Phi_{\Delta_3}^{\epsilon_3} (x_3)  \rangle  &=  \left(  \prod_{i=1}^3 \int_{0}^{+\infty}  \frac{d\omega_i}{2 \pi} \omega_i^{\Delta_i - 1}  e^{-i \epsilon_i \omega_i u_i } \right) \int_{Flat} d^4 X   e^{-i \sum_{i=1}^3 p^\mu_i X_\mu} \\
    &=   \left(  \prod_{i=1}^3 \int_{0}^{+\infty}  \frac{d\omega_i}{2 \pi} \omega_i^{\Delta_i - 1}  e^{-i \epsilon_i \omega_i u_i } \right) (2\pi)^4 \delta^{(4)}(p_1 + p _2 + p_3),  
\end{split}
\end{equation} again, up to an overall constant. By comparing this expression to \eqref{modified mellin transform}, this corresponds to the definition of the three-point Carrollian amplitude associated with the momentum space contact diagram $\mathcal{A}_3 =  \delta^{(4)}(p_1 + p _2 + p_3)$. The above example can be easily extended to any $n$-point contact diagram. For completeness, we also display the four-point contact diagram: 
\begin{multline}
\label{4pointcontact}
    \langle \Phi_{\Delta_1}^{\epsilon_1} (x_1) \Phi_{\Delta_2}^{\epsilon_2} (x_2) \Phi_{\Delta_3}^{\epsilon_3} (x_3) \Phi_{\Delta_4}^{\epsilon_4} (x_4)  \rangle_c \\
    =  \int_{Flat} d^4 X \mathcal{G}^{Flat, \Delta_1}_{Bb,\epsilon_1}(x_1;X) \mathcal{G}^{Flat, \Delta_2}_{Bb,\epsilon_2} (x_2;X)\mathcal{G}^{Flat, \Delta_3}_{Bb,\epsilon_3}(x_3;X) \mathcal{G}^{Flat, \Delta_4}_{Bb,\epsilon_4}(x_4;X) .
\end{multline}

The four-point exchange diagram coming from the cubic interaction $\phi^3$ can also be expressed by
\begin{align}
\label{flat space4ptexch}
&\langle \Phi_{\Delta_1}^{\epsilon_1} (x_1) \Phi_{\Delta_2}^{\epsilon_2} (x_2) \Phi_{\Delta_3}^{\epsilon_3} (x_3) \Phi_{\Delta_4}^{\epsilon_4} (x_4)  \rangle_e \\ &=   \int_{Flat} d^4 X d^4Y \mathcal{G}^{Flat,\Delta_1}_{Bb,\epsilon_1} (x_1;X) \mathcal{G}^{Flat,\Delta_2}_{Bb,\epsilon_2} (x_2;X) \mathcal{G}^{Flat}_{BB}(X,Y) \mathcal{G}^{Flat,\Delta_3}_{Bb,\epsilon_3} (x_3;Y) 
    \mathcal{G}^{Flat,\Delta_4}_{Bb,\epsilon_4} (x_4;Y) .\nonumber 
\end{align}  Using the integral representation of the bulk-to-bulk \eqref{fourierBulktoBulk} and bulk-to-boundary \eqref{bulktoboundaryflat} propagators, we can rewrite this expression as
\begin{align}
&\langle \Phi_{\Delta_1}^{\epsilon_1} (x_1) \Phi_{\Delta_2}^{\epsilon_2} (x_2) \Phi_{\Delta_3}^{\epsilon_3} (x_3) \Phi_{\Delta_4}^{\epsilon_4} (x_4)  \rangle_e \nonumber\\
     &=  \prod_{i=1}^4 \left( \int_{0}^{+\infty}   \frac{d\omega_i}{2 \pi} \omega_i^{\Delta_i-1} e^{-i \epsilon_i \omega_i u_i } \right) \int \frac{d^4 p}{(2\pi)^4} \int d^4 X d^4Y \frac{e^{-ip^\mu (X-Y)_\mu}}{p^2} e^{-i (p_1 + p_2)^\mu X_\mu} e^{-i (p_3 + p_4)^\mu Y_\mu} \nonumber \\
     &=   \prod_{i=1}^4 \left( \int_{0}^{+\infty}   \frac{d\omega_i}{2 \pi} \omega_i^{\Delta_i-1} e^{-i \epsilon_i \omega_i u_i } \right)  \int d^4 p \frac{(2\pi)^8}{p^2} \delta^{(4)} (p+ p_1 + p_2) \delta^{(4)} (p- p_3 - p_4) \nonumber \\
     &=   \prod_{i=1}^4 \left( \int_{0}^{+\infty}   \frac{d\omega_i}{2 \pi} \omega_i^{\Delta_i-1} e^{-i \epsilon_i \omega_i u_i } \right)\frac{(2\pi)^8}{(p_1 +p_2)^2} \delta^{(4)} \Big(\sum_{i=1}^4 p_i \Big) \ .
\end{align}
Comparing with \eqref{modified mellin transform}, this corresponds to the definition of the four-point Carrollian amplitude associated with the $s$-channel exchange diagram $\mathcal{A}_{4,e} =  \frac{1}{s} \delta^{(4)}( p_1 + p _2 + p_3 + p_4)$. The $t$- and $u$-channels can be obtained similarly. The general proof of the representation of Carrollian amplitudes in terms of Feynman diagrams can be found in \cite{Liu:2024nfc}.

\subsection{Differential equations}

Differential equations satisfied by correlation functions are essential elements in CFT. Some famous examples are Belavin-Polyakov-Zamolodchikov (BPZ) equations \cite{Belavin:1984vu}, Knizhnik-Zamolodchikov (KZ) equations \cite{Knizhnik:1984nr} in two dimensions, and differential equations of conformal partial waves by Dolan and Osborn \cite{Dolan:2011dv} in higher dimensions. These differential equations place powerful constraints on the correlation functions. For example, the four-point correlators in minimal models and in the WZW model can be determined by solving differential equations subject to physically sensible monodromy properties \cite{Belavin:1984vu,Knizhnik:1984nr,DiFrancesco:1997nk}.

MHV Carrollian gluon and graviton amplitudes have been obtained explicitly in \cite{Mason:2023mti}. In this section, we write some
linear differential equations satisfied by these amplitudes \cite{Ruzziconi:2024zkr}, which are expected to play an important role in the dynamical description of the dual theory. These differential equations were first written in the framework of celestial holography \cite{Banerjee:2020vnt,Hu:2021lrx}, and their origin can be understood as BCFW shifts in momentum space. Their Carrollian counterpart was obtained in \cite{Ruzziconi:2024zkr} by using the Carrollian/celestial correspondence, see Section \ref{sec:From Carrollian to celestial holography}. Here we just display the final equations.

A generic $n$-point color-ordered MHV Carrollian gluon amplitude $\mathcal{C}_n (1^{J_1} , \ldots , n^{J_n})$, obtained from the celebrated Parke-Taylor formula \cite{Parke:1986gb} by applying definition \eqref{Carrollian S-matrix} \cite{Mason:2023mti}, can be shown to satisfy the differential equation \cite{Ruzziconi:2024zkr}
\begin{align}
&\left( \partial_i -\frac{1+u_i\partial_{u_i}}{z_{i-1,i}} -\frac{1}{z_{i+1,i}}\right)\partial_{u_{i-1}}\mathcal C_n(1,\ldots,n) \nonumber\\
&+ \left(  \frac{1+u_{i-1}\partial_{u_{i-1}} -J_{i-1}+\bar{z}_{i-1,i} \, \bar{\partial}_{i-1}}{z_{i-1,i}} \right) \partial_{u_{i}} \mathcal C_n(1,\ldots,n) = 0  ,\label{eq:CarrollBG}
\end{align}
where particle $i$ is a gluon with positive helicity $J_i$. There is another set of equations that can be obtained from Equation \eqref{eq:CarrollBG} by the exchange $i-1 \leftrightarrow i+1$. One can easily obtain the $\overline{\text{MHV}}$ Carrollian gluon amplitudes from \eqref{eq:CarrollBG} by switching $z$ and $\bar{z}$, and flipping $J$ to $-J$,
\begin{align}
&\left( \bar{\partial}_i -\frac{1+u_i\partial_{u_i}}{\bar{z}_{i-1,i}} -\frac{1}{\bar{z}_{i+1,i}}\right)\partial_{u_{i-1}}\mathcal C_{n}(1 , \ldots , n) \nonumber\\
&+ \left(  \frac{1+u_{i-1}\partial_{u_{i-1}} +J_{i-1}+{z}_{i-1,i} \, \partial_{i-1}}{\bar{z}_{i-1,i}} \right) \partial_{u_{i}} \mathcal C_{n}(1,\ldots ,n) = 0  ,\label{eq:CarrollBGMHVbar}
\end{align}
together with the one obtained by the exchange $i-1\leftrightarrow i+1$. For completeness, we can also write the differential equations for Carrollian correlators with arbitrary $\Delta_i$ \eqref{modified mellin transform}, associated to MHV gluon amplitudes. We find
\begin{align}
&\left(\partial_i -\frac{\Delta_i +u_i\partial_{u_i}}{z_{i-1,i}}-\frac{1}{z_{i+1,i}}\right)\partial_{u_{i-1}}\mathcal{C}_n^{\Delta_1\dots \Delta_n} (1 , \ldots , n) \nonumber\\
&+ \left(\frac{\Delta_{i-1} +u_{i-1}\partial_{u_{i-1}}-J_{i-1}+\bar{z}_{i-1,i}\bar{\partial}_{i-1}}{z_{i-1,i}} \right)\partial_{u_i}\mathcal{C}_n^{\Delta_1\dots \Delta_n} (1 , \ldots , n) = 0  . \label{eq:CBGu2}
\end{align}
Besides the MHV amplitudes, interesting solutions of these equations can be explicitly written, and describe e.g. scattering amplitudes on some dilaton background. We refer to \cite{Ruzziconi:2024zkr} for more details. 

Finally, differential equations can also be written for MHV Carrollian graviton amplitudes. We have explicitly
\begin{align}
&-\sum_{i=1}^{n-1} \frac{(1+u_i\partial_{u_i}-J_i) (\bar{z}_i-\bar{z}_n) +(\bar{z}_i-\bar{z}_n)^2\bar{\partial}_i}{z_i-z_n} \partial_{u_n} \mathcal{C}_n(1,\cdots,n)  \nonumber\\
&+(2+u_n\partial_{u_n} -J_n) \sum_{i=1}^{n-1} \frac{\bar{z}_i-\bar{z}_n}{z_i-z_n} \partial_{u_i} \mathcal{C}_n(1,\cdots,n) =0  , \label{eq:Cgradiff1}
\end{align} together with
\begin{align}
&\left( \partial_n \partial_{u_n} - \sum_{i=1}^{n-1} \frac{(1+u_i\partial_{u_i} -J_i) +2(\bar{z}_i-\bar{z}_n) \bar{\partial}_i}{z_i-z_n} \, \partial_{u_n}\right)\mathcal{C}_n(1,\cdots,n) \nonumber\\
&+ \left( (2+u_n\partial_{u_n}) \sum_{i=1}^{n-1} \frac{1}{z_i-z_n} \partial_{u_i} + \bar{\partial}_n \sum_{i=1}^{n-1} \frac{\bar{z}_i-\bar{z}_n}{z_i-z_n} \partial_{u_i}\right)\mathcal{C}_n(1,\cdots,n) = 0  . \label{eq:Cgradiff2}
\end{align}
The covariance of these differential equations under the global conformal Carrollian transformations \eqref{boundary action of poincare} can be explicitly checked \cite{Ruzziconi:2024zkr}.

\section{From Carrollian to celestial holography}
\label{sec:From Carrollian to celestial holography}

As discussed in the introduction, a seemingly different approach to flat space holography, known as celestial holography, has been developed over the past few years and has achieved notable success in recasting flat space scattering amplitudes in the language of a two-dimensional CFT living at the boundary \cite{deBoer:2003vf,He:2015zea,Pasterski:2016qvg,Stieberger:2018onx} (see also \cite{Pasterski:2021rjz,Raclariu:2021zjz,Pasterski:2021raf} for reviews). In this section, following \cite{Donnay:2022aba, Bagchi:2022emh, Donnay:2022wvx}, we explain the relation between Carrollian holography, presented in the previous section, and celestial holography. As we shall see, for massless scattering these two proposals are in fact equivalent.

\subsection{A primer on celestial holography}

\subsubsection{Conformal primary wave functions}

The starting point of celestial holography consists of re-expressing the scattering amplitudes in flat space in a boost eigenstate basis. Let us focus on the massless case. The conformal primary wave functions are defined through the Mellin transform of the usual plane waves through \cite{deBoer:2003vf,Pasterski:2016qvg,Pasterski:2017kqt}
\begin{equation}
    \begin{split}
        V_{I}^{*\alpha}(\Delta,w,\bar w|X) = \varepsilon^{*\alpha}_I(w,\bar w) \int_0^{+\infty} d\omega\, \omega^{\Delta-1}\, e^{i \omega q^{\mu} X_{\mu} -\epsilon\omega} = \varepsilon^{*\alpha}_I(w,\bar w) \frac{ i^\Delta \Gamma[\Delta]}{(q^{\mu} X_{\mu} + i \epsilon)^\Delta} .
    \end{split} \label{mellin of plane wave}
\end{equation}
This implements the change of basis between energy eigenstates and boost eigenstates. Using the identity
\begin{equation}
    \int_0^{+\infty}d\omega\, \omega^{i\nu-1} = 2\pi\,\delta(\nu), \label{delta represented in mellin}
\end{equation}
one can show \cite{Pasterski:2017kqt} that the statement that plane waves form a delta-function normalizable basis     
for the Klein-Gordon inner product translates, after Mellin transform, into requiring that $\Delta$ lies on the principal continuous series of the irreducible unitary representations of the Lorentz group, i.e. $\Delta = 1+i\nu$, $\nu \in \mathbb{R}$.

Instead of the plane wave expansion of the spin-$s$ massless bulk field \eqref{pre-fourier}, we can use the new basis \eqref{mellin of plane wave} to write 
\begin{equation}
    \phi^{(s)}_I (X) 
    = \frac{K_{(s)}}{32\pi^4} \sum_{\alpha=\pm}\int d \nu\, d^2 w \left[a^{(s)}_{2-\Delta,\alpha}(w,\bar w) V_I^{*\alpha}(\Delta,w,\bar w|X) + a^{(s)}_{2-\Delta,\alpha}(w,\bar w)^\dagger V^{\alpha}_I (\Delta,w,\bar w|X) \right]
    \label{Phi expanded in Mellin bulk}
\end{equation} 
where we defined the ladder operators in the Mellin basis as in \cite{Pasterski:2021dqe} 
\begin{equation}
\begin{aligned} 
a_{\Delta, \alpha}^{(s)}(w,\bar w) \equiv  \int_0^{+\infty} d \omega\,\omega^{\Delta-1}\,a^{(s)}_\alpha(\omega,w,\bar w), \qquad
    a_{\Delta, \alpha}^{(s)}(w,\bar w)^\dagger \equiv \int_0^{+\infty} d \omega\,\omega^{\Delta-1}\,a^{(s)}_\alpha(\omega,w,\bar w)^\dagger
     \label{mellin O}
\end{aligned}
\end{equation} and these relations can be easily inverted as
\begin{equation}
    \begin{split}
        a^{(s)}_\alpha(\omega, w,\bar w) = \frac{1}{2\pi} \int_{-\infty}^{+\infty}d\nu \, \omega^{-\Delta} \, a^{(s)}_{\Delta, \alpha}(w,\bar w) , \qquad a^{(s)}_\alpha(\omega, w,\bar w)^\dagger = \frac{1}{2\pi} \int_{-\infty}^{+\infty}d \nu \, \omega^{-\Delta} \, a^{(s)}_{\Delta, \alpha}(w,\bar w)^\dagger .
    \end{split}
\label{ladder op inverse mellin}
\end{equation}
Crucially, applying the Mellin transforms \eqref{mellin O} to \eqref{transfo a poincare}, these operators can be shown to transform as
\begin{equation}
a'^{(s)}_{\Delta,\alpha}(w',\bar w') = \left(\frac{\partial w'}{\partial w}\right)^{ -\frac{\Delta +J}{2}}\left(\frac{\partial \bar w'}{\partial \bar w}\right)^{-\frac{\Delta -J}{2}} a^{(s)}_{\Delta,\alpha}(w,\bar w)
\label{a delta transfo lorentz finite}
\end{equation}
under the action of the Lorentz group, which is precisely the definition of a $2d$ conformal primary field of dimension $\Delta$ and spin $J=\alpha s$, $\alpha = \pm 1$. The action of the Poincaré translations in the Mellin basis is given by \cite{Stieberger:2018onx}
\begin{equation}
a_{\Delta,\alpha}'^{(s)}(w',\bar w') = e^{-i q^\mu(w,\bar w)t_\mu \hat\partial_{\Delta}}a^{(s)}_{\Delta,\alpha}(w,\bar w) = \sum_{n=0}^{+\infty} \frac{\left(-i q^\mu(w,\bar w)t_\mu\right)^n}{n!} a^{(s)}_{\Delta+n,\alpha}(w,\bar w) \label{transfo a mellin translations}
\end{equation}
where we defined the discrete derivative operator $\hat\partial_\Delta F(\Delta) \equiv F(\Delta+1)$. The infinitesimal action of Poincaré transformations in the Mellin basis reads as
\begin{equation}
\begin{split}
    &\delta_{\xi(\mathcal T,\mathcal Y, \bar{\mathcal{Y}})} a^{(s)}_{\Delta,\alpha}(w,\bar w) = \left(-i\mathcal T\hat\partial_{\Delta} + \mathcal{Y} \partial + \bar{\mathcal Y} \bar \partial + \frac{\Delta+ J}{2} \partial \mathcal{Y}  + \frac{\Delta- J}{2} \bar \partial \bar{\mathcal Y} \right) a^{(s)}_{\Delta,\alpha}(w,\bar w).
\end{split}
\label{a delta transfo poincare inf}
\end{equation}

\subsubsection{Celestial amplitudes}

In the same spirit as what we discussed in Section \ref{sec:Definition of Carrollian amplitudes}, one can express $\mathcal S$-matrix correlators in this Mellin basis \cite{Pasterski:2016qvg,Pasterski:2017kqt}. The asymptotic states in this basis are given by
\begin{equation}
\bra{\Delta,z,\bar z,s,\alpha}= -\frac{iK_{(s)}}{4\pi} \bra{0}a_{\Delta,\alpha}^{(s)\,\text{out}}(z,\bar z)  = \int_0^{+\infty} d\omega \,\omega^{\Delta-1} \bra{\omega,z,\bar z,s,\alpha}  \label{outgoing celestial}
\end{equation}
for outgoing particles, and
\begin{equation}
\ket{\Delta,z,\bar z,s,\alpha} =   \frac{iK_{(s)}}{4\pi} a_{\Delta,\alpha}^{(s)\,\text{in}}(z,\bar z)^\dagger \ket{0} = \int_0^{+\infty} d\omega \,\omega^{\Delta-1} \ket{\omega,z,\bar z,s,\alpha} \label{incoming celestial}
\end{equation}
for incoming particles. Because of \eqref{a delta transfo lorentz finite}, states defined as \eqref{outgoing celestial}--\eqref{incoming celestial} are indeed boost eigenstates. The so-called ``celestial amplitudes'' $\mathcal M_n = \braket{\text{out}|\text{in}}_{\text{boost}}$ involving $n$ inserted particles on the celestial sphere are now obtained as 
\begin{multline}
 \mathcal M_n ( \{\Delta_1,z_1,\bar z_1 \}^{\epsilon_1}_{J_1}, \dots , \{\Delta_n,z_n,\bar z_n\}^{\epsilon_n}_{J_n}) \\
 = \prod_{k=1}^n \int_0^{+\infty} d\omega_k \, \omega_k^{\Delta_k-1} \mathcal A_n ( \{\omega_1,z_1,\bar z_1 \}^{\epsilon_1}_{J_1}, \dots , \{\omega_n,z_n,\bar z_n\}^{\epsilon_n}_{J_n})
    \label{Celestial S-matrix}
\end{multline} where the momentum space amplitude was defined in \eqref{usual momentum space amplitude}. Due to their transformation law, the ladder operators can be identified with CFT primaries on the celestial sphere, 
\begin{equation} \label{celestial primary}
\mathbb{O}^{\epsilon = +1}_{\Delta, J} (z, \bar z) = -\frac{iK_{(s)}}{4\pi} a_{\Delta,\alpha}^{(s)\,\text{out}}(z,\bar z), \qquad \mathbb{O}^{\epsilon = -1}_{\Delta, J} (z, \bar z) = \frac{iK_{(s)}}{4\pi} a_{\Delta,\alpha}^{(s)\,\text{in}}(z,\bar z)^\dagger .
\end{equation}
These are called celestial primaries, and they transform as
\begin{equation} \label{Poincare transfo celestiql primary}
  \delta_{\xi(\mathcal T,\mathcal Y, \bar{\mathcal{Y}})} \mathbb{O}^{\epsilon}_{\Delta, J} (z, \bar z) = -   i \epsilon \mathcal T(z,\bar z)\hat\partial_{\Delta}\mathbb O^{\epsilon}_{(\Delta,J)}(z,\bar z) + \left(\mathcal Y \partial + \bar{\mathcal Y} \bar \partial  + h\,\partial \mathcal Y + \bar h \, \bar \partial \bar{\mathcal Y}  \right) \mathbb O^{\epsilon}_{(\Delta,J)}(z,\bar z)
\end{equation} under Poincaré transformations, where 
\begin{equation}
    h  = \frac{\Delta + \epsilon J}{2}, \qquad \bar h = \frac{\Delta - \epsilon J}{2}
\end{equation} are the $2d$ conformal weights. This is a direct consequence of \eqref{a delta transfo poincare inf} and \eqref{celestial primary}. Notice the peculiar action of spacetime translations, shifting the conformal dimension by one. This is unusual from a $2d$ CFT perspective and will have drastic consequences for the form of the correlation functions. In terms of these primaries, the celestial amplitudes \eqref{Celestial S-matrix} can be re-interpreted as correlation functions in a $2d$ CFT living on the celestial sphere
\begin{equation}
    \langle \mathbb{O}^{\epsilon_1}_{\Delta_1, J_1} (z_1, \bar z_{1}) \ldots   \mathbb{O}^{\epsilon_n}_{\Delta_n, J_n} (z_n, \bar z_{n}) \rangle =  \mathcal M_n ( \{\Delta_1,z_1,\bar z_1 \}^{\epsilon_1}_{J_1}, \dots , \{\Delta_n,z_n,\bar z_n\}^{\epsilon_n}_{J_n})
\end{equation} which justifies the terminology of celestial amplitudes. By construction, these correlators satisfy the global conformal Ward identities. As we shall discuss in the next section, they also satisfy the local conformal Ward identities, which further reinforces the holographic CFT interpretation.

\subsection{From Carrollian to celestial amplitudes}

\subsubsection{Integral transform}

\begin{figure}[ht!]
    \centering
    \begin{tikzpicture}
        \tikzmath{\a = 2.5; \b = sqrt(3)*\a;} 
        \tikzstyle{nb} = [rectangle,draw,fill=white,rounded corners,outer sep=3pt,text width=3.7cm,align=center,inner sep=5pt];
        \coordinate (A) at (-\a, 0);
        \coordinate (B) at ( \a, 0);
        \coordinate (C) at ( 0, \b);
        \draw[opacity=0] ($(A)+(-4,-1.5)$) -- ($(B)+(+4,-1.5)$) -- ($(B)+(0,\b)+(+4,1)$) -- ($(A)+(0,\b)+(-4,1)$) -- cycle;
        \node[nb,left] (An) at (A) {Position space $\Phi^\epsilon_J(u,z,\bar z)$};
        \node[nb,right] (Bn) at (B) {Mellin space $\mathbb{O}^{\epsilon}_{\Delta, J}(z,\bar z)$};
        \node[nb] (Cn) at (C) {Momentum space \\
        $a_J^{\epsilon}(\omega,z,\bar z)$};
        \draw[Latex-Latex] (An) -- (Bn);
        \draw[Latex-Latex] (Bn) -- (Cn);
        \draw[Latex-Latex] (An) -- (Cn);
        \node[below,align=center,text width=3.5cm] at ($(A)!0.5!(B)-(0,0.2)$) {Carrollian/celestial \\
        \eqref{from Carroll to celestial}};
\node[align=center,anchor=south east] at ($(An)!0.45!(Cn)$) {Fourier transform\\ \eqref{outgoing spin s}--\eqref{fourier transform formula at scri}};
        \node[align=center,anchor=south west] at ($(Bn)!0.45!(Cn)$) {Mellin transform\\ \eqref{mellin O}--\eqref{ladder op inverse mellin}};
    \end{tikzpicture}
    \caption{Interplay between three scattering bases in flat spacetime \cite{Donnay:2022wvx}.}
    \label{fig:FBM}
\end{figure}
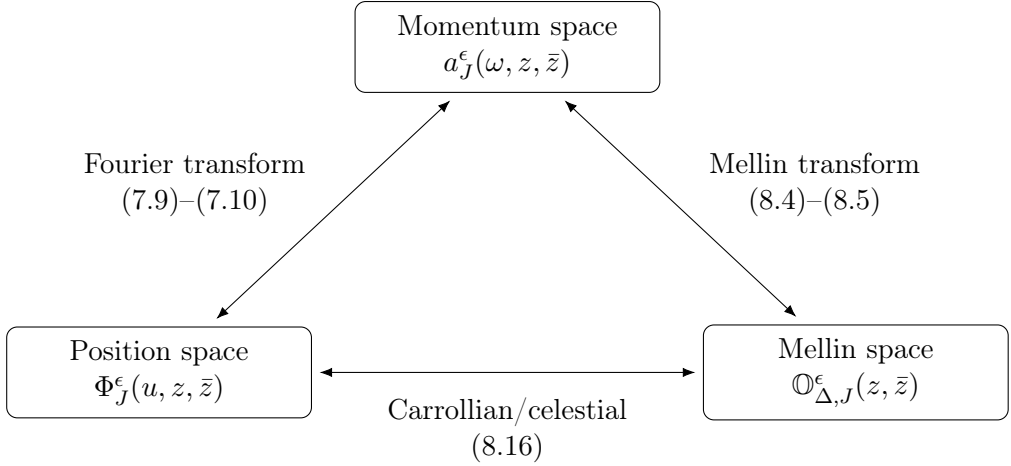

Let us summarize the situation. The usual scattering amplitudes \eqref{usual momentum space amplitude} are expressed in momentum space, where the operators defining the $\mathcal{S}$-matrix elements are creation and annihilation operators $a^{\epsilon}_J(\omega,z,\bar z)$ (here $a^{+1}_J = a_J$ and $a^{-1}_J = a_J^\dagger$). Then we have defined the Carrollian amplitudes \eqref{Carrollian S-matrix} as amplitudes written in position space at null infinity. The associated position space operators \eqref{outgoing spin s} have been identified with Carrollian primaries $\Phi^\epsilon_J (u, z, \bar z)$ in \eqref{identification operators}. Finally, the celestial amplitudes have been defined as the amplitudes written in Mellin space \eqref{Celestial S-matrix}, and the associated ladder operators are celestial primaries $\mathbb{O}_J^\epsilon (z, \bar z)$ identified in \eqref{celestial primary}. The relation between the three bases is illustrated in Figure \ref{fig:FBM}. We refer, e.g., to \cite{Donnay:2022sdg,Freidel:2022skz} for more discussion on scattering amplitudes in these three different choices of bases. 

Hence, the relation between Carrollian and celestial primaries is simply obtained by closing the triangle in Figure \ref{fig:FBM}, and combining the inverse Fourier transform, together with the Mellin transform. This transform has been worked out in \cite{Donnay:2022aba, Donnay:2022wvx} and we have explicitly
\begin{equation}
\mathbb{O}^{\epsilon}_{(\Delta,J)} (z, \bar{z})  = \kappa_\Delta^\epsilon \int^{+\infty}_{-\infty} \frac{d u}{(u+i \epsilon\varepsilon)^{\Delta}}\, \Phi^{\epsilon}_{J}(u,z,\bar z)
\label{from Carroll to celestial}
\end{equation} with $\kappa_\Delta^\epsilon= 4 \pi (i \epsilon)^{\Delta + 1} \Gamma (\Delta)$. This can be used to map $3d$ Carrollian CFT correlators to $2d$ celestial CFT correlators as 
\begin{multline} \label{carroll-celestial transform}
   \langle \mathbb{O}^{\epsilon_1}_{(\Delta_1,J_1)} (z_1, \bar{z}_1) \ldots \mathbb{O}^{\epsilon_n}_{(\Delta_n,J_n)} (z_n, \bar{z}_n)    \rangle \\ = 
   \prod_{i=1}^n \left( \kappa_{\Delta_i}^{\epsilon_i} \int^{+\infty}_{-\infty} \frac{d u_i}{(u_i+i \epsilon_i\varepsilon)^{\Delta_i}} \right) \langle \Phi^{\epsilon_1}_{J_1} (u_1, z_1, \bar{z}_1) \ldots \Phi^{\epsilon_n}_{J_n} (u_n, z_n, \bar{z}_n)    \rangle .
\end{multline}
A few comments are in order. First, we see that the integral transform \eqref{from Carroll to celestial} trades the $u$-dependence of the Carrollian primary $\Phi^{\epsilon}_{J}(u,z,\bar z)$ for the conformal dimension $\Delta$ of the celestial primary $\mathbb{O}^{\epsilon}_{(\Delta,J)} (z, \bar{z})$, which explains why the spectrum of the celestial CFT is continuous. Second, one can easily check that \eqref{from Carroll to celestial} is compatible with the transformations of Carrollian and celestial primaries. Indeed, starting from \eqref{Carrollian primary 3d} with Carrollian weights fixed as in \eqref{fixed weight}, one recovers \eqref{Poincare transfo celestiql primary}. Third, since the integral transform \eqref{from Carroll to celestial} combines the (inverse) Fourier and Mellin transforms, it is invertible. It implies that the Carrollian CFT correlators as defined in \eqref{Carrollian S-matrix} are in one-to-one correspondence with the celestial CFT correlators, which fully encode the $\mathcal{S}$-matrix for massless scattering. This establishes the equivalence between Carrollian and celestial holography, at least for massless scattering.

In Section \ref{sec:Definition of Carrollian amplitudes} we also defined Carrollian correlators involving primaries with arbitrary conformal dimension using a modified Mellin transform \eqref{modified mellin transform}. Although these correlators are redundant to encode the $\mathcal{S}$-matrix, we argued that they were useful to get rid of the IR divergences, as well as to relate with celestial holography. Indeed, one can exploit the redundancy of \eqref{modified mellin transform} and set $u_i = 0$ for all $i$ to recover the Mellin transform \eqref{Celestial S-matrix} defining the celestial amplitudes. This observation is very convenient for practical purposes, and was indeed used in \cite{Bagchi:2022emh} to discuss the relation between Carrollian and celestial amplitudes. More explicitly, we have 
\begin{equation}
     \mathbb{O}^\epsilon_{\Delta, J}(z,\bar z) \equiv 2 \pi \,  \Phi^\epsilon_{(\Delta , J)} (u = 0, z, \bar z)
\end{equation} where the right-hand side is a Carrollian primary appearing in the left-hand side of \eqref{modified mellin transform}. Similarly, the conformal Carrollian primary wavefunction discussed in Section \ref{sec:Conformal Carrollian primary wave functions} reduces to the conformal primary wavefunction \eqref{mellin of plane wave}.

The non-trivial interplay between $3d$ Carrollian physics, $2d$ CFT, and physics in $4d$ asymptotically flat spacetimes has had an important impact on the field and has generated a large body of work. In the following, we illustrate the Carrollian/celestial correspondence with examples at the level of correlators, soft theorems, and OPEs.

\subsubsection{Examples of celestial amplitudes}
\label{sec:Examples of celestial amplitudes}

Let us give a few examples of celestial amplitudes, starting from their Carrollian counterparts discussed in Section \ref{sec:Example of Carrollian amplitudes}. We follow \cite{Kulkarni:2025qcx} and refer to \cite{Banerjee:2019prz, Mason:2023mti} for earlier references. We start with the $2$-point Carrollian amplitude \eqref{2point News} and take $u_1, u_2 \to 0$:
\begin{equation}
    \begin{split}
\label{eq:2ptcelamp}
    \langle \mathbb{O}^{-1}_{\Delta_1, J_1} (z_1, \bar{z}_1) \mathbb{O}^{+1}_{\Delta_2, J_2} (z_2, \bar{z}_2) \rangle  &=\frac{K^2_{J_1,J_2}}{4\pi}  \left[ \lim_{\varepsilon, u_i \to 0}\frac{\Gamma\left(\Delta_1 + \Delta_2 - 2\right)}{\left(u_{12}+i  \varepsilon\right)^{\Delta_1 + \Delta_2-2}}\right] \delta_{J_1, J_2} \\
    &= - \frac{i}{2} K^2_{J_1,J_2}  \delta^{(2)}\left(z_{12}\right) \delta\left(\Delta_1 + \Delta_2-2\right) \delta_{J_1, J_2} .
    \end{split}
\end{equation}
In arriving at this formula, we have made use of the distributional identity $\lim_{\nu \to 0} \Gamma\left(x\right)\nu^{-x} = -2\pi i \delta\left(x\right).$ This corresponds to the celestial $2$-point function. Notice that this is not the usual $2$-point function in $2d$ CFT. This contact term ensures compatibility with bulk translation invariance, which acts peculiarly by shifting the conformal dimension, as mentioned below \eqref{Poincare transfo celestiql primary}. This is one of the motivations for Carrollian holography: translations act naturally as spacetime symmetries on Carrollian primaries, and the presence of the spatial delta function is a feature shared by all electric Carrollian CFTs, rather than something exotic.

Following a similar procedure, the three- and four-point celestial amplitudes can be extracted from the $u_i \to 0$ limit of \eqref{eq:3ptwithTheta} and \eqref{C4general}, respectively. The three-point function reads as 
\begin{equation}
	\begin{split}
	\langle \mathbb{O}_1 \mathbb{O}_2 \mathbb{O}_3 \rangle &= \Theta\left(-\frac{z_{13}}{z_{23}}\epsilon_1\epsilon_2\right)\Theta\left(\frac{z_{12}}{z_{23}}\epsilon_1\epsilon_3\right) \delta (\Delta_1+ \Delta_2 + \Delta_3 + J_1 + J_2 + J_3 - 4)\\
	&\times \delta(\bz_{12}) \delta(\bz_{13}) \, (z_{12})^{h_3 - h_1 - h_2}\,(z_{23})^{h_1 - h_2 - h_3}\,(z_{13})^{h_2 - h_1 - h_3}
	\end{split}
	\end{equation} up to an overall factor. This matches the results of \cite{Pasterski:2017ylz} obtained directly by applying the Mellin transform on the three-point amplitude \eqref{spinhel3pt}. The four-point celestial amplitude reads as 
\begin{multline}
 \langle \mathbb{O}_1 \mathbb{O}_2 \mathbb{O}_3 \mathbb{O}_4 \rangle =   \Theta\left(-z\left|\frac{z_{24}}{z_{12}}\right|^2 \epsilon_1 \epsilon_4  \right)\Theta\left(\frac{1-z}{z} \left|\frac{z_{34}}{z_{23}}\right|^2\epsilon_2 \epsilon_4  \right)\Theta\left(-\frac{1}{1-z}\left|\frac{z_{14}}{z_{13}}\right|^2\epsilon_3 \epsilon_4 \right)\\ \times \delta\left(\D_1+\D_2+\D_3+\D_4-4\right) \, \prod_{i<j} \frac{G(z)\delta(z-\bz)}{(z_{ij})^{a_{ij}} (\bz_{ij})^{\bar{a}_{ij}}}
\end{multline} up to an overall factor. $G(z)$ was given in \eqref{Gz for gluon and gravitons} for gluons and gravitons, and $a_{ij}$ and $\bar a_{ij}$ can be read from \eqref{aijcft}. This matches the result of \cite{Pasterski:2017ylz} for gluons, and \cite{Stieberger:2018edy,Puhm:2019zbl} for gravitons. Again, the distributional nature of the three- and four-point functions is not standard for a $2d$ CFT, and constitutes an important obstruction to applying bootstrap methods to the celestial CFT.

\subsubsection{Conformally soft theorems}

As discussed in Section \ref{sec:Soft limit and Carrollian CFT stress tensor Ward identities}, the leading and subleading soft graviton theorems can be written in position space at null infinity and have a natural interpretation in terms of Carrollian CFT stress tensor Ward identities. One could instead rewrite the soft theorems in a celestial basis by applying the Mellin transform \eqref{Celestial S-matrix} on \eqref{leading soft omega} and \eqref{subleading soft omega}. In this basis, the soft theorems become ``conformally soft theorems'' and have a natural interpretation in terms of $2d$ CFT Ward identities \cite{Strominger:2013jfa,He:2014laa,Kapec:2014opa, Kapec:2016jld, Adamo:2019ipt, Donnay:2018neh,Puhm:2019zbl}. Alternatively, one could also obtain these expressions using the Carrollian/celestial dictionary discussed above. Indeed, applying \eqref{carroll-celestial transform} on \eqref{leading soft them position} and \eqref{subleading soft graviton theorem}, one can obtain
\begin{multline}
\label{conformally soft 1}
    \langle i S^+_0 (z,\bar{z}) \mathbb{O}^{\epsilon_1}_{\Delta_1, J_1} (z_1, \bz_1) \ldots \mathbb{O}^{\epsilon_n}_{\Delta_n, J_n} (z_n, \bz_n)   \rangle \\
    = \sum_{i=1}^n    \frac{\bar z- \bar z_i}{z- z_i}   \langle  \mathbb{O}^{\epsilon_1}_{\Delta_1, J_1} (z_1, \bar z_1) \ldots \mathbb{O}^{\epsilon_i}_{\Delta_i+1, J_i} (z_i, \bar z_i)  \ldots \mathbb{O}^{\epsilon_n}_{\Delta_n, J_n} (z_n, \bar z_n) \rangle  
\end{multline} and 
\begin{multline} 
\label{conformally soft 2}
    \langle i S_1^+ (z,\bar{z})  \mathbb{O}^{\epsilon_1}_{\Delta_1, J_1} (z_1, \bz_1) \ldots \mathbb{O}^{\epsilon_n}_{\Delta_n, J_n} (z_n, \bz_n)   \rangle  \\
    =  \sum_{i=1}^n \frac{(\bar z-\bar z_i)^2 \bar \partial_i+ 2 \bar h_i  (\bar z_i- \bar z) }{z - z_i}   \langle \mathbb{O}^{\epsilon_1}_{\Delta_1, J_1} (z_1, \bz_1) \ldots \mathbb{O}^{\epsilon_n}_{\Delta_n, J_n} (z_n, \bz_n)    \rangle . 
\end{multline} Introducing the operator
\begin{equation}
    P(z, \bar z) = i \bar \partial S^+_0 ( z, \bar z) ,
\end{equation} Equation \eqref{conformally soft 1} can be rewritten as 
\begin{multline}
\label{conformally soft 12}
    \langle P (z,\bar{z}) \mathbb{O}^{\epsilon_1}_{\Delta_1, J_1} (z_1, \bz_1) \ldots \mathbb{O}^{\epsilon_n}_{\Delta_n, J_n} (z_n, \bz_n)   \rangle \\
    = \sum_{i=1}^n    \frac{1}{z- z_i}   \langle  \mathbb{O}_{\Delta_1, J_1} (z_1, \bar z_1) \ldots \mathbb{O}_{\Delta_i+1, J_i} (z_i, \bar z_i)  \ldots \mathbb{O}_{\Delta_n, J_n} (z_n, \bar z_n) \rangle  , 
\end{multline} which corresponds to a $2d$ CFT Ward identity for a $U(1)$ Kac-Moody current $P(z, \bar{z})$ with conformal weights $(\frac{3}{2}, \frac{1}{2})$ \cite{Strominger:2013jfa,He:2014laa,Donnay:2018neh,Puhm:2019zbl}. Furthermore, defining
\begin{equation}
    \bar{T}(\bar z) = \frac{i}{2\pi} \int d^2 w \frac{1}{(\bar z - \bar w)} \bar \partial^3 S^+_1 (w, \bar{w}), 
\label{celestial stress tensor}
\end{equation} as the celestial CFT stress tensor \cite{Kapec:2016jld}, Equation \eqref{conformally soft 2} can be rewritten as
\begin{multline} 
    \langle \bar T (\bar z) \mathbb{O}^{\epsilon_1}_{\Delta_1, J_1} (z_1, \bz_1) \ldots \mathbb{O}^{\epsilon_n}_{\Delta_n, J_n} (z_n, \bz_n)   \rangle  \\
    =  \sum_{i=1}^n \left[ \frac{1}{\bar z - \bar z_i} \bar \partial_i + \frac{\bar h_i}{(\bar z - \bar z_i)^2} \right]   \langle \mathbb{O}^{\epsilon_1}_{\Delta_1, J_1} (z_1, \bz_1) \ldots \mathbb{O}^{\epsilon_n}_{\Delta_n, J_n} (z_n, \bz_n)    \rangle . 
\end{multline}  These correspond to $2d$ CFT Ward identities, and show that the local conformal transformations in $2d$ are also symmetries of the celestial CFT. This result was one of the strong hints pointing towards celestial holography. 

Notice that improvements of the celestial stress tensor \eqref{celestial stress tensor} have been discussed in the literature. Adding quadratic soft terms to the expression of the celestial stress tensor \cite{He:2017fsb} has been shown to capture the one-loop correction to the subleading soft graviton theorem \cite{Bern:2014oka}. An improvement of this prescription was proposed in \cite{Donnay:2022hkf,Pasterski:2022djr}, motivated by consistency with the split between hard and soft gravitational phase space at null infinity \cite{Donnay:2021wrk}. The latter proposal has also been shown \cite{Agrawal:2023zea} to be related to the $\log \omega$ corrections to the subleading soft graviton theorem arising at one loop \cite{Sahoo:2018lxl,Saha:2019tub}; see also \cite{Choi:2024ygx,Alessio:2024onn,Geiller:2024ryw,Sen:2024qzb,Choi:2024ajz} for related discussions. The leading soft theorem being tree-level exact, and the subleading soft graviton theorem being one-loop exact, these discussions ensure that supertranslations and superrotations are symmetries of the celestial CFT, and therefore of the gravitational $\mathcal{S}$-matrix, at all orders of perturbation.

\subsubsection{Celestial OPEs}

A notion of holomorphic OPEs can be defined in the $2d$ celestial CFT from the collinear limit of scattering amplitudes \cite{Fan:2019emx, Himwich:2021dau}, in a completely analogous way to what we have presented for Carrollian OPEs in Section \ref{sec:Collinear limit and holomorphic Carrollian OPEs}, by applying the Mellin transform on the collinear factorization \eqref{eq:collinear limit}. Alternatively, the celestial OPEs can be obtained by applying the Carrollian/celestial correspondence formula \eqref{from Carroll to celestial} to the Carrollian OPE formula \eqref{eq:carrollOPEformula1} \cite{Mason:2023mti}. We find
\begin{align}
    &\mathbb{O}_{\D_1, J_1} \left(z_1, \zb_1 \right)\mathbb{O}_{\D_2, J_2} \left(z_2, \zb_2 \right)\nonumber \\
    &\hspace{3.5cm} \sim  -\sum_p \frac{\kappa_{J_1, J_2, -J}}{2\pi}\frac{\zb_{12}^{p}}{z_{12}} (-i)^{\D_1+\D_2} \Gamma (\D_1) \Gamma(\D_2) \int_{-\infty}^{\infty} du_1 \, du_2 \, u_1^{-\D_1} \, u_2^{-\D_2} \\
    &\nonumber \hspace{4.0cm} \times\int_0^1 dt\,  t^{J_2-J-1} (1-t)^{J_1-J-1} \,  \left(\frac{\partial}{\partial u_2}\right)^{p}\Phi_{J}\left(u_2 + t u_{12}, z_2, \zb_2 + t \zb_{12}\right) . 
\end{align}
We can deform the $u_i$ integrals to pick up the discontinuity across the branch cut on the negative $u_i$ axis and reduce them to integrals over the positive real line. We further define $\tilde u_1 = t u_1, \tilde u_2 = (1-t)u_2$, yielding
\begin{align}
    &\mathbb{O}_{\D_1, J_1} \left(z_1, \zb_1 \right)\mathbb{O}_{\D_2, J_2} \left(z_2, \zb_2 \right) \sim \sum_p \frac{2\pi(-i)^{\D_1+\D_2}\kappa_{J_1, J_2, -J}}{\Gamma (1-\D_1) \Gamma(1-\D_2) }\frac{\zb_{12}^{p}}{z_{12}} \int_0^1 dt\,  t^{2\hb_1+p-1} (1-t)^{2\hb_2-1} \,  \\
    &\nonumber \hspace{5.5cm} (-1)^p\times\int_{0}^{\infty} d\tilde u_1 \, d\tilde u_2 \, \tilde u_1^{-\D_1} \, \tilde u_2^{-\D_2} \left(\frac{\partial}{\partial \tilde u_2}\right)^{p}\Phi_{J}\left(\tilde u_1+\tilde u_2, z_2, \zb_2 + t \zb_{12}\right) .
\end{align}
Integrating the $u_2$ derivative by parts and then making a final change of variables $\tilde u_1 = x u, \tilde u_2 = (1-x) u$ allows us to factorize the integrals. The integral over $x$ is then readily performed and the expression simplifies considerably to 
\begin{align}
    &\mathbb{O}_{\D_1, J_1} \left(z_1, \zb_1 \right)\mathbb{O}_{\D_2, J_2} \left(z_2, \zb_2 \right) \nonumber \\
    &\hspace{3.5cm}\sim  -\sum_p \kappa_{J_1, J_2, -J}(-1)^p\frac{\text{sin} \pi(\D_2+p)}{\text{sin} \pi\D_2 }\frac{\zb_{12}^{p}}{z_{12}}  \int_0^1 dt\,  t^{2\hb_1+p-1} (1-t)^{2\hb_2+p-1}\\
    & \hspace{4.0cm} \times \frac{2\pi(-i)^{\D_1+\D_2+1}}{\Gamma(2-\D_1-\D_2-p)} \, \int_0^{\infty} du \, u^{1-\D_1-\D_2-p} \Phi_{J}\left(\tilde u_1 + \tilde u_2, z_2, \zb_2 + t \zb_{12}\right) .\nonumber 
\end{align}
Identifying the expression on the final line with the celestial operator (after a contour deformation) and noting that $\frac{\text{sin} \pi(\D_2+p)}{\text{sin} \pi\D_2 } = (-1)^p$, we arrive at the celestial OPE formula \cite{Fan:2019emx, Himwich:2021dau}
\begin{align} \label{celestial OPE formula}
    &\mathbb{O}_{\D_1, J_1} \left(z_1, \zb_1 \right)\mathbb{O}_{\D_2, J_2} \left(z_2, \zb_2 \right)  \nonumber \\ &\hspace{2.5cm}\sim  - \sum_p \kappa_{J_1, J_2, -J}\frac{\zb_{12}^{p}}{z_{12}}  \int_0^1 dt\,  t^{2\hb_1+p-1} (1-t)^{2\hb_2+p-1}\mathbb{O}_{\D_1+\D_2+p-1, J} \left(z_2, \zb_2 + t \zb_{12}\right).
\end{align} 

\subsection{Celestial symmetries}

Besides its original ambitions for flat space holography, we have seen that celestial holography is a powerful framework to reformulate scattering amplitudes in terms of a $2d$ CFT language. Using the Mellin basis, one may make certain properties of the scattering amplitudes manifest. One example of this is the discovery of the celestial symmetries \cite{Guevara:2021abz,Strominger:2021mtt}, which arguably constitutes one of the most important achievements of the celestial holography program. These symmetry algebras have been obtained from the above celestial OPE formulae \eqref{celestial OPE formula} by taking conformally soft limits and defining the commutator from the OPE, as in standard $2d$ CFT. In the case of Yang–Mills theory, the celestial symmetries form the $S$-algebra:
\begin{equation}
  \bigl[ S^{p,a}_{\bar m,\,m},\, S^{q,b}_{\bar n,\,n} \bigr]
= - i\, f^{abc}\, S^{p+q-1,\,c}_{\bar m+ \bar n,\, m+n}
\end{equation} where $a,b$ are $SU(N)$ color indices, $f^{abc}$ are the structure constants, $p = 1, \frac{3}{2}, 2, \frac{5}{2}, \ldots$, $m+p \in \mathbb{Z}$ and $\bar m + p \in \mathbb{Z}$ in the $\bar m$-wedge $|\bar{m}| < p$. The modes $p=1$ encode the leading soft gluon operator \cite{He:2015zea}. In the case of gravity, they form the $Lw_{1+\infty}$ algebra (or more precisely, $LHam (\mathbb C^2)$, i.e. the loop algebra of the algebra of canonical transformations of $\mathbb{C}^2$ \cite{Bittleston:2023bzp}):
\begin{equation}
    \{ w^{p}_{\bar m,\, m},\, w^{q}_{\bar n,\, n} \}
= 2\bigl(\bar m (q-1) - \bar n (p-1)\bigr)\,
w^{p+q-2}_{\bar m + \bar n,\, m + n}
\end{equation} where $p = 1, \frac{3}{2}, 2, \frac{5}{2}, \ldots$, $m+p \in \mathbb{Z}$ and $\bar m + p \in \mathbb{Z}$ in the $\bar m$-wedge $|\bar{m}| < p$. The generators $p=1$ are central in this algebra. The modes $p=\frac{3}{2}$ and $p = 2$ are related to the leading \cite{Strominger:2013jfa,He:2014laa} and subleading \cite{Kapec:2014opa} soft gravitons, respectively. They can be identified with a subset of some chiral version of the generalized BMS algebra discussed in \cite{Banerjee:2020zlg}, see also Section 3.4 of \cite{Ruzziconi:2024kzo}. Remarkably, the $Lw_{1+\infty}$ symmetries already appeared in Penrose's non-linear graviton construction of self-dual spacetimes \cite{Penrose:1976jq,Penrose:1976js}, and find a natural interpretation on twistor space in terms of Poisson diffeomorphisms \cite{Adamo:2021lrv,Mason:2022hly}.

The celestial symmetries have also been shown to appear in the phase space of self-dual Yang-Mills \cite{Freidel:2023gue} and gravity \cite{Freidel:2021ytz, Geiller:2024bgf} at null infinity, where their canonical generators have been constructed using heuristic methods based on recursion relations. This line of work has been further discussed in \cite{Cresto:2024fhd,Cresto:2024mne,Cresto:2025bfo,Cresto:2025ubl}; see \cite{Cresto:2025fbc} for a comprehensive review. The interpretation of these symmetries from a spacetime perspective remains unclear and is suspected to be related to overleading gauge transformations or diffeomorphisms \cite{Nagy:2022xxs,Nagy:2024dme,Nagy:2024jua,Diaz-Jaramillo:2025gxw,Nagy:2025hip}. However, as mentioned above, they admit a natural geometric interpretation on twistor space. Exploiting this fact, the celestial charges have been derived from first principles by starting from twistor space actions for self-dual gravity \cite{Kmec:2024nmu} and self-dual Yang-Mills \cite{Kmec:2025ftx} and applying standard phase space methods \cite{Iyer:1994ys,Barnich:2001jy}. The resulting charges were then translated to spacetime expressions at null infinity using the non-local correspondence between spacetime and twistor space and matched with those obtained via heuristic approaches. Finally, spacetime expressions for the $Lw_{1+\infty}$ charges have been constructed in the phase space of a null hypersurface at finite distance, such as a black hole or a cosmological horizon, both using heuristic methods \cite{Ruzziconi:2025fct,Ruzziconi:2025fuy} and first-principles derivations from twistor space \cite{Kmec:2026dis}.

The celestial symmetries also appear in the context of twisted holography \cite{Costello:2022wso,Costello:2022jpg,Costello:2023hmi,Bittleston:2024efo}, where they are realized as generalized $2d$ chiral algebras on the celestial sphere and play a crucial role in constructing top-down models for celestial holography. The relation between the holographic celestial OPE approach, the twistor approach, and the spacetime approach to these symmetries has been clarified in \cite{Kmec:2025ftx}. These symmetries can also be embedded into a Carrollian framework, as discussed e.g. in \cite{Mason:2023mti, Saha:2023abr, Donnay:2024qwq, Ruzziconi:2024kzo}, using the Carrollian/celestial correspondence described above. Finally, the celestial symmetries have recently been extended to AdS spacetimes \cite{Taylor:2023ajd, Bittleston:2024rqe}, where they are interpreted as light-ray operator algebras in the boundary CFT$_3$ \cite{Sheta:2025oep}.

An interesting question is to understand the precise implications of these symmetries beyond the self-dual sector. In scattering theory, they are known to organize the MHV expansion, with interactions beyond self-duality appearing as controlled deformations constrained by the underlying celestial symmetry algebra; see e.g. \cite{Heuveline:2025nmb} for a review. Remarkably, in gravity, the $Lw_{1+\infty}$ charges have been shown to be related to the multipoles appearing in the metric expansion in full general relativity \cite{Compere:2022zdz, Kmec:2026dis}, which constitute important observables for gravitational wave physics.

\section{Flat space/Carrollian limit of AdS/CFT}
\label{sec:Flat space/Carrollian limit of AdS/CFT}

As stated in the introduction, AdS/CFT is the framework in which holography has been most extensively studied and is best understood, due to its derivation within string theory and to the remarkable fact that the dual theory is a CFT, a well-studied class of QFT. A natural approach to flat space holography is to take the flat space limit in the bulk, going from AdS to flat space by sending the AdS radius to infinity \cite{Susskind:1998vk,Polchinski:1999ry,Giddings:1999jq}. In practice, this limit is very subtle and has generated a substantial body of work (see e.g. \cite{Penedones:2010ue,Fitzpatrick:2011hu,Marotta:2024sce} and references therein). Only recently has it been understood as a Carrollian limit in the dual theory, offering a potential route towards a top-down realization of flat space holography by implementing a limit on both sides of the duality. Here, we present recent progress in this direction, notably at the level of holographic correlators. We will mostly follow \cite{Alday:2024yyj,Lipstein:2025jfj,Kulkarni:2025qcx} and refer to \cite{deGioia:2022fcn,deGioia:2023cbd,Bagchi:2023fbj,Bagchi:2023cen,Marotta:2024sce,Kraus:2024gso,deGioia:2024yne,deGioia:2025mwt,Adamo:2025bfr,Poulias:2025eck,Marotta:2025qjh} for closely related works. Let us emphasize that the approach presented here is based on a series of works in the context of classical general relativity, where the flat space limit of Einstein’s equations has been reinterpreted as a Carrollian limit of conservation laws at the boundary \cite{Barnich:2012aw,Ciambelli:2018wre,Poole:2018koa,Compere:2019bua,Compere:2020lrt,Campoleoni:2023fug}. The key strategy in this procedure is to take the limit in position space on both sides of the duality.

\subsection{Coordinates in AdS}

The (rectangular) Bondi coordinates, $(u,r,z, \bar z)$, with $u,r\in \mathbb R, z \in \mathbb C$, discussed in Section \ref{sec:Minkowski spacetime in Bondi coordinates}, also exist in AdS \cite{Barnich:2012aw,Poole:2018koa,Compere:2019bua,Compere:2020lrt}. In these coordinates, the AdS metric takes the form
\begin{equation}
    ds^2_{AdS} = - \frac{r^2}{\ell^2} du^2 - 2 du dr + 2 r^2 dz d\bar{z} ,
    \label{AdSBondi}
\end{equation} where the dimensions of length are $\ell \sim L$, $u \sim L$, $r \sim L$, $z \sim L^0$, $\bar z \sim L^0$. One can indeed check explicitly that the metric has constant curvature, i.e. $R_{\mu\nu\rho\sigma} = -\frac{1}{\ell^2}( g_{\mu\rho}g_{\nu\sigma}- g_{\mu\sigma} g_{\nu\rho})$. The advantage of working in Bondi coordinates is twofold: first they admit a well-defined flat space limit by taking the following dimensionless quantity to zero:
\begin{equation} \label{flat space limit def}
   \text{Flat space limit:} \qquad \frac{r}{\ell} \to 0 .
\end{equation} Indeed, in this limit, the line element \eqref{AdSBondi} consistently reduces to the Minkowski metric \eqref{Mink best}, $ds^2_{AdS} \to ds^2_{Flat}$. Notice that, in practice, we can just formally take $\ell \to \infty$ while keeping $r$ fixed. Furthermore, as we shall explain in more detail later, the Bondi coordinates cover two Poincaré patches: one with $r>0$ and the other with $r<0$. The separation hypersurface $r=0$ between the two patches is null, see Figure \ref{Fig:Poincare}. This will allow us to define a notion of incoming and outgoing propagator in Lorentzian AdS.

We now give a basic argument to illustrate the correspondence between the flat space limit in the bulk and the Carrollian limit at the boundary. The boundary metric in AdS is obtained from \eqref{AdSBondi} by restricting the metric on a $r=\bar{r} = \text{constant}$ hypersurface, rescaling the metric by $\bar{r}^{-2}$ to remove the double pole, and then sending $\bar{r} \to \epsilon\infty$ to reach the boundary of the first Poincaré patch ($\epsilon = +1$) or the second Poincaré patch ($\epsilon = -1$). We have 
\begin{equation}
    ds^2_{\partial AdS} =  \lim_{\bar{r} \to \epsilon \infty} \bar{r}^{-2}   \left( ds^2_{AdS} \right)\Big|_{r= \bar{r}} = - \frac{1}{\ell^2} du^2 + 2 dz d\bar{z}   .
\label{boundary metric AdS}
\end{equation} One can now draw the following commuting diagram:
\begin{equation}
	\begin{tikzpicture}[baseline=(current bounding box.center)]
		\tikzset{nodx/.style = {draw=black,outer sep=4pt,inner sep=7pt,text width=1.25cm,align=center}
}
		\tikzset{nodrelax/.style = {draw=black,outer sep=4pt,inner sep=7pt}
}
		\def\h{1.5};\def\l{3};
		\coordinate (A) at (-\l, \h);
		\coordinate (B) at ( \l, \h);
		\coordinate (C) at ( \l,-\h);
		\coordinate (D) at (-\l,-\h);
		\node[nodx] (a) at (A) {$ds^2_{{AdS}}$};
		\node[nodx] (b) at (B) {$ds^2_{{Flat}}$};
		\node[nodx] (d) at (D) {$ds^2_{\partial{AdS}}$};
		\node[nodx] (c) at (C) {$ds^2_{\partial{Flat}}$};
		\draw[-Latex] (a) -- (b) node[midway,above] {$\ell\to \infty$};
		\draw[-Latex] (d) -- (c) node[midway,below] {$c\to 0$};
		\draw[-Latex] (a) -- (d) node[midway,left] {$r\to \infty$};
		\draw[-Latex] (b) -- (c) node[midway,right] {$r\to \infty$};
		\node[outer sep=4pt] (e) at ($(A)!.5!(B)$) {};
		\node[outer sep=4pt] (f) at ($(C)!.5!(D)$) {};
		\node[nodrelax] (o) at (0,0) {$c_{{Boundary}}\equiv \frac{1}{\ell_{{Bulk}}}$};
		\draw[double, double distance=4pt] (o.north) -- (e.south);
		\draw[double, double distance=4pt] (o.south) -- (f.north);
	\end{tikzpicture} \label{c1L}
\end{equation}
 In the lower-right box, we recover the degenerate metric which is part of the Carrollian geometry at null infinity \cite{Henneaux:1979vn,1977asst.conf....1G,Duval:2014uva,Ashtekar:2014zsa}, see Section \ref{sec:Conformal Carrollian geometry of null infinity}, and in particular Equation \eqref{degenerate metric at scri}. From this, we see that the flat space limit in the bulk induces a Carrollian limit at the boundary:
\begin{equation}
    \text{Carrollian limit:} \qquad \frac{du}{\ell dz} \to 0 .
\end{equation} Indeed, by doing the formal identification $c_{Boundary}\equiv \ell^{-1}_{Bulk}$, keeping the coordinates fixed, the lower arrow induces exactly the Carrollian limit on the boundary metric, as discussed around Equation \eqref{Carrollian limit on the metric}. 

In Section~\ref{sec:Flat space limit of Witten diagrams}, we will show that the bulk flat space limit of position-space Witten diagrams, used to compute holographic CFT correlators, reproduces the Feynman diagrams for Carrollian amplitudes discussed in Section~\ref{sec:Feynman rules for Carrollian amplitudes}. This holds diagram by diagram. Then, in Section~\ref{sec:Carrollian limit of holographic correlators}, we will show that the Carrollian limit of holographic CFT correlators in position space reduces exactly to the Carrollian amplitudes. Hence, these two independent computations match and establish the flat space/Carrollian limit correspondence at the level of correlators; see Figure \ref{fig:diag}.

Following \cite{Alday:2024yyj}, we introduce the embedding space coordinates and discuss their relation with intrinsic coordinate systems: global, Poincaré and Bondi coordinates. Lorentzian AdS$_4$ can be described as a hyperboloid embedded in $\mathbb{R}^{3,2}$ with coordinates 
$X^I = \left(X^+, X^-, X^0, X^1, X^2\right)$ and metric
\begin{equation}
    \label{eq:embeddingmetric}
    G_{IJ} dX^I dX^J =  -dX^+dX^- -(dX^0)^2 + (dX^1)^2 + (dX^2)^2 .
\end{equation}
Indeed, AdS${}_{4}$ is the hyperboloid defined by the universal cover of 
\begin{equation}
    \label{eq:LAdSdef}
    X \cdot X = -X^{+}X^{-} -(X^0)^2 + (X^1)^2 + (X^2)^2 = -\ell^2  . 
\end{equation} We typically work with the universal cover of this quadric since the latter has closed timelike loops. This can be seen by temporarily switching from light-cone coordinates to ordinary ones via $X^{\pm} = X^3 \pm X^4$. This gives
\begin{equation}
    (X^3)^2 + (X^0)^2
    = \ell^2 + (X^1)^2 + (X^2)^2 + (X^4)^2 .
\end{equation}
We will also use Euclidean AdS$_4$ which is defined as the quadric
\begin{equation}
    \label{eq:EAdSdef}
    X \cdot X = -X^{+}X^{-} +(X^0)^2 + (X^1)^2 + (X^2)^2 = -\ell^2 \ , \qquad X^+ + X^- >0  ,
\end{equation}
in the embedding space $\mathbb{R}^{4,1}$. This is related to the Lorentzian one by analytic continuation.  In the rest of this section, we will work with Lorentzian AdS$_4$. We introduce three intrinsic coordinate systems.

\begin{itemize}
\item Global coordinates in AdS are obtained by choosing the following parametrization: 
\begin{equation}
    (X^3,X^0)=\frac{\ell}{\cos R}
    \left(\cos\tau,\sin\tau\right),
    \quad
    (X^1,X^2,X^4)=\ell\tan R
    \left(
        \frac{z+\bar z}{1+z\bar z},
        \frac{-i(z-\bar z)}{1+z\bar z},
        \frac{1-z\bar z}{1+z\bar z}
    \right).
\end{equation}
The metric reads 
\begin{equation}
    ds^2 = \frac{\ell^2}{\cos^2 R} [-d\tau^2 + d R^2 + \sin^2R \, d\Omega^2_{S} ]  ,
    \label{global coordinate}
\end{equation} where $d\Omega^2_{S} = \frac{4dz d\bar z}{(1+ z\bar z)^2}$ is the round sphere metric. From this embedding, the range of coordinates is $\tau \in [-\pi, \pi]$ and $R \in [0,\frac{\pi}{2}]$, see Figure \ref{Fig:Poincare}. As mentioned above, to get the universal cover of the quadric, we take instead $\tau \in \mathbb R$.    
\item Poincar{\' e} coordinates $(\rho,x^\mu)$ are obtained by choosing the following parametrization:
\begin{equation}
\label{eq:Poincare}
    X = \frac{\ell}{\rho}\left(1, \rho^2+ x \cdot x, x^0, x^1, x^2\right) ,
\end{equation}
where $x \cdot x = \eta_{\mu \nu} x^{\mu} x^{\nu}$ with $\eta$ being the metric on $\mathbb{R}^{1,2}$. Here we chose $\rho, x^\mu$ to be dimensionless and $\ell$ to have dimensions of length. The metric in these coordinates is 
\begin{equation}
    ds^2_{AdS} = \frac{\ell^2}{\rho^2} \Big(d\rho^2 - (dx^0)^2 +(dx^1)^2 + (dx^2)^2 \Big)  .
\end{equation} 
The conformal boundary is located at $\rho = 0$ and the region where $\rho > 0$ describes a Poincar{\'e} patch, see Figure \ref{Fig:Poincare} (the other Poincar\'e patch corresponds to the region where $\rho < 0$). Boundary Poincar{\'e} coordinates are denoted by
\begin{equation}
    x = \left(x^0, x^1, x^2 \right)
\end{equation}
and the boundary metric is 
\begin{equation}
    ds^2_{\partial AdS} = -(dx^0)^2 +(dx^1)^2 + (dx^2)^2  .
\end{equation} 
\item Bondi coordinates $(u,r,z,\bar z)$ are obtained via the parametrization
\begin{equation}
    X = r\left(1, \frac{2u}{r}-\frac{u^2}{\ell^2}+2 z \bar z, -\frac{\ell}{r}+\frac{u}{\ell}, \frac{z+\bar z}{\sqrt{2}},  \frac{z-\bar z}{\sqrt{2}i}\right) .
\end{equation}
The relation between Poincar{\'e} and Bondi coordinates is 
\begin{equation}
    \label{eq:ptob}
    \rho = \frac{\ell}{r}  , \quad x^0 = -\frac{\ell}{r}+\frac{u}{\ell} , \quad x^1 = \frac{z+\bar z}{\sqrt{2}}  , \quad x^2 = \frac{z-\bar z}{\sqrt{2}i}  .
\end{equation}
The interplay between Bondi coordinates and Poincar\'e patch is discussed in Figure \ref{Fig:Poincare}. One can check that the metric in Bondi coordinates is indeed given by \eqref{AdSBondi}. The conformal boundary is reached by letting $r \to \pm \infty$. After a suitable rescaling, the boundary coordinates are
\begin{equation}
      \left(1, -\frac{u^2}{\ell^2}+2 z \bar z, \frac{u}{\ell}, \frac{z+\bar z}{\sqrt{2}},  \frac{z-\bar z}{\sqrt{2}i}\right) \ ,
\end{equation}
and the boundary metric reproduces \eqref{boundary metric AdS}.  
\end{itemize}

\begin{figure}[ht!]
\centering
\begin{tikzpicture}
	\def\h{3};\def\l{2};\def\dy{0.5};\def\ofx{1.7};\def\ofy{0.2};
	\coordinate (bl) at (-\l,-\h);
	\coordinate (br) at ( \l,-\h);
	\coordinate (cl) at (-\l,  0);
	\coordinate (cr) at ( \l,  0);
	\coordinate (cc) at (  0,  0);
	\coordinate (tl) at (-\l, \h);
	\coordinate (tr) at ( \l, \h);
	\coordinate (al) at ($(cl)+(0,1.5)$);
	\coordinate (ar) at ($(al)+(2*\l,-2*\l)$);
	\coordinate (bracetopfin) at ($(tl)+(-\ofx,-\ofy)$);
	\coordinate (bracetopini) at ($(cl)+(-\ofx, \ofy)$);
	\coordinate (bracebotini) at ($(bl)+(-\ofx, \ofy)$);
	\coordinate (bracebotfin) at ($(cl)+(-\ofx,-\ofy)$);
	\draw[red,opacity=0] ($(bl)+(-3.1,-1)$) -- ($(tl)+(-3.1,1)$) -- ($(tr)+(3.1,1)$) -- ($(br)+(3.1,-1)$) -- cycle;
	\draw[] (bl)node[left]{$\tau=-\pi$} -- (cl)node[left]{$\tau=0$} -- (tl)node[left]{$\tau=+\pi$};
	\draw[] (br) -- (tr)node[anchor=north west]{$\mathscr I_{\text{AdS}}$};
	\draw[] (bl) arc[x radius=\l, y radius=\dy, start angle=-180, end angle=0];
	\draw[densely dashed] (bl) arc[x radius=\l, y radius=\dy, start angle=180, end angle=0];
	\draw[] (cl) arc[x radius=\l, y radius=\dy, start angle=-180, end angle=0];
	\draw[densely dashed] (cl) arc[x radius=\l, y radius=\dy, start angle=180, end angle=0];
	\draw[] (tl) arc[x radius=\l, y radius=\dy, start angle=-180, end angle=180];
	\node[left] at (al) {\scriptsize $(u,z,\bar z)$};
	\node[right] at (ar) {\scriptsize $(u,z,\bar z)$};
	\draw[decorate, decoration={brace}] (bracetopini) -- (bracetopfin);
	\draw[decorate, decoration={brace}] (bracebotini) --(bracebotfin);
	\node[left,outer sep=3pt] at ($(bracetopini)!0.5!(bracetopfin)$) {$u>0$};
	\node[left,outer sep=3pt] at ($(bracebotini)!0.5!(bracebotfin)$) {$u<0$};
	\draw[navyblue] (cr) to[out=180,in=-80] (tl);
	\draw[navyblue,dashed] (cr) to[out=100,in=0] (tl);
	\draw[navyblue,dashed] (cr) to[out=180,in=80] (bl);
	\draw[navyblue] (cr) to[out=-100,in=0] (bl);
	\fill[navyblue,opacity=0.2] (tl) to[in=100,out=0] (cr) to[out=-100,in=0] (bl);
	\fill[navyblue,opacity=0.5] (tl) to[in=180,out=-80]  (cr) to[out=-100,in=0] (bl);
	\draw[->,navyblue] ($(cr)+(0.2,0)$) -- ($(cr)+(1,0)$)node[right]{\scriptsize Poincaré patch};
	\node[rotate=-45,anchor=south west,red!75!black] at (al) {\scriptsize $r\to+\infty$};
	\node[rotate=-45,anchor=north east,red!75!black] at (ar) {\scriptsize $r\to-\infty\,$};
	\draw[Latex-Latex,red!75!black] (al)node[circle,fill,inner sep=1pt]{} -- (ar)node[circle,fill,inner sep=1pt]{};
\end{tikzpicture}
\caption{The blue region represents a Poincar\'e patch ($\rho>0$) in AdS. To cover the other half of the cylinder, we can use a second Poincar\'e patch corresponding to $\rho <0$. In particular, this naturally divides the boundary into two regions. The Bondi coordinates also cover the region outside of the patch (this follows from \eqref{eq:ptob} and $r\in \mathbb R$). In red is a null ray defined by $(u,z,\bar z) = \text{constant}$ in Bondi coordinates. The point obtained by taking $r\to +\infty$ is inside of the patch, while the point $r\to -\infty$ is outside of the patch \cite{Alday:2024yyj}.} \label{Fig:Poincare}
\end{figure}
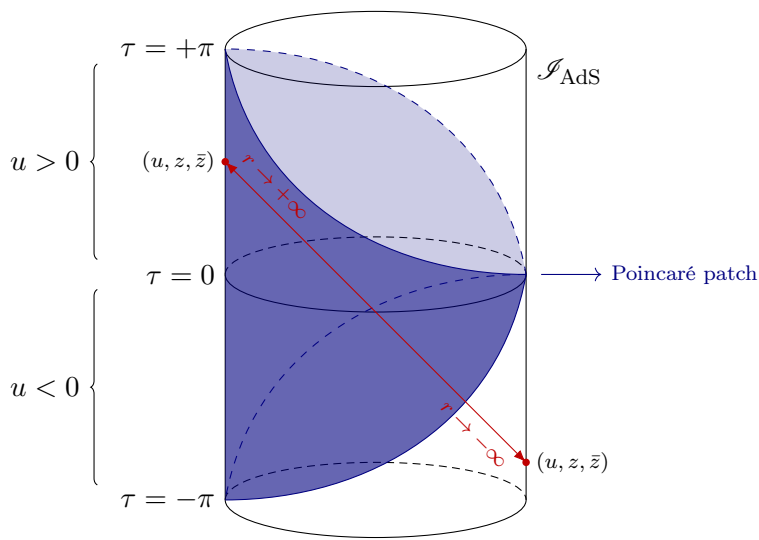

The chordal distance between two points in AdS${}_4$ appears frequently in computations. This is defined as
\begin{equation}
    \xi_{12}^{AdS} = (X_1 - X_2)\cdot (X_1 - X_2) = \frac{\ell^2}{\rho_1 \rho_2} \Big(\rho^2_{12}+x_{12}^2 \Big) \label{eq:chordalp} \ ,
\end{equation} 
where $x_{12}^2 = \eta_{\mu \nu} (x_1 - x_2)^{\mu} (x_1 - x_2 )^{\nu}$ and the first expression for $\xi^{AdS}_{12}$ is in terms of the embedding coordinates while the second in terms of Poincar{\'e} coordinates. This can also be written in Bondi coordinates as 
\begin{equation}
     \xi_{12}^{AdS} = - \frac{1}{\ell^2} r_1 r_2 u_{12}^2  -2  r_{12} u_{12} + 2 r_1 r_2 |z_{12}|^2  = \xi^{Flat}_{12} - \frac{1}{\ell^2}r_1 r_2 u_{12}^2 \ . \label{eq:chordalb} 
\end{equation}
A virtue of Bondi coordinates is that the chordal distance between two points reduces to the flat space distance \eqref{distance flat} in the limit $\ell \to \infty$.

\subsection{Flat space limit of Witten diagrams}
\label{sec:Flat space limit of Witten diagrams}

\subsubsection{Propagators} 

The two key ingredients in the computation of Witten diagrams in AdS are the bulk-to-bulk and bulk-to-boundary propagators. For scalars, the bulk-to-bulk propagator $\mathcal{G}^{AdS,\Delta}_{BB}$ can be obtained by solving the sourced Klein-Gordon equation in AdS$_4$ 
\begin{equation}
    (\Box_{X_1} + m^2) \mathcal{G}^{AdS,\Delta}_{BB}(X_1,X_2) = \frac{1}{\sqrt{-g}}\delta^{(4)} \left(X_{12}\right) \ ,
\label{KG equation AdS}
\end{equation} with $m^2 \ell^2 = \Delta (3-\Delta)$. In Bondi coordinates, the d'Alembert operator reads explicitly as
\begin{equation}
\label{eq:wavebondi}
    \Box = \Big[  \frac{2}{r^2} \partial_z {\partial}_{\bar z} + \frac{r}{\ell^2} (4 \partial_r + r \partial_r^2) - \frac{2}{r} \partial_u - 2 \partial_u \partial_r   \Big]
\end{equation} and Equation \eqref{KG equation AdS} consistently reduces to the flat space sourced Klein-Gordon Equation \eqref{box equation flat} when $\ell \to \infty$. We will assume that $\D$ does not rescale with $\ell$ as we take the flat space limit, so that we go from a generically massive scalar field in AdS to a massless scalar field in flat space.

The second order equation \eqref{KG equation AdS} admits two independent solutions, and we will only keep one of them by imposing regularity in the interior \cite{Witten:1998qj}. Expressing this solution in the Bondi parametrization and taking the flat space limit, we recover the Feynman propagator for a massless scalar field in Minkowski space \eqref{bulktobulkflat}:
\begin{equation}
   \mathcal{G}^{AdS,\Delta}_{BB}(X_1,X_2)  \xrightarrow[]{\ell \to \infty} 
   \mathcal{G}^{Flat}_{BB}(X_1,X_2)  .
\label{flatlimitBB}
\end{equation} We refer to \cite{Alday:2024yyj} for more details. In particular, the parameter $\Delta$ completely disappears in the limit. This is consistent with the observation made below \eqref{eq:wavebondi} where the massive Klein-Gordon equation in AdS exactly reduces to the massless one in flat space.

\begin{figure}[h!]
\begin{center}
\begin{tikzpicture}[scale=2.2]
    \tikzset{blok/.style={draw,outer sep=2pt,inner sep=5pt}};
    \tikzset{->-/.style={decoration={markings, mark=at position .53 with {\arrow{stealth}}}, postaction={decorate}}};
    \def\l{3};
    \def\shrink{2};

    \begin{scope}
        \node[blok] (topleft) at (0,\l)
            {$\mathcal{G}_{BB}^{AdS,\Delta}(X_1,X_2)$};

        \node[blok] (botleft) at (0,\l/\shrink)
            {$\mathcal{G}_{Bb,+1}^{AdS,\Delta}(x_1;X_2)$};

        \node[blok] (topcent) at (\l,\l)
            {$\mathcal{G}_{BB}^{Flat}(X_1,X_2)$};

        \node[blok] (botcent) at (\l,\l/\shrink)
            {$\mathcal{G}_{Bb,+1}^{Flat,\Delta}(x_1;X_2)$};

        \draw[->-] (topleft) -- (botleft);
        \draw[->-] (topcent) -- (botcent);
        \draw[->-] (topleft) -- (topcent);
        \draw[->-] (botleft) -- (botcent);

        \node[left] at ($(topleft)!.5!(botleft)$) {$r_1 \to + \infty$};
        \node[right] at ($(topcent)!.5!(botcent)$) {$r_1\to+\infty,\ \partial_u^{\Delta-1}$};
        \node[above] at ($(topleft)!.5!(topcent)$) {$\ell\to+\infty$};
        \node[below] at ($(botleft)!.5!(botcent)$)
            {$\ell \to \infty$};
    \end{scope}
\end{tikzpicture}

\caption{Relations between bulk-to-bulk and bulk-to-boundary propagators in AdS and flat spacetime.}
\label{fig:limits}
\end{center}
\end{figure}
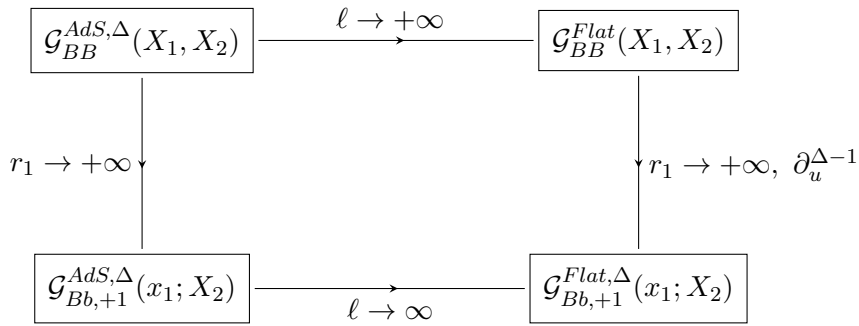

Analogously to the flat space discussion in Section \ref{sec:Propagators}, the bulk-to-boundary propagator in AdS can be obtained from the bulk-to-bulk propagator by sending one point to infinity, and multiplying by the appropriate power of $r$ to obtain a finite result:
\begin{equation}
    \begin{split}
   \mathcal{G}_{Bb, +1}^{AdS,\Delta} \left(x_1;X_2\right)  
   &\equiv \ell \lim_{r_1 \to \infty} \left(\frac{r_1}{\ell}\right)^{\D} \,\mathcal{G}_{BB}^{AdS,\Delta} (X_1,X_2) \\
   &=  C (\Delta) \ell^{-1}   \Big(\frac{-4\ell}{- \frac{1}{\ell^2} r_2 u_{12}^2  -2 u_{12} + 2 r_2 |z_{12}|^2 + i \epsilon \varepsilon }\Big)^{\D} ,
    \end{split} \label{eq:bboundary}
\end{equation} where $C (\Delta)$ is a fixed constant that does not depend on $\ell$.\footnote{We pulled out the factor of $\ell^{-2}$ in the definition of $C_2 (\Delta)$ given in Equation (3.23) of \cite{Alday:2024yyj}, i.e. $C^{here}(\Delta) = \ell^2 C_2^{there}(\Delta)$.} The factors of $r$ allow us to extract the finite piece, and the factor of $\ell$ is fixed by consistency with the units (see e.g. \cite{Penedones:2010ue}). As in flat space, in Bondi coordinates, we can distinguish between outgoing and incoming bulk-to-boundary propagators in AdS, corresponding to $r_1 \to +\infty$ (the case above) and $r_2 \to - \infty$, respectively (see Figure \ref{fig:flatlimit}). Hence the outgoing ($\epsilon = +1$) and incoming ($\epsilon = -1$) bulk-to-boundary propagators are given by 
\begin{equation}
\mathcal{G}_{Bb,\epsilon}^{AdS,\Delta} \left(x;X\right) =  C (\Delta)   \Big(\frac{-4\ell}{- \frac{1}{\ell^2} r_X u_{xX}^2  -2 u_x - 2 q_x \cdot X + i \epsilon \varepsilon }\Big)^{\D}  .
\end{equation} For $\epsilon = +1$ (resp. $\epsilon = -1)$, the point $x$ is at the boundary of the first (resp. second) Poincaré patch. In the flat space limit, these bulk-to-boundary propagators reproduce those of Minkowski spacetime written in Equation \eqref{Bb propagator with D},
\begin{equation}
   \label{flat limit Bb}\mathcal{G}^{AdS,\Delta}_{Bb,\epsilon}(x;X)  \xrightarrow[]{\ell \to \infty} 
   \alpha (\Delta) \ell^{\Delta - 1}\mathcal{G}^{Flat,\Delta}_{Bb, \epsilon}(x;X) ,
\end{equation} where $\alpha (\Delta)$ is a fixed constant that does not depend on $\ell$. In particular, for $\Delta = 1$, the expression for the bulk-to-boundary propagator reduces exactly to the one in Minkowski space \eqref{bulktoboundaryflat} without extra factor, i.e. $\alpha (\Delta = 1) = 1$. As discussed previously, the other values of $\Delta$ are redundant to encode the $\mathcal{S}$-matrix in flat space, and the bulk-to-boundary propagators with integer values $\Delta = 1, 2, 3, \ldots$ can be obtained by taking $\partial_u$-derivatives on the bulk-to-boundary propagator obtained by extrapolation of the Feynman propagator to null infinity; see below \eqref{Bb propagator with D}. In Figure \ref{fig:limits}, we summarise the connection between the various propagators in AdS and flat space.

\subsubsection{Witten diagrams}
\label{sec:AdS Witten diagram}

In the previous section, we showed that the massive bulk-to-bulk and bulk-to-boundary propagators reduce respectively to the massless bulk-to-bulk and bulk-to-boundary propagators in Minkowski spacetime in the flat limit $\ell \to \infty$. These propagators in AdS being the building blocks of the Witten diagrams, we are somehow guaranteed that the latter will also have a well-defined limit. In this section, we show that this is indeed the case and that we directly end up with the flat space Carrollian amplitudes expressed in terms of Feynman rules in Section \ref{sec:Diagrams}. Let us emphasize that the limit is taken entirely in position space \cite{Alday:2024yyj}.

\begin{figure}[ht!]
\centering
\begin{tikzpicture}[scale=.9]
	\def\h{3};\def\l{2};\def\dy{0.5};\def\ofx{1.7};\def\ofy{0.2};
	\coordinate (bl) at (-\l,-\h);
	\coordinate (br) at ( \l,-\h);
	\coordinate (cl) at (-\l,  0);
	\coordinate (cr) at ( \l,  0);
	\coordinate (cc) at (  0,  0);
	\coordinate (tl) at (-\l, \h);
	\coordinate (tr) at ( \l, \h);
	\coordinate (bbl) at (-\l,-2*\h);
	\coordinate (bbr) at ( \l,-2*\h);
	\coordinate (al) at ($(cl)+(0,1.5)$);
	\coordinate (ar) at ($(al)+(2*\l,-2*\l)$);
	\coordinate (bracetopfin) at ($(tl)+(-\ofx,-\ofy)$);
	\coordinate (bracetopini) at ($(cl)+(-\ofx, \ofy)$);
	\coordinate (bracebotini) at ($(bl)+(-\ofx, \ofy)$);
	\coordinate (bracebotfin) at ($(cl)+(-\ofx,-\ofy)$);
	\draw[red,opacity=0] ($(bl)+(-1.5,-4)$) -- ($(tl)+(-1.5,1)$) -- ($(tr)+(13,1)$) -- ($(br)+(13,-4)$) -- cycle;
	\draw[] (bl) -- (cl) -- (tl);
	\draw[] (br) -- (tr)node[anchor=north west]{$\mathscr I_{\text{AdS}}$};
	\draw[] (bl) arc[x radius=\l, y radius=\dy, start angle=-180, end angle=0];
	\draw[densely dashed] (bl) arc[x radius=\l, y radius=\dy, start angle=180, end angle=0];
	\draw[] (cl) arc[x radius=\l, y radius=\dy, start angle=-180, end angle=0];
	\draw[densely dashed] (cl) arc[x radius=\l, y radius=\dy, start angle=180, end angle=0];
	\draw[] (tl) arc[x radius=\l, y radius=\dy, start angle=-180, end angle=180];
	\draw[navyblue] (cr) to[out=180,in=-80] (tl);
	\draw[navyblue,dashed] (cr) to[out=100,in=0] (tl);
	\draw[navyblue,dashed] (cr) to[out=180,in=80] (bl);
	\draw[navyblue] (cr) to[out=-100,in=0] (bl);
	\fill[navyblue,opacity=0.2] (tl) to[in=100,out=0] (cr) to[out=-100,in=0] (bl);
	\fill[navyblue,opacity=0.5] (tl) to[in=180,out=-80]  (cr) to[out=-100,in=0] (bl);
	\draw (bl) -- (bbl);
	\draw (br) -- (bbr);
	\draw[densely dashed] (bbl) arc[x radius=\l, y radius=\dy, start angle=180, end angle=0];
	\draw (bbl) arc[x radius=\l, y radius=\dy, start angle=-180, end angle=0];
	\draw[orange,dashed] (bbr) to[out=100,in=0] (bl);
	\draw[orange] (bbr) to[out=180,in=-80] (bl);
	\fill[orange,opacity=0.5] (bbr) to[in=-80,out=180]  (bl) to[out=0,in=-100] (cr);
	\fill[orange,opacity=0.2] (bbr) to[in=-80,out=180]  (bl) to[out=0,in=100] (bbr);
	\node[navyblue,rotate=90,anchor=south east] at (cl) {out $(\epsilon=+1)$};
	\node[orange,rotate=-90,anchor=south west] at (br) {in $(\epsilon=-1)$};
	\def\del{1.5};
	\coordinate (nexus) at ($(cr)!0.5!(br)$);
	\coordinate (inarr) at ($(nexus)+(\del,0)$);
	\coordinate (fiarr) at ($(inarr)+(2*\del,0)$);
	\draw[very thick,black!70,-Latex] (inarr) -- (fiarr);
	\node[black!70,above,outer sep=4pt] at ($(inarr)!0.5!(fiarr)$) {$\ell\to+\infty$};
	\def\are{3};
	\coordinate (L) at ($(fiarr)+(\del,0)$);
	\coordinate (B) at ($(L)+(\are,-\are)$);
	\coordinate (R) at ($(B)+(\are, \are)$);
	\coordinate (T) at ($(R)-(\are,-\are)$);
	\draw (L) -- (B) -- (R) -- (T) -- cycle;
	\draw[] (L) to[out=-20,in=-160] (R);
	\draw[densely dashed] (L) to[out=20,in=160] (R);
	\fill[orange,opacity=0.5] (L) to[out=20,in=160] (R) -- (B) -- (L);
	\fill[navyblue,opacity=0.5] (L) to[out=-20,in=-160] (R) -- (T) -- (L);
	\node[anchor=south west] at ($(T)!0.5!(R)$) {$\mathscr I^+$};
	\node[anchor=north west] at ($(B)!0.5!(R)$) {$\mathscr I^-$};
	\node[below] at (B) {$i^-$};
	\node[above] at (T) {$i^+$};
	\node[right] at (R) {$i^0$};
	\node[left] at (L) {$i^0$};
	\node[circle,inner sep=1.5pt,fill] at (B) {};
	\node[circle,inner sep=1.5pt,fill] at (T) {};
	\node[circle,inner sep=1.5pt,fill] at (R) {};
	\node[circle,inner sep=1.5pt,fill] at (L) {};
	\node[navyblue,rotate=45,above] at ($(T)!0.5!(L)$) {out $(\epsilon=+1)$};
	\node[orange,rotate=-45,below] at ($(B)!0.5!(L)$) {in $(\epsilon=-1)$};
	\node[anchor=south west,outer sep=10pt] (toparrowL) at ($(cr)!0.3!(tr)$) {};
	\node[anchor=south east,outer sep=10pt] (toparrowR) at ($(T)!0.5!(L)$) {};
	\draw[line width=3pt,navyblue!20,-Latex] (toparrowL) to[out=20,in=145] (toparrowR);
	\node[anchor=north west,outer sep=10pt] (downarrowL) at ($(br)!0.3!(bbr)$) {};
	\node[anchor=north east,outer sep=10pt] (downarrowR) at ($(B)!0.5!(L)$) {};
	\draw[line width=3pt,orange!20,-Latex] (downarrowL) to[out=-20,in=-145] (downarrowR);
	\node[rotate=-45,anchor=south west,red!75!black] at (al) {\scriptsize $r\to+\infty$};
	\node[rotate=-45,anchor=north east,red!75!black] at (ar) {\scriptsize $r\to-\infty\,$};
	\draw[red!75!black,Latex-Latex] (al)node[circle,fill,inner sep=1pt]{} -- (ar)node[circle,fill,inner sep=1pt]{};
	\node[left] at (al) {\scriptsize $(u,z,\bar z)$};
	\node[right] at (ar) {\scriptsize $(u,z,\bar z)$};
	\coordinate (tipr) at ($(L)!0.3!(T)$);
	\coordinate (orir) at ($(B)!0.3!(R)$);
	\draw[red!75!black,Latex-Latex] (tipr)node[circle,fill,inner sep=1pt]{} -- (orir)node[circle,fill,inner sep=1pt]{};
	\node[below,rotate=45] at (orir) {\scriptsize $(u,z,\bar z)$};
	\node[rotate=-45,anchor=south west,red!75!black] at (tipr) {\scriptsize $r\to+\infty$};
	\node[rotate=-45,anchor=north east,red!75!black] at (orir) {\scriptsize $r\to-\infty\,$};
\end{tikzpicture}
\caption{The left figure represents two Poincar\'e patches in AdS: the blue one is covered by Bondi coordinates with $r>0$, the orange one is covered by Bondi coordinates with $r<0$ (see also Figure \ref{Fig:Poincare}). These two patches provide a natural division of the AdS boundary into two regions characterised by $r\to +\infty$ and $r\to -\infty$, respectively. These two regions will be the locus for the insertion of operators creating outgoing ($\epsilon = +1$) and incoming ($\epsilon = -1$) propagators. The right figure corresponds to flat space in Bondi coordinates (see also Figure \ref{FigureBondi}). The complete figure summarizes the effect of taking the flat limit $\ell \to \infty$ in the bulk: the boundary of the blue patch is sent to $\mathscr I^+$ and the boundary of the orange region is sent to $\mathscr I^-$. The upper/lower boundary limit of the blue (resp. orange) region is mapped to $\mathscr{I}^+_+/\mathscr{I}^+_-$ (resp. $\mathscr{I}^-_+/\mathscr{I}^-_-$). Importantly, there is no kinematic restriction before taking the limit (the operators can be inserted anywhere on the two patches) \cite{Alday:2024yyj}.} \label{fig:flatlimit}
\end{figure}
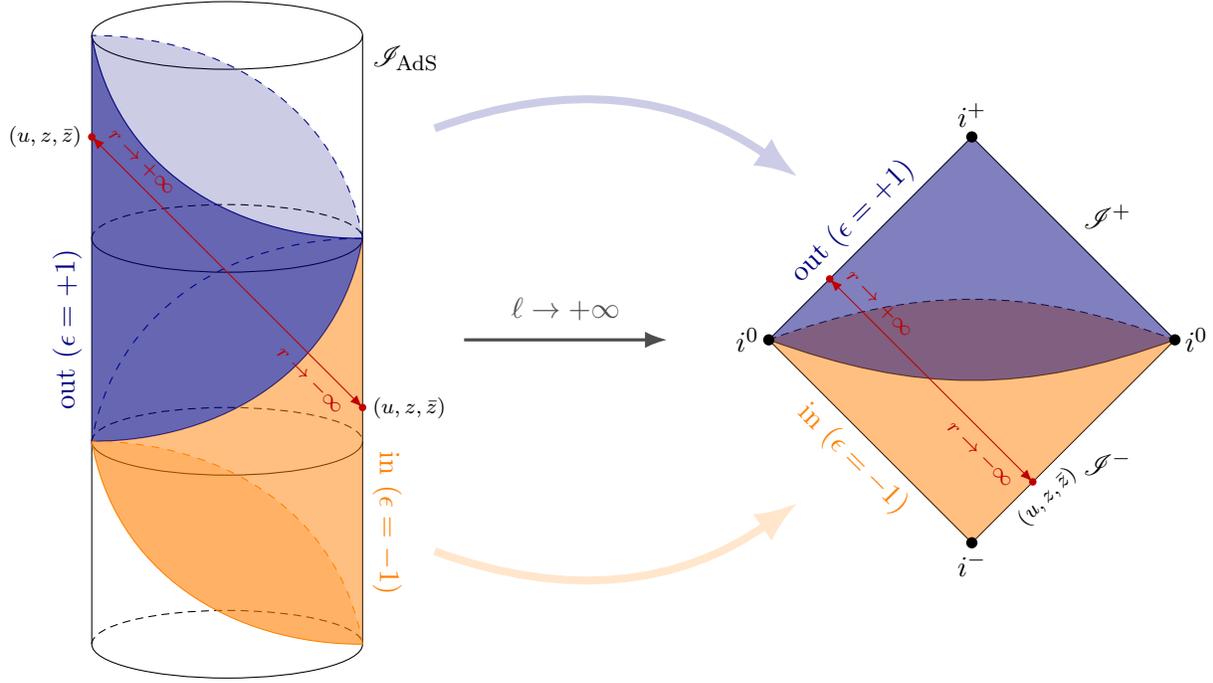

Correlators of operators inserted at the boundary of AdS$_4$ can be computed explicitly using Witten diagrams \cite{Witten:1998qj,Freedman:1998tz,Liu:1998bu,DHoker:1998ecp,DHoker:1998bqu}. These correlators are associated with CFT correlators at strong coupling via the AdS/CFT correspondence. We denote by $\mathcal{O}^\epsilon_\Delta (x)$ the boundary operators corresponding to CFT primaries of spin $s=0$ and conformal dimension $\Delta$. The extra index $\epsilon = \pm 1$ labels operators inserted at the boundary of the first/second Poincaré patch, respectively (see Figure \ref{fig:flatlimit}), and will be associated with outgoing/incoming operators.

The Lorentzian two-point correlation function can be computed by starting from $\mathcal{G}_{Bb,+1}^{AdS,\Delta}$ in \eqref{eq:bboundary} and taking $r_2 \to -\infty$. With the appropriate rescaling, we find
\begin{equation}
      \langle \mathcal{O}^{+1}_{\Delta} (x_1) \mathcal{O}^{-1}_{\Delta} (x_2) \rangle = \ell \lim_{r_2 \to - \infty } \left(\frac{r_2}{\ell}\right)^{\D}\mathcal{G}_{Bb,+1}^{AdS,\Delta}(x_1,X_2) = \mathcal{N}_2  \frac{1}{(-\frac{1}{\ell^2} u_{12}^2 + 2 |z_{12}|^2 - i \varepsilon)^\Delta}  ,
\label{2ptAdSWitten}
\end{equation} where $\mathcal{N}_2$ is a constant that does not scale with $\ell$. Alternatively, the CFT $2$-point function can be computed via the following integrated product of bulk-to-boundary propagators: 
\begin{equation}
     \langle \mathcal{O}^{+1}_{\Delta} (x_1) \mathcal{O}^{-1}_{\Delta} (x_2) \rangle = \int_{AdS} d^4 X \, \mathcal G_{Bb, +1}^{AdS , \Delta} (x_1; X) \mathcal G_{Bb, -1}^{AdS , \Delta} (x_2; X) 
     \label{2 point in AdS Wd}
\end{equation} up to an overall constant that does not scale with $\ell$ (see \cite{Alday:2024yyj} for a consistent choice of normalization). We can take the flat space limit of this expression by taking the limit inside of the integral and using \eqref{flat limit Bb}. However, there is a subtle point here: recall that strictly speaking, the flat space limit is defined as in \eqref{flat space limit def}. Since there is an integral running over all values of $r$ in \eqref{2 point in AdS Wd}, the correct way to take this limit is to introduce an IR cut-off in the $r$-integral, then take $\ell \to \infty$ and remove the cut-off at the very end. By applying this procedure, we find that this expression reduces exactly to the Feynman diagram for the $2$-point Carrollian amplitude in \eqref{2point carrollian diagram}, up to some overall factors. Re-absorbing these global factors into a redefinition of the primaries in the limit, we find 
\begin{equation}
    \langle \mathcal{O}^{+1}_{\Delta} (x_1) \mathcal{O}^{-1}_{\Delta} (x_2) \rangle \xrightarrow[]{\ell \to \infty}   \langle \Phi^{+1}_{\Delta} (x_1) \Phi^{-1}_{\Delta} (x_2) \rangle, \qquad  \mathcal{O}^\epsilon_\Delta (x) \sim \ell^{\Delta-1} \Phi^\epsilon_\Delta (x) . \label{electric scaling in the flat limit}
\end{equation} On the left-hand side, $x= (u,z, \bar z)$ corresponds to a point at the boundary of AdS, $\mathscr{I}_{AdS}$, while on the right-hand side, this corresponds to a point at the boundary of flat space, $\mathscr I^\pm$. The scaling of $\mathcal{O}_\Delta^{\epsilon}$ with $\ell$ in the limit is fixed by the scaling of the bulk-to-boundary propagator in \eqref{flat limit Bb}. As stated before, the conformal dimension $\Delta$ in AdS becomes redundant in the flat space limit to encode the bulk $\mathcal{S}$-matrix.

The story is very similar for the $3$-point function, which is computed through 
\begin{equation}
    \langle \mathcal{O}^{\epsilon_1}_{\Delta_1} (x_1) \mathcal{O}^{\epsilon_2}_{\Delta_2} (x_2) \mathcal{O}^{\epsilon_3}_{\Delta_3} (x_3) \rangle = \int_{AdS} d^4X \mathcal{G}^{AdS,\Delta_1}_{Bb,\epsilon_1}(x_1;X) \mathcal{G}^{AdS,\Delta_2}_{Bb,\epsilon_2} (x_2;X)\mathcal{G}^{AdS,\Delta_3}_{Bb,\epsilon_3}(x_3;X) .
\label{3ptWd}
\end{equation} Using Equation \eqref{flat limit Bb} and the procedure described above, we see that \eqref{3ptWd} reduces precisely to \eqref{3pointdiagram} provided the CFT primaries are related to the Carrollian CFT primaries via the same scaling as for the $2$-point function in Equation \eqref{electric scaling in the flat limit}.

Similarly, the AdS correlator associated with the four-point contact diagram reads as 
\begin{multline}
\label{fourpointAdS}
    \langle \mathcal{O}^{\epsilon_1}_{\Delta_1} (x_1) \mathcal{O}^{\epsilon_2}_{\Delta_2} (x_2) \mathcal{O}^{\epsilon_3}_{\Delta_3} (x_3) \mathcal{O}^{\epsilon_4}_{\Delta_4} (x_4) \rangle_c \\
    =   \int_{AdS} d^4X \mathcal{G}^{AdS,\Delta_1}_{Bb,\epsilon_1}(x_1;X) \mathcal{G}^{AdS,\Delta_2}_{Bb,\epsilon_2} (x_2;X)\mathcal{G}^{AdS,\Delta_3}_{Bb,\epsilon_3}(x_3;X) \mathcal{G}^{AdS,\Delta_4}_{Bb,\epsilon_4}(x_4;X) 
\end{multline} and reduces to \eqref{4pointcontact} after the rescaling \eqref{electric scaling in the flat limit}. Finally, the four-point exchange diagram can be computed as 
\begin{align}
&\langle \mathcal{O}^{\epsilon_1}_{\Delta_1} (x_1) \mathcal{O}^{\epsilon_2}_{\Delta_2} (x_2) \mathcal{O}^{\epsilon_3}_{\Delta_3} (x_3) \mathcal{O}^{\epsilon_4}_{\Delta_4} (x_4) \rangle_e \\ &\qquad=  \int_{AdS} d^4 X d^4Y \mathcal{G}^{AdS,\Delta_1}_{Bb,\epsilon_1} (x_1;X) \mathcal{G}^{AdS,\Delta_2}_{Bb,\epsilon_2} (x_2;X) \mathcal{G}^{AdS,\Delta}_{BB}(X,Y) \mathcal{G}^{AdS,\Delta_3}_{Bb,\epsilon_3} (x_3;Y) 
    \mathcal{G}^{AdS,\Delta_4}_{Bb,\epsilon_4} (x_4;Y) \nonumber
\end{align} and maps perfectly to the corresponding flat space Carrollian amplitude \eqref{flat space4ptexch} after taking \eqref{flatlimitBB} and \eqref{flat limit Bb} into account, as well as the rescaling \eqref{electric scaling in the flat limit}.

\subsection{Carrollian limit of holographic correlators}
\label{sec:Carrollian limit of holographic correlators}

In this section, we present a purely boundary computation, which consists of taking the Carrollian limit of the holographic CFT correlators and showing that these reduce consistently to the Carrollian amplitudes encoding the flat-space $\mathcal{S}$-matrix.

\subsubsection{General procedure}

The procedure for obtaining the Carrollian amplitudes for massless scalars $\an{\Phi_{\D_1} \dots \Phi_{\D_n}}$ from holographic CFT correlators is performed following these steps \cite{Alday:2024yyj}:
\begin{itemize}
    \item The holographic CFT correlators, when known in closed form, are usually expressed in Euclidean signature. Following our prescription for taking the limit, the first step is to analytically continue these correlators to Lorentzian/Kleinian signature. This analytic continuation is very subtle and not unique; it corresponds to the different possibilities of operator insertions at the boundary of either of the two Poincaré patches in AdS (see Figure~\ref{Fig:Poincare}). Hence, only certain of these analytic continuations lead to a non-vanishing result in the flat space limit.

    \item The second step is to restore the speed of light in the correlator expressions, by simply using the coordinates $x = (c u, z, \bar{z})$, which coincide with those induced by the bulk Bondi coordinates through the formal identification $c_{Boundary} \equiv \ell^{-1}$ explained in Equation \eqref{c1L}. The coupling constants may also be rescaled by some power of $c$, as explained in Section \ref{sec:Coupling constants}. 
    
    \item The third and last step consists of implementing the $c\to 0$ limit. As we shall explain, similarly to what happens for the $c\to 0$ limit at the level of the Lagrangians discussed in Section~\ref{sec:Scalar field}, there are different branches of correlation functions appearing in the limit, which are of electric or magnetic type. In particular, these are consistent with the Carrollian correlation functions obtained in Section~\ref{sec:Ward identities} by solving the Carrollian CFT Ward identities. We will be interested in the electric type of branch, as it is leading in the $c \to 0$ limit and coincides with the Carrollian amplitudes. More precisely, we will compute
    \begin{equation}
\lim_{c \to 0}  c^{\sum_i \Delta_i - n}
\left\langle \mathcal{O}_{\Delta_1} \dots \mathcal{O}_{\Delta_n} \right\rangle
 \end{equation}
by keeping track of distributional terms, and identify the rescaled operators as Carrollian primaries,
\begin{equation} \label{scaling with c}
\Phi_\Delta (x) \sim c^{\Delta - 1} \mathcal{O}_{\Delta}(x) 
\end{equation}
up to an overall normalization. Notice that this ``electric scaling'' is compatible with the scaling in $\ell$ considered in \eqref{electric scaling in the flat limit} via the formal identification \eqref{c1L}.

\end{itemize}
We outline below how this procedure works for $2$-, $3$- and $4$-point correlators, and refer to \cite{Alday:2024yyj} for more details. Again, we stress that our Carrollian limit procedure is taken entirely in position space, and refer to \cite{Marotta:2025qjh} for the analogous discussion in momentum space.

\subsubsection{Two-point function}

 The two-point function is completely fixed by conformal symmetry. After analytic continuation to Lorentzian signature \cite{Duffin} it is given by
\begin{align}
    \an{\mo_{\D} \left(x_1\right) \mo_{\D} \left(x_2\right)}  = \frac{\mathcal{N}_2}{(x_{12}^2+ i \e)^{\D}},
\end{align}
where $\mathcal{N}_2$ is a normalization and $x_{ij}^2 = -c^2 u_{ij}^2 + 2 |z_{ij}|^2$. This is consistent with the extrapolate result \eqref{2ptAdSWitten}, after taking into account the formal identification \eqref{c1L}. Following the procedure outlined above, we compute 
\begin{multline}
     \label{limit 2pt}
\an{\mo_{\D}\left(x_1\right)\mo_{\D}\left(x_2\right)} = \frac{\mathcal{N}_2}{\left(-c^2 u_{12}^2 + 2 |z_{12}|^2+i\varepsilon \right)^{\D}} \\ \xrightarrow[]{c \to 0} c^0 \underbrace{\frac{  \mathcal{N}_2}{2^{\D} |z_{12}|^{2\D} }}_{\text{Magnetic}}  + c^{2- 2\Delta}  \underbrace{\frac{\pi \mathcal{N}_2}{2\left(\D-1\right)} \frac{\delta^{(2)}\left(z_{12}\right) }{\left(-u_{12}^2+i \varepsilon \right)^{\D-1}}}_{\text{Electric}}  . 
\end{multline}
The naive result when taking $c\to 0$ yields the time independent result, which coincides with the magnetic Carrollian $2$-point function discussed in \eqref{two-point magnetic} and is of order $c^0$. However, on the support of $|z_{12}| = 0$, one has to be careful when taking the $c\to 0$ limit. Indeed, a $u$-dependent distributional branch appears, which corresponds to the electric Carrollian $2$-point function \eqref{two-point electric}. It scales like $c^{2-2\Delta}$, which is leading for the conformal dimensions of interest in Carrollian correlators where $\Delta \ge 1$, see Section \ref{sec:Definition of Carrollian amplitudes}. For the case $\Delta = 1$, the $2$-point function diverges, which is in agreement with the $2$-point Carrollian amplitude, see discussion below \eqref{2point News}. The formula \eqref{limit 2pt} can be seen as a generalization of the well-known formula $\frac{1}{x \pm i\varepsilon}
= \frac{1}{x} \mp i\pi\delta(x)$. The precise factor in front of the $\delta$-function can be easily computed by requiring the normalization $\int d^2 z \delta^{(2)}(z_{12}) = 1$ and applying this on the left-hand side.

Hence the scaling of the primary $\mathcal{O}_\Delta$ with $c$ tells us on which branch we land in the limit. For our purpose, as stated above, we focus on the leading electric branch, which is found by taking the scaling \eqref{scaling with c}. Putting all together, we have
\begin{equation}
   \lim_{c \to 0} c^{2\D-2}  \an{\mo_{\D} \left(x_1\right) \mo_{\D} \left(x_2\right)} =  \frac{\pi \mathcal{N}_2\, \delta^{2}(z_{12})}{2(\Delta -1) (-u_{12}+i \varepsilon)^{2\Delta - 2}} = \an{\Phi_{\D}(u_1, z_1, \bar z_1) \Phi_{\D} (u_2, z_2, \bar z_2) }
\end{equation} 
 where the last equality matches exactly \eqref{2point News} after appropriate normalization and setting $\e_1 = -\e_2 = -1$.

\subsubsection{Three-point function}
 
As discussed in Section \ref{sec:Three-point function}, it is convenient to work in Klein signature in the bulk since it allows for non-trivial $3$-point amplitudes. This amounts to treating $z_i, \zb_i$ as real and independent. After analytic continuation, the time-ordered correlator with $z_i, \zb_i$ real and independent is
 \begin{align}
\an{\mo_{\D_1}\left(x_1\right)\mo_{\D_2}\left(x_2\right)\mo_{\D_3}\left(x_3\right)}_K = \frac{\mathcal{N}_3}{c}\frac{1}{\left(x_{12}^2 + i \varepsilon\right)^{\D_{12}} \left(x_{23}^2 + i \varepsilon\right)^{\D_{23}} \left(x_{13}^2+ i \varepsilon\right)^{\D_{13}}} 
\end{align} where $\mathcal{N}_3$ is a normalization that does not depend on $c$ and $\D_{ij} = \D_i + \D_j - \frac{1}{2}\sum_{k=1}^3 \D_k$. Applying the procedure outlined above, and focusing on the relevant leading electric branch $\sim \delta (\bar{z}_{12}) \delta (\bar z_{23})$, we get 
\begin{align}
      & \lim_{c\to 0} c^{ \sum_{j=1}^3 \Delta_j-3}\langle \mathcal{O}_{\Delta_1}(u_1, z_1, \bar z_1) \mathcal{O}_{\Delta_2}(u_2, z_2, \bar z_2) \mathcal{O}_{\Delta_3}(u_3, z_3, \bar z_3)   \rangle \\
      & \nonumber \qquad= \frac{ \tilde{\mathcal{N}_3}\, \delta (\bar z_{12}) \delta (\bar{z}_{23})\Theta\left(z_{12}z_{32}\right)\Theta\left(z_{13}z_{23}\right) z_{12}^{\Delta_3-2}z_{23}^{\Delta_1-2}z_{13}^{\Delta_2-2}}{\left(u_1 z_{23}+u_2 z_{31}+u_3 z_{12}+i \text{sign }z_{23} \varepsilon \right)^{\sum_{j=1}^3 \Delta_j-4}} =  \langle \Phi_{\Delta_1} \Phi_{\Delta_2} \Phi_{\Delta_3} \rangle.
  \end{align} We refer the reader to \cite{Alday:2024yyj} for the normalization factor $\tilde{\mathcal{N}}_3$ and details. Again the precise functional dependence is found by requiring the normalization of the $\delta$-function and using the explicit form of the CFT $3$-point function. This result coincides with the $3$-point Carrollian scalar amplitude with $\e_1 = -\e_2 = -\e_3 = 1$ displayed in \eqref{eq:3ptwithThetascal}.  

\subsubsection{Four-point function}

\paragraph{Four-point contact diagram:} We focus on the contact diagram in the bulk, whose expression as a CFT correlator in Euclidean signature takes the form
\begin{equation}
    \langle \mathcal{O}_{\Delta_1}(x_1) \mathcal{O}_{\Delta_2}(x_2) \mathcal{O}_{\Delta_3}(x_3) \mathcal{O}_{\Delta_4}(x_4) \rangle^c_E = \mathcal{N}_4 \frac{\left(x_{14}^2\right)^{\frac{1}{2}\Sigma_{\D}-\D_1-\D_4}\left(x_{34}^2\right)^{\frac{1}{2}\Sigma_{\D}-\D_3-\D_4}}{\left(x_{13}^2\right)^{\frac{1}{2}\Sigma_{\D}-\D_4}\left(x_{24}^2\right)^{\D_2}}\bar{D}_{\D_1, \D_2, \D_3, \D_4}\left(U,V\right)  
\end{equation} 
where $\Sigma_\Delta = \sum_{j=1}^4 \Delta_j$, $\mathcal{N}_4$ is a constant that does not scale with $c$, and  
\begin{equation} \label{cross ratios 3d}
    U = \frac{x_{12}^2x_{34}^2}{x_{13}^2 x_{24}^2} = Z \Zb  , \qquad V = \frac{x_{23}^2x_{14}^2}{x_{13}^2x_{24}^2} =(1-Z)(1-\Zb)
\end{equation} 
are the $3d$ conformal cross ratios, which can also be expressed in terms of the complex variables $Z$ and $\bar Z$. The definition of $\bar{D}$ functions and various useful properties can be found in Appendix D of \cite{Arutyunov:2002fh}. The analytic continuation from Euclidean to Lorentzian signature requires treating $Z$ and $\Zb$ as complex independent variables. In Euclidean signature, the $\bar{D}$ function is regular as $Z \to \Zb$. However, it can become singular upon analytic continuation \cite{Gary:2009ae,Maldacena:2015iua} and its leading singularity is \cite{Alday:2024yyj}
\begin{align}
\label{eq:lsgen}
     \bar{D}_{\D_1, \D_2, \D_3, \D_4}\left(U,V\right) \xrightarrow[]{Z \to \Zb} \hat{\Phi}^{l.s}_{\D_1, \D_2, \D_3, \D_4} \equiv \mathcal{K}_\D \frac{Z^{\D_{3}+\D_{4}-2}(1-Z)^{\D_{1}+\D_{4}-2}}{(Z-\bar{Z})^{\sum_{i=1}^4 \Delta_i-3}}
\end{align} where $\mathcal{K}_\Delta$ is a constant. This is referred to as the leading ``bulk-point singularity''. Furthermore, we have $Z - \bar Z = z - \bar{z} + \mathcal{O}(c^2)$ where $z = \frac{z_{12} z_{34}}{z_{13}z_{24}}$ is the $2d$ cross ratio on the celestial sphere. Analogously to case of the $2$- and $3$-point functions, the electric scaling of the $4$-point function will arise on the most singular term in the $c \to 0$ limit, coming from the leading singularity on the support $z = \bar z$:
 \begin{align} \label{deltaz-zb}
     \lim_{c \to 0} c^{ \sum_\Delta-4}\hat{\Phi}^{l.s}_{\D_1, \D_2, \D_3, \D_4} = \mathcal{R}\left(u_i, z_i\right) \delta (z-\zb ),
 \end{align}
 where $\mathcal{R}\left(u_i, z_i\right)$ is a complicated function of the coordinates obtained by requiring normalization of the $\delta$-function, and whose explicit expression can be found in \cite{Alday:2024yyj}. Using these results, we can show that applying the procedure outlined at the beginning of this section to the correlator corresponding to the four-point contact diagram results in
\begin{equation}
    \begin{split}
        &\lim_{c\to 0} c^{\sum_\Delta - 4}  \langle \mathcal{O}_{\Delta_1}(x_1) \mathcal{O}_{\Delta_2}(x_2) \mathcal{O}_{\Delta_3}(x_3) \mathcal{O}_{\Delta_4}(x_4) \rangle  \\
        &=  \mathcal{N}'_4\,\left(\frac{\left|z_{23}\right|^2}{\left|z_{34}\right|^2\left|z_{24}\right|^2}\right)^{\frac{4-\Sigma_{\D}}{2}} \frac{z^{2-\D_1-\D_2}\left(1-z\right)^{\D_1+\D_4-2} \delta\left(z-\zb\right)}{(F_{1234})^{\sum_{i=1}^4 \D_i -4}} \propto \langle \Phi_{\Delta_1}^{\e_1} \Phi_{\Delta_2}^{\e_2} \Phi_{\Delta_3}^{\e_3} \Phi_{\Delta_4}^{\e_4} \rangle,
    \end{split}
\end{equation}
where $F_{1234}$ is the $u$-dependent combination introduced in \eqref{F1234}. Depending on the details of the analytic continuation, we can have $z<0$, $0<z<1$ or $z>1$. This coincides with the allowed values of $z$ for which the Carrollian amplitude is non-zero. Focussing on $0<z<1$, the proportionality can be turned into an equality after properly taking into account the normalization of the fields and setting $\e_1 = \e_2 = -\e_3 = -\e_4 = 1$. This reproduces the Carrollian correlator \eqref{4pt contact diagram Carroll} associated to a four-point contact diagram in flat space.

\paragraph{Four-point exchange diagram:} In Euclidean signature, the four-point correlator for a bulk scalar exchange is
\begin{align} \an{\mo_{\D_1}\left(x_1\right)\mo_{\D_2}\left(x_2\right)\mo_{\D_3}\left(x_3\right)\mo_{\D_4}\left(x_4\right)}_E^e =I_s + I_t + I_u , 
\end{align}
where $I_s$ represents the $s$-channel exchange of an operator with conformal dimension $\Delta$,
\begin{align} I_s = \int_{AdS_4} d^4X \, d^4Y \, \prod_{i=1}^2 \, \mathcal G^{\D_i}_{Bb}\left(x_i; X\right) \, \prod_{i=3}^4 \, \mathcal G^{\D_i}_{Bb}\left(x_i; Y\right) \mathcal G^{\D}_{BB}\left(X,Y\right) . \end{align}
The $t$- and $u$-channel contributions are obtained via $2 \leftrightarrow 3$ and $2 \leftrightarrow 4$, respectively. For generic dimensions, exchange diagrams cannot be expressed in terms of elementary functions in the CFT. However, when $-\Delta + \Delta_3 + \Delta_4 \in 2\mathbb{Z}^+$, the $s$-channel diagram reduces to a finite sum of contact diagrams:
\begin{align} \label{eq:finitesumexchange} I_s &= \sum_{k=k_{\text{min}}}^{k_{\text{max}}} a_k \left(x_{34}^2\right)^{k-\D_4} D_{\D_1, \D_2, \D_3-\D_4+k,k} \\ &\nonumber = \frac{\left(x_{14}^2\right)^{\frac{1}{2}\Sigma_{\D}-\D_1-\D_4}\left(x_{34}^2\right)^{\frac{1}{2}\Sigma_{\D}-\D_3-\D_4}}{\left(x_{13}^2\right)^{\frac{1}{2}\Sigma_{\D}-\D_4}\left(x_{24}^2\right)^{\D_2}} \sum_{k=k_{\text{min}}}^{k_{\text{max}}} a_k \mathcal{N}''_4 \bar{D}_{\D_1, \D_2, \D_3-\D_4+k,k} , 
\end{align}
with $\mathcal{N}_4''$ a constant, $k_{\text{min}} = \frac{1}{2}(\Delta - \Delta_3 + \Delta_4)$, $k_{\text{max}} = \Delta_4 - 1$, and coefficients $a_k$ satisfying
\begin{align}
a_{k-1} = \frac{\left(k-\frac{\Delta}{2} + \frac{\Delta_3 - \Delta_4}{2}\right)\left(k-\frac{3}{2} + \frac{\Delta}{2} + \frac{\Delta_3 - \Delta_4}{2}\right)}{(k-1)(k-1+\Delta_3-\Delta_4)} a_k , \qquad a_{k_{\text{max}}} = \frac{1}{4(\Delta_3 - \frac{3}{2})(\Delta_4 - \frac{3}{2})} .
\end{align}
A relevant example is $\Delta_1 = \Delta_2 = \Delta_3 = \Delta_4 = \Delta = 2$, for which
\begin{align}
I_s = \frac{\ell^2}{4\pi^6 (x_{13}^2)^2 (x_{24}^2)^2}  \bar{D}_{2,2,1,1} ,
\end{align}
with $I_t$ and $I_u$ obtained by permuting $2 \leftrightarrow 3$ and $2 \leftrightarrow 4$, respectively.

It follows from \eqref{flatlimitBB} that the bulk-to-bulk propagator in AdS reduces to the Feynman propagator in flat space for any value of $\Delta$. This observation motivates the following strategy to analyze the Carrollian limit of a general scalar exchange diagram. Focusing on the $s$-channel, we choose the external conformal dimensions $\Delta_i$ such that $\Delta_3 + \Delta_4 - \Delta \in 2\mathbb{Z}^+$. In this case, the exchange diagram reduces to a finite sum of contact diagrams (Equation \eqref{eq:finitesumexchange}), allowing us to compute the Carrollian limit of each term using the methods presented above. The result matches the corresponding Carrollian amplitude, is analytic in the external dimensions $\Delta_i$, and is independent of the exchanged dimension $\Delta$. Consequently, the restriction on $\Delta_i$ can be relaxed, extending the result to generic exchange diagrams. The computation of the Carrollian limit of a finite sum of contact diagrams follows the same logic as the one presented around \eqref{deltaz-zb}, which focuses on the leading term in the expansion around $Z = \bar{Z}$. As $Z \to \bar{Z}$, we have
\begin{align} I_s \xrightarrow[]{Z \to \Zb} \frac{\left(x_{14}^2\right)^{\frac{1}{2}\Sigma_{\D}-\D_1-\D_4}\left(x_{34}^2\right)^{\frac{1}{2}\Sigma_{\D}-\D_3-\D_4}}{\left(x_{13}^2\right)^{\frac{1}{2}\Sigma_{\D}-\D_4}\left(x_{24}^2\right)^{\D_2}} \sum_{k=k_{\text{min}}}^{k_{\text{max}}} a_k\, \mathcal{N}''_4\, \Phi^{ls}_{\D_1, \D_2, \D_3-\D_4+k,k} . 
\end{align}
The leading singularity comes from $k = k_{\text{max}} = \Delta_4 - 1$, since $\Phi^{ls}_{\Delta_1, \Delta_2, \Delta_3, \Delta_4} \propto (Z-\bar{Z})^{3-\Sigma_\Delta}$. The Carrollian limit is then computed exactly as before, yielding
\begin{align}
\lim_{c \to 0}  c^{\sum_\Delta-4} I_s = \langle \Phi_1 \Phi_2 \Phi_3 \Phi_4  \rangle_e ,
\end{align}
where $\langle \Phi_1 \Phi_2 \Phi_3 \Phi_4  \rangle_e$ was written down in \eqref{four-point exchange}. This result is independent of $\Delta$ and analytic in $\Delta_i$, confirming that the Carrollian limit of a scalar exchange diagram coincides with the corresponding Carrollian amplitude in full generality.

Let us emphasize that the Carrollian limit discussed in this section is taken intrinsically in the CFT, without referring to the bulk spacetime. Hence we have proven the equivalence between taking the flat space limit of Witten diagrams in the bulk and taking the Carrollian limit of the corresponding holographic CFT correlators at the boundary. We refer to Figure \ref{fig:diag} for a summary.

\begin{center}
\begin{figure}[h]
\centering
\begin{tikzpicture}[
  every node/.style={font=\small},
  box/.style={draw, rounded corners, minimum width=3.8cm, minimum height=1cm, align=center,inner sep=3pt,outer sep=3pt},
  arrow/.style={-{Latex}, thick},
  barrow/.style={Latex-Latex, thick}  
]

\node[box] (witten) at (0,0) {Witten diagrams \\ in $\text{AdS}_4$};
\node[box] (carrollAmp) at (7,0) {Feynman diagrams for \\ Carrollian amplitudes in $4d$ \\
(Sec. \ref{sec:Diagrams})};

\node[box] (holoCFT) at (0,-2.5) {Holographic $\text{CFT}_{3}$ \\ correlators};
\node[box] (carrollCFT) at (7,-2.5) {Carrollian $\text{CFT}_{3}$ \\ correlators \\
(Sec. \ref{sec:Definition of Carrollian amplitudes})};

\node[box] (celestialCFT) at (7,-5) {Celestial CFT$_{2}$ \\ correlators\\
(Sec. \ref{sec:Examples of celestial amplitudes})};

\draw[arrow] (witten) -- (carrollAmp) node[above,midway] {$\ell \to \infty$} node[below,midway] {(Sec. \ref{sec:Flat space limit of Witten diagrams})};
\draw[arrow] (holoCFT) -- (carrollCFT) node[above,midway] {$c \to 0$} node[below,midway] {(Sec. \ref{sec:Carrollian limit of holographic correlators})} ;
\draw[barrow] (holoCFT.south) |- (celestialCFT.west) node[midway,anchor=south west]{$\qquad\qquad\quad$\cite{deGioia:2024yne}} ;
\draw[barrow] (witten) -- (holoCFT) node[left,midway] {AdS/CFT}{};
\draw[barrow] (carrollAmp) -- node[right] {Carrollian holography (Sec. \ref{sec:Carrollian and celestial amplitudes})} (carrollCFT){};
\draw[barrow] (carrollCFT) -- (celestialCFT) node[right,midway] {Carroll/celestial correspondence (Sec. \ref{sec:From Carrollian to celestial holography})};

\end{tikzpicture}
\caption{Summary of the interplay between the flat space limit / Carrollian limit of AdS/CFT correlators \cite{Alday:2024yyj, Kulkarni:2025qcx}.}
\label{fig:diag}
\end{figure}
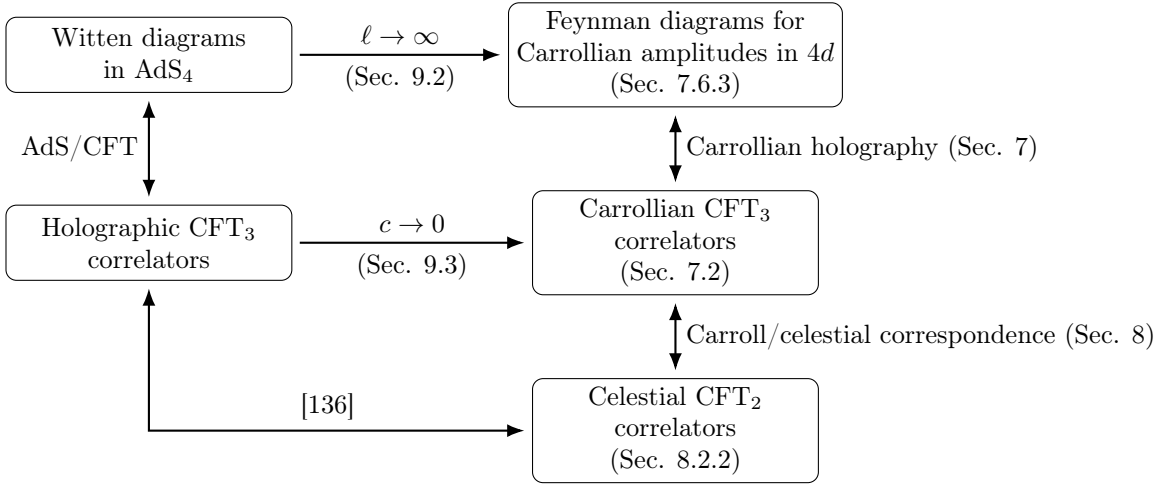
\end{center}

\subsection{Towards a top-down flat space hologram from AdS/CFT}
\label{sec:Towards a top-down flat space hologram from AdS/CFT}

The above discussion has been presented for a general scalar subsector of AdS/CFT in $4d$ bulk. In this section, we apply these ideas to a specific realization of AdS$_4$/CFT$_3$: M-theory on AdS${}_4 \times$S$^7/\mathbb{Z}_{k_{CS}}$ dual to a $3d$ superconformal Chern-Simons matter theory on $\mathbb{R}^{1,2}$ known as the ABJM theory \cite{Aharony:2008ug}. The idea consists of taking the limit on both sides of the duality by using the flat space / Carrollian limit correspondence. We will mostly follow \cite{Lipstein:2025jfj}.

\subsubsection{Starting point}

In AdS${}_4 \times$S$^7/\mathbb{Z}_{k_{CS}}$, the AdS${}_4$ has radius $\ell$ and the $S^7$ has radius $2\ell$. The dual ABJM theory is a superconformal Chern-Simons matter theory with gauge group $U(N)_{k_{CS}} \times U(N)_{-k_{CS}}$, where $k_{CS}$ is the Chern-Simons level and the matter fields are in the bifundamental representation of the gauge group. The theory has a Lagrangian description with $\mathcal{N}=6$ supersymmetry \cite{Benna:2008zy,Bandres:2008ry}, but for $k_{CS}=1,2$, the quantum theory has maximal $\mathcal{N}=8$ supersymmetry \cite{Kapustin:2010xq}. We will only consider the case where the Chern-Simons level is $k_{CS} = 1$.

The central charge $c_T$ is defined as the coefficient of the stress tensor two-point function. When $N \gg k_{CS}$, the relationship between $c_T$ and $N$ is \cite{Klebanov:1996un,Aharony:2008ug}
\begin{align}
    \label{eq:ctNrel}
c_T = \frac{64}{3\pi} \sqrt{2k_{CS}} N^{\frac{3}{2}} +
\mathcal{O}\left(N^{1/2}\right).
\end{align}
Moreover, when $N \gg k_{CS}^5$, the bulk is described by supergravity on AdS$_4 \times$S$^7$ and we have 
\begin{align}
    \label{eq:Nlrel}
    \frac{\ell^6}{\ell_{11}^6} = \left(\frac{3\pi c_T k_{CS}}{2^{11}}\right)^{\frac{2}{3}} + \mathcal{O}\left(c_T^0\right) = \frac{N k_{CS}}{8} + \mathcal{O}\left(N^0\right),
\end{align}
where $\ell_{11}$ is the 11-dimensional Planck length. 

We will focus on correlators of scalar operators which are $1/2$-BPS (i.e. annihilated by half of the supersymmetry generators) and are dual to KK modes on the 7-sphere. These operators take the form $\mo_k^{I_1 \dots I_k}$, where $I_1, \dots , I_k$ are $SO(8)$ $R$-symmetry indices and $\mo_k^{I_1 \dots I_k}$ is symmetric trace-free. 
To make this property manifest, we will contract the indices with complex null vectors $t_I$ 
\begin{align}
\label{eq:contractedop}
    \mo_k\left(x, t\right) \equiv \mo_k^{I_1 \dots I_k}t_{I_1} \dots t_{I_k},
\end{align}
where the subscript $k$ denotes the R-charge of the conformal primary (not to be confused with the Carrollian weights appearing e.g. in \eqref{Carrollian primary 3d}). The scaling dimensions of these operators are protected and an operator with R-charge $k$ has conformal dimension $\D_k = \frac{k}{2}$. For the minimal value $k=2$, these operators belong to the stress tensor multiplet. 

The 't Hooft coupling is $\lambda=N/k_{CS}$ and the planar limit corresponds to taking $k_{CS}$ and $N$ to infinity while holding $\lambda$ fixed. In this limit, the theory becomes integrable (see \cite{Klose:2010ki} for a review). On the other hand, the enhancement of supersymmetry at $k_{CS}=1,2$ arises from non-perturbative effects involving monopole operators. In this regime, the operators $\mathcal{O}_k$ are quantum operators which are not constructed directly out of the fields in the Lagrangian and have no classical analogue \cite{Klebanov:2009sg,Lambert:2019khh}. As a result, their correlation functions have been mainly computed using superconformal bootstrap methods \cite{Rastelli:2017udc,Chester:2018aca,Alday:2020dtb,Alday:2021ymb,Alday:2022rly,Chester:2024esn,Chester:2024bij}.

\subsubsection{Carrollian limit of ABJM}
\label{sec:Carrollian limit of ABJM}

Let us first consider the boundary theory. At the level of symmetries, the Carrollian limit of superconformal algebras has been studied in \cite{Bagchi:2022owq, Lipstein:2025jfj} and leads to finite-dimensional global superconformal Carrollian algebras. As discussed in Section~\ref{sec:Infinite-dimensional enhancement}, these algebras can be further enhanced to infinite-dimensional ones. Remarkably, this enhancement also extends to the fermionic generators \cite{Fotopoulos:2020bqj,Henneaux:2020ekh,Fuentealba:2023hzq}, resulting in genuine superconformal Carrollian algebras. The R-symmetries and their associated coordinates survive the Carrollian limit \cite{Lipstein:2025jfj}, provided the corresponding generators are not scaled with $c$, which is the choice we adopt here. We therefore expect one of these superconformal Carrollian algebras to be a symmetry of the theory obtained by taking the Carrollian limit of ABJM theory.

The Carrollian limit of ABJM theory can be taken either at the level of the action, following steps similar to those presented in Section~\ref{sec:Carrollian field theories}, or directly at the level of correlators, using the procedure described in Section~\ref{sec:Carrollian limit of holographic correlators}. We first adopt the former approach and then turn to the latter. The Carrollian limit of Chern–Simons theory has been studied in detail in \cite{Miskovic:2023zfz}. In addition, the Carrollian limit of a Chern-Simons theory with gauge group $SU(N)\times SU(M)$ coupled to bifundamental scalar matter was analyzed in \cite{Bagchi:2024efs}. The relativistic action reads
\begin{align}
S=\int du \, d^2 \vec{x} & \left\{ \frac{ik}{8\pi}\epsilon^{abc}\operatorname{Tr}_{N}\left(A_{a}\partial_{b}A_{c}+\frac{2i}{3}A_{a}A_{b}A_{c}\right)\right. \nonumber\\
& -\frac{i k}{8 \pi} \epsilon^{abc} \operatorname{Tr}_{M}\left(B_{a} \partial_{b} B_{c}+\frac{2 i}{3} B_{a} B_{b} B_{c}\right) \left.+\operatorname{Tr}_{M}\left[\left(D_{a} \phi\right)^{\dagger}\left(D_{a} \phi\right)\right]\right\}
\end{align}
where $k_{N}=-k_{M}=k$ is the Chern-Simons level, $\phi$ is a scalar field in the $(N, \bar{M})$ representation of $\operatorname{SU}(N) \times \operatorname{SU}(M)$, $A_{a}$ and $B_{a}$ are $SU(N)$ and $SU(M)$ gauge fields, respectively, and
\begin{equation}
D_{a} \phi=\partial_{a} \phi-i A_{a} \phi+i \phi B_{a}.
\end{equation}
This model is closely related to ABJM theory, which additionally requires the inclusion of fermions and scalar potential terms. Following a procedure analogous to that described in Section~\ref{sec:Carrollian field theories}, the electric Carrollian limit is obtained by appropriately rescaling the fields and taking $c \to 0$. This is the relevant limit for describing massless scattering amplitudes, since we have shown above that the Carrollian amplitudes are of electric type. The electric Carrollian contraction leads to
\begin{align}
S_{e}=\int du \, d^2 \vec{x} & \left\{\frac{i k}{8 \pi} \epsilon^{abc} \operatorname{Tr}_{N}\left(a_{a} \partial_{b} a_{c}+\frac{2 i}{3} a_{a} a_{b} a_{c}\right)\right. \nonumber \\
& -\frac{i k}{8 \pi} \epsilon^{abc} \operatorname{Tr}_{M}\left(b_{a} \partial_{b} b_{c}+\frac{2 i}{3} b_{a} b_{b} b_{c}\right) \left.-\operatorname{Tr}_{M}\left[\left(D_{u} \phi\right)^{\dagger}  D_{u} \phi\right]\right\},
\end{align}
where $a_a$ and $b_a$ are the rescaled gauge fields and $D_u \phi = \partial_u \phi - i a_u \phi + i \phi b_u$ \cite{Bagchi:2024efs}. This theory can be shown to possess an infinite-dimensional conformal Carrollian symmetry algebra and, as such, admits a natural interpretation as a theory living at null infinity. At this stage, however, the action is purely classical, and quantization is required in order to obtain a candidate dual Carrollian CFT. As emphasized in Section~\ref{sec:Quantization}, directly quantizing electric Carrollian theories by starting from the Lagrangian is a highly subtle issue \cite{Cotler:2024xhb}, and the problem of obtaining non-trivial correlators such as the Carrollian amplitudes remains open.

Another possibility is to take the Carrollian limit directly at the level of the correlators, as presented in Section \ref{sec:Carrollian limit of holographic correlators}. This road has been investigated in \cite{Lipstein:2025jfj}, following the general procedure discussed in \cite{Alday:2024yyj}. The $2$-point function in ABJM reads as
\begin{equation}
    \label{eq:ABJM2ptfunc} 
    \an{\mo_{k_1}\left(x_1, t_1\right) \mo_{k_2} \left(x_2, t_2\right) } = \frac{\delta_{k_1, k_2}\, t_{12}^{k_1}}{x_{12}^{k_1}}
\end{equation} where $t_{ij} = t_{i} \cdot t_j$ and $x^2_{ij} = -c^2 u_{ij}^2 + 2 z_{ij} \zb_{ij}$. This is the usual form of the CFT $2$-point function with $\Delta_k = \frac{k}{2}$ and with R-symmetry space decoration. The electric limit of \eqref{eq:ABJM2ptfunc} after analytic continuation leads to 
\begin{align}
\label{eq:2ptcarrolllim}
   \lim_{c\to 0} \frac{\an{\mo_{k_1}\left(x_1\right) \mo_{k_2}\left(x_2\right)}}{c^{2 - \frac{k_1+k_2}{2} } \sigma_{k_1} \sigma_{k_2}} = \an{\Phi_{k_1} \Phi_{k_2}} = \frac{\delta_{k_1, k_2}}{4\pi}\frac{(-1)^{\frac{k_1}{2}-1} \Gamma(k_1-2)}{\left(u_{12}-i \e\right)^{k_1-2} } \,t_{12}^{k_1}\, \delta^2\left(z_{12}\right), 
\end{align} where we chose the scaling $\mo_k =\sigma_k \, c^{1 - \frac{k}{2}}  \, \Phi_k$ with $\sigma_k =  \frac{2\pi}{\sqrt{\Gamma\left(k-1\right)}}$. This is the same electric scaling as the one we chose for the general discussion in \eqref{scaling with c}.

Similarly the ABJM three-point function reads as 
\begin{align}
   \label{eq:ABJM3ptfunc}
\an{\mo_{k_1}\,\mo_{k_2}\,\mo_{k_3}} = c^{-1} R_{k_1, k_2, k_3} \frac{t_{12}^{\alpha_3}t_{23}^{\alpha_1}t_{13}^{\alpha_2}}{x_{12}^{\alpha_3}x_{23}^{\alpha_1}x_{13}^{\alpha_2}},
\end{align} where $\alpha_i = \frac{1}{2}\sum_{j=1}^3 k_j -k_i$ and $R_{k_1, k_2, k_3}$ is a constant that does not depend on R-symmetry space coordinates and whose explicit expression can be found in \cite{Lipstein:2025jfj}. There is a subtlety regarding the global scaling with $c$, which we re-absorbed in $R_{k_1, k_2, k_3}$, and we refer to that reference for details. Up to this subtlety, the $c\to 0$ limit yields
\begin{align}
    \label{eq:carroll3ptabjm}
    \lim_{c \to 0} \frac{\an{\mo_{k_1}\mo_{k_2}\mo_{k_3}}}{c^{3-\frac{k_1+k_2+k_3}{2}}\prod_{i=1}^3 \sigma_{k_i}} =& \frac{\Gamma\left(\frac{\alpha}{2}-2\right)\Theta\left(z_{12}z_{31}\right)\Theta\left(z_{13}z_{23}\right)}{\Gamma\left(\frac{\alpha_1}{2}\right)\Gamma\left(\frac{\alpha_2}{2}\right)\Gamma\left(\frac{\alpha_3}{2}\right)}\pi^2 R_{k_1, k_2, k_3}\prod_{i=1}^3 \frac{1}{\sigma_{k_i}} \\
    &\nonumber \qquad \times \delta (\bar z_{12}) \delta (\bar{z}_{23})\frac{t_{12}^{\alpha_3}\,t_{23}^{\alpha_1}\,t_{13}^{\alpha_2}z_{12}^{\frac{k_3}{2}-2}z_{23}^{\frac{k_1}{2}-2}z_{13}^{\frac{k_2}{2}-2}}{\left(u_1 z_{23}+u_2 z_{31}+u_3 z_{12}+i \varepsilon \right)^{\frac{k_1+k_2+k_3}{2}-4}} .
\end{align}

Remarkably, the $4$-point functions of half-BPS operators have been computed in the large $N$ limit in Mellin space variables, using superconformal bootstrap methods \cite{Rastelli:2017udc,Chester:2018aca,Alday:2020dtb,Alday:2021ymb,Alday:2022rly,Chester:2024esn,Chester:2024bij}. Following \cite{Lipstein:2025jfj}, these correlators can be re-expressed in position space as a finite sum of $\bar{D}$-functions, on which we know how to implement the limit, as explained in Section \ref{sec:Carrollian limit of holographic correlators}. Four-point functions of the superconformal primaries in ABJM can be written as \cite{Dolan:2004mu,Alday:2020dtb}
\begin{align}
\label{eq:4ptcorr}
    \an{\mo_{k_1}\dots \mo_{k_4}} = \prod_{i<j} \left(\frac{t_{ij}^2}{x_{ij}^{2}}\right)^{\frac{\gamma_{ij}^0}{2}} \left(\frac{t_{12}^2 t_{34}^2}{x_{12}^{2}x_{34}^{2}}\right)^{\frac{\mathcal{E}}{2}} \mathcal{G}_{k_1, \dots, k_4}\left(U, V,\sigma, \tau\right).
\end{align}
Here $U,V$ are the $3d$ cross ratios defined in \eqref{cross ratios 3d} and $\sigma = \frac{t_{13} t_{24}}{t_{12} t_{34}} = \alpha \bar{\alpha}$, $\tau = \frac{t_{14} t_{23}}{t_{12} t_{34}}= \left(1-\alpha\right) \left(1-\bar{\alpha}\right)$ are the R-symmetry space cross ratios.
The extremality $\mathcal{E}$ is
\begin{align}
\label{eq:extremality}
     \mathcal{E} = \begin{cases}
        \frac{k_1+k_2+k_3-k_4}{2} \qquad\qquad &\text{Case I}: k_1+k_4 \geq k_2+k_3,\\
        k_1  \qquad\qquad  &\text{Case II}: k_1+k_4 < k_2+k_3,
    \end{cases}
\end{align}
and the exponents $\gamma_{ij}^0$ are  given by
\begin{align}
    \label{eq:gammas}
    &\gamma_{12}^0 = \gamma_{13}^0 = 0, \qquad \gamma_{34}^0 = \frac{\kappa_s}{2}, \qquad \gamma_{24}^0 = \frac{\kappa_u}{2},\\
    &\nonumber \text{Case I: } \gamma_{14}^0 = \frac{\kappa_t}{2}, \qquad \gamma_{23}^0 = 0, \qquad \text{Case II: } \gamma_{14}^0 = 0, \qquad \gamma_{23}^0 = \frac{\kappa_t}{2} ,
\end{align}
where $\kappa_s = \left|k_1 + k_2 - k_3 - k_4\right|$, $\kappa_t = \left|k_1 + k_4 - k_2 - k_3\right|$, $\kappa_u = \left|k_2 + k_4 - k_1 - k_3\right|$. The Carrollian limit of this expression leads to 
\begin{align}
 \label{G2kkkk Carrollian}
     \lim_{c \to 0} \frac{\an{\mo_{k_1} \dots \mo_{k_4}}}{\prod_{i=1}^4 c^{1-\frac{k_i}{2}}\sigma_k}= & \tilde{\mathcal{N}}\, \left[\prod_{i<j} \left(\frac{t_{ij}}{\left|z_{ij}\right|}\right)^{\gamma_{ij}^0}\left(\frac{t_{12}t_{34}}{\left|z_{12}\right|\left|z_{34}\right|}\right)^{\mathcal{E}} \left(\frac{\left|z_{23}\right|^2}{\left|z_{34}\right|^2 \left|z_{24}\right|^2}\right)^{1-\frac{\sum_i k_i}{4}}\right] \\
     &\qquad \times \left[z^{-2a_s} \left(1-z\right)^{\frac{\sum_i k_i}{2}-2-2a_t} \frac{\delta\left(z-\zb \right)}{(F_{1234})^{\frac{\sum_i k_i}{2}-2}}\right] \times \left[\left(1-\alpha z\right)^2 \left(1-\bar{\alpha}z\right)^2 \mathcal{P}_{k_i}\left(\sigma,\tau\right)\right]\nonumber,
 \end{align} 
 where 
\begin{align}
\label{eq:polyk}
    \mathcal{P}_{k_i} \left(\sigma, \tau\right) = \underset{\underset{0\leq i,j,k\leq \mathcal{E}-2}{i+j+k = \mathcal{E}-2}}{\sum} \frac{\left(\mathcal{E}-2\right)! \sigma^i \tau^j}{i!  j! k! \left(i + \frac{\kappa_u}{2}\right)! \left(j + \frac{\kappa_t}{2}\right)! \left(k+\frac{\kappa_u}{2}\right)! }.
\end{align} Again, the final expression \eqref{G2kkkk Carrollian} is compatible with the expected result for the Carrollian $4$-point function \eqref{electric 4pt}, with some R-symmetry space decoration. 

In summary, the electric Carrollian limit of ABJM theory can, in principle, be taken either at the level of the Lagrangian or directly at the level of correlation functions. In the former case, one obtains an electric Carrollian Lagrangian, but there is currently no known quantization procedure that yields non-Gaussian correlation functions. In contrast, the latter approach leads to non-trivial correlators, which take the expected form of Carrollian CFT correlators. This naturally raises two questions: does this set of Carrollian ABJM correlators define a consistent theory, and what is their bulk interpretation? We address the latter question in the next section and defer the former to the discussion in Section~\ref{sec:Through the looking glass}.

\subsubsection{Bulk interpretation}
\label{sec:Bulk interpretation}

In the previous section, we obtained a set of correlators living in a $3d$ Carrollian CFT. We expect to match these results with bulk scattering amplitudes in flat space. However, once we consider the flat space limit of an explicit realization of AdS/CFT, one has to address the fate of the internal space. More explicitly, the metric on AdS$_4 \times S^7$ is given by
\begin{equation}
    ds^2_{AdS_4 \times S^7}  = ds^2_{AdS_4} + 4 \ell^2 ds^2_{S^7}
\end{equation} where $ds^2_{AdS_4}$ is the $AdS_4$ metric, which can be expressed in Bondi coordinates as in \eqref{AdSBondi}, and $ds^2_{S^7}$ is the unit $S^7$ metric. As emphasised above, the radius of the sphere is precisely twice the AdS radius, ensuring that this background metric is a supergravity solution. In the flat space limit, $\ell \to \infty$, the $S^7$ radius becomes infinite and the internal space decompactifies, leading to $11d$ Minkowski space:
\begin{equation} \label{from4to11}
    ds^2_{AdS_4 \times S^7} \xrightarrow[]{\ell \to \infty} ds^2_{Mink_{11}} . 
\end{equation} In the large $N$ regime we are considering, the resulting theory in the flat space limit is $\mathcal{N}=1$ eleven-dimensional supergravity, whose natural observables are, for example, $11d$ tree-level graviton amplitudes. This leads to an apparent paradox: from a purely boundary computation, we obtain Carrollian ABJM correlators in $3d$, whereas we expect to be describing an $11d$ flat space theory, whose boundary is $10d$. There are two possible ways to proceed:
\begin{itemize}
\item The first option consists of starting from the graviton amplitudes in $11d$ Minkowski space and restricting the kinematics to a $4d$ hyperplane $\mathbb{R}^{1,3}$, such that all external momenta lie in this $4d$ hyperplane while the graviton polarizations remain transverse to it. Hence, from the $4d$ perspective, this corresponds to effective scattering amplitudes of massless scalar fields, which we can then match with the $3d$ Carrollian CFT correlators through the Carrollian holography dictionary described in the previous sections. This is the procedure we will implement here, following \cite{Lipstein:2025jfj}; see Figure \ref{fig:cube}.

\item The second option, if at all possible, would consist of changing the way the limit is taken in the dual theory to end up with a $10d$ Carrollian CFT living at the boundary of $11d$ Minkowski space, instead of a $3d$ Carrollian CFT. This would, in principle, allow one to capture the full $11d$ graviton amplitude with generic kinematics, and thereby reconstruct the complete flat space theory. Such a procedure cannot be a naive Carrollian limit, since the latter does not change the number of dimensions, and would require understanding the effect of decompactifying the $S^7$ from the dual theory perspective. We will comment on this intriguing possibility in the discussion section \ref{sec:Through the looking glass}.

\end{itemize}

\begin{figure}[h]
\begin{center}

	\begin{tikzpicture}[baseline=(current bounding box.center)]
		\tikzset{nodx/.style = {draw=black,outer sep=4pt,inner sep=7pt,text width=3.3cm,align=center,minimum height=1.2cm}
}
		\tikzset{nodrelax/.style = {draw=black,outer sep=4pt,inner sep=7pt}
}
		\def\h{3};\def\l{3.8};
		\coordinate (A) at (-\l, \h);
		\coordinate (B) at ( \l, \h);
		\coordinate (C) at ( \l,-\h);
		\coordinate (D) at (-\l,-\h);
		\coordinate (E) at ($(B)!.5!(C)$);
		\node[nodx] (a) at (A) {M-theory on $\text{AdS}_4\times S^7$};
		\node[nodx] (b) at (B) {M-theory on $11d$ flat space};
		\node[nodx] (d) at (D) {ABJM theory in $3d$};
		\node[nodx] (c) at (C) {Carrollian ABJM theory in $3d$};
		\node[nodx] (e) at (E) {Kinematic \\restriction to $\mathbb R^{1,3}$};
		\draw[-Latex] (a) -- (b) node[midway,above] {$\ell\to \infty$} node[midway,below] {(Eq. \eqref{from4to11})};
		\draw[-Latex] (d) -- (c) node[midway,above] {$c\to 0$} node[midway,below] {(Sec. \ref{sec:Carrollian limit of ABJM})};
		\draw[Latex-Latex] (a) -- (d) node[midway,left] {AdS/CFT};
		\draw[-Latex] (b) -- (e) node[midway,right] {(Eq. \eqref{restrictkin})};
		\draw[Latex-Latex] (e) -- (c) node[midway,right] {(Eq. \eqref{carrAbjmdic})};
	\end{tikzpicture}
    \caption{Summary of the flat space / Carrollian limit of AdS$_4$ / CFT$_3$ using the first approach described in Section~\ref{sec:Bulk interpretation}, which consists in reproducing the flat space amplitudes with restricted kinematics in the limit.}

    \label{fig:summarylast}
\end{center}
\end{figure}
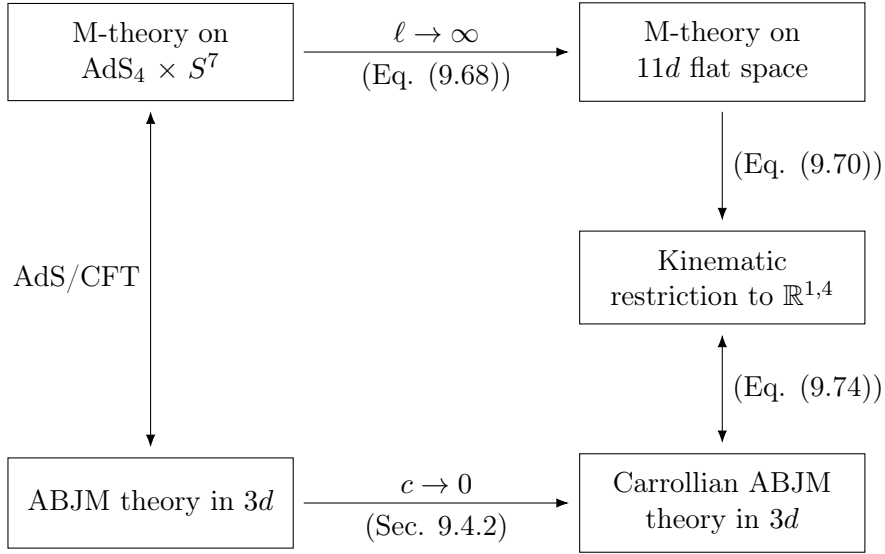

\begin{center}
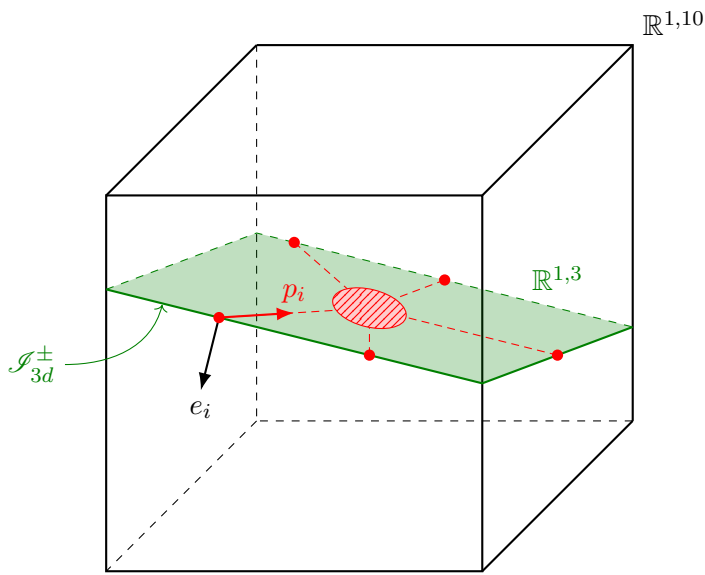
\begin{figure}[h]
\centering
\begin{tikzpicture}
	\def\l{2.5}; \def\q{2};
	\coordinate (A) at (-\l, \l);
	\coordinate (B) at ( \l, \l);
	\coordinate (C) at ( \l,-\l);
	\coordinate (D) at (-\l,-\l);
	\coordinate (a) at ($(A)-(\q,\q)$);
	\coordinate (b) at ($(B)-(\q,\q)$);
	\coordinate (c) at ($(C)-(\q,\q)$);
	\coordinate (d) at ($(D)-(\q,\q)$);
	\coordinate (Q1) at ($(b)!.5!(c)$);
	\coordinate (Q2) at ($(a)!.25!(d)$);
	\coordinate (Q3) at ($(A)!.5!(D)$);
	\coordinate (Q4) at ($(C)!.25!(B)$);
	\coordinate (I1) at ($(Q1)!.7!(Q2)$);
	\coordinate (I2) at ($(Q1)!.3!(Q2)$);
	\coordinate (I3) at ($(Q1)!.5!(Q4)$);
	\coordinate (I4) at ($(Q3)!.5!(Q4)$);
	\coordinate (I5) at ($(Q3)!.1!(Q4)$);	
	\coordinate (O) at ($(Q2)!.5!(Q4)$);
	
	\draw[opacity=0] ($(B)+(1.5,.5)$) rectangle ($(d)-(1.5,.5)$);
	
	\draw[dashed] (D) -- (d);
	\draw[thick] (A) -- (B) -- (C);
	\draw[dashed] (C) -- (D) -- (A);
	\draw[thick] (a) -- (b) -- (c) -- (d) -- cycle;

	\fill[opacity=0.25,green!50!black] 	(Q2) -- (Q1) -- (Q4) -- (Q3) -- cycle;
	
	\draw[green!50!black,thick] (Q2) -- (Q1) -- (Q4);	
	\draw[green!50!black,dashed] (Q2) -- (Q3) -- (Q4);
	
	\draw[densely dashed,red] (I1) -- (O);
	\draw[densely dashed,red] (I2) -- (O);
	\draw[densely dashed,red] (I3) -- (O);
	\draw[densely dashed,red] (I4) -- (O);
	\draw[densely dashed,red] (I5) -- (O);
	\node[preaction={fill=red!20},pattern=north east lines, pattern color=red, draw=red, ellipse, rotate=-14, minimum width=1cm, minimum height=0.5cm] at (O) {};
	
	\draw[-Latex,thick] (I1) -- ($(I1)-({sin(14)},{cos(14)})$) node[below]{$e_i$};
	\draw[-Latex,red,thick] (I1) -- ($(I1)!.5!(O)$) node[above] {$p_i$};
	
	\node[circle,inner sep=1.5pt,fill=red] at (I1) {};
	\node[circle,inner sep=1.5pt,fill=red] at (I2) {};
	\node[circle,inner sep=1.5pt,fill=red] at (I3) {};
	\node[circle,inner sep=1.5pt,fill=red] at (I4) {};
	\node[circle,inner sep=1.5pt,fill=red] at (I5) {};
	
	\draw[thick] (A) -- (a);
	\draw[thick] (B) -- (b);
	\draw[thick] (C) -- (c);
	
	\node[anchor=south west] at (B) {$\mathbb R^{1,10}$};
	
	\node[outer sep=0pt,inner sep=0pt] (point) at ($(Q2)!.5!(I1)$) {};
	\node[green!50!black] (orig) at ($(Q2)-(1,1)$) {$\mathscr I^\pm_{3d}$};
	\draw[green!50!black,->] (orig) to[out=0,in=-104] (point);
	\node[green!50!black,above,yshift=3pt] at ($(Q4)!.2!(Q3)$) {$\mathbb R^{1,3}$};
\end{tikzpicture}
\caption{Graviton amplitude in $\mathbb{R}^{1,10}$ with restricted kinematics in $\mathbb{R}^{1,3}$. Here $p_i$ denotes the momentum of the $i^{th}$ graviton longitudinal to $\mathbb{R}^{1,3}$, and $e_i$ its polarization transverse to $\mathbb{R}^{1,3}$. In position space, the boundary of the restricted kinematic region $\mathbb{R}^{1,3}$ is $\mathscr{I}^\pm_{3d}$, which is where the Carrollian ABJM correlators derived in Section \ref{sec:Carrollian limit of ABJM} are living.}
\label{fig:cube}
\end{figure}
\end{center}
We start from the tree-level graviton amplitudes in $11d$ $\mathcal{N}=1$ supergravity, which can be written as
\begin{equation}
\mathcal{A}_n({ P_i, e_i })
\end{equation}
where $P_i$ and $e_i$ denote the $11d$ momentum and polarization of the $i^{\rm th}$ graviton, respectively. Following the first option described above, we split $\mathbb{R}^{1,10} = \mathbb{R}^{1,3} \times \mathbb{R}^7$ and restrict ourselves to the specific kinematics 
\begin{align} \label{restrictkin}
   P_i^{\alpha} = \left(p_i^{\mu}, 0\right), \qquad p_i \in \mathbb{R}^{1,3}, \qquad e_i^{\alpha} = \left(0 , \xi_i^I\right), \qquad \xi_i \in \mathbb{C}^7 .
\end{align} This situation is depicted in Figure \ref{fig:cube}. We can embed the complex $7d$ null vector $\xi_i$ into an $8d$ vector and identify $t_i \equiv \left(0, \xi_i\right) \in \mathbb{C}^8$ where $t_i$ is the R-symmetry null vector. This has the property $e_i \cdot e_j = t_i \cdot t_j \equiv t_{ij}$ and $e_i \cdot P_j = 0$. One can then use scattering states adapted to this decomposition; they are product of plane waves in the $\mathbb{R}^{1,3}$ space and polynomials of degree $k-2$ in the complexified $\mathbb{C}^7$ space. Furthermore, one can show that they satisfy the massless spin-$2$ field equations of motion in $11d$, but are not normalizable. We refer to \cite{Lipstein:2025jfj} for details. Decomposing the amplitudes in these scattering states with restricted kinematics, we find massless scattering amplitudes for scalar fields labelled by $k$ taking place in the $\mathbb{R}^{1,3}$ plane: 
\begin{align}
    \label{eq:2ptfromZ}
    \mathcal{A}_{k_1, k_2} = \mathcal{N}_2\frac{t_{12}^{k_1}\delta_{k_1, k_2}}{\om_1} \, \delta_{\e_1, -\e_2}\delta\left(\om_1 - \om_2\right) \delta^2\left(z_{12}\right),
\end{align}
\begin{align}
    \label{eq:3ptamp}
    \mathcal{A}_{k_1, k_2, k_3} = \mathcal{N}_3 t_{12}^{\alpha_3}t_{23}^{\alpha_1}t_{13}^{\alpha_2} \,\delta^{(4)}\left(p_1+p_2+p_3\right),
\end{align}
 \begin{align}
    \mathcal{A}_{k_1,k_2,k_3,k_4} = \mathcal{N}_4 \ell^9_{11} \prod_{i<j} t_{ij}^{\gamma_{ij}^0}\frac{\left(t_{12}t_{34}\right)^{\mathcal{E}}}{s\, t\, u}\left(t \, u + s \, u\, \sigma + s \, t \, \tau\right)^2 \mathcal{P}_{k_i}\left(\sigma, \tau\right) \delta^{(4)}\left(p_1+p_2+p_3 +p_4\right) .
\end{align} Computing the corresponding Carrollian amplitudes via \eqref{modified mellin transform} with fixed $\Delta_j = \frac{k_j}{2}$, i.e.
\begin{equation} \label{carrAbjmdic}
    \langle \Phi_{k_1} \ldots \Phi_{k_n}  \rangle = \int \prod_{j=1}^n \frac{d\om_j}{2\pi} \om_j^{\D_j-1} e^{-i \, u_j \om_j \e_j } \mathcal{A}_{k_1, \ldots , k_n} 
\end{equation} we can match them exactly with the Carrollian ABJM correlators \eqref{eq:2ptcarrolllim}, \eqref{eq:carroll3ptabjm} and \eqref{G2kkkk Carrollian} computed from an intrinsic boundary perspective. Our discussion is summarized in Figure \ref{fig:summarylast}. 

\section{Through the looking glass...}
\label{sec:Through the looking glass}

In this review, we have surveyed recent developments in Carrollian physics within the emerging framework of flat space holography. Our focus has been on the scattering problem in four-dimensional asymptotically flat spacetimes, motivated and strengthened by its close connection to celestial holography \cite{Donnay:2022aba, Bagchi:2022emh, Donnay:2022wvx}. The current status of the program may be summarized as follows. Assuming that a holographic principle applies to asymptotically flat spacetimes, and adopting a bottom-up perspective, the geometry and symmetry structure at null infinity naturally point toward a putative dual description in terms of a theory with BMS symmetries, or equivalently a Carrollian CFT. Indeed, we have shown how Carrollian physics arises intrinsically on null hypersurfaces. We reviewed the general properties of Carrollian field theories and explained how explicit examples can be constructed by taking Carrollian limits of relativistic field theories.

With the goal of establishing a flat space holographic dictionary, we showed that massless scattering amplitudes in flat spacetime can be recast as correlators at null infinity obeying Carrollian CFT Ward identities, which we referred to as Carrollian amplitudes. Their explicit form provides strong dynamical constraints on any would-be dual theory. In this spirit, we reformulated several familiar statements about flat-space scattering amplitudes in terms of Carrollian physics at null infinity. This allowed us to extract structural information about the putative dual theory, including aspects of its operator product expansion, the role of stress tensor Ward identities, and differential equations constraining its dynamics.

An important point in favor of the Carrollian approach to flat space holography is its direct and concrete relation to AdS/CFT. In particular, we have shown that taking the flat space limit in the bulk of AdS is closely tied to taking a Carrollian limit at the boundary. This observation raises the prospect of obtaining explicit realizations of Carrollian CFTs and of achieving a top-down construction of flat space holography by taking suitable limits on both sides of the AdS/CFT duality. We have taken initial steps in this direction by applying this procedure to a concrete example of AdS$_4$/CFT$_3$. This framework makes it possible, at least for restricted kinematics, to compute flat space scattering amplitudes from purely boundary data: starting from CFT correlators (for instance obtained via bootstrap methods), taking their Carrollian limit, and finally reinterpreting the result as flat space amplitudes using the Carrollian holography dictionary. This constitutes a tangible and encouraging advance in the broader flat space holography program.

Despite this progress, a fundamental question remains open. It is not yet clear whether the Carrollian CFTs that arise either as limits of holographic CFTs or as reformulations of flat space observables at null infinity admit a consistent intrinsic definition as quantum field theories. In other words, can one define a genuine quantum Carrollian CFT that captures flat space physics in its own right? Or should flat space holography instead be viewed as a singular limit of AdS/CFT—well defined only as a limiting procedure applied to observables, but not as a standalone theory? Resolving this issue is, in our view, one of the central challenges for the future development of the subject.

We now outline a few intriguing directions for the program, some of which may offer concrete ways to address this crucial question.

As emphasized throughout this review, the quantization of Carrollian field theories is a delicate and subtle issue. In particular, electric Carrollian field theories, which are relevant for massless scattering in flat spacetime, do not appear to give rise to non-Gaussian correlators when quantized using a lattice regulator \cite{Cotler:2024xhb}. By contrast, as we have discussed, taking the Carrollian limit directly at the level of correlation functions does lead to non-trivial electric Carrollian correlators. This observation suggests that the speed of light $c$ itself may play the role of a natural regulator for the quantization of Carrollian field theories. From this perspective, one may view a Carrollian theory as arising from a relativistic parent theory, which we know how to quantize, by tuning the value of $c$. However, starting from a given Carrollian field theory, there are in general multiple inequivalent ways to uplift it to a relativistic theory by restoring a finite speed of light. It would be interesting to explore whether there exists a preferred uplift of this kind, and whether this requirement could provide a principled prescription for regulating and quantizing Carrollian field theories.

An important caveat of the current Carrollian holography dictionary is that it incorporates only massless scattering amplitudes in flat spacetime in terms of Carrollian CFT data. While celestial holography provides a prescription for encoding massive states, the corresponding construction in the Carrollian framework remains an open problem and would be essential for completing the holographic dictionary. One proposal is to encode massive scattering data at timelike infinity rather than at null infinity \cite{Have:2024dff}, since timelike geodesics naturally terminate there. However, the Carrollian geometry at timelike infinity is four-dimensional \cite{Figueroa-OFarrill:2021sxz} and therefore does not lead to a holographic reduction in the number of dimensions, weakening its appeal from a holographic perspective. It would therefore be highly interesting to explore whether massive states can instead be encoded at null infinity. Although timelike geodesics do not reach this boundary, this does not, in principle, preclude massive degrees of freedom from being represented there. In this respect, the situation could be analogous to AdS/CFT: while timelike geodesics do not reach the AdS boundary, massive states are nonetheless fully encoded in boundary observables. It would also be valuable to better understand the behavior of massive states in the flat space limit, for which we expect their conformal dimension to scale as $\Delta \sim \ell$, and to clarify their fate in the Carrollian framework.

We have seen that the Carrollian limit of the conformal algebra leads to the global conformal Carrollian algebra, which admits an infinite-dimensional enhancement through the inclusion of supertranslations, and potentially superrotations. These symmetries have been connected to the infrared sector of massless scattering \cite{Strominger:2017zoo}, which is intimately tied to soft modes. In Carrollian language, there are strong indications that this soft sector is described by a magnetic Carrollian CFT, since its correlation functions are time independent (see, for instance, \cite{Himwich:2020rro, Fiorucci:2023lpb, Jorstad:2024yzm}) and the actions governing its dynamics are effectively two-dimensional \cite{He:2024vlp}. This stands in contrast to hard scattering amplitudes which, as we have explained, are naturally described by an electric Carrollian CFT. It would be interesting to revisit the infrared sector of flat spacetime from a Carrollian perspective and to understand how this structure emerges in the flat space limit of AdS/CFT.

In taking the limit of an explicit realization of AdS/CFT, a key step is to understand what happens with the internal space. As we have explained in Section \ref{sec:Towards a top-down flat space hologram from AdS/CFT}, there are essentially two options: either the flat space limit / Carrollian limit only allows to reproduce amplitudes with restricted kinematics, which is the approach we have adopted here (see Figures \ref{fig:cube} and \ref{fig:summarylast}), or there is a subtle way to take the limit to reproduce the full flat space amplitudes, with unrestricted kinematics. Important insights along these lines were recently discussed in \cite{Fontanella:2025tbs} by implementing a Carrollian contraction on both sides of the AdS/CFT duality. It would be extremely interesting to understand the decompactification of $S^7$ in the dual theory, via a specific R-symmetry scaling in the Carrollian limit.

As we have emphasized, one of the central open problems of the program is to define Carrollian CFTs intrinsically, without appealing to a limiting procedure from relativistic theories. Despite some interesting progress outlined in this report, the state-operator correspondence and the structure of conformal block decompositions remain largely unexplored, and may well differ substantially from their relativistic counterparts. Moreover, the intricate and nonstandard OPE structure of Carrollian CFTs may obstruct a straightforward extraction of the data required to define such theories in the usual axiomatic sense. Recent works even suggest that the theory describing massless scattering in flat space may be non-local, which drastically reshapes the axiomatic definition of what we mean by a Carrollian CFT \cite{Cotler:2025npu}. Interestingly, closely related developments exist for non-relativistic CFTs \cite{Niederer:1972zz,Hagen:1972pd,Henkel:1993sg,Mehen:1999nd,Nishida:2007pj,Nishida:2010tm,Golkar:2014mwa,Goldberger:2014hca,Pal:2018idc}, which have also found notable applications in holography \cite{Son:2008ye,Balasubramanian:2008dm,Goldberger:2008vg,Maldacena:2008wh,Taylor:2008tg,Volovich:2009yh}. This program has been remarkably successful, despite facing challenges closely analogous to those encountered in the Carrollian setting: a causal structure distinct from that of relativistic CFTs, unconventional forms of correlation functions, and a different organization of primaries and descendants. Motivated by these results, it would be natural to investigate whether some of the techniques developed in the non-relativistic context, can be adapted to Carrollian CFTs.

Of course, one of the main motivations behind the development of the holographic principle is to understand the microscopic origin of the Bekenstein-Hawking entropy formula \cite{Bekenstein:1973ur,Hawking:1975vcx},
\[
S = \frac{A}{4G},
\]
where $A$ is the area of the black hole horizon. Indeed, the universality of this formula, independently of the value of $\Lambda$, suggests that holography should hold beyond AdS spacetimes. Although such a microscopic counting has been achieved in certain situations \cite{Strominger:1996sh,  Strominger:1997eq,Guica:2008mu, Castro:2010fd}, it remains an open problem to account for this formula in the simplest and perhaps most physically relevant cases, namely Schwarzschild and non-extremal Kerr black holes in asymptotically flat spacetimes. These are of major interest since they model the black holes observed in the universe. The long-term hope is that progress in flat space holography will ultimately shed light on this fundamental question in quantum gravity.

It is clear that the tools developed in the pursuit of Carrollian holography have a scope that extends well beyond flat space holography itself. As emphasized in the introduction, Carrollian physics is naturally associated with null surfaces, which arise ubiquitously across theoretical physics, including black hole and cosmological horizons, as well as null defects in relativistic CFTs \cite{Erramilli:2025pfh}. More broadly, Carrollian structures have already appeared in diverse contexts ranging from cosmology \cite{deBoer:2021jej} and gravitational waves \cite{Duval:2017els}, to condensed matter systems \cite{Bidussi:2021nmp,Bagchi:2022eui,Figueroa-OFarrill:2023vbj}, and the dynamics of extremal fluids \cite{deBoer:2017ing,Ciambelli:2018xat,Bagchi:2023ysc,Petkou:2022bmz,Bagchi:2023rwd, Shukla:2026xig}. These developments strongly suggest that Carrollian physics provides a unifying language for phenomena involving null hypersurfaces and, more generally, extreme regimes in which conventional notions of causality break down.

In the course of developing Carrollian holography, we have encountered not only new technical tools but also a number of conceptual questions that invite a reassessment of familiar notions of locality, geometry, dynamics, and quantum field theory. The program has already generated a substantial body of ideas while remaining at an early stage of development. What ultimately lies beyond the looking glass is not yet clear, but this openness leaves considerable room for further progress and exploration.

\section*{Acknowledgements}

It is my pleasure to thank all my collaborators on the topic of Carrollian physics and holography. I am particularly grateful to Jordan Cotler, Adrien Fiorucci, Daniel Grumiller, Simon Heuveline, Walker Melton, Kyrill Michaelsen and Atul Sharma for their useful feedback on the manuscript. I am deeply indebted to Stephan Stieberger for inviting me to write this report and for his continual encouragement throughout the process. Finally, I would like to thank the organizers and participants of the Celestial Holography Summer School 2024 (Simons Collaboration) at the Perimeter Institute, where part of this material was presented. This work is supported by the European Union’s Horizon Europe research and innovation programme under the Marie Skłodowska-Curie grant agreement No. 101104845 (UniFlatHolo), hosted at Harvard University and École Polytechnique.

\section*{References}\addcontentsline{toc}{section}{References}
\bibliographystyle{style}
\renewcommand\refname{\vskip -1.3cm}
\bibliography{references}

@misc{Duffin,
  author        = {David Simmons-Duffin},
  title         = {TASI Lectures on Conformal Field Theory
in Lorentzian Signature},
  year          = {2019}
}

@article{Freedman:1998tz,
    author = "Freedman, Daniel Z. and Mathur, Samir D. and Matusis, Alec and Rastelli, Leonardo",
    title = "{Correlation functions in the CFT(d) / AdS(d+1) correspondence}",
    eprint = "hep-th/9804058",
    archivePrefix = "arXiv",
    reportNumber = "MIT-CTP-2727",
    doi = "10.1016/S0550-3213(99)00053-X",
    journal = "Nucl. Phys. B",
    volume = "546",
    pages = "96--118",
    year = "1999"
}

@article{deBoer:2023fnj,
    author = "de Boer, Jan and Hartong, Jelle and Obers, Niels A. and Sybesma, Watse and Vandoren, Stefan",
    title = "{Carroll stories}",
    eprint = "2307.06827",
    archivePrefix = "arXiv",
    primaryClass = "hep-th",
    reportNumber = "NORDITA-2023-036",
    doi = "10.1007/JHEP09(2023)148",
    journal = "JHEP",
    volume = "09",
    pages = "148",
    year = "2023"
}

@article{Bittleston:2024rqe,
    author = "Bittleston, Roland and Bogna, Giuseppe and Heuveline, Simon and Kmec, Adam and Mason, Lionel and Skinner, David",
    title = "{On AdS$_4$ deformations of celestial symmetries}",
    eprint = "2403.18011",
    archivePrefix = "arXiv",
    primaryClass = "hep-th",
    month = "3",
    year = "2024"
}

@article{Taylor:2023ajd,
    author = "Taylor, Tomasz R. and Zhu, Bin",
    title = "{w1+\ensuremath{\infty} Algebra with a Cosmological Constant and the Celestial Sphere}",
    eprint = "2312.00876",
    archivePrefix = "arXiv",
    primaryClass = "hep-th",
    doi = "10.1103/PhysRevLett.132.221602",
    journal = "Phys. Rev. Lett.",
    volume = "132",
    number = "22",
    pages = "221602",
    year = "2024"
}

@article{DHoker:1998bqu,
    author = "D'Hoker, Eric and Freedman, Daniel Z.",
    title = "{Gauge boson exchange in AdS(d+1)}",
    eprint = "hep-th/9809179",
    archivePrefix = "arXiv",
    reportNumber = "UCLA-98-TEP-33, MIT-CTP-2781",
    doi = "10.1016/S0550-3213(98)00852-9",
    journal = "Nucl. Phys. B",
    volume = "544",
    pages = "612--632",
    year = "1999"
}

@article{DHoker:1998ecp,
    author = "D'Hoker, Eric and Freedman, Daniel Z.",
    title = "{General scalar exchange in AdS(d+1)}",
    eprint = "hep-th/9811257",
    archivePrefix = "arXiv",
    reportNumber = "UCLA-98-TEP-34, MIT-CTP-2795",
    doi = "10.1016/S0550-3213(99)00169-8",
    journal = "Nucl. Phys. B",
    volume = "550",
    pages = "261--288",
    year = "1999"
}

@article{Liu:1998bu,
    author = "Liu, Hong and Tseytlin, Arkady A.",
    title = "{D = 4 superYang-Mills, D = 5 gauged supergravity, and D = 4 conformal supergravity}",
    eprint = "hep-th/9804083",
    archivePrefix = "arXiv",
    reportNumber = "IMPERIAL-TP-97-98-39, NSF-ITP-98-049",
    doi = "10.1016/S0550-3213(98)00443-X",
    journal = "Nucl. Phys. B",
    volume = "533",
    pages = "88--108",
    year = "1998"
}

@article{Freidel:2022skz,
    author = "Freidel, Laurent and Pranzetti, Daniele and Raclariu, Ana-Maria",
    title = "{A discrete basis for celestial holography}",
    eprint = "2212.12469",
    archivePrefix = "arXiv",
    primaryClass = "hep-th",
    doi = "10.1007/JHEP02(2024)176",
    journal = "JHEP",
    volume = "02",
    pages = "176",
    year = "2024"
}

@article{deGioia:2024yne,
    author = "de Gioia, Leonardo Pipolo and Raclariu, Ana-Maria",
    title = "{Celestial amplitudes from conformal correlators with bulk-point kinematics}",
    eprint = "2405.07972",
    archivePrefix = "arXiv",
    primaryClass = "hep-th",
    month = "5",
    year = "2024"
}

@article{Ara:2024fbr,
    author = "Ara, Nisa and Banerjee, Aritra and Basu, Rudranil and Krishnan, Bhagya",
    title = "{Flat bands and compact localised states: A Carrollian roadmap}",
    eprint = "2412.18965",
    archivePrefix = "arXiv",
    primaryClass = "hep-th",
    doi = "10.21468/SciPostPhys.19.2.046",
    journal = "SciPost Phys.",
    volume = "19",
    pages = "046",
    year = "2025"
}

@article{Rodriguez:2021tcz,
    author = "Rodr{\'\i}guez, Pablo and Tempo, David and Troncoso, Ricardo",
    title = "{Mapping relativistic to ultra/non-relativistic conformal symmetries in 2D and finite $ \sqrt{T\overline{T}} $ deformations}",
    eprint = "2106.09750",
    archivePrefix = "arXiv",
    primaryClass = "hep-th",
    reportNumber = "CECS-PHY-20/03",
    doi = "10.1007/JHEP11(2021)133",
    journal = "JHEP",
    volume = "11",
    pages = "133",
    year = "2021"
}

@article{Tempo:2022ndz,
    author = "Tempo, David and Troncoso, Ricardo",
    title = "{Nonlinear automorphism of the conformal algebra in 2D and continuous $ \sqrt{T\overline{T}} $ deformations}",
    eprint = "2210.00059",
    archivePrefix = "arXiv",
    primaryClass = "hep-th",
    reportNumber = "CECS-PHY-22/05",
    doi = "10.1007/JHEP12(2022)129",
    journal = "JHEP",
    volume = "12",
    pages = "129",
    year = "2022"
}

@article{Parekh:2023xms,
    author = "Parekh, Pulastya and Tempo, David and Troncoso, Ricardo",
    title = "{BMS$_{3}$ (Carrollian) field theories from a bound in the coupling of current-current deformations of CFT$_{2}$}",
    eprint = "2307.06367",
    archivePrefix = "arXiv",
    primaryClass = "hep-th",
    doi = "10.1007/JHEP09(2023)083",
    journal = "JHEP",
    volume = "09",
    pages = "083",
    year = "2023"
}

@article{Banerjee:2026qyc,
    author = "Banerjee, Aritra and Parekh, Pulastya and Raj, Robin",
    title = "{On $\sqrt{T\overline{T}}$ deformed pathways: CFT to CCFT}",
    eprint = "2601.15376",
    archivePrefix = "arXiv",
    primaryClass = "hep-th",
    month = "1",
    year = "2026"
}

@article{Kasikci:2023tvs,
    author = "Kasikci, Oguzhan and Ozkan, Mehmet and Pang, Yi",
    title = "{Carrollian origin of spacetime subsystem symmetry}",
    eprint = "2304.11331",
    archivePrefix = "arXiv",
    primaryClass = "hep-th",
    doi = "10.1103/PhysRevD.108.045020",
    journal = "Phys. Rev. D",
    volume = "108",
    number = "4",
    pages = "045020",
    year = "2023"
}

@article{Blitz:2025spc,
    author = "Blitz, Samuel and McNutt, David and Nurowski, Pawel",
    title = "{Unique Carrollian manifolds emerging from Einstein spacetimes}",
    doi = "10.1088/1361-6382/adbb4e",
    journal = "Class. Quant. Grav.",
    volume = "42",
    number = "7",
    pages = "075006",
    year = "2025"
}

@article{Koutrolikos:2023evq,
    author = "Koutrolikos, Konstantinos and Najafizadeh, Mojtaba",
    title = "{Super-Carrollian and Super-Galilean Field Theories}",
    eprint = "2309.16786",
    archivePrefix = "arXiv",
    primaryClass = "hep-th",
    reportNumber = "IPM/P-2023/61",
    doi = "10.1103/PhysRevD.108.125014",
    journal = "Phys. Rev. D",
    volume = "108",
    number = "12",
    pages = "125014",
    year = "2023"
}

@article{Adami:2023fbm,
    author = "Adami, H. and Parvizi, A. and Sheikh-Jabbari, M. M. and Taghiloo, V. and Yavartanoo, H.",
    title = "{Hydro {\&} thermo dynamics at causal boundaries, examples in 3d gravity}",
    eprint = "2305.01009",
    archivePrefix = "arXiv",
    primaryClass = "hep-th",
    doi = "10.1007/JHEP07(2023)038",
    journal = "JHEP",
    volume = "07",
    pages = "038",
    year = "2023"
}

@article{Adami:2023wbe,
    author = "Adami, H. and Parvizi, A. and Sheikh-Jabbari, M. M. and Taghiloo, V. and Yavartanoo, H.",
    title = "{Carrollian structure of the null boundary solution space}",
    eprint = "2311.03515",
    archivePrefix = "arXiv",
    primaryClass = "hep-th",
    doi = "10.1007/JHEP02(2024)073",
    journal = "JHEP",
    volume = "02",
    pages = "073",
    year = "2024"
}

@article{Adami:2024mtu,
    author = "Adami, H. and Parvizi, A. and Sheikh-Jabbari, M. M. and Taghiloo, V.",
    title = "{Heisenberg soft hair on Robinson-Trautman spacetimes}",
    eprint = "2402.17658",
    archivePrefix = "arXiv",
    primaryClass = "hep-th",
    doi = "10.1007/JHEP05(2024)191",
    journal = "JHEP",
    volume = "05",
    pages = "191",
    year = "2024"
}

@article{Adami:2024rkr,
    author = "Adami, H. and Sheikh-Jabbari, M. M. and Taghiloo, V.",
    title = "{Gravitational stress tensor and current at null infinity in three dimensions}",
    eprint = "2405.00149",
    archivePrefix = "arXiv",
    primaryClass = "hep-th",
    doi = "10.1016/j.physletb.2024.138835",
    journal = "Phys. Lett. B",
    volume = "855",
    pages = "138835",
    year = "2024"
}

@article{Ekiz:2025hdn,
    author = "Ekiz, Ertu{\u{g}}rul and Kahya, Emre Onur and Zorba, Utku",
    title = "{Quantization of Carrollian fermions}",
    eprint = "2502.05645",
    archivePrefix = "arXiv",
    primaryClass = "hep-th",
    doi = "10.1103/PhysRevD.111.105019",
    journal = "Phys. Rev. D",
    volume = "111",
    number = "10",
    pages = "105019",
    year = "2025"
}

@article{Capone:2021ouo,
    author = "Capone, Federico",
    title = "{General null asymptotics and superrotation-compatible configuration spaces in $d\ge4$}",
    eprint = "2108.01203",
    archivePrefix = "arXiv",
    primaryClass = "hep-th",
    doi = "10.1007/JHEP10(2021)158",
    journal = "JHEP",
    volume = "10",
    pages = "158",
    year = "2021",
    note = "[Erratum: JHEP 02, 113 (2022)]"
}

@article{Capone:2019aiy,
    author = "Capone, Federico and Taylor, Marika",
    title = "{Cosmic branes and asymptotic structure}",
    eprint = "1904.04265",
    archivePrefix = "arXiv",
    primaryClass = "hep-th",
    doi = "10.1007/JHEP10(2019)138",
    journal = "JHEP",
    volume = "10",
    pages = "138",
    year = "2019"
}

@article{Gomis:2020wxp,
    author = "Gomis, Joaquim and Hidalgo, Diego and Salgado-Rebolledo, Patricio",
    title = "{Non-relativistic and Carrollian limits of Jackiw-Teitelboim gravity}",
    eprint = "2011.15053",
    archivePrefix = "arXiv",
    primaryClass = "hep-th",
    doi = "10.1007/JHEP05(2021)162",
    journal = "JHEP",
    volume = "05",
    pages = "162",
    year = "2021"
}

@article{Bunster:2012hm,
    author = "Bunster, Claudio and Henneaux, Marc",
    title = "{Duality Invariance Implies Poincar{\'e} Invariance}",
    eprint = "1208.6302",
    archivePrefix = "arXiv",
    primaryClass = "hep-th",
    doi = "10.1103/PhysRevLett.110.011603",
    journal = "Phys. Rev. Lett.",
    volume = "110",
    number = "1",
    pages = "011603",
    year = "2013"
}

@article{Li:2008dq,
    author = "Li, Wei and Song, Wei and Strominger, Andrew",
    title = "{Chiral Gravity in Three Dimensions}",
    eprint = "0801.4566",
    archivePrefix = "arXiv",
    primaryClass = "hep-th",
    doi = "10.1088/1126-6708/2008/04/082",
    journal = "JHEP",
    volume = "04",
    pages = "082",
    year = "2008"
}

@article{Bagchi:2012yk,
    author = "Bagchi, Arjun and Detournay, Stephane and Grumiller, Daniel",
    title = "{Flat-Space Chiral Gravity}",
    eprint = "1208.1658",
    archivePrefix = "arXiv",
    primaryClass = "hep-th",
    doi = "10.1103/PhysRevLett.109.151301",
    journal = "Phys. Rev. Lett.",
    volume = "109",
    pages = "151301",
    year = "2012"
}

@article{Arenas-Henriquez:2025rpt,
    author = "Arenas-Henriquez, Gabriel and Ciambelli, Luca and Diaz, Felipe and Jia, Weizhen and Rivera-Betancour, David",
    title = "{Radiation in fluid/gravity and the flat limit}",
    eprint = "2508.01446",
    archivePrefix = "arXiv",
    primaryClass = "hep-th",
    doi = "10.1007/JHEP01(2026)086",
    journal = "JHEP",
    volume = "01",
    pages = "086",
    year = "2026"
}

@article{Ciambelli:2024kre,
    author = "Ciambelli, Luca and Pasterski, Sabrina and Tabor, Elisa",
    title = "{Radiation in holography}",
    eprint = "2404.02146",
    archivePrefix = "arXiv",
    primaryClass = "hep-th",
    doi = "10.1007/JHEP09(2024)124",
    journal = "JHEP",
    volume = "09",
    pages = "124",
    year = "2024"
}

@article{Poole:2018koa,
    author = "Poole, Aaron and Skenderis, Kostas and Taylor, Marika",
    title = "{(A)dS$\mathbf{_4}$ in Bondi gauge}",
    eprint = "1812.05369",
    archivePrefix = "arXiv",
    primaryClass = "hep-th",
    doi = "10.1088/1361-6382/ab117c",
    journal = "Class. Quant. Grav.",
    volume = "36",
    number = "9",
    pages = "095005",
    year = "2019"
}

@article{Liu:2024nfc,
    author = "Liu, Wen-Bin and Long, Jiang and Ye, Xiao-Quan",
    title = "{Feynman rules and loop structure of Carrollian amplitudes}",
    eprint = "2402.04120",
    archivePrefix = "arXiv",
    primaryClass = "hep-th",
    doi = "10.1007/JHEP05(2024)213",
    journal = "JHEP",
    volume = "05",
    pages = "213",
    year = "2024"
}

@article{Mason:2023mti,
    author = "Mason, Lionel and Ruzziconi, Romain and Yelleshpur Srikant, Akshay",
    title = "{Carrollian amplitudes and celestial symmetries}",
    eprint = "2312.10138",
    archivePrefix = "arXiv",
    primaryClass = "hep-th",
    doi = "10.1007/JHEP05(2024)012",
    journal = "JHEP",
    volume = "05",
    pages = "012",
    year = "2024"
}

@article{Hansen:1978jz,
	author = {Hansen, R. O. and Newman, E. T. and Penrose, R. and Tod, K. P.},
	doi = {10.1098/rspa.1978.0177},
	journal = {Proc. Roy. Soc. Lond.},
	pages = {445-468},
	title = {{The Metric and Curvature Properties of H Space}},
	volume = {A363},
	year = {1978}}

@article{Bidussi:2021nmp,
    author = "Bidussi, Leo and Hartong, Jelle and Have, Emil and Musaeus, J\o{}rgen and Prohazka, Stefan",
    title = "{Fractons, dipole symmetries and curved spacetime}",
    eprint = "2111.03668",
    archivePrefix = "arXiv",
    primaryClass = "hep-th",
    doi = "10.21468/SciPostPhys.12.6.205",
    journal = "SciPost Phys.",
    volume = "12",
    number = "6",
    pages = "205",
    year = "2022"
}

@article{Figueroa-OFarrill:2023vbj,
    author = "Figueroa-O'Farrill, Jos\'e and P\'erez, Alfredo and Prohazka, Stefan",
    title = "{Carroll/fracton particles and their correspondence}",
    eprint = "2305.06730",
    archivePrefix = "arXiv",
    primaryClass = "hep-th",
    reportNumber = "EMPG-23-09",
    doi = "10.1007/JHEP06(2023)207",
    journal = "JHEP",
    volume = "06",
    pages = "207",
    year = "2023"
}

@article{Chen:2023pqf,
    author = "Chen, Bin and Liu, Reiko and Sun, Haowei and Zheng, Yu-fan",
    title = "{Constructing Carrollian field theories from null reduction}",
    eprint = "2301.06011",
    archivePrefix = "arXiv",
    primaryClass = "hep-th",
    doi = "10.1007/JHEP11(2023)170",
    journal = "JHEP",
    volume = "11",
    pages = "170",
    year = "2023"
}

@article{Bagchi:2017cpu,
    author = "Bagchi, Arjun and Gary, Mirah and Zodinmawia",
    title = "{The nuts and bolts of the BMS Bootstrap}",
    eprint = "1705.05890",
    archivePrefix = "arXiv",
    primaryClass = "hep-th",
    doi = "10.1088/1361-6382/aa8003",
    journal = "Class. Quant. Grav.",
    volume = "34",
    number = "17",
    pages = "174002",
    year = "2017"
}

@article{Merbis:2019wgk,
    author = "Merbis, Wout and Riegler, Max",
    title = "{Geometric actions and flat space holography}",
    eprint = "1912.08207",
    archivePrefix = "arXiv",
    primaryClass = "hep-th",
    doi = "10.1007/JHEP02(2020)125",
    journal = "JHEP",
    volume = "02",
    pages = "125",
    year = "2020"
}

@article{Adamo:2021lrv,
    author = "Adamo, Tim and Mason, Lionel and Sharma, Atul",
    title = "{Celestial $w_{1+\infty}$ Symmetries from Twistor Space}",
    eprint = "2110.06066",
    archivePrefix = "arXiv",
    primaryClass = "hep-th",
    doi = "10.3842/SIGMA.2022.016",
    journal = "SIGMA",
    volume = "18",
    pages = "016",
    year = "2022"
}

@article{Freidel:2021ytz,
    author = "Freidel, Laurent and Pranzetti, Daniele and Raclariu, Ana-Maria",
    title = "{Higher spin dynamics in gravity and w1+{\ensuremath{\infty}} celestial symmetries}",
    eprint = "2112.15573",
    archivePrefix = "arXiv",
    primaryClass = "hep-th",
    doi = "10.1103/PhysRevD.106.086013",
    journal = "Phys. Rev. D",
    volume = "106",
    number = "8",
    pages = "086013",
    year = "2022"
}

@article{Adamo:2014yya,
    author = "Adamo, Tim and Casali, Eduardo and Skinner, David",
    title = "{Perturbative gravity at null infinity}",
    eprint = "1405.5122",
    archivePrefix = "arXiv",
    primaryClass = "hep-th",
    reportNumber = "DAMTP-2014-29",
    doi = "10.1088/0264-9381/31/22/225008",
    journal = "Class. Quant. Grav.",
    volume = "31",
    number = "22",
    pages = "225008",
    year = "2014"
}

@article{Geiller:2022vto,
    author = "Geiller, Marc and Zwikel, C\'eline",
    title = "{The partial Bondi gauge: Further enlarging the asymptotic structure of gravity}",
    eprint = "2205.11401",
    archivePrefix = "arXiv",
    primaryClass = "hep-th",
    doi = "10.21468/SciPostPhys.13.5.108",
    journal = "SciPost Phys.",
    volume = "13",
    pages = "108",
    year = "2022"
}

@article{Petkou:2022bmz,
    author = "Petkou, Anastasios C. and Petropoulos, P. Marios and Betancour, David Rivera and Siampos, Konstantinos",
    title = "{Relativistic fluids, hydrodynamic frames and their Galilean versus Carrollian avatars}",
    eprint = "2205.09142",
    archivePrefix = "arXiv",
    primaryClass = "hep-th",
    reportNumber = "CPHT-RR021.042022",
    doi = "10.1007/JHEP09(2022)162",
    journal = "JHEP",
    volume = "09",
    pages = "162",
    year = "2022"
}

@article{Pasterski:2022djr,
    author = "Pasterski, Sabrina",
    title = "{A Comment on Loop Corrections to the Celestial Stress Tensor}",
    eprint = "2205.10901",
    archivePrefix = "arXiv",
    primaryClass = "hep-th",
    month = "5",
    year = "2022"
}

@article{Adami:2021kvx,
    author = "Adami, H. and Sheikh-Jabbari, M. M. and Taghiloo, V. and Yavartanoo, H.",
    title = "{Null surface thermodynamics}",
    eprint = "2110.04224",
    archivePrefix = "arXiv",
    primaryClass = "hep-th",
    doi = "10.1103/PhysRevD.105.066004",
    journal = "Phys. Rev. D",
    volume = "105",
    number = "6",
    pages = "066004",
    year = "2022"
}

@article{Campoleoni:2020ejn,
    author = "Campoleoni, Andrea and Francia, Dario and Heissenberg, Carlo",
    title = "{On asymptotic symmetries in higher dimensions for any spin}",
    eprint = "2011.04420",
    archivePrefix = "arXiv",
    primaryClass = "hep-th",
    reportNumber = "NORDITA 2020-103",
    doi = "10.1007/JHEP12(2020)129",
    journal = "JHEP",
    volume = "12",
    pages = "129",
    year = "2020"
}

@article{Mangano:1990by,
    author = "Mangano, Michelangelo L. and Parke, Stephen J.",
    title = "{Multiparton amplitudes in gauge theories}",
    eprint = "hep-th/0509223",
    archivePrefix = "arXiv",
    reportNumber = "FERMILAB-PUB-90-113-T",
    doi = "10.1016/0370-1573(91)90091-Y",
    journal = "Phys. Rept.",
    volume = "200",
    pages = "301--367",
    year = "1991"
}

@article{Barnich:2012rz,
    author = "Barnich, Glenn and Gomberoff, Andr\'es and Gonz\'alez, Hern\'an A.",
    title = "{Three-dimensional Bondi-Metzner-Sachs invariant two-dimensional field theories as the flat limit of Liouville theory}",
    eprint = "1210.0731",
    archivePrefix = "arXiv",
    primaryClass = "hep-th",
    doi = "10.1103/PhysRevD.87.124032",
    journal = "Phys. Rev. D",
    volume = "87",
    number = "12",
    pages = "124032",
    year = "2013"
}

@article{Barnich:2017jgw,
    author = "Barnich, Glenn and Gonzalez, Hernan A. and Salgado-Rebolledo, Patricio",
    title = "{Geometric actions for three-dimensional gravity}",
    eprint = "1707.08887",
    archivePrefix = "arXiv",
    primaryClass = "hep-th",
    doi = "10.1088/1361-6382/aa9806",
    journal = "Class. Quant. Grav.",
    volume = "35",
    number = "1",
    pages = "014003",
    year = "2018"
}

@article{Jiang:2017ecm,
    author = "Jiang, Hongliang and Song, Wei and Wen, Qiang",
    title = "{Entanglement Entropy in Flat Holography}",
    eprint = "1706.07552",
    archivePrefix = "arXiv",
    primaryClass = "hep-th",
    doi = "10.1007/JHEP07(2017)142",
    journal = "JHEP",
    volume = "07",
    pages = "142",
    year = "2017"
}

@article{Bekaert:2015xua,
    author = "Bekaert, Xavier and Morand, Kevin",
    title = "{Connections and dynamical trajectories in generalised Newton-Cartan gravity II. An ambient perspective}",
    eprint = "1505.03739",
    archivePrefix = "arXiv",
    primaryClass = "hep-th",
    doi = "10.1063/1.5030328",
    journal = "J. Math. Phys.",
    volume = "59",
    number = "7",
    pages = "072503",
    year = "2018"
}

@article{Henneaux:2021yzg,
    author = "Henneaux, Marc and Salgado-Rebolledo, Patricio",
    title = "{Carroll contractions of Lorentz-invariant theories}",
    eprint = "2109.06708",
    archivePrefix = "arXiv",
    primaryClass = "hep-th",
    doi = "10.1007/JHEP11(2021)180",
    journal = "JHEP",
    volume = "11",
    pages = "180",
    year = "2021"
}

@article{Henneaux:1979vn,
    author = "Henneaux, Marc",
    title = "{Geometry of Zero Signature Space-times}",
    reportNumber = "PRINT-79-0606 (PRINCETON)",
    journal = "Bull. Soc. Math. Belg.",
    volume = "31",
    pages = "47--63",
    year = "1979"
}

@inproceedings{Pasterski:2021raf,
    author = "Pasterski, Sabrina and Pate, Monica and Raclariu, Ana-Maria",
    title = "{Celestial Holography}",
    booktitle = "{2022 Snowmass Summer Study}",
    eprint = "2111.11392",
    archivePrefix = "arXiv",
    primaryClass = "hep-th",
    month = "11",
    year = "2021"
}

@article{Bagchi:2014iea,
    author = "Bagchi, Arjun and Basu, Rudranil and Grumiller, Daniel and Riegler, Max",
    title = "{Entanglement entropy in Galilean conformal field theories and flat holography}",
    eprint = "1410.4089",
    archivePrefix = "arXiv",
    primaryClass = "hep-th",
    reportNumber = "TUW-14-14",
    doi = "10.1103/PhysRevLett.114.111602",
    journal = "Phys. Rev. Lett.",
    volume = "114",
    number = "11",
    pages = "111602",
    year = "2015"
}

@article{Bagchi:2015wna,
    author = "Bagchi, Arjun and Grumiller, Daniel and Merbis, Wout",
    title = "{Stress tensor correlators in three-dimensional gravity}",
    eprint = "1507.05620",
    archivePrefix = "arXiv",
    primaryClass = "hep-th",
    reportNumber = "MIT-CTP-4694, TUW-15-13",
    doi = "10.1103/PhysRevD.93.061502",
    journal = "Phys. Rev. D",
    volume = "93",
    number = "6",
    pages = "061502",
    year = "2016"
}

@article{Compere:2016hzt,
    author = "Comp\`ere, Geoffrey and Long, Jiang",
    title = "{Classical static final state of collapse with supertranslation memory}",
    eprint = "1602.05197",
    archivePrefix = "arXiv",
    primaryClass = "gr-qc",
    doi = "10.1088/0264-9381/33/19/195001",
    journal = "Class. Quant. Grav.",
    volume = "33",
    number = "19",
    pages = "195001",
    year = "2016"
}

@article{Mann:2005yr,
    author = "Mann, Robert B. and Marolf, Donald",
    title = "{Holographic renormalization of asymptotically flat spacetimes}",
    eprint = "hep-th/0511096",
    archivePrefix = "arXiv",
    doi = "10.1088/0264-9381/23/9/010",
    journal = "Class. Quant. Grav.",
    volume = "23",
    pages = "2927--2950",
    year = "2006"
}

@article{Bagchi:2016geg,
    author = "Bagchi, Arjun and Gary, Mirah and Zodinmawia",
    title = "{Bondi-Metzner-Sachs bootstrap}",
    eprint = "1612.01730",
    archivePrefix = "arXiv",
    primaryClass = "hep-th",
    doi = "10.1103/PhysRevD.96.025007",
    journal = "Phys. Rev. D",
    volume = "96",
    number = "2",
    pages = "025007",
    year = "2017"
}

@article{Arcioni:2003xx,
    author = "Arcioni, Giovanni and Dappiaggi, Claudio",
    title = "{Exploring the holographic principle in asymptotically flat space-times via the BMS group}",
    eprint = "hep-th/0306142",
    archivePrefix = "arXiv",
    reportNumber = "SPIN-03-18, ITP-UU-03-27",
    doi = "10.1016/j.nuclphysb.2003.09.051",
    journal = "Nucl. Phys. B",
    volume = "674",
    pages = "553--592",
    year = "2003"
}

@article{Giddings:1999jq,
    author = "Giddings, Steven B.",
    title = "{Flat space scattering and bulk locality in the AdS / CFT correspondence}",
    eprint = "hep-th/9907129",
    archivePrefix = "arXiv",
    doi = "10.1103/PhysRevD.61.106008",
    journal = "Phys. Rev. D",
    volume = "61",
    pages = "106008",
    year = "2000"
}

@article{Polchinski:1999ry,
    author = "Polchinski, Joseph",
    title = "{S-matrices from AdS space-time}",
    eprint = "hep-th/9901076",
    archivePrefix = "arXiv",
    reportNumber = "NST-ITP-99-02",
    month = "1",
    year = "1999"
}

@article{Susskind:1998vk,
    author = "Susskind, Leonard",
    editor = "Burgess, C. P. and Myers, Robert C.",
    title = "{Holography in the flat space limit}",
    eprint = "hep-th/9901079",
    archivePrefix = "arXiv",
    doi = "10.1063/1.1301570",
    journal = "AIP Conf. Proc.",
    volume = "493",
    number = "1",
    pages = "98--112",
    year = "1999"
}

@article{Atanasov:2021oyu,
    author = "Atanasov, Alexander and Ball, Adam and Melton, Walker and Raclariu, Ana-Maria and Strominger, Andrew",
    title = "{(2, 2) Scattering and the celestial torus}",
    eprint = "2101.09591",
    archivePrefix = "arXiv",
    primaryClass = "hep-th",
    doi = "10.1007/JHEP07(2021)083",
    journal = "JHEP",
    volume = "07",
    pages = "083",
    year = "2021"
}

@article{Brown:1986nw,
    author = "Brown, J. David and Henneaux, M.",
    title = "{Central Charges in the Canonical Realization of Asymptotic Symmetries: An Example from Three-Dimensional Gravity}",
    doi = "10.1007/BF01211590",
    journal = "Commun. Math. Phys.",
    volume = "104",
    pages = "207--226",
    year = "1986"
}

@article{Duval:2014lpa,
    author = "Duval, C. and Gibbons, G. W. and Horvathy, P. A.",
    title = "{Conformal Carroll groups}",
    eprint = "1403.4213",
    archivePrefix = "arXiv",
    primaryClass = "hep-th",
    doi = "10.1088/1751-8113/47/33/335204",
    journal = "J. Phys. A",
    volume = "47",
    number = "33",
    pages = "335204",
    year = "2014"
}

@article{Detournay:2014fva,
    author = {Detournay, Stephane and Grumiller, Daniel and Sch\"oller, Friedrich and Sim\'on, Joan},
    title = "{Variational principle and one-point functions in three-dimensional flat space Einstein gravity}",
    eprint = "1402.3687",
    archivePrefix = "arXiv",
    primaryClass = "hep-th",
    reportNumber = "TUW-14-02",
    doi = "10.1103/PhysRevD.89.084061",
    journal = "Phys. Rev. D",
    volume = "89",
    number = "8",
    pages = "084061",
    year = "2014"
}

@article{Bagchi:2012xr,
    author = "Bagchi, Arjun and Detournay, St\'ephane and Fareghbal, Reza and Sim\'on, Joan",
    title = "{Holography of 3D Flat Cosmological Horizons}",
    eprint = "1208.4372",
    archivePrefix = "arXiv",
    primaryClass = "hep-th",
    reportNumber = "EMPG-12-18",
    doi = "10.1103/PhysRevLett.110.141302",
    journal = "Phys. Rev. Lett.",
    volume = "110",
    number = "14",
    pages = "141302",
    year = "2013"
}

@article{Barnich:2012xq,
    author = "Barnich, Glenn",
    title = "{Entropy of three-dimensional asymptotically flat cosmological solutions}",
    eprint = "1208.4371",
    archivePrefix = "arXiv",
    primaryClass = "hep-th",
    reportNumber = "ULB-TH-12-14",
    doi = "10.1007/JHEP10(2012)095",
    journal = "JHEP",
    volume = "10",
    pages = "095",
    year = "2012"
}

@article{Barnich:2012aw,
    author = "Barnich, Glenn and Gomberoff, Andres and Gonzalez, Hernan A.",
    title = "{The Flat limit of three dimensional asymptotically anti-de Sitter spacetimes}",
    eprint = "1204.3288",
    archivePrefix = "arXiv",
    primaryClass = "gr-qc",
    doi = "10.1103/PhysRevD.86.024020",
    journal = "Phys. Rev. D",
    volume = "86",
    pages = "024020",
    year = "2012"
}

@article{Duval:2014uva,
    author = "Duval, C. and Gibbons, G. W. and Horvathy, P. A.",
    title = "{Conformal Carroll groups and BMS symmetry}",
    eprint = "1402.5894",
    archivePrefix = "arXiv",
    primaryClass = "gr-qc",
    doi = "10.1088/0264-9381/31/9/092001",
    journal = "Class. Quant. Grav.",
    volume = "31",
    pages = "092001",
    year = "2014"
}

@article{Bagchi:2022owq,
    author = "Bagchi, Arjun and Grumiller, Daniel and Nandi, Poulami",
    title = "{Carrollian superconformal theories and super BMS}",
    eprint = "2202.01172",
    archivePrefix = "arXiv",
    primaryClass = "hep-th",
    doi = "10.1007/JHEP05(2022)044",
    journal = "JHEP",
    volume = "05",
    pages = "044",
    year = "2022"
}

@article{Dolan:2004mu,
    author = "Dolan, Francis A. and Gallot, Laurent and Sokatchev, Emery",
    title = "{On four-point functions of 1/2-BPS operators in general dimensions}",
    eprint = "hep-th/0405180",
    archivePrefix = "arXiv",
    reportNumber = "DAMTP-04-45, LAPTH-1047-04",
    doi = "10.1088/1126-6708/2004/09/056",
    journal = "JHEP",
    volume = "09",
    pages = "056",
    year = "2004"
}

@article{He:2024vlp,
    author = "He, Temple and Raclariu, Ana-Maria and Zurek, Kathryn M.",
    title = "{An infrared on-shell action and its implications for soft charge fluctuations in asymptotically flat spacetimes}",
    eprint = "2408.01485",
    archivePrefix = "arXiv",
    primaryClass = "hep-th",
    reportNumber = "CALT-TH 2024-028",
    doi = "10.1088/1751-8121/adc4a2",
    journal = "J. Phys. A",
    volume = "58",
    number = "16",
    pages = "165402",
    year = "2025"
}

@article{Niederer:1972zz,
    author = "Niederer, U.",
    title = "{The maximal kinematical invariance group of the free Schrodinger equation.}",
    doi = "10.5169/seals-114417",
    journal = "Helv. Phys. Acta",
    volume = "45",
    pages = "802--810",
    year = "1972"
}

@article{Hagen:1972pd,
    author = "Hagen, C. R.",
    title = "{Scale and conformal transformations in galilean-covariant field theory}",
    doi = "10.1103/PhysRevD.5.377",
    journal = "Phys. Rev. D",
    volume = "5",
    pages = "377--388",
    year = "1972"
}

@article{Henkel:1993sg,
    author = "Henkel, Malte",
    title = "{Schrodinger invariance in strongly anisotropic critical systems}",
    eprint = "hep-th/9310081",
    archivePrefix = "arXiv",
    reportNumber = "OUTP-93-33-S, UGVA-DPT-1993-09-833",
    doi = "10.1007/BF02186756",
    journal = "J. Statist. Phys.",
    volume = "75",
    pages = "1023--1061",
    year = "1994"
}

@article{Mehen:1999nd,
    author = "Mehen, Thomas and Stewart, Iain W. and Wise, Mark B.",
    title = "{Conformal invariance for nonrelativistic field theory}",
    eprint = "hep-th/9910025",
    archivePrefix = "arXiv",
    reportNumber = "CALT-68-2242, UCSD-PTH-99-14",
    doi = "10.1016/S0370-2693(00)00006-X",
    journal = "Phys. Lett. B",
    volume = "474",
    pages = "145--152",
    year = "2000"
}

@article{Nishida:2007pj,
    author = "Nishida, Yusuke and Son, Dam T.",
    title = "{Nonrelativistic conformal field theories}",
    eprint = "0706.3746",
    archivePrefix = "arXiv",
    primaryClass = "hep-th",
    reportNumber = "INT-PUB-07-16",
    doi = "10.1103/PhysRevD.76.086004",
    journal = "Phys. Rev. D",
    volume = "76",
    pages = "086004",
    year = "2007"
}

@article{Son:2008ye,
    author = "Son, D. T.",
    title = "{Toward an AdS/cold atoms correspondence: A Geometric realization of the Schrodinger symmetry}",
    eprint = "0804.3972",
    archivePrefix = "arXiv",
    primaryClass = "hep-th",
    reportNumber = "INT-PUB-08-08",
    doi = "10.1103/PhysRevD.78.046003",
    journal = "Phys. Rev. D",
    volume = "78",
    pages = "046003",
    year = "2008"
}

@article{Nishida:2010tm,
    author = "Nishida, Yusuke and Son, Dam Thanh",
    title = "{Unitary Fermi gas, epsilon expansion, and nonrelativistic conformal field theories}",
    eprint = "1004.3597",
    archivePrefix = "arXiv",
    primaryClass = "cond-mat.quant-gas",
    reportNumber = "MIT-CTP-4141, INT-PUB-10-017",
    doi = "10.1007/978-3-642-21978-8_7",
    journal = "Lect. Notes Phys.",
    volume = "836",
    pages = "233--275",
    year = "2012"
}

@article{Golkar:2014mwa,
    author = "Golkar, Siavash and Son, Dam T.",
    title = "{Operator Product Expansion and Conservation Laws in Non-Relativistic Conformal Field Theories}",
    eprint = "1408.3629",
    archivePrefix = "arXiv",
    primaryClass = "hep-th",
    reportNumber = "EFI-14-27",
    doi = "10.1007/JHEP12(2014)063",
    journal = "JHEP",
    volume = "12",
    pages = "063",
    year = "2014"
}

@article{Balasubramanian:2008dm,
    author = "Balasubramanian, Koushik and McGreevy, John",
    title = "{Gravity duals for non-relativistic CFTs}",
    eprint = "0804.4053",
    archivePrefix = "arXiv",
    primaryClass = "hep-th",
    doi = "10.1103/PhysRevLett.101.061601",
    journal = "Phys. Rev. Lett.",
    volume = "101",
    pages = "061601",
    year = "2008"
}

@article{Goldberger:2014hca,
    author = "Goldberger, Walter D. and Khandker, Zuhair U. and Prabhu, Siddharth",
    title = "{OPE convergence in non-relativistic conformal field theories}",
    eprint = "1412.8507",
    archivePrefix = "arXiv",
    primaryClass = "hep-th",
    doi = "10.1007/JHEP12(2015)048",
    journal = "JHEP",
    volume = "12",
    pages = "048",
    year = "2015"
}

@article{Pal:2018idc,
    author = "Pal, Sridip",
    title = "{Unitarity and universality in nonrelativistic conformal field theory}",
    eprint = "1802.02262",
    archivePrefix = "arXiv",
    primaryClass = "hep-th",
    doi = "10.1103/PhysRevD.97.105031",
    journal = "Phys. Rev. D",
    volume = "97",
    number = "10",
    pages = "105031",
    year = "2018"
}

@article{Maldacena:2008wh,
    author = "Maldacena, Juan and Martelli, Dario and Tachikawa, Yuji",
    title = "{Comments on string theory backgrounds with non-relativistic conformal symmetry}",
    eprint = "0807.1100",
    archivePrefix = "arXiv",
    primaryClass = "hep-th",
    doi = "10.1088/1126-6708/2008/10/072",
    journal = "JHEP",
    volume = "10",
    pages = "072",
    year = "2008"
}

@article{Taylor:2008tg,
    author = "Taylor, Marika",
    title = "{Non-relativistic holography}",
    eprint = "0812.0530",
    archivePrefix = "arXiv",
    primaryClass = "hep-th",
    reportNumber = "ITFA-2008-48",
    month = "12",
    year = "2008"
}

@article{Goldberger:2008vg,
    author = "Goldberger, Walter D.",
    title = "{AdS/CFT duality for non-relativistic field theory}",
    eprint = "0806.2867",
    archivePrefix = "arXiv",
    primaryClass = "hep-th",
    doi = "10.1088/1126-6708/2009/03/069",
    journal = "JHEP",
    volume = "03",
    pages = "069",
    year = "2009"
}

@article{Erramilli:2025pfh,
    author = "Erramilli, Rajeev S. and Kulp, Justin and Popov, Fedor K.",
    title = "{Do null defects dream of conformal symmetry?}",
    eprint = "2509.04578",
    archivePrefix = "arXiv",
    primaryClass = "hep-th",
    month = "9",
    year = "2025"
}

@article{Bagchi:2023rwd,
    author = "Bagchi, Arjun and Kolekar, Kedar S. and Mandal, Taniya and Shukla, Ashish",
    title = "{Heavy-ion collisions, Gubser flow, and Carroll hydrodynamics}",
    eprint = "2310.03167",
    archivePrefix = "arXiv",
    primaryClass = "hep-th",
    reportNumber = "CPHT-RR046.072023",
    doi = "10.1103/PhysRevD.109.056004",
    journal = "Phys. Rev. D",
    volume = "109",
    number = "5",
    pages = "056004",
    year = "2024"
}

@article{Despontin:2025dog,
    author = "Despontin, Emilie and Detournay, Stephane and Dutta, Sudipta and Fontaine, Dima",
    title = "{Anisotropic conformal Carroll field theories and their gravity duals}",
    eprint = "2505.23755",
    archivePrefix = "arXiv",
    primaryClass = "hep-th",
    doi = "10.1007/JHEP09(2025)056",
    journal = "JHEP",
    volume = "09",
    pages = "056",
    year = "2025"
}

@article{Afshar:2024llh,
    author = "Afshar, Hamid and Bekaert, Xavier and Najafizadeh, Mojtaba",
    title = "{Classification of conformal carroll algebras}",
    eprint = "2409.19953",
    archivePrefix = "arXiv",
    primaryClass = "hep-th",
    reportNumber = "IPM/P-2024/31",
    doi = "10.1007/JHEP12(2024)148",
    journal = "JHEP",
    volume = "12",
    pages = "148",
    year = "2024"
}

@article{Aharony:1999ti,
    author = "Aharony, Ofer and Gubser, Steven S. and Maldacena, Juan Martin and Ooguri, Hirosi and Oz, Yaron",
    title = "{Large N field theories, string theory and gravity}",
    eprint = "hep-th/9905111",
    archivePrefix = "arXiv",
    reportNumber = "CERN-TH-99-122, HUTP-99-A027, LBNL-43113, RU-99-18, UCB-PTH-99-16, LBL-43113",
    doi = "10.1016/S0370-1573(99)00083-6",
    journal = "Phys. Rept.",
    volume = "323",
    pages = "183--386",
    year = "2000"
}

@article{Barnich:2021dta,
    author = "Barnich, Glenn and Ruzziconi, Romain",
    title = "{Coadjoint representation of the BMS group on celestial Riemann surfaces}",
    eprint = "2103.11253",
    archivePrefix = "arXiv",
    primaryClass = "gr-qc",
    doi = "10.1007/JHEP06(2021)079",
    journal = "JHEP",
    volume = "06",
    pages = "079",
    year = "2021"
}

@article{Strominger:2021mtt,
    author = "Strominger, Andrew",
    title = "{$w_{1+\infty}$ Algebra and the Celestial Sphere: Infinite Towers of Soft Graviton, Photon, and Gluon Symmetries}",
    doi = "10.1103/PhysRevLett.127.221601",
    journal = "Phys. Rev. Lett.",
    volume = "127",
    number = "22",
    pages = "221601",
    year = "2021"
}

@article{Newman:1961qr,
    author = "Newman, Ezra and Penrose, Roger",
    title = "{An Approach to gravitational radiation by a method of spin coefficients}",
    doi = "10.1063/1.1724257",
    journal = "J. Math. Phys.",
    volume = "3",
    pages = "566--578",
    year = "1962"
}

@article{Brown:1992br,
    author = "Brown, J. David and York, Jr., James W.",
    title = "{Quasilocal energy and conserved charges derived from the gravitational action}",
    eprint = "gr-qc/9209012",
    archivePrefix = "arXiv",
    reportNumber = "IFP-423-UNC, TAR-009-UNC",
    doi = "10.1103/PhysRevD.47.1407",
    journal = "Phys. Rev. D",
    volume = "47",
    pages = "1407--1419",
    year = "1993"
}

@book{DiFrancesco:1997nk,
    author = "Di Francesco, P. and Mathieu, P. and Senechal, D.",
    title = "{Conformal Field Theory}",
    doi = "10.1007/978-1-4612-2256-9",
    isbn = "978-0-387-94785-3, 978-1-4612-7475-9",
    publisher = "Springer-Verlag",
    address = "New York",
    series = "Graduate Texts in Contemporary Physics",
    year = "1997"
}

@article{Ashtekar:2014zsa,
    author = "Ashtekar, Abhay",
    title = "{Geometry and Physics of Null Infinity}",
    eprint = "1409.1800",
    archivePrefix = "arXiv",
    primaryClass = "gr-qc",
    reportNumber = "IGC-14-9-1",
    month = "9",
    year = "2014"
}

@article{Freidel:2021fxf,
    author = "Freidel, Laurent and Oliveri, Roberto and Pranzetti, Daniele and Speziale, Simone",
    title = "{The Weyl BMS group and Einstein\textquoteright{}s equations}",
    eprint = "2104.05793",
    archivePrefix = "arXiv",
    primaryClass = "hep-th",
    doi = "10.1007/JHEP07(2021)170",
    journal = "JHEP",
    volume = "07",
    pages = "170",
    year = "2021"
}

@article{Ciambelli:2020ftk,
    author = "Ciambelli, Luca and Marteau, Charles and Petropoulos, P. Marios and Ruzziconi, Romain",
    title = "{Fefferman-Graham and Bondi Gauges in the Fluid/Gravity Correspondence}",
    eprint = "2006.10083",
    archivePrefix = "arXiv",
    primaryClass = "hep-th",
    reportNumber = "CPHT-PC031.052020",
    doi = "10.22323/1.376.0154",
    journal = "PoS",
    volume = "CORFU2019",
    pages = "154",
    year = "2020"
}

@article{Hu:2021lrx,
    author = "Hu, Yangrui and Ren, Lecheng and Srikant, Akshay Yelleshpur and Volovich, Anastasia",
    title = "{Celestial dual superconformal symmetry, MHV amplitudes and differential equations}",
    eprint = "2106.16111",
    archivePrefix = "arXiv",
    primaryClass = "hep-th",
    doi = "10.1007/JHEP12(2021)171",
    journal = "JHEP",
    volume = "12",
    pages = "171",
    year = "2021"
}

@article{Ciambelli:2020eba,
    author = "Ciambelli, Luca and Marteau, Charles and Petropoulos, P. Marios and Ruzziconi, Romain",
    title = "{Gauges in Three-Dimensional Gravity and Holographic Fluids}",
    eprint = "2006.10082",
    archivePrefix = "arXiv",
    primaryClass = "hep-th",
    reportNumber = "CPHT-RR021.032020",
    doi = "10.1007/JHEP11(2020)092",
    journal = "JHEP",
    volume = "11",
    pages = "092",
    year = "2020"
}

@article{Adami:2021nnf,
    author = "Adami, H. and Grumiller, D. and Sheikh-Jabbari, M. M. and Taghiloo, V. and Yavartanoo, H. and Zwikel, C.",
    title = "{Null boundary phase space: slicings, news \& memory}",
    eprint = "2110.04218",
    archivePrefix = "arXiv",
    primaryClass = "hep-th",
    doi = "10.1007/JHEP11(2021)155",
    journal = "JHEP",
    volume = "11",
    pages = "155",
    year = "2021"
}

@article{Donnay:2021wrk,
    author = "Donnay, Laura and Ruzziconi, Romain",
    title = "{BMS flux algebra in celestial holography}",
    eprint = "2108.11969",
    archivePrefix = "arXiv",
    primaryClass = "hep-th",
    doi = "10.1007/JHEP11(2021)040",
    journal = "JHEP",
    volume = "11",
    pages = "040",
    year = "2021"
}

@article{Newman:1962cia,
    author = "Newman, Ezra T. and Unti, Theodore W. J.",
    title = "{Behavior of Asymptotically Flat Empty Spaces}",
    doi = "10.1063/1.1724303",
    journal = "J. Math. Phys.",
    volume = "3",
    number = "5",
    pages = "891",
    year = "1962"
}

@article{Himwich:2020rro,
    author = "Himwich, Elizabeth and Narayanan, Sruthi A. and Pate, Monica and Paul, Nisarga and Strominger, Andrew",
    title = "{The Soft $\mathcal{S}$-Matrix in Gravity}",
    eprint = "2005.13433",
    archivePrefix = "arXiv",
    primaryClass = "hep-th",
    doi = "10.1007/JHEP09(2020)129",
    journal = "JHEP",
    volume = "09",
    pages = "129",
    year = "2020"
}

@article{Ciambelli:2018ojf,
    author = "Ciambelli, Luca and Marteau, Charles",
    title = "{Carrollian conservation laws and Ricci-flat gravity}",
    eprint = "1810.11037",
    archivePrefix = "arXiv",
    primaryClass = "hep-th",
    reportNumber = "CPHT-RR101.102018",
    doi = "10.1088/1361-6382/ab0d37",
    journal = "Class. Quant. Grav.",
    volume = "36",
    number = "8",
    pages = "085004",
    year = "2019"
}

@article{Pasterski:2021dqe,
    author = "Pasterski, Sabrina and Puhm, Andrea and Trevisani, Emilio",
    title = "{Revisiting the conformally soft sector with celestial diamonds}",
    eprint = "2105.09792",
    archivePrefix = "arXiv",
    primaryClass = "hep-th",
    reportNumber = "CPHT-RR037.052021",
    doi = "10.1007/JHEP11(2021)143",
    journal = "JHEP",
    volume = "11",
    pages = "143",
    year = "2021"
}

@article{Fitzpatrick:2011hu,
    author = "Fitzpatrick, A. Liam and Kaplan, Jared",
    title = "{Analyticity and the Holographic S-Matrix}",
    eprint = "1111.6972",
    archivePrefix = "arXiv",
    primaryClass = "hep-th",
    reportNumber = "SLAC-PUB-14841",
    doi = "10.1007/JHEP10(2012)127",
    journal = "JHEP",
    volume = "10",
    pages = "127",
    year = "2012"
}

@article{Penedones:2010ue,
    author = "Penedones, Joao",
    title = "{Writing CFT correlation functions as AdS scattering amplitudes}",
    eprint = "1011.1485",
    archivePrefix = "arXiv",
    primaryClass = "hep-th",
    doi = "10.1007/JHEP03(2011)025",
    journal = "JHEP",
    volume = "03",
    pages = "025",
    year = "2011"
}

@article{Gary:2009ae,
    author = "Gary, Mirah and Giddings, Steven B. and Penedones, Joao",
    title = "{Local bulk S-matrix elements and CFT singularities}",
    eprint = "0903.4437",
    archivePrefix = "arXiv",
    primaryClass = "hep-th",
    reportNumber = "CERN-PH-TH-2009-035, NSF-KITP-09-35",
    doi = "10.1103/PhysRevD.80.085005",
    journal = "Phys. Rev. D",
    volume = "80",
    pages = "085005",
    year = "2009"
}

@article{tHooft:1993dmi,
    author = "'t Hooft, Gerard",
    title = "{Dimensional reduction in quantum gravity}",
    eprint = "gr-qc/9310026",
    archivePrefix = "arXiv",
    reportNumber = "THU-93-26",
    journal = "Conf. Proc. C",
    volume = "930308",
    pages = "284--296",
    year = "1993"
}

@article{Susskind:1994vu,
    author = "Susskind, Leonard",
    title = "{The World as a hologram}",
    eprint = "hep-th/9409089",
    archivePrefix = "arXiv",
    reportNumber = "SU-ITP-94-33",
    doi = "10.1063/1.531249",
    journal = "J. Math. Phys.",
    volume = "36",
    pages = "6377--6396",
    year = "1995"
}

@article{Chandrasekaran:2021hxc,
    author = "Chandrasekaran, Venkatesa and Flanagan, Eanna E. and Shehzad, Ibrahim and Speranza, Antony J.",
    title = "{Brown-York charges at null boundaries}",
    eprint = "2109.11567",
    archivePrefix = "arXiv",
    primaryClass = "hep-th",
    doi = "10.1007/JHEP01(2022)029",
    journal = "JHEP",
    volume = "01",
    pages = "029",
    year = "2022"
}

@article{Donnelly:2020xgu,
    author = "Donnelly, William and Freidel, Laurent and Moosavian, Seyed Faroogh and Speranza, Antony J.",
    title = "{Gravitational edge modes, coadjoint orbits, and hydrodynamics}",
    eprint = "2012.10367",
    archivePrefix = "arXiv",
    primaryClass = "hep-th",
    doi = "10.1007/JHEP09(2021)008",
    journal = "JHEP",
    volume = "09",
    pages = "008",
    year = "2021"
}

@article{Bergshoeff:2017btm,
    author = "Bergshoeff, Eric and Gomis, Joaquim and Rollier, Blaise and Rosseel, Jan and ter Veldhuis, Tonnis",
    title = "{Carroll versus Galilei Gravity}",
    eprint = "1701.06156",
    archivePrefix = "arXiv",
    primaryClass = "hep-th",
    doi = "10.1007/JHEP03(2017)165",
    journal = "JHEP",
    volume = "03",
    pages = "165",
    year = "2017"
}

@article{Maldacena:2015iua,
    author = "Maldacena, Juan and Simmons-Duffin, David and Zhiboedov, Alexander",
    title = "{Looking for a bulk point}",
    eprint = "1509.03612",
    archivePrefix = "arXiv",
    primaryClass = "hep-th",
    doi = "10.1007/JHEP01(2017)013",
    journal = "JHEP",
    volume = "01",
    pages = "013",
    year = "2017"
}

@article{Costello:2022upu,
    author = "Costello, Kevin and Paquette, Natalie M.",
    title = "{Associativity of One-Loop Corrections to the Celestial Operator Product Expansion}",
    eprint = "2204.05301",
    archivePrefix = "arXiv",
    primaryClass = "hep-th",
    doi = "10.1103/PhysRevLett.129.231604",
    journal = "Phys. Rev. Lett.",
    volume = "129",
    number = "23",
    pages = "231604",
    year = "2022"
}

@article{Guica:2008mu,
    author = "Guica, Monica and Hartman, Thomas and Song, Wei and Strominger, Andrew",
    title = "{The Kerr/CFT Correspondence}",
    eprint = "0809.4266",
    archivePrefix = "arXiv",
    primaryClass = "hep-th",
    doi = "10.1103/PhysRevD.80.124008",
    journal = "Phys. Rev. D",
    volume = "80",
    pages = "124008",
    year = "2009"
}

@article{Bagchi:2019xfx,
    author = "Bagchi, Arjun and Mehra, Aditya and Nandi, Poulami",
    title = "{Field Theories with Conformal Carrollian Symmetry}",
    eprint = "1901.10147",
    archivePrefix = "arXiv",
    primaryClass = "hep-th",
    doi = "10.1007/JHEP05(2019)108",
    journal = "JHEP",
    volume = "05",
    pages = "108",
    year = "2019"
}

@article{Bagchi:2016bcd,
    author = "Bagchi, Arjun and Basu, Rudranil and Kakkar, Ashish and Mehra, Aditya",
    title = "{Flat Holography: Aspects of the dual field theory}",
    eprint = "1609.06203",
    archivePrefix = "arXiv",
    primaryClass = "hep-th",
    doi = "10.1007/JHEP12(2016)147",
    journal = "JHEP",
    volume = "12",
    pages = "147",
    year = "2016"
}

@article{Ciambelli:2018wre,
    author = "Ciambelli, Luca and Marteau, Charles and Petkou, Anastasios C. and Petropoulos, P. Marios and Siampos, Konstantinos",
    title = "{Flat holography and Carrollian fluids}",
    eprint = "1802.06809",
    archivePrefix = "arXiv",
    primaryClass = "hep-th",
    reportNumber = "CPHT-RR049.082017, CERN-TH-2017-229",
    doi = "10.1007/JHEP07(2018)165",
    journal = "JHEP",
    volume = "07",
    pages = "165",
    year = "2018"
}

@article{Hartong:2015xda,
    author = "Hartong, Jelle",
    title = "{Gauging the Carroll Algebra and Ultra-Relativistic Gravity}",
    eprint = "1505.05011",
    archivePrefix = "arXiv",
    primaryClass = "hep-th",
    doi = "10.1007/JHEP08(2015)069",
    journal = "JHEP",
    volume = "08",
    pages = "069",
    year = "2015"
}

@article{Chandrasekaran:2018aop,
    author = "Chandrasekaran, Venkatesa and Flanagan, Eanna E. and Prabhu, Kartik",
    title = "{Symmetries and charges of general relativity at null boundaries}",
    eprint = "1807.11499",
    archivePrefix = "arXiv",
    primaryClass = "hep-th",
    doi = "10.1007/JHEP11(2018)125",
    journal = "JHEP",
    volume = "11",
    pages = "125",
    year = "2018"
}

@article{Strominger:1997eq,
    author = "Strominger, Andrew",
    title = "{Black hole entropy from near horizon microstates}",
    eprint = "hep-th/9712251",
    archivePrefix = "arXiv",
    reportNumber = "HUTP-97-A106",
    doi = "10.1088/1126-6708/1998/02/009",
    journal = "JHEP",
    volume = "02",
    pages = "009",
    year = "1998"
}

@article{Fotopoulos:2020bqj,
    author = "Fotopoulos, Angelos and Stieberger, Stephan and Taylor, Tomasz R. and Zhu, Bin",
    title = "{Extended Super BMS Algebra of Celestial CFT}",
    eprint = "2007.03785",
    archivePrefix = "arXiv",
    primaryClass = "hep-th",
    month = "7",
    year = "2020"
}

@article{Banerjee:2019prz,
    author = "Banerjee, Shamik and Ghosh, Sudip and Pandey, Pranjal and Saha, Arnab Priya",
    title = "{Modified celestial amplitude in Einstein gravity}",
    eprint = "1909.03075",
    archivePrefix = "arXiv",
    primaryClass = "hep-th",
    doi = "10.1007/JHEP03(2020)125",
    journal = "JHEP",
    volume = "03",
    pages = "125",
    year = "2020"
}

@article{Strominger:1996sh,
    author = "Strominger, Andrew and Vafa, Cumrun",
    title = "{Microscopic origin of the Bekenstein-Hawking entropy}",
    eprint = "hep-th/9601029",
    archivePrefix = "arXiv",
    reportNumber = "HUTP-96-A002, RU-96-01",
    doi = "10.1016/0370-2693(96)00345-0",
    journal = "Phys. Lett. B",
    volume = "379",
    pages = "99--104",
    year = "1996"
}

@article{Barnich:2014cwa,
      author         = "Barnich, Glenn and Donnay, Laura and Matulich, Javier and
                        Troncoso, Ricardo",
      title          = "{Asymptotic symmetries and dynamics of three-dimensional
                        flat supergravity}",
      journal        = "JHEP",
      volume         = "08",
      year           = "2014",
      pages          = "071",
      doi            = "10.1007/JHEP08(2014)071",
      eprint         = "1407.4275",
      archivePrefix  = "arXiv",
      primaryClass   = "hep-th",
      reportNumber   = "CECS-PHY-14-02",
      SLACcitation   = "%%CITATION = ARXIV:1407.4275;%%"
}

@article{Donnay:2019jiz,
      author         = "Donnay, Laura and Marteau, Charles",
      title          = "{Carrollian Physics at the Black Hole Horizon}",
      journal        = "Class. Quant. Grav.",
      volume         = "36",
      year           = "2019",
      number         = "16",
      pages          = "165002",
      doi            = "10.1088/1361-6382/ab2fd5",
      eprint         = "1903.09654",
      archivePrefix  = "arXiv",
      primaryClass   = "hep-th",
      SLACcitation   = "%%CITATION = ARXIV:1903.09654;%%"
}

@article{Ciambelli:2019lap,
    author = "Ciambelli, Luca and Leigh, Robert G. and Marteau, Charles and Petropoulos, P. Marios",
    title = "{Carroll Structures, Null Geometry and Conformal Isometries}",
    eprint = "1905.02221",
    archivePrefix = "arXiv",
    primaryClass = "hep-th",
    reportNumber = "CPHT-RR025.052019, CPHT-RR010.022019",
    doi = "10.1103/PhysRevD.100.046010",
    journal = "Phys. Rev. D",
    volume = "100",
    number = "4",
    pages = "046010",
    year = "2019"
}

@article{Campoleoni:2018ltl,
    author = "Campoleoni, Andrea and Ciambelli, Luca and Marteau, Charles and Petropoulos, P. Marios and Siampos, Konstantinos",
    title = "{Two-dimensional fluids and their holographic duals}",
    eprint = "1812.04019",
    archivePrefix = "arXiv",
    primaryClass = "hep-th",
    reportNumber = "CPHT-RR078.082018, CERN-TH-2018-231",
    doi = "10.1016/j.nuclphysb.2019.114692",
    journal = "Nucl. Phys. B",
    volume = "946",
    pages = "114692",
    year = "2019"
}

@article{Guevara:2021abz,
    author = "Guevara, Alfredo and Himwich, Elizabeth and Pate, Monica and Strominger, Andrew",
    title = "{Holographic symmetry algebras for gauge theory and gravity}",
    eprint = "2103.03961",
    archivePrefix = "arXiv",
    primaryClass = "hep-th",
    doi = "10.1007/JHEP11(2021)152",
    journal = "JHEP",
    volume = "11",
    pages = "152",
    year = "2021"
}

@article{Barnich:2015uva,
    author = "Barnich, Glenn and Oblak, Blagoje",
    title = "{Notes on the BMS group in three dimensions: II. Coadjoint representation}",
    eprint = "1502.00010",
    archivePrefix = "arXiv",
    primaryClass = "hep-th",
    doi = "10.1007/JHEP03(2015)033",
    journal = "JHEP",
    volume = "03",
    pages = "033",
    year = "2015"
}

@article{Ciambelli:2018xat,
    author = "Ciambelli, Luca and Marteau, Charles and Petkou, Anastasios C. and Petropoulos, P. Marios and Siampos, Konstantinos",
    title = "{Covariant Galilean versus Carrollian hydrodynamics from relativistic fluids}",
    eprint = "1802.05286",
    archivePrefix = "arXiv",
    primaryClass = "hep-th",
    reportNumber = "CPHT-RR048.082017, CERN-TH-2017-228",
    doi = "10.1088/1361-6382/aacf1a",
    journal = "Class. Quant. Grav.",
    volume = "35",
    number = "16",
    pages = "165001",
    year = "2018"
}

@article{Compere:2020lrt,
    author = "Comp\`ere, Geoffrey and Fiorucci, Adrien and Ruzziconi, Romain",
    title = "{The $\Lambda$-BMS$_4$ charge algebra}",
    eprint = "2004.10769",
    archivePrefix = "arXiv",
    primaryClass = "hep-th",
    doi = "10.1007/JHEP10(2020)205",
    journal = "JHEP",
    volume = "10",
    pages = "205",
    year = "2020"
}

@article{Compere:2018aar,
    author = "Comp\`ere, Geoffrey and Fiorucci, Adrien",
    archivePrefix = "arXiv",
    eprint = "1801.07064",
    month = "1",
    primaryClass = "hep-th",
    title = "{Advanced Lectures on General Relativity}",
    year = "2018"
}

@article{Banerjee:2020zlg,
    author = "Banerjee, Shamik and Ghosh, Sudip and Paul, Partha",
    title = "{MHV graviton scattering amplitudes and current algebra on the celestial sphere}",
    eprint = "2008.04330",
    archivePrefix = "arXiv",
    primaryClass = "hep-th",
    doi = "10.1007/JHEP02(2021)176",
    journal = "JHEP",
    volume = "02",
    pages = "176",
    year = "2021"
}

@article{Maldacena:1997re,
    author = "Maldacena, Juan Martin",
    title = "{The Large N limit of superconformal field theories and supergravity}",
    eprint = "hep-th/9711200",
    archivePrefix = "arXiv",
    reportNumber = "HUTP-97-A097, HUTP-98-A097",
    doi = "10.1023/A:1026654312961",
    journal = "Adv. Theor. Math. Phys.",
    volume = "2",
    pages = "231--252",
    year = "1998"
}

@article{Campiglia:2020qvc,
    author = "Campiglia, Miguel and Peraza, Javier",
    title = "{Generalized BMS charge algebra}",
    eprint = "2002.06691",
    archivePrefix = "arXiv",
    primaryClass = "gr-qc",
    doi = "10.1103/PhysRevD.101.104039",
    journal = "Phys. Rev. D",
    volume = "101",
    number = "10",
    pages = "104039",
    year = "2020"
}

@article{Puhm:2019zbl,
    author = "Puhm, Andrea",
    title = "{Conformally Soft Theorem in Gravity}",
    eprint = "1905.09799",
    archivePrefix = "arXiv",
    primaryClass = "hep-th",
    reportNumber = "CPHT-RR021.052019",
    doi = "10.1007/JHEP09(2020)130",
    journal = "JHEP",
    volume = "09",
    pages = "130",
    year = "2020"
}

@article{Pate:2019lpp,
      author         = "Pate, Monica and Raclariu, Ana-Maria and Strominger,
                        Andrew and Yuan, Ellis Ye",
      title          = "{Celestial Operator Products of Gluons and Gravitons}",
      year           = "2019",
      eprint         = "1910.07424",
      archivePrefix  = "arXiv",
      primaryClass   = "hep-th",
      SLACcitation   = "%%CITATION = ARXIV:1910.07424;%%"
}

@article{Fan:2019emx,
      author         = "Fan, Wei and Fotopoulos, Angelos and Taylor, Tomasz R.",
      title          = "{Soft Limits of Yang-Mills Amplitudes and Conformal
                        Correlators}",
      journal        = "JHEP",
      volume         = "05",
      year           = "2019",
      pages          = "121",
      doi            = "10.1007/JHEP05(2019)121",
      eprint         = "1903.01676",
      archivePrefix  = "arXiv",
      primaryClass   = "hep-th",
      SLACcitation   = "%%CITATION = ARXIV:1903.01676;%%"
}

@article{Donnay:2018neh,
      author         = "Donnay, Laura and Puhm, Andrea and Strominger, Andrew",
      title          = "{Conformally Soft Photons and Gravitons}",
      journal        = "JHEP",
      volume         = "01",
      year           = "2019",
      pages          = "184",
      doi            = "10.1007/JHEP01(2019)184",
      eprint         = "1810.05219",
      archivePrefix  = "arXiv",
      primaryClass   = "hep-th",
      SLACcitation   = "%%CITATION = ARXIV:1810.05219;%%"
}

@article{Himwich:2021dau,
    author = "Himwich, Elizabeth and Pate, Monica and Singh, Kyle",
    title = "{Celestial operator product expansions and w$_{1+\infty}$ symmetry for all spins}",
    eprint = "2108.07763",
    archivePrefix = "arXiv",
    primaryClass = "hep-th",
    doi = "10.1007/JHEP01(2022)080",
    journal = "JHEP",
    volume = "01",
    pages = "080",
    year = "2022"
}

@article{Banerjee:2022wht,
    author = "Banerjee, Shamik and Pasterski, Sabrina",
    title = "{Revisiting the Shadow Stress Tensor in Celestial CFT}",
    eprint = "2212.00257",
    archivePrefix = "arXiv",
    primaryClass = "hep-th",
    month = "11",
    year = "2022"
}

@article{Banerjee:2020vnt,
    author = "Banerjee, Shamik and Ghosh, Sudip",
    title = "{MHV gluon scattering amplitudes from celestial current algebras}",
    eprint = "2011.00017",
    archivePrefix = "arXiv",
    primaryClass = "hep-th",
    doi = "10.1007/JHEP10(2021)111",
    journal = "JHEP",
    volume = "10",
    pages = "111",
    year = "2021"
}

@article{Saha:2023hsl,
    author = "Saha, Amartya",
    title = "{Carrollian approach to 1 + 3D flat holography}",
    eprint = "2304.02696",
    archivePrefix = "arXiv",
    primaryClass = "hep-th",
    doi = "10.1007/JHEP06(2023)051",
    journal = "JHEP",
    volume = "06",
    pages = "051",
    year = "2023"
}

@article{Stieberger:2018edy,
      author         = "Stieberger, Stephan and Taylor, Tomasz R.",
      title          = "{Strings on Celestial Sphere}",
      journal        = "Nucl. Phys.",
      volume         = "B935",
      year           = "2018",
      pages          = "388-411",
      doi            = "10.1016/j.nuclphysb.2018.08.019",
      eprint         = "1806.05688",
      archivePrefix  = "arXiv",
      primaryClass   = "hep-th",
      reportNumber   = "MPP-2018-136",
      SLACcitation   = "%%CITATION = ARXIV:1806.05688;%%"
}

@article{Iyer:1994ys,
    author = "Iyer, Vivek and Wald, Robert M.",
    archivePrefix = "arXiv",
    doi = "10.1103/PhysRevD.50.846",
    eprint = "gr-qc/9403028",
    journal = "Phys. Rev. D",
    pages = "846--864",
    title = "{Some properties of Noether charge and a proposal for dynamical black hole entropy}",
    volume = "50",
    year = "1994"
}

@article{Wald:1999wa,
      author         = "Wald, Robert M. and Zoupas, Andreas",
      title          = "{A General definition of `conserved quantities' in
                        general relativity and other theories of gravity}",
      journal        = "Phys. Rev.",
      volume         = "D61",
      year           = "2000",
      pages          = "084027",
      doi            = "10.1103/PhysRevD.61.084027",
      eprint         = "gr-qc/9911095",
      archivePrefix  = "arXiv",
      primaryClass   = "gr-qc",
      SLACcitation   = "%%CITATION = GR-QC/9911095;%%"
}

@article{Oblak:2015qia,
      author         = "Oblak, Blagoje",
      title          = "{From the Lorentz Group to the Celestial Sphere}",
      url            = "http://inspirehep.net/record/1386700/files/arXiv:1508.00920.pdf",
      year           = "2015",
      eprint         = "1508.00920",
      archivePrefix  = "arXiv",
      primaryClass   = "math-ph",
      SLACcitation   = "%%CITATION = ARXIV:1508.00920;%%"
}

@article{Weinberg:1965nx,
      author         = "Weinberg, Steven",
      title          = "{Infrared photons and gravitons}",
      journal        = "Phys. Rev.",
      volume         = "140",
      year           = "1965",
      pages          = "B516-B524",
      doi            = "10.1103/PhysRev.140.B516",
      SLACcitation   = "%%CITATION = PHRVA,140,B516;%%"
}

@article{Witten:1998qj,
      author         = "Witten, Edward",
      title          = "{Anti-de Sitter space and holography}",
      journal        = "Adv. Theor. Math. Phys.",
      volume         = "2",
      year           = "1998",
      pages          = "253-291",
      doi            = "10.4310/ATMP.1998.v2.n2.a2",
      eprint         = "hep-th/9802150",
      archivePrefix  = "arXiv",
      primaryClass   = "hep-th",
      reportNumber   = "IASSNS-HEP-98-15",
      SLACcitation   = "%%CITATION = HEP-TH/9802150;%%"
}

@article{Mack:1969rr,
      author         = "Mack, G. and Salam, Abdus",
      title          = "{Finite component field representations of the conformal
                        group}",
      journal        = "Annals Phys.",
      volume         = "53",
      year           = "1969",
      pages          = "174-202",
      doi            = "10.1016/0003-4916(69)90278-4",
      reportNumber   = "IC-68-68",
      SLACcitation   = "%%CITATION = APNYA,53,174;%%"
}

@inproceedings{Penedones:2016voo,
      author         = "Penedones, Joao",
      title          = "{TASI lectures on AdS/CFT}",
      booktitle      = "{Proceedings, Theoretical Advanced Study Institute in
                        Elementary Particle Physics: New Frontiers in Fields and
                        Strings (TASI 2015): Boulder, CO, USA, June 1-26, 2015}",
      year           = "2017",
      url            = "http://inspirehep.net/record/1481834/files/arXiv:1608.04948.pdf",
      pages          = "75-136",
      doi            = "10.1142/9789813149441_0002",
      eprint         = "1608.04948",
      archivePrefix  = "arXiv",
      primaryClass   = "hep-th",
      SLACcitation   = "%%CITATION = ARXIV:1608.04948;%%"
}

@article{Barnich:2017ubf,
      author         = "Barnich, Glenn",
      title          = "{Centrally extended BMS4 Lie algebroid}",
      journal        = "JHEP",
      volume         = "06",
      year           = "2017",
      pages          = "007",
      doi            = "10.1007/JHEP06(2017)007",
      eprint         = "1703.08704",
      archivePrefix  = "arXiv",
      primaryClass   = "hep-th",
      SLACcitation   = "%%CITATION = ARXIV:1703.08704;%%"
}

@article{Barnich:2016lyg,
      author         = "Barnich, Glenn and Troessaert, CÃ©dric",
      title          = "{Finite BMS transformations}",
      journal        = "JHEP",
      volume         = "03",
      year           = "2016",
      pages          = "167",
      doi            = "10.1007/JHEP03(2016)167",
      eprint         = "1601.04090",
      archivePrefix  = "arXiv",
      primaryClass   = "gr-qc",
      SLACcitation   = "%%CITATION = ARXIV:1601.04090;%%"
}

@article{Barnich:2013yka,
      author         = "Barnich, Glenn and Gonzalez, Hernan A.",
      title          = "{Dual dynamics of three dimensional asymptotically flat
                        Einstein gravity at null infinity}",
      journal        = "JHEP",
      volume         = "05",
      year           = "2013",
      pages          = "016",
      doi            = "10.1007/JHEP05(2013)016",
      eprint         = "1303.1075",
      archivePrefix  = "arXiv",
      primaryClass   = "hep-th",
      reportNumber   = "ULB-TH-12-24",
      SLACcitation   = "%%CITATION = ARXIV:1303.1075;%%"
}

@article{Barnich:2011mi,
      author         = "Barnich, Glenn and Troessaert, Cedric",
      title          = "{BMS charge algebra}",
      journal        = "JHEP",
      volume         = "12",
      year           = "2011",
      pages          = "105",
      doi            = "10.1007/JHEP12(2011)105",
      eprint         = "1106.0213",
      archivePrefix  = "arXiv",
      primaryClass   = "hep-th",
      reportNumber   = "ULB-TH-11-10",
      SLACcitation   = "%%CITATION = ARXIV:1106.0213;%%"
}

@article{Barnich:2010eb,
      author         = "Barnich, Glenn and Troessaert, Cedric",
      title          = "{Aspects of the BMS/CFT correspondence}",
      journal        = "JHEP",
      volume         = "05",
      year           = "2010",
      pages          = "062",
      doi            = "10.1007/JHEP05(2010)062",
      eprint         = "1001.1541",
      archivePrefix  = "arXiv",
      primaryClass   = "hep-th",
      reportNumber   = "ULB-TH-09-28",
      SLACcitation   = "%%CITATION = ARXIV:1001.1541;%%"
}

@article{Barnich:2011ct,
      author         = "Barnich, Glenn and Troessaert, Cedric",
      title          = "{Supertranslations call for superrotations}",
      booktitle      = "{Proceedings, Satellite Workshop on Non Commutative Field
                        Theory and Gravity : 10th Hellenic School and Workshops on
                        Elementary Particle Physics and Gravity (CORFU2010-NC):
                        Corfu 2010, Greece, September 8-12, 2010}",
      journal        = "PoS",
      volume         = "CNCFG",
      year           = "2010",
      pages          = "010",
      note           = "[Ann. U. Craiova Phys.21,S11(2011)]",
      eprint         = "1102.4632",
      archivePrefix  = "arXiv",
      primaryClass   = "gr-qc",
      reportNumber   = "ULB-TH-11-02",
      SLACcitation   = "%%CITATION = ARXIV:1102.4632;%%"
}

@article{Barnich:2009se,
      author         = "Barnich, Glenn and Troessaert, Cedric",
      title          = "{Symmetries of asymptotically flat 4 dimensional
                        spacetimes at null infinity revisited}",
      journal        = "Phys. Rev. Lett.",
      volume         = "105",
      year           = "2010",
      pages          = "111103",
      doi            = "10.1103/PhysRevLett.105.111103",
      eprint         = "0909.2617",
      archivePrefix  = "arXiv",
      primaryClass   = "gr-qc",
      reportNumber   = "ULB-TH-09-24",
      SLACcitation   = "%%CITATION = ARXIV:0909.2617;%%"
}

@article{Barnich:2001jy,
      author         = "Barnich, Glenn and Brandt, Friedemann",
      title          = "{Covariant theory of asymptotic symmetries, conservation
                        laws and central charges}",
      journal        = "Nucl. Phys.",
      volume         = "B633",
      year           = "2002",
      pages          = "3-82",
      doi            = "10.1016/S0550-3213(02)00251-1",
      eprint         = "hep-th/0111246",
      archivePrefix  = "arXiv",
      primaryClass   = "hep-th",
      reportNumber   = "ULB-TH-01-19, MPI-MIS-94-2001",
      SLACcitation   = "%%CITATION = HEP-TH/0111246;%%"
}

@article{Ashtekar:1981bq,
      author         = "Ashtekar, A. and Streubel, M.",
      title          = "{Symplectic Geometry of Radiative Modes and Conserved
                        Quantities at Null Infinity}",
      journal        = "Proc. Roy. Soc. Lond.",
      volume         = "A376",
      year           = "1981",
      pages          = "585-607",
      doi            = "10.1098/rspa.1981.0109",
      SLACcitation   = "%%CITATION = PRSLA,A376,585;%%"
}

@article{Bondi:1962px,
      author         = "Bondi, H. and van der Burg, M. G. J. and Metzner, A. W.
                        K.",
      title          = "{Gravitational waves in general relativity. 7. Waves from
                        axisymmetric isolated systems}",
      journal        = "Proc. Roy. Soc. Lond.",
      volume         = "A269",
      year           = "1962",
      pages          = "21",
      doi            = "10.1098/rspa.1962.0161",
      SLACcitation   = "%%CITATION = PRSLA,A269,21;%%"
}

@article{Sachs:1962wk,
      author         = "Sachs, R. K.",
      title          = "{Gravitational waves in general relativity. 8. Waves in
                        asymptotically flat space-times}",
      journal        = "Proc. Roy. Soc. Lond.",
      volume         = "A270",
      year           = "1962",
      pages          = "103-126",
      doi            = "10.1098/rspa.1962.0206",
      SLACcitation   = "%%CITATION = PRSLA,A270,103;%%"
}

@article{Sachs:1962zza,
      author         = "Sachs, R.",
      title          = "{Asymptotic symmetries in gravitational theory}",
      journal        = "Phys. Rev.",
      volume         = "128",
      year           = "1962",
      pages          = "2851-2864",
      doi            = "10.1103/PhysRev.128.2851",
      SLACcitation   = "%%CITATION = PHRVA,128,2851;%%"
}

@article{Dolan:2011dv,
      author         = "Dolan, F. A. and Osborn, H.",
      title          = "{Conformal Partial Waves: Further Mathematical Results}",
      year           = "2011",
      eprint         = "1108.6194",
      archivePrefix  = "arXiv",
      primaryClass   = "hep-th",
      reportNumber   = "DAMTP-11-64, CCTP-2011-32",
      SLACcitation   = "%%CITATION = ARXIV:1108.6194;%%"
}

@article{deBoer:2003vf,
      author         = "de Boer, Jan and Solodukhin, Sergey N.",
      title          = "{A Holographic reduction of Minkowski space-time}",
      journal        = "Nucl. Phys.",
      volume         = "B665",
      year           = "2003",
      pages          = "545-593",
      doi            = "10.1016/S0550-3213(03)00494-2",
      eprint         = "hep-th/0303006",
      archivePrefix  = "arXiv",
      primaryClass   = "hep-th",
      reportNumber   = "ITFA-2003-11",
      SLACcitation   = "%%CITATION = HEP-TH/0303006;%%"
}

@article{Campiglia:2015yka,
      author         = "Campiglia, Miguel and Laddha, Alok",
      title          = "{New symmetries for the Gravitational S-matrix}",
      journal        = "JHEP",
      volume         = "04",
      year           = "2015",
      pages          = "076",
      doi            = "10.1007/JHEP04(2015)076",
      eprint         = "1502.02318",
      archivePrefix  = "arXiv",
      primaryClass   = "hep-th",
      SLACcitation   = "%%CITATION = ARXIV:1502.02318;%%"
}

@article{Campiglia:2014yka,
      author         = "Campiglia, Miguel and Laddha, Alok",
      title          = "{Asymptotic symmetries and subleading soft graviton
                        theorem}",
      journal        = "Phys. Rev.",
      volume         = "D90",
      year           = "2014",
      number         = "12",
      pages          = "124028",
      doi            = "10.1103/PhysRevD.90.124028",
      eprint         = "1408.2228",
      archivePrefix  = "arXiv",
      primaryClass   = "hep-th",
      SLACcitation   = "%%CITATION = ARXIV:1408.2228;%%"
}

@article{Kapec:2016jld,
      author         = "Kapec, Daniel and Mitra, Prahar and Raclariu, Ana-Maria
                        and Strominger, Andrew",
      title          = "{2D Stress Tensor for 4D Gravity}",
      journal        = "Phys. Rev. Lett.",
      volume         = "119",
      year           = "2017",
      number         = "12",
      pages          = "121601",
      doi            = "10.1103/PhysRevLett.119.121601",
      eprint         = "1609.00282",
      archivePrefix  = "arXiv",
      primaryClass   = "hep-th",
      SLACcitation   = "%%CITATION = ARXIV:1609.00282;%%"
}

@article{Pasterski:2017ylz,
      author         = "Pasterski, Sabrina and Shao, Shu-Heng and Strominger,
                        Andrew",
      title          = "{Gluon Amplitudes as 2d Conformal Correlators}",
      journal        = "Phys. Rev.",
      volume         = "D96",
      year           = "2017",
      number         = "8",
      pages          = "085006",
      doi            = "10.1103/PhysRevD.96.085006",
      eprint         = "1706.03917",
      archivePrefix  = "arXiv",
      primaryClass   = "hep-th",
      SLACcitation   = "%%CITATION = ARXIV:1706.03917;%%"
}

@book{Strominger:2017zoo,
      author         = "Strominger, Andrew",
      title          = "{Lectures on the Infrared Structure of Gravity and Gauge
                        Theory}",
      year           = "2018",
      publisher			 = "{Princeton University Press}",
      eprint         = "1703.05448",
      archivePrefix  = "arXiv",
      primaryClass   = "hep-th",
      SLACcitation   = "%%CITATION = ARXIV:1703.05448;%%"
}

@article{Adamo:2019ipt,
    author = "Adamo, Tim and Mason, Lionel and Sharma, Atul",
    title = "{Celestial amplitudes and conformal soft theorems}",
    eprint = "1905.09224",
    archivePrefix = "arXiv",
    primaryClass = "hep-th",
    reportNumber = "IMPERIAL-TP-TA-2019-02",
    doi = "10.1088/1361-6382/ab42ce",
    journal = "Class. Quant. Grav.",
    volume = "36",
    number = "20",
    pages = "205018",
    year = "2019"
}

@article{Costello:2022jpg,
    author = "Costello, Kevin and Paquette, Natalie M. and Sharma, Atul",
    title = "{Top-Down Holography in an Asymptotically Flat Spacetime}",
    eprint = "2208.14233",
    archivePrefix = "arXiv",
    primaryClass = "hep-th",
    doi = "10.1103/PhysRevLett.130.061602",
    journal = "Phys. Rev. Lett.",
    volume = "130",
    number = "6",
    pages = "061602",
    year = "2023"
}

@article{Fontanella:2025tbs,
    author = "Fontanella, Andrea and Payne, Oliver",
    title = "{A Carroll Limit of AdS/CFT: A Triality with Flat Space Holography?}",
    eprint = "2508.10085",
    archivePrefix = "arXiv",
    primaryClass = "hep-th",
    month = "8",
    year = "2025"
}

@article{Shukla:2026xig,
    author = "Shukla, Ashish and Singh, Rajeev and Soni, Pushkar",
    title = "{Carroll hydrodynamics with spin}",
    eprint = "2601.15023",
    archivePrefix = "arXiv",
    primaryClass = "hep-th",
    month = "1",
    year = "2026"
}

@article{Majumdar:2025juk,
    author = "Majumdar, Sucheta",
    title = "{Carroll theories from Lorentzian light-cone theories}",
    eprint = "2507.03081",
    archivePrefix = "arXiv",
    primaryClass = "hep-th",
    month = "7",
    year = "2025"
}

@article{Kolekar:2024cfg,
    author = "Kolekar, Kedar S. and Mandal, Taniya and Shukla, Ashish and Soni, Pushkar",
    title = "{Hydrodynamics in the Carrollian Regime}",
    eprint = "2409.18763",
    archivePrefix = "arXiv",
    primaryClass = "hep-th",
    reportNumber = "CPHT-RR061.082024",
    doi = "10.1142/S0217751X26500168",
    journal = "Int. J. Mod. Phys. A",
    volume = "41",
    number = "02",
    pages = "2650016",
    year = "2026"
}

@article{Ahmadi-Jahmani:2025iqc,
    author = "Ahmadi-Jahmani, M. M. and Parvizi, A.",
    title = "{Fracton and non-Lorentzian particle duality: gauge field couplings and geometric implications}",
    eprint = "2503.21660",
    archivePrefix = "arXiv",
    primaryClass = "hep-th",
    doi = "10.1007/JHEP08(2025)157",
    journal = "JHEP",
    volume = "08",
    pages = "157",
    year = "2025"
}

@article{Pasterski:2017kqt,
      author         = "Pasterski, Sabrina and Shao, Shu-Heng",
      title          = "{Conformal basis for flat space amplitudes}",
      journal        = "Phys. Rev.",
      volume         = "D96",
      year           = "2017",
      number         = "6",
      pages          = "065022",
      doi            = "10.1103/PhysRevD.96.065022",
      eprint         = "1705.01027",
      archivePrefix  = "arXiv",
      primaryClass   = "hep-th",
      SLACcitation   = "%%CITATION = ARXIV:1705.01027;%%"
}

@article{Freidel:2022bai,
    author = "Freidel, Laurent and Jai-akson, Puttarak",
    title = "{Carrollian hydrodynamics from symmetries}",
    eprint = "2209.03328",
    archivePrefix = "arXiv",
    primaryClass = "hep-th",
    doi = "10.1088/1361-6382/acb194",
    journal = "Class. Quant. Grav.",
    volume = "40",
    number = "5",
    pages = "055009",
    year = "2023"
}

@article{Freidel:2022vjq,
    author = "Freidel, Laurent and Jai-akson, Puttarak",
    title = "{Carrollian hydrodynamics and symplectic structure on stretched horizons}",
    eprint = "2211.06415",
    archivePrefix = "arXiv",
    primaryClass = "gr-qc",
    reportNumber = "RIKEN-iTHEMS-Report-22",
    doi = "10.1007/JHEP05(2024)135",
    journal = "JHEP",
    volume = "05",
    pages = "135",
    year = "2024"
}

@article{Baiguera:2022lsw,
    author = "Baiguera, Stefano and Oling, Gerben and Sybesma, Watse and S{\o}gaard, Benjamin T.",
    title = "{Conformal Carroll scalars with boosts}",
    eprint = "2207.03468",
    archivePrefix = "arXiv",
    primaryClass = "hep-th",
    reportNumber = "NORDITA 2022-047",
    doi = "10.21468/SciPostPhys.14.4.086",
    journal = "SciPost Phys.",
    volume = "14",
    number = "4",
    pages = "086",
    year = "2023"
}

@article{Hansen:2021fxi,
    author = "Hansen, Dennis and Obers, Niels A. and Oling, Gerben and S\o{}gaard, Benjamin T.",
    title = "{Carroll Expansion of General Relativity}",
    eprint = "2112.12684",
    archivePrefix = "arXiv",
    primaryClass = "hep-th",
    reportNumber = "NORDITA 2021-156",
    doi = "10.21468/SciPostPhys.13.3.055",
    journal = "SciPost Phys.",
    volume = "13",
    number = "3",
    pages = "055",
    year = "2022"
}

@article{Hartong:2015usd,
    author = "Hartong, Jelle",
    title = "{Holographic Reconstruction of 3D Flat Space-Time}",
    eprint = "1511.01387",
    archivePrefix = "arXiv",
    primaryClass = "hep-th",
    doi = "10.1007/JHEP10(2016)104",
    journal = "JHEP",
    volume = "10",
    pages = "104",
    year = "2016"
}

@article{Ciambelli:2017wou,
    author = "Ciambelli, Luca and Petkou, Anastasios C. and Petropoulos, P. Marios and Siampos, Konstantinos",
    title = "{The Robinson-Trautman spacetime and its holographic fluid}",
    eprint = "1707.02995",
    archivePrefix = "arXiv",
    primaryClass = "hep-th",
    reportNumber = "CPHT-PC037.062017",
    doi = "10.22323/1.292.0076",
    journal = "PoS",
    volume = "CORFU2016",
    pages = "076",
    year = "2017"
}

@article{Armas:2023dcz,
    author = "Armas, Jay and Have, Emil",
    title = "{Carrollian Fluids and Spontaneous Breaking of Boost Symmetry}",
    eprint = "2308.10594",
    archivePrefix = "arXiv",
    primaryClass = "hep-th",
    doi = "10.1103/PhysRevLett.132.161606",
    journal = "Phys. Rev. Lett.",
    volume = "132",
    number = "16",
    pages = "161606",
    year = "2024"
}

@article{Basu:2018dub,
    author = "Basu, Rudranil and Chowdhury, Udit Narayan",
    title = "{Dynamical structure of Carrollian Electrodynamics}",
    eprint = "1802.09366",
    archivePrefix = "arXiv",
    primaryClass = "hep-th",
    doi = "10.1007/JHEP04(2018)111",
    journal = "JHEP",
    volume = "04",
    pages = "111",
    year = "2018"
}

@article{Dautcourt:1997hb,
    author = "Dautcourt, G.",
    editor = "Demianski, M. and Kopczynski, W.",
    title = "{On the ultrarelativistic limit of general relativity}",
    eprint = "gr-qc/9801093",
    archivePrefix = "arXiv",
    journal = "Acta Phys. Polon. B",
    volume = "29",
    pages = "1047--1055",
    year = "1998"
}

@article{Campoleoni:2022ebj,
    author = "Campoleoni, Andrea and Henneaux, Marc and Pekar, Simon and P{\'e}rez, Alfredo and Salgado-Rebolledo, Patricio",
    title = "{Magnetic Carrollian gravity from the Carroll algebra}",
    eprint = "2207.14167",
    archivePrefix = "arXiv",
    primaryClass = "hep-th",
    doi = "10.1007/JHEP09(2022)127",
    journal = "JHEP",
    volume = "09",
    pages = "127",
    year = "2022"
}

@article{Islam:2023rnc,
    author = "Islam, Minhajul",
    title = "{Carrollian Yang-Mills theory}",
    eprint = "2301.00953",
    archivePrefix = "arXiv",
    primaryClass = "hep-th",
    doi = "10.1007/JHEP05(2023)238",
    journal = "JHEP",
    volume = "05",
    pages = "238",
    year = "2023"
}

@article{Isham:1975ur,
    author = "Isham, C. J.",
    title = "{Some Quantum Field Theory Aspects of the Superspace Quantization of General Relativity}",
    reportNumber = "Print-76-0255 (KING S COLL.)",
    doi = "10.1098/rspa.1976.0138",
    journal = "Proc. Roy. Soc. Lond. A",
    volume = "351",
    pages = "209--232",
    year = "1976"
}

@article{Ecker:2024czh,
    author = "Ecker, Florian and Fiorucci, Adrien and Grumiller, Daniel",
    title = "{Tantum gravity}",
    eprint = "2501.00095",
    archivePrefix = "arXiv",
    primaryClass = "hep-th",
    doi = "10.1103/PhysRevD.111.L021901",
    journal = "Phys. Rev. D",
    volume = "111",
    number = "2",
    pages = "L021901",
    year = "2025"
}

@article{Mars:1993mj,
    author = "Mars, Marc and Senovilla, Jose M. M.",
    title = "{Geometry of general hypersurfaces in space-time: Junction conditions}",
    eprint = "gr-qc/0201054",
    archivePrefix = "arXiv",
    doi = "10.1088/0264-9381/10/9/026",
    journal = "Class. Quant. Grav.",
    volume = "10",
    pages = "1865--1897",
    year = "1993"
}

@article{Belinsky:1970ew,
    author = "Belinsky, V. A. and Khalatnikov, I. M. and Lifshitz, E. M.",
    title = "{Oscillatory approach to a singular point in the relativistic cosmology}",
    doi = "10.1080/00018737000101171",
    journal = "Adv. Phys.",
    volume = "19",
    pages = "525--573",
    year = "1970"
}

@inproceedings{Teitelboim:1978wv,
    author = "Teitelboim, Claudio",
    title = "{SURFACE DEFORMATIONS, THEIR SQUARE ROOT AND THE SIGNATURE OF SPACE-TIME}",
    booktitle = "{7th International Group Theory Colloquium: The Integrative Conference on Group Theory and Mathematical Physics}",
    reportNumber = "Print-78-1134 (PRINCETON)",
    month = "12",
    year = "1978"
}

@article{OConnor:2024rku,
    author = "O'Connor, Josh A. and Pekar, Simon",
    title = "{A note on non-Lorentzian duality symmetries}",
    eprint = "2409.12279",
    archivePrefix = "arXiv",
    primaryClass = "hep-th",
    doi = "10.1007/JHEP03(2025)084",
    journal = "JHEP",
    volume = "03",
    pages = "084",
    year = "2025"
}

@article{Ciambelli:2023mir,
    author = "Ciambelli, Luca and Freidel, Laurent and Leigh, Robert G.",
    title = "{Null Raychaudhuri: canonical structure and the dressing time}",
    eprint = "2309.03932",
    archivePrefix = "arXiv",
    primaryClass = "hep-th",
    doi = "10.1007/JHEP01(2024)166",
    journal = "JHEP",
    volume = "01",
    pages = "166",
    year = "2024"
}

@phdthesis{Ruzziconi:2020cjt,
    author = "Ruzziconi, Romain",
    title = "{On the Various Extensions of the BMS Group}",
    eprint = "2009.01926",
    archivePrefix = "arXiv",
    primaryClass = "hep-th",
    school = "U. Brussels",
    year = "2020"
}

@article{Geroch:1973am,
    author = "Geroch, Robert P. and Held, A. and Penrose, R.",
    title = "{A space-time calculus based on pairs of null directions}",
    doi = "10.1063/1.1666410",
    journal = "J. Math. Phys.",
    volume = "14",
    pages = "874--881",
    year = "1973"
}

@INPROCEEDINGS{1977asst.conf....1G,
       author = {{Geroch}, Robert},
        title = "{Asymptotic Structure of Space-Time}",
    booktitle = {Asymptotic Structure of Space-Time},
         year = 1977,
       editor = {{Esposito}, F. Paul and {Witten}, Louis},
        month = jan,
        pages = {1},
       adsurl = {https://ui.adsabs.harvard.edu/abs/1977asst.conf....1G}
}

@article{Ruzziconi:2020wrb,
    author = "Ruzziconi, Romain and Zwikel, C\'eline",
    title = "{Conservation and Integrability in Lower-Dimensional Gravity}",
    eprint = "2012.03961",
    archivePrefix = "arXiv",
    primaryClass = "hep-th",
    doi = "10.1007/JHEP04(2021)034",
    journal = "JHEP",
    volume = "04",
    pages = "034",
    year = "2021"
}

@article{Bagchi:2023fbj,
    author = "Bagchi, Arjun and Dhivakar, Prateksh and Dutta, Sudipta",
    title = "{AdS Witten diagrams to Carrollian correlators}",
    eprint = "2303.07388",
    archivePrefix = "arXiv",
    primaryClass = "hep-th",
    doi = "10.1007/JHEP04(2023)135",
    journal = "JHEP",
    volume = "04",
    pages = "135",
    year = "2023"
}

@article{deGioia:2022fcn,
    author = "de Gioia, Leonardo Pipolo and Raclariu, Ana-Maria",
    title = "{Eikonal approximation in celestial CFT}",
    eprint = "2206.10547",
    archivePrefix = "arXiv",
    primaryClass = "hep-th",
    doi = "10.1007/JHEP03(2023)030",
    journal = "JHEP",
    volume = "03",
    pages = "030",
    year = "2023"
}

@article{deGioia:2023cbd,
    author = "de Gioia, Leonardo Pipolo and Raclariu, Ana-Maria",
    title = "{Celestial sector in CFT: Conformally soft symmetries}",
    eprint = "2303.10037",
    archivePrefix = "arXiv",
    primaryClass = "hep-th",
    doi = "10.21468/SciPostPhys.17.1.002",
    journal = "SciPost Phys.",
    volume = "17",
    number = "1",
    pages = "002",
    year = "2024"
}

@article{Campoleoni:2023fug,
    author = "Campoleoni, Andrea and Delfante, Arnaud and Pekar, Simon and Petropoulos, P. Marios and Rivera-Betancour, David and Vilatte, Matthieu",
    title = "{Flat from anti de Sitter}",
    eprint = "2309.15182",
    archivePrefix = "arXiv",
    primaryClass = "hep-th",
    reportNumber = "CPHT-RR054.082023",
    doi = "10.1007/JHEP12(2023)078",
    journal = "JHEP",
    volume = "12",
    pages = "078",
    year = "2023"
}

@article{Banerjee:2018fgd,
    author = "Banerjee, Shamik",
    title = "{Symmetries of free massless particles and soft theorems}",
    eprint = "1804.06646",
    archivePrefix = "arXiv",
    primaryClass = "hep-th",
    doi = "10.1007/s10714-019-2609-z",
    journal = "Gen. Rel. Grav.",
    volume = "51",
    number = "9",
    pages = "128",
    year = "2019"
}

@article{Compere:2018ylh,
    author = "Comp\`ere, Geoffrey and Fiorucci, Adrien and Ruzziconi, Romain",
    archivePrefix = "arXiv",
    doi = "10.1007/JHEP11(2018)200",
    eprint = "1810.00377",
    journal = "JHEP",
    pages = "200",
    primaryClass = "hep-th",
    title = "{Superboost transitions, refraction memory and super-Lorentz charge algebra}",
    volume = "11",
    year = "2018"
}

@article{Campoleoni:2022wmf,
    author = "Campoleoni, Andrea and Ciambelli, Luca and Delfante, Arnaud and Marteau, Charles and Petropoulos, P. Marios and Ruzziconi, Romain",
    title = "{Holographic Lorentz and Carroll frames}",
    eprint = "2208.07575",
    archivePrefix = "arXiv",
    primaryClass = "hep-th",
    doi = "10.1007/JHEP12(2022)007",
    journal = "JHEP",
    volume = "12",
    pages = "007",
    year = "2022"
}

@article{Nguyen:2023vfz,
    author = "Nguyen, Kevin and West, Peter",
    title = "{Carrollian Conformal Fields and Flat Holography}",
    eprint = "2305.02884",
    archivePrefix = "arXiv",
    primaryClass = "hep-th",
    doi = "10.3390/universe9090385",
    journal = "Universe",
    volume = "9",
    number = "9",
    pages = "385",
    year = "2023"
}

@article{Cotler:2024cia,
    author = "Cotler, Jordan and Jensen, Kristan and Prohazka, Stefan and Riegler, Max and Salzer, Jakob",
    title = "{Soft gravitons in three dimensions}",
    eprint = "2411.13633",
    archivePrefix = "arXiv",
    primaryClass = "hep-th",
    doi = "10.1007/JHEP07(2025)002",
    journal = "JHEP",
    volume = "07",
    pages = "002",
    year = "2025"
}

@article{Lipstein:2025jfj,
    author = "Lipstein, Arthur and Ruzziconi, Romain and Yelleshpur Srikant, Akshay",
    title = "{Towards a flat space Carrollian hologram from AdS$_{4}$/CFT$_{3}$}",
    eprint = "2504.10291",
    archivePrefix = "arXiv",
    primaryClass = "hep-th",
    doi = "10.1007/JHEP06(2025)073",
    journal = "JHEP",
    volume = "06",
    pages = "073",
    year = "2025"
}

@article{Belavin:1984vu,
    author = "Belavin, A. A. and Polyakov, Alexander M. and Zamolodchikov, A. B.",
    editor = "Khalatnikov, I. M. and Mineev, V. P.",
    title = "{Infinite Conformal Symmetry in Two-Dimensional Quantum Field Theory}",
    reportNumber = "CERN-TH-3827",
    doi = "10.1016/0550-3213(84)90052-X",
    journal = "Nucl. Phys. B",
    volume = "241",
    pages = "333--380",
    year = "1984"
}

@article{Knizhnik:1984nr,
    author = "Knizhnik, V. G. and Zamolodchikov, A. B.",
    editor = "Khalatnikov, I. M. and Mineev, V. P.",
    title = "{Current Algebra and Wess-Zumino Model in Two-Dimensions}",
    doi = "10.1016/0550-3213(84)90374-2",
    journal = "Nucl. Phys. B",
    volume = "247",
    pages = "83--103",
    year = "1984"
}

@article{Nguyen:2025sqk,
    author = "Nguyen, Kevin and Salzer, Jakob",
    title = "{Operator product expansion in Carrollian CFT}",
    eprint = "2503.15607",
    archivePrefix = "arXiv",
    primaryClass = "hep-th",
    doi = "10.1007/JHEP07(2025)193",
    journal = "JHEP",
    volume = "07",
    pages = "193",
    year = "2025"
}

@article{Henneaux:2020ekh,
    author = "Henneaux, Marc and Matulich, Javier and Neogi, Turmoli",
    title = "{Asymptotic realization of the super-BMS algebra at spatial infinity}",
    eprint = "2004.07299",
    archivePrefix = "arXiv",
    primaryClass = "hep-th",
    doi = "10.1103/PhysRevD.101.126016",
    journal = "Phys. Rev. D",
    volume = "101",
    number = "12",
    pages = "126016",
    year = "2020"
}

@article{Alday:2024yyj,
    author = "Alday, Luis F. and Nocchi, Maria and Ruzziconi, Romain and Yelleshpur Srikant, Akshay",
    title = "{Carrollian amplitudes from holographic correlators}",
    eprint = "2406.19343",
    archivePrefix = "arXiv",
    primaryClass = "hep-th",
    doi = "10.1007/JHEP03(2025)158",
    journal = "JHEP",
    volume = "03",
    pages = "158",
    year = "2025"
}

@article{Ruzziconi:2024zkr,
    author = "Ruzziconi, Romain and Stieberger, Stephan and Taylor, Tomasz R. and Zhu, Bin",
    title = "{Differential equations for Carrollian amplitudes}",
    eprint = "2407.04789",
    archivePrefix = "arXiv",
    primaryClass = "hep-th",
    doi = "10.1007/JHEP09(2024)149",
    journal = "JHEP",
    volume = "09",
    pages = "149",
    year = "2024"
}

@article{Arutyunov:2002fh,
    author = "Arutyunov, G. and Dolan, F. A. and Osborn, H. and Sokatchev, E.",
    title = "{Correlation functions and massive Kaluza-Klein modes in the AdS / CFT correspondence}",
    eprint = "hep-th/0212116",
    archivePrefix = "arXiv",
    reportNumber = "AEI-2002-096, DAMTP-02-148, LAPTH-953-02",
    doi = "10.1016/S0550-3213(03)00448-6",
    journal = "Nucl. Phys. B",
    volume = "665",
    pages = "273--324",
    year = "2003"
}

@article{Bagchi:2024efs,
    author = "Bagchi, Arjun and Lipstein, Arthur and Mandlik, Mangesh and Mehra, Aditya",
    title = "{3d Carrollian Chern-Simons theory {\&} 2d Yang-Mills}",
    eprint = "2407.13574",
    archivePrefix = "arXiv",
    primaryClass = "hep-th",
    doi = "10.1007/JHEP11(2024)006",
    journal = "JHEP",
    volume = "11",
    pages = "006",
    year = "2024"
}

@article{Miskovic:2023zfz,
    author = "Miskovic, Olivera and Olea, Rodrigo and Petropoulos, P. Marios and Rivera-Betancour, David and Siampos, Konstantinos",
    title = "{Chern-Simons action and the Carrollian Cotton tensors}",
    eprint = "2310.19929",
    archivePrefix = "arXiv",
    primaryClass = "hep-th",
    reportNumber = "CPHT-RR048.072022",
    doi = "10.1007/JHEP12(2023)130",
    journal = "JHEP",
    volume = "12",
    pages = "130",
    year = "2023"
}

@article{SenGupta:1966qer,
    author = "Sen Gupta, N. D.",
    title = "{On an analogue of the Galilei group}",
    doi = "10.1007/BF02740871",
    journal = "Nuovo Cim. A",
    volume = "44",
    number = "2",
    pages = "512--517",
    year = "1966"
}

@article{Fuentealba:2023hzq,
    author = "Fuentealba, Oscar and Henneaux, Marc",
    title = "{Simplifying (super-)BMS algebras}",
    eprint = "2309.07600",
    archivePrefix = "arXiv",
    primaryClass = "hep-th",
    doi = "10.1007/JHEP11(2023)108",
    journal = "JHEP",
    volume = "11",
    pages = "108",
    year = "2023"
}

@article{Awada:1985by,
    author = "Awada, M. A. and Gibbons, G. W. and Shaw, W. T.",
    title = "{Conformal supergravity, twistors and the super BMS group}",
    reportNumber = "Print-85-0937 (CAMBRIDGE)",
    doi = "10.1016/S0003-4916(86)80023-9",
    journal = "Annals Phys.",
    volume = "171",
    pages = "52",
    year = "1986"
}

@online{Kaplan_AdSCFT_from_bottom_up,
    author  = "Jared Kaplan",
    title   = "Lectures on AdS/CFT from the Bottom Up",
    year    = "2016",
    url     = "https://sites.krieger.jhu.edu/jared-kaplan/files/2016/05/AdSCFTCourseNotesCurrentPublic.pdf",
    note    = "Lecture notes, Johns Hopkins University"
}

@article{Anninos:2025zgr,
    author = "Anninos, Dionysios and Galante, Dami{\'a}n A. and Georgescu, Silvia and Maneerat, Chawakorn and Svesko, Andrew",
    title = "{The Stretched Horizon Limit}",
    eprint = "2512.16738",
    archivePrefix = "arXiv",
    primaryClass = "hep-th",
    month = "12",
    year = "2025"
}

@article{Cotler:2025npu,
    author = "Cotler, Jordan and Dhivakar, Prateksh and Jensen, Kristan",
    title = "{Carrollian holographic duals are non-local}",
    eprint = "2512.05072",
    archivePrefix = "arXiv",
    primaryClass = "hep-th",
    month = "12",
    year = "2025"
}

@article{Bagchi:2026wcu,
    author = "Bagchi, Arjun and Banerjee, Aritra and Chatterjee, Ritankar and Pandit, Priyadarshini",
    title = "{The Tensionless Lives of Null Strings}",
    eprint = "2601.20959",
    archivePrefix = "arXiv",
    primaryClass = "hep-th",
    month = "1",
    year = "2026"
}

@article{Long:2026cpq,
    author = "Long, Jiang and Yang, Jing-Long",
    title = "{Constraining bulk-to-boundary correlators in the theories with Poincar{\'e} symmetry}",
    eprint = "2601.18461",
    archivePrefix = "arXiv",
    primaryClass = "hep-th",
    month = "1",
    year = "2026"
}

@article{Adamo:2015fwa,
    author = "Adamo, Tim and Casali, Eduardo",
    title = "{Perturbative gauge theory at null infinity}",
    eprint = "1504.02304",
    archivePrefix = "arXiv",
    primaryClass = "hep-th",
    reportNumber = "DAMTP-2015-19",
    doi = "10.1103/PhysRevD.91.125022",
    journal = "Phys. Rev. D",
    volume = "91",
    number = "12",
    pages = "125022",
    year = "2015"
}

@article{Ryu:2006bv,
    author = "Ryu, Shinsei and Takayanagi, Tadashi",
    title = "{Holographic derivation of entanglement entropy from AdS/CFT}",
    eprint = "hep-th/0603001",
    archivePrefix = "arXiv",
    reportNumber = "NSF-KITP-06-11, NSF-KITP-06-11",
    doi = "10.1103/PhysRevLett.96.181602",
    journal = "Phys. Rev. Lett.",
    volume = "96",
    pages = "181602",
    year = "2006"
}

@article{Skenderis:2002wp,
    author = "Skenderis, Kostas",
    editor = "de Wit, B. and Vandoren, S.",
    title = "{Lecture notes on holographic renormalization}",
    eprint = "hep-th/0209067",
    archivePrefix = "arXiv",
    reportNumber = "PUTP-2047",
    doi = "10.1088/0264-9381/19/22/306",
    journal = "Class. Quant. Grav.",
    volume = "19",
    pages = "5849--5876",
    year = "2002"
}

@article{Geiller:2024bgf,
    author = "Geiller, Marc",
    title = "{Celestial $w_{1+\infty}$ charges and the subleading structure of asymptotically-flat spacetimes}",
    eprint = "2403.05195",
    archivePrefix = "arXiv",
    primaryClass = "hep-th",
    doi = "10.21468/SciPostPhys.18.1.023",
    journal = "SciPost Phys.",
    volume = "18",
    number = "1",
    pages = "023",
    year = "2025"
}

@article{Kmec:2024nmu,
    author = "Kmec, Adam and Mason, Lionel and Ruzziconi, Romain and Yelleshpur Srikant, Akshay",
    title = "{Celestial $Lw_{1+\infty}$ charges from a twistor action}",
    eprint = "2407.04028",
    archivePrefix = "arXiv",
    primaryClass = "hep-th",
    doi = "10.1007/JHEP10(2024)250",
    journal = "JHEP",
    volume = "10",
    pages = "250",
    year = "2024"
}

@article{Costello:2022wso,
    author = "Costello, Kevin and Paquette, Natalie M.",
    title = "{Celestial holography meets twisted holography: 4d amplitudes from chiral correlators}",
    eprint = "2201.02595",
    archivePrefix = "arXiv",
    primaryClass = "hep-th",
    doi = "10.1007/JHEP10(2022)193",
    journal = "JHEP",
    volume = "10",
    pages = "193",
    year = "2022"
}

@article{Costello:2023hmi,
    author = "Costello, Kevin and Paquette, Natalie M. and Sharma, Atul",
    title = "{Burns space and holography}",
    eprint = "2306.00940",
    archivePrefix = "arXiv",
    primaryClass = "hep-th",
    doi = "10.1007/JHEP10(2023)174",
    journal = "JHEP",
    volume = "10",
    pages = "174",
    year = "2023"
}

@article{Stieberger:2022zyk,
    author = "Stieberger, Stephan and Taylor, Tomasz R. and Zhu, Bin",
    title = "{Celestial Liouville theory for Yang-Mills amplitudes}",
    eprint = "2209.02724",
    archivePrefix = "arXiv",
    primaryClass = "hep-th",
    doi = "10.1016/j.physletb.2022.137588",
    journal = "Phys. Lett. B",
    volume = "836",
    pages = "137588",
    year = "2023"
}

@article{Penrose:1976js,
    author = "Penrose, R.",
    title = "{Nonlinear Gravitons and Curved Twistor Theory}",
    doi = "10.1007/BF00762011",
    journal = "Gen. Rel. Grav.",
    volume = "7",
    pages = "31--52",
    year = "1976"
}

@article{Penrose:1976jq,
    author = "Penrose, R.",
    title = "{The Nonlinear Graviton}",
    doi = "10.1007/BF00763433",
    journal = "Gen. Rel. Grav.",
    volume = "7",
    pages = "171--176",
    year = "1976"
}

@article{Cresto:2024fhd,
    author = "Cresto, Nicolas and Freidel, Laurent",
    title = "{Asymptotic higher spin symmetries I: covariant wedge algebra in gravity}",
    eprint = "2409.12178",
    archivePrefix = "arXiv",
    primaryClass = "hep-th",
    doi = "10.1007/s11005-025-01921-4",
    journal = "Lett. Math. Phys.",
    volume = "115",
    number = "2",
    pages = "39",
    year = "2025"
}

@article{Cresto:2024mne,
    author = "Cresto, Nicolas and Freidel, Laurent",
    title = "{Asymptotic Higher Spin Symmetries II: Noether Realization in Gravity}",
    eprint = "2410.15219",
    archivePrefix = "arXiv",
    primaryClass = "hep-th",
    month = "10",
    year = "2024"
}

@article{Cresto:2025bfo,
    author = "Cresto, Nicolas",
    title = "{Asymptotic higher spin symmetries III: Noether realization in Yang{\textendash}Mills theory}",
    eprint = "2501.08856",
    archivePrefix = "arXiv",
    primaryClass = "hep-th",
    doi = "10.1007/s11005-025-02027-7",
    journal = "Lett. Math. Phys.",
    volume = "115",
    number = "6",
    pages = "133",
    year = "2025"
}

@article{Cresto:2025ubl,
    author = "Cresto, Nicolas and Freidel, Laurent",
    title = "{Asymptotic higher spin symmetries. Part IV. Einstein-Yang-Mills theory}",
    eprint = "2505.04327",
    archivePrefix = "arXiv",
    primaryClass = "hep-th",
    doi = "10.1007/JHEP12(2025)097",
    journal = "JHEP",
    volume = "12",
    pages = "097",
    year = "2025"
}

@phdthesis{Cresto:2025fbc,
    author = "Cresto, Nicolas",
    title = "{Asymptotic Higher Spin Symmetries: Noether Realization {\&} Algebraic Structure in Einstein-Yang-Mills Theory}",
    eprint = "2509.17137",
    archivePrefix = "arXiv",
    primaryClass = "hep-th",
    school = "U. Waterloo (main)",
    year = "2025"
}

@article{Nagy:2022xxs,
    author = "Nagy, Silvia and Peraza, Javier",
    title = "{Radiative phase space extensions at all orders in r for self-dual Yang-Mills and gravity}",
    eprint = "2211.12991",
    archivePrefix = "arXiv",
    primaryClass = "hep-th",
    doi = "10.1007/JHEP02(2023)202",
    journal = "JHEP",
    volume = "02",
    pages = "202",
    year = "2023"
}

@article{Nagy:2024dme,
    author = "Nagy, Silvia and Peraza, Javier and Pizzolo, Giorgio",
    title = "{General hierarchy of charges at null infinity via the Todd polynomials}",
    eprint = "2405.06629",
    archivePrefix = "arXiv",
    primaryClass = "hep-th",
    doi = "10.1103/PhysRevD.111.L061903",
    journal = "Phys. Rev. D",
    volume = "111",
    number = "6",
    pages = "L061903",
    year = "2025"
}

@article{Nagy:2024jua,
    author = "Nagy, Silvia and Peraza, Javier and Pizzolo, Giorgio",
    title = "{Infinite-dimensional hierarchy of recursive extensions for all sub$^{n}$-leading soft effects in Yang-Mills}",
    eprint = "2407.13556",
    archivePrefix = "arXiv",
    primaryClass = "hep-th",
    doi = "10.1007/JHEP12(2024)068",
    journal = "JHEP",
    volume = "12",
    pages = "068",
    year = "2024"
}

@article{Diaz-Jaramillo:2025gxw,
    author = "D{\'\i}az-Jaramillo, Felipe and Nagy, Silvia and Pizzolo, Giorgio",
    title = "{Homotopy kinematic algebras at null infinity}",
    eprint = "2503.21035",
    archivePrefix = "arXiv",
    primaryClass = "hep-th",
    doi = "10.1103/p945-xjlk",
    journal = "Phys. Rev. D",
    volume = "112",
    number = "8",
    pages = "086002",
    year = "2025"
}

@article{Nagy:2025hip,
    author = "Nagy, Silvia and Peraza, Javier and Pizzolo, Giorgio",
    title = "{Boundary Actions and Loop Groups: A Geometric Picture of Gauge Symmetries at Null Infinity}",
    eprint = "2509.08725",
    archivePrefix = "arXiv",
    primaryClass = "hep-th",
    month = "9",
    year = "2025"
}

@article{Kmec:2025ftx,
    author = "Kmec, Adam and Mason, Lionel and Ruzziconi, Romain and Sharma, Atul",
    title = "{S-algebra in gauge theory: twistor, spacetime and holographic perspectives}",
    eprint = "2506.01888",
    archivePrefix = "arXiv",
    primaryClass = "hep-th",
    doi = "10.1088/1361-6382/ae0673",
    journal = "Class. Quant. Grav.",
    volume = "42",
    number = "19",
    pages = "195008",
    year = "2025"
}

@article{Ruzziconi:2025fuy,
    author = "Ruzziconi, Romain and Zwikel, C{\'e}line",
    title = "{Celestial $Lw_{1+\infty}$ Symmetries and Subleading Phase Space of Null Hypersurfaces}",
    eprint = "2511.07525",
    archivePrefix = "arXiv",
    primaryClass = "hep-th",
    month = "11",
    year = "2025"
}

@article{Ruzziconi:2025fct,
    author = "Ruzziconi, Romain and Zwikel, C{\'e}line",
    title = "{Celestial Symmetries of Black Hole Horizons}",
    eprint = "2504.08027",
    archivePrefix = "arXiv",
    primaryClass = "hep-th",
    month = "4",
    year = "2025"
}

@article{Sheta:2025oep,
    author = "Sheta, Ahmed and Strominger, Andrew and Tropper, Adam and Wei, Hongji",
    title = "{Soft Algebras in AdS$_4$ from Light Ray Operators in CFT$_3$}",
    eprint = "2601.00096",
    archivePrefix = "arXiv",
    primaryClass = "hep-th",
    month = "12",
    year = "2025"
}

@article{Bittleston:2024efo,
    author = "Bittleston, Roland and Costello, Kevin and Zeng, Keyou",
    title = "{Self-Dual Gauge Theory from the Top Down}",
    eprint = "2412.02680",
    archivePrefix = "arXiv",
    primaryClass = "hep-th",
    month = "12",
    year = "2024"
}

@article{Stieberger:2023fju,
    author = "Stieberger, Stephan and Taylor, Tomasz R. and Zhu, Bin",
    title = "{Yang-Mills as a Liouville theory}",
    eprint = "2308.09741",
    archivePrefix = "arXiv",
    primaryClass = "hep-th",
    doi = "10.1016/j.physletb.2023.138229",
    journal = "Phys. Lett. B",
    volume = "846",
    pages = "138229",
    year = "2023"
}

@article{Have:2024dff,
    author = "Have, Emil and Nguyen, Kevin and Prohazka, Stefan and Salzer, Jakob",
    title = "{Massive carrollian fields at timelike infinity}",
    eprint = "2402.05190",
    archivePrefix = "arXiv",
    primaryClass = "hep-th",
    reportNumber = "UWThPh 2024-6",
    doi = "10.1007/JHEP07(2024)054",
    journal = "JHEP",
    volume = "07",
    pages = "054",
    year = "2024"
}

@article{Adamo:2024mqn,
    author = "Adamo, Tim and Bu, Wei and Tourkine, Piotr and Zhu, Bin",
    title = "{Eikonal amplitudes on the celestial sphere}",
    eprint = "2405.15594",
    archivePrefix = "arXiv",
    primaryClass = "hep-th",
    doi = "10.1007/JHEP10(2024)192",
    journal = "JHEP",
    volume = "10",
    pages = "192",
    year = "2024"
}

@article{Banerjee:2024hvb,
    author = "Banerjee, Sourish and Basu, Rudranil and Atul Bhatkar, Sayali",
    title = "{Light transformation: a celestial and Carrollian perspective}",
    eprint = "2407.08379",
    archivePrefix = "arXiv",
    primaryClass = "hep-th",
    doi = "10.1007/JHEP12(2024)122",
    journal = "JHEP",
    volume = "12",
    pages = "122",
    year = "2024"
}

@article{Kraus:2024gso,
    author = "Kraus, Per and Myers, Richard M.",
    title = "{Carrollian partition functions and the flat limit of AdS}",
    eprint = "2407.13668",
    archivePrefix = "arXiv",
    primaryClass = "hep-th",
    doi = "10.1007/JHEP01(2025)183",
    journal = "JHEP",
    volume = "01",
    pages = "183",
    year = "2025"
}

@article{Jorstad:2024yzm,
    author = "J{\o}rstad, Eivind and Pasterski, Sabrina",
    title = "{A Comment on Boundary Correlators: Soft Omissions and the Massless S-Matrix}",
    eprint = "2410.20296",
    archivePrefix = "arXiv",
    primaryClass = "hep-th",
    month = "10",
    year = "2024"
}

@article{Kulp:2024scx,
    author = "Kulp, Justin and Pasterski, Sabrina",
    title = "{Multiparticle states for the flat hologram}",
    eprint = "2501.00462",
    archivePrefix = "arXiv",
    primaryClass = "hep-th",
    doi = "10.1007/JHEP08(2025)091",
    journal = "JHEP",
    volume = "08",
    pages = "091",
    year = "2025"
}

@article{Kraus:2025wgi,
    author = "Kraus, Per and Myers, Richard M.",
    title = "{Carrollian partition function for bulk Yang-Mills theory}",
    eprint = "2503.00916",
    archivePrefix = "arXiv",
    primaryClass = "hep-th",
    doi = "10.1007/JHEP08(2025)180",
    journal = "JHEP",
    volume = "08",
    pages = "180",
    year = "2025"
}

@article{Bergshoeff:2023vfd,
    author = "Bergshoeff, Eric A. and Campoleoni, Andrea and Fontanella, Andrea and Mele, Lea and Rosseel, Jan",
    title = "{Carroll fermions}",
    eprint = "2312.00745",
    archivePrefix = "arXiv",
    primaryClass = "hep-th",
    doi = "10.21468/SciPostPhys.16.6.153",
    journal = "SciPost Phys.",
    volume = "16",
    number = "6",
    pages = "153",
    year = "2024"
}

@article{Duval:2017els,
    author = "Duval, C. and Gibbons, G. W. and Horvathy, P. A. and Zhang, P. -M.",
    title = "{Carroll symmetry of plane gravitational waves}",
    eprint = "1702.08284",
    archivePrefix = "arXiv",
    primaryClass = "gr-qc",
    doi = "10.1088/1361-6382/aa7f62",
    journal = "Class. Quant. Grav.",
    volume = "34",
    number = "17",
    pages = "175003",
    year = "2017"
}

@article{Bagchi:2023ysc,
    author = "Bagchi, Arjun and Kolekar, Kedar S. and Shukla, Ashish",
    title = "{Carrollian Origins of Bjorken Flow}",
    eprint = "2302.03053",
    archivePrefix = "arXiv",
    primaryClass = "hep-th",
    reportNumber = "CPHT079.122022",
    doi = "10.1103/PhysRevLett.130.241601",
    journal = "Phys. Rev. Lett.",
    volume = "130",
    number = "24",
    pages = "241601",
    year = "2023"
}

@article{Geiller:2024amx,
    author = "Geiller, Marc and Zwikel, C{\'e}line",
    title = "{The partial Bondi gauge: Gauge fixings and asymptotic charges}",
    eprint = "2401.09540",
    archivePrefix = "arXiv",
    primaryClass = "hep-th",
    doi = "10.21468/SciPostPhys.16.3.076",
    journal = "SciPost Phys.",
    volume = "16",
    number = "3",
    pages = "076",
    year = "2024"
}

@article{Gubser:1998bc,
    author = "Gubser, S. S. and Klebanov, Igor R. and Polyakov, Alexander M.",
    title = "{Gauge theory correlators from noncritical string theory}",
    eprint = "hep-th/9802109",
    archivePrefix = "arXiv",
    reportNumber = "PUPT-1767",
    doi = "10.1016/S0370-2693(98)00377-3",
    journal = "Phys. Lett. B",
    volume = "428",
    pages = "105--114",
    year = "1998"
}

@article{Aharony:2008ug,
    author = "Aharony, Ofer and Bergman, Oren and Jafferis, Daniel Louis and Maldacena, Juan",
    title = "{N=6 superconformal Chern-Simons-matter theories, M2-branes and their gravity duals}",
    eprint = "0806.1218",
    archivePrefix = "arXiv",
    primaryClass = "hep-th",
    reportNumber = "WIS-12-08-JUN-DPP",
    doi = "10.1088/1126-6708/2008/10/091",
    journal = "JHEP",
    volume = "10",
    pages = "091",
    year = "2008"
}

@article{Cotler:2025dau,
    author = "Cotler, Jordan and Dhivakar, Prateksh and Jensen, Kristan",
    title = "{A finite Carrollian critical point}",
    eprint = "2504.12289",
    archivePrefix = "arXiv",
    primaryClass = "hep-th",
    doi = "10.1007/JHEP08(2025)172",
    journal = "JHEP",
    volume = "08",
    pages = "172",
    year = "2025"
}

@article{Chen:2024voz,
    author = "Chen, Bin and Sun, Haowei and Zheng, Yu-fan",
    title = "{Quantization of Carrollian conformal scalar theories}",
    eprint = "2406.17451",
    archivePrefix = "arXiv",
    primaryClass = "hep-th",
    doi = "10.1103/PhysRevD.110.125010",
    journal = "Phys. Rev. D",
    volume = "110",
    number = "12",
    pages = "125010",
    year = "2024"
}

@article{Cotler:2024xhb,
    author = "Cotler, Jordan and Jensen, Kristan and Prohazka, Stefan and Raz, Amir and Riegler, Max and Salzer, Jakob",
    title = "{Quantizing Carrollian field theories}",
    eprint = "2407.11971",
    archivePrefix = "arXiv",
    primaryClass = "hep-th",
    doi = "10.1007/JHEP10(2024)049",
    journal = "JHEP",
    volume = "10",
    pages = "049",
    year = "2024"
}

@article{Campoleoni:2016vsh,
    author = "Campoleoni, Andrea and Gonzalez, Hernan A. and Oblak, Blagoje and Riegler, Max",
    editor = "Brink, Lars and Henneaux, Marc and Vasiliev, Mikhail A.",
    title = "{BMS Modules in Three Dimensions}",
    eprint = "1603.03812",
    archivePrefix = "arXiv",
    primaryClass = "hep-th",
    doi = "10.1142/S0217751X16500688",
    journal = "Int. J. Mod. Phys. A",
    volume = "31",
    number = "12",
    pages = "1650068",
    year = "2016"
}

@phdthesis{Damour:1979wya,
    author = "Damour, Thibaut",
    title = "{Quelques proprietes mecaniques, electromagnet iques, thermodynamiques et quantiques des trous noir}",
    school = "Paris U., VI-VII",
    year = "1979"
}

@article{Damour:1978cg,
    author = "Damour, T.",
    title = "{Black Hole Eddy Currents}",
    doi = "10.1103/PhysRevD.18.3598",
    journal = "Phys. Rev. D",
    volume = "18",
    pages = "3598--3604",
    year = "1978"
}

@BOOK{1986bhmp.book.....T,
       author = {{Thorne}, Kip S. and {Price}, Richard H. and {MacDonald}, Douglas A.},
        title = "{Black holes: The membrane paradigm}",
         year = 1986,
       adsurl = {https://ui.adsabs.harvard.edu/abs/1986bhmp.book.....T}
}

@article{Price:1986yy,
    author = "Price, R. H. and Thorne, K. S.",
    title = "{Membrane Viewpoint on Black Holes: Properties and Evolution of the Stretched Horizon}",
    doi = "10.1103/PhysRevD.33.915",
    journal = "Phys. Rev. D",
    volume = "33",
    pages = "915--941",
    year = "1986"
}

@article{Freidel:2024emv,
    author = "Freidel, Laurent and Jai-akson, Puttarak",
    title = "{Geometry of Carrollian stretched horizons}",
    eprint = "2406.06709",
    archivePrefix = "arXiv",
    primaryClass = "gr-qc",
    reportNumber = "RIKEN-iTHEMS-Report-24",
    doi = "10.1088/1361-6382/adaf6e",
    journal = "Class. Quant. Grav.",
    volume = "42",
    number = "6",
    pages = "065010",
    year = "2025"
}

@article{Kulkarni:2025qcx,
    author = "Kulkarni, Harshal and Ruzziconi, Romain and Yelleshpur Srikant, Akshay",
    title = "{On Carrollian and celestial correlators in general dimensions}",
    eprint = "2508.06602",
    archivePrefix = "arXiv",
    primaryClass = "hep-th",
    doi = "10.1007/JHEP10(2025)187",
    journal = "JHEP",
    volume = "10",
    pages = "187",
    year = "2025"
}

@phdthesis{Vilatte:2024jjr,
    author = "Vilatte, Matthieu Bernard Jean",
    title = "{Adventures in Thermal Wonderland:~Aspects of Carrollian physics, asymptotically flat spacetimes and thermal field theory}",
    reportNumber = "2024IPPAX065, tel-04791687",
    school = "Thessaloniki U",
    month = "11",
    year = "2024"
}

@article{Hartong:2024hvs,
    author = "Hartong, Jelle and Palumbo, Giandomenico and Pekar, Simon and P{\'e}rez, Alfredo and Prohazka, Stefan",
    title = "{Fractons on curved spacetime in 2 + 1 dimensions}",
    eprint = "2409.04525",
    archivePrefix = "arXiv",
    primaryClass = "hep-th",
    reportNumber = "DIAS-STP-24-16",
    doi = "10.21468/SciPostPhys.18.1.022",
    journal = "SciPost Phys.",
    volume = "18",
    number = "1",
    pages = "022",
    year = "2025"
}

@article{Bagchi:2024ikw,
    author = "Bagchi, Arjun and Banerjee, Aritra and Mondal, Saikat and Sarkar, Sayantan",
    title = "{Carroll in Shallow Water}",
    eprint = "2411.04190",
    archivePrefix = "arXiv",
    primaryClass = "hep-th",
    month = "11",
    year = "2024"
}

@article{Bagchi:2025vri,
    author = "Bagchi, Arjun and Banerjee, Aritra and Dhivakar, Prateksh and Mondal, Saikat and Shukla, Ashish",
    title = "{The Carrollian Kaleidoscope}",
    eprint = "2506.16164",
    archivePrefix = "arXiv",
    primaryClass = "hep-th",
    month = "6",
    year = "2025"
}

@article{Fiorucci:notes,
    author = "Fiorucci, Adrien",
    title = "{Carroll Physics and Geometry (to appear)}"
}

@article{Bekenstein:1973ur,
    author = "Bekenstein, Jacob D.",
    title = "{Black holes and entropy}",
    doi = "10.1103/PhysRevD.7.2333",
    journal = "Phys. Rev. D",
    volume = "7",
    pages = "2333--2346",
    year = "1973"
}

@article{Chrusciel:2000cu,
    author = "Chrusciel, Piotr T. and Delay, Erwann and Galloway, Gregory J. and Howard, Ralph",
    title = "{The Area theorem}",
    eprint = "gr-qc/0001003",
    archivePrefix = "arXiv",
    doi = "10.1007/PL00001029",
    journal = "Annales Henri Poincare",
    volume = "2",
    pages = "109--178",
    year = "2001"
}

@article{Hawking:1971vc,
    author = "Hawking, S. W.",
    title = "{Black holes in general relativity}",
    doi = "10.1007/BF01877517",
    journal = "Commun. Math. Phys.",
    volume = "25",
    pages = "152--166",
    year = "1972"
}

@article{Hawking:1975vcx,
    author = "Hawking, S. W.",
    editor = "Gibbons, G. W. and Hawking, S. W.",
    title = "{Particle Creation by Black Holes}",
    doi = "10.1007/BF02345020",
    journal = "Commun. Math. Phys.",
    volume = "43",
    pages = "199--220",
    year = "1975",
    note = "[Erratum: Commun.Math.Phys. 46, 206 (1976)]"
}

@article{Kmec:2026dis,
    author = "Kmec, Adam and Mason, Lionel and Ruzziconi, Romain",
    title = "{Quasi-Local Celestial Charges and Multipoles}",
    eprint = "2604.13362",
    archivePrefix = "arXiv",
    primaryClass = "hep-th",
    month = "4",
    year = "2026"
}

@article{Geiller:2025dqe,
    author = "Geiller, Marc and Mao, Pujian and Vincenti, Antoine",
    title = "{Twisting asymptotically-flat spacetimes}",
    eprint = "2511.13814",
    archivePrefix = "arXiv",
    primaryClass = "gr-qc",
    month = "11",
    year = "2025"
}

@article{Fiorucci:2024ndw,
    author = "Fiorucci, Adrien and Matulich, Javier and Ruzziconi, Romain",
    title = "{Superrotations at spacelike infinity}",
    eprint = "2404.02197",
    archivePrefix = "arXiv",
    primaryClass = "hep-th",
    doi = "10.1103/PhysRevD.110.L061502",
    journal = "Phys. Rev. D",
    volume = "110",
    number = "6",
    pages = "L061502",
    year = "2024"
}

@article{Flanagan:2015pxa,
    author = "Flanagan, {\'E}anna {\'E}. and Nichols, David A.",
    title = "{Conserved charges of the extended Bondi-Metzner-Sachs algebra}",
    eprint = "1510.03386",
    archivePrefix = "arXiv",
    primaryClass = "hep-th",
    doi = "10.1103/PhysRevD.95.044002",
    journal = "Phys. Rev. D",
    volume = "95",
    number = "4",
    pages = "044002",
    year = "2017",
    note = "[Erratum: Phys.Rev.D 108, 069902 (2023)]"
}

@article{Fuentealba:2022xsz,
    author = "Fuentealba, Oscar and Henneaux, Marc and Troessaert, C{\'e}dric",
    title = "{Logarithmic supertranslations and supertranslation-invariant Lorentz charges}",
    eprint = "2211.10941",
    archivePrefix = "arXiv",
    primaryClass = "hep-th",
    doi = "10.1007/JHEP02(2023)248",
    journal = "JHEP",
    volume = "02",
    pages = "248",
    year = "2023"
}

@article{Ecker:2024czx,
    author = "Ecker, Florian and Grumiller, Daniel and Henneaux, Marc and Salgado-Rebolledo, Patricio",
    title = "{Carroll swiftons}",
    eprint = "2403.00544",
    archivePrefix = "arXiv",
    primaryClass = "hep-th",
    reportNumber = "TUW-24-01",
    doi = "10.1103/PhysRevD.110.L041901",
    journal = "Phys. Rev. D",
    volume = "110",
    number = "4",
    pages = "L041901",
    year = "2024"
}

@book{Elvang:2015rqa,
    author = "Elvang, Henriette and Huang, Yu-tin",
    title = "{Scattering Amplitudes in Gauge Theory and Gravity}",
    isbn = "978-1-316-19142-2, 978-1-107-06925-1",
    publisher = "Cambridge University Press",
    month = "4",
    year = "2015"
}

\end{document}